\documentclass{article}
\usepackage{fancybox}
\usepackage[dvips]{graphicx}  
\makeatletter
\@addtoreset{equation}{section}

\makeatother

\begin{document}
\begin{flushright}
hep-th/0103003

March 2001

KUNS-1709
\end{flushright}

\begin{center}
 {\Large \bf Investigation of Matrix Theory via Super Lie
  Algebra}\footnote{This paper is based on the master's dissertation
  submitted to the Department of Physics, Faculty of Science, Kyoto
  University on February 2001.}\\
\vspace{5mm}
Takehiro Azuma \footnote{e-mail address :
  azuma@gauge.scphys.kyoto-u.ac.jp}\\
\vspace{5mm}
\textit{Department of Physics, Kyoto University, Kyoto 606-8502,
  Japan}
\end{center}

 \begin{abstract}
  Superstring theory is the most powerful candidate for the theory
 unifying Standard model and gravity, and this field has been  rigorously
 researched. The discovery of a BPS object named D-brane has lead to
 the idea that different kinds of superstring theory - type IIA, type
 IIB, type I, $SO(32)$ heterotic and $E_{8} \times E_{8}$ heterotic
 theory - are related with each other by duality. Now, the formulation
 of superstring theory has been realized only 
 perturbatively. However, if we succeed in the formulation of the
 constructive definition of superstring theory, which does not depend
 on the perturbation, this will be a genuine unified theory describing 
 all interactions. 'Matrix model' is regarded as the most powerful
 arena to describe the nonperturbative superstring theory. Here, we
 mean 'matrix model' by a model in which the theory is described in
 terms of $N \times N$ matrices and superstring theory is reproduced
 in the limit $N \to \infty$. This belief is based on the series of 
 works in the late 1980's. These works have provided us with the
 computation of the exact
 solution of the nonperturbative bosonic string theory in less than 1
 dimensional spacetime by describing the bosonic string in terms of $N 
 \times N$ matrices.  Many proposals for the constructive definition of
 superstring theory  
  have been hitherto made, and the most successful existing proposal
  is IKKT model. This is a dimensional reduction of 10 dimensional
  ${\cal N}=1$ SYM to 0 dimension, and is identical to the matrix
  regularization of the Green-Schwarz action of type IIB superstring
  theory. Yet, it is an exciting issue to pursue a model  exceeding IKKT
  model, and we investigate the matrix model  proposed by L. Smolin
  \cite{0002009} \cite{0006137}. He proposed a cubic matrix model
  in which both the bosons and the fermions are embedded in one
  multiplet.  He proposed two Lie algebras for the framework of this
  cubic matrix model. One is $osp(1|32,R)$. This is a natural arena in 
  that this is the maximal simple Lie algebra of the symmetry of 11
  dimensional M-theory. The other is the $u(1|16,16)$, which is
  suggested as an extension of $osp(1|32,R)$.
  This paper reports the research of the formulation of the matrix
  model based on Smolin's proposal. We discuss the original proposal
  $osp(1|32,R)$ super Lie algebra, and $gl(1|32,R)$, which is the analytic
  continuation of the gauged $u(1|16,16)$ theory. We discuss the
  relationship with the existing matrix model and the supersymmetry
  for these two models. This paper is based on the collaboration with
  S. Iso, H. Kawai and Y. Ohwashi \cite{virginal}.
 \end{abstract}

\newpage
\tableofcontents

\newpage
\section{Introduction}
     One of the main themes in elementary particle  physics is
  unification. Four kinds of interactions are known - weak,
  strong, electromagnetic and gravitational interactions. Much effort
  has been  
  hitherto made to understand these interactions by means of a unified 
  quantized theory. In 1967, the Glashow-Weinberg-Salam model succeeded in
  unifying the electromagnetic and weak interactions in terms of
  $SU(2) \times U(1)$ gauge theory. The ensuing success is the
  emergence of Grand Unified Theory (GUT) in the early 1970's, which
  further unifies the strong interactions. This theory possesses the
  gauge group $SU(3) \times SU(2) \times U(1)$, and another name is
  {\it Standard Model}. 
  \begin{figure}[htbp]
   \begin{center}
    \scalebox{.5}{\includegraphics{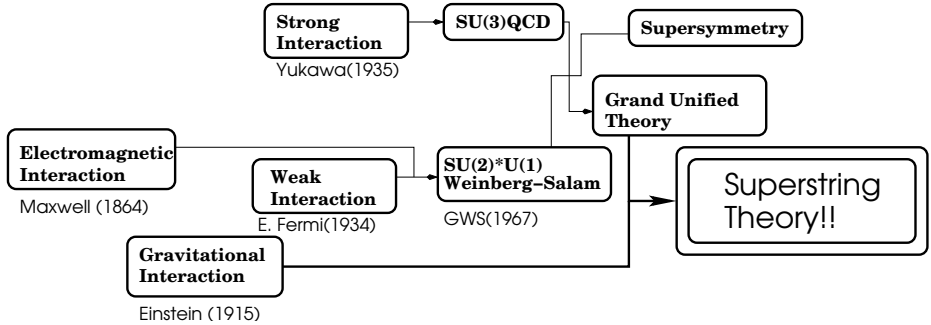} }
   \end{center}
   \caption{The chronological table of the development of particle
     physics. All four interactions are believed to be unified by
     superstring theory.}
  \label{Unification}
  \end{figure}

  The last and most difficult theme is the unification of the
  gravity. Unlike the other interactions, the gravitational
  interaction cannot be renormalized due to the intense
  divergence. Then we require another formalism than the gauge theory
  to complete a consistent quantum theory which unifies the Standard
  Model and the gravity. The most promising candidate is the
  superstring theory. 

  Superstring theory possesses many splendid properties.
  Superstring theory naturally contains not only matter and gauge
  fields but also  gravitational fields. Every consistent superstring
  theory contains a massless spin-2 state, which corresponds to
  graviton. It possesses a sufficiently large gauge group to include the
  conventional Standard model. $E_{8} \times E_{8}$ heterotic
  superstring compactified on Calabi-Yau manifold is strikingly
  similar to the $SU(3) \times SU(2) \times U(1)$ Standard Model. 

  The research of superstring theory has advanced at an astonishing
  rate in recent years. One of the significant discovery of 
  superstring theory is the BPS object named D-brane, defined as an
  object on which a superstring can terminate. D-brane has made a
  remarkable contribution in the unification of five kinds of
  superstring which seems to differ in the perturbative framework -
  type I, type IIA, type IIB, $SO(32)$  heterotic and  $E_{8} \times
  E_{8}$ heterotic superstring theory. The discovery of D-brane
  enables us to relate these theories by duality, and these theories
  are now regarded as a limit of one unified theory. And if we succeed 
  in finding the
  constructive definition, which describes the nonperturbative behavior 
  of the superstring, of any one of these superstring theories, the
  rest of the theories are described by duality, perturbatively or
  nonperturbatively. If such a theory is found, this may become
  'Theory of Everything'. The word 'Theory of Everything' means the
  theory which describes all interactions and phenomena in our whole
  universe. We believe that all interactions in the whole universe are 
  distinguished by the above-mentioned four interactions, and if we obtain 
  a theory which unifies all of them, this can be called 'Theory of
  Everything'. The last and the biggest dream of particle physics is to 
  find this ultimate theory.

   \begin{figure}[htbp]
   \begin{center}
    \scalebox{.5}{\includegraphics{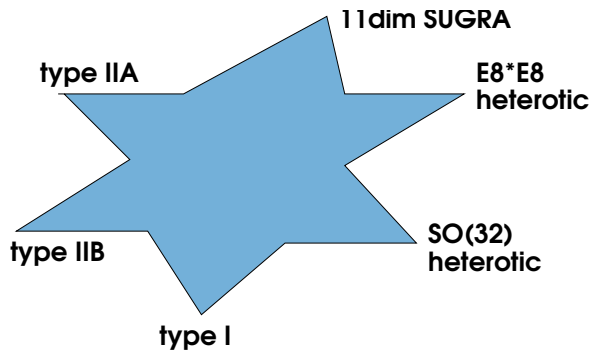} }
   \end{center}
   \caption{All string theories, and M-theory, are conjectured to be a 
limit of one theory.}
  \label{Mtheory}
  \end{figure}

  Now, many string theorists believe the conjecture that the unified
  theory may be described by a matrix model. The first proposal was
  made by Banks, Fischler, Shenker and Susskind  \cite{9610043}. Their 
  matrix model was obtained by the dimensional reduction of the 10
  dimensional super-Yang-Mills(SYM) theory to 1 dimension. This model are
  deeply related to type IIA superstring theory, and type IIA SUGRA
  is induced by one-loop effect. In this theory, the physical
  quantities are described by $N \times N$ matrices, and when we take
  the size of these matrices to an infinity, this theory gives a
  microscopic second-quantized description of M-theory in light-cone
  coordinates.  

  Another proposal for a matrix model was made by
  Ishibashi, Kawai, Kitazawa and Tsuchiya \cite{9612115}
  \footnote{for a review, see \cite{9908038} or \cite{shino}} . This is
  IKKT (IIB matrix) model and is the most powerful candidate for the
  constructive definition of superstring theory.  This theory is the
  dimensional reduction  of the 10 dimensional SYM theory to 0
  dimension.  The action of IKKT model is surprisingly simple :
  \begin{eqnarray}
   S = - \frac{1}{g^{2}} Tr ( \frac{1}{4} [ A_{i} , A_{j}][A^{i} , 
   A^{j} ] + \frac{1}{2} {\bar \psi} \Gamma^{i} [ A_{i} , \psi]). 
   \label{1AZMikkt} \end{eqnarray}
  As the name of this theory indicates, this theory is deeply related
  to type IIB superstring theory. This theory has splendid properties, 
  which we explain in Sec. 3. 
   These properties gives us a confidence that IKKT model
   may be a successful constructive definition of the superstring
   theory. 

  Then, here comes one simple but fundamental question.
   \begin{center}
   {\it Why do we stick to the large N matrix theory ?}
   \end{center}
  The background of this belief dates back to the late
  1980's \footnote{for a review of the progress of old matrix theory
  (2D quantum gravity), see \cite{9304011}},  long 
  before the discovery of D-brane by J. Polchinski. Grand Unified
  Theory (GUT) was a great success in unifying three of the four
  fundamental interactions in elementary particle physics. Another
  success is the description of the nonperturbative behavior of
  Quantum Chromodynamics(QCD). There are two ways to describe the
  strong-coupling region - large $N$ expansion and lattice gauge
  theory. It has been speculated that the same may be true of string
  theory, and many attempts have been made to describe
  string theory 
  in terms of large $N$ matrix theory. Brezin and Kazakov
  \cite{brezinkazakov} succeeded in solving exactly the behavior of
  bosonic string theory of less than 1 spacetime dimension by means of 
  orthogonal polynomial method, and their analysis well reproduced the 
  behavior of string theory. These works, although they do not give so 
  many clues technically, support strongly the belief that the
  nonperturbative behavior of superstring theory should be described
  by large $N$ matrix theory.

  Based on this philosophy, we speculate that the Ariadne's thread to
  'Theory of Everything' should lie in large $N$ matrix theory. Our
  research is a pursuit of another matrix model than IKKT model,
  expecting this to exceed IKKT model, and
  the clue to the new matrix model was proposed by L. Smolin
  \cite{0002009} \cite{0006137}. He proposed a cubic matrix model
  with the multiplet belonging to super Lie algebra $osp(1|32,R)$ or
  $u(1|16,16)$\footnote{Throughout this paper, we denote the Lie
  groups by the capital letters, and the Lie algebras by the small
  letters.}. This super Lie algebra $osp(1|32,R)$ has been known
  as the maximal super Lie algebra possessing the symmetry of 11
  dimensional M-theory, and this super Lie algebra is a natural arena
  for describing a constructive definition of superstring theory. We
  will explain in the subsequent section our motivation to follow the
  idea of L. Smolin and pursue this cubic matrix model.

  This paper is organized as follows.
  \begin{itemize}
   \item{Sec. 2. is devoted to a  brief review of the success of old
       matrix theory. Although this knowledge does not provide one with 
       the technical hint of modern matrix model, these series of
       works are of importance in that they give us the belief that
       the constructive definition of superstring theory should be
       described by matrix model.}
  \item{Sec. 3. is a brief review of IKKT model. We review the
      successful aspects of this model, and especially introduce a
      knowledge we inherit in our research of the new cubic matrix model.}
  \item{Sec. 4. is based on our research of $osp(1|32,R)$ cubic matrix  model
      \cite{virginal}. We especially investigate the relationship of
      this cubic matrix model with the existing proposal of matrix
      model. We compare the supersymmetry of this cubic matrix model
      with IKKT model, and consider how IKKT model is induced from the 
      cubic matrix model.}
  \item{Sec. 5. introduces another version of this cubic matrix
      model, called 'gauged action'. We treat the theory with the
      gauge symmetry vastly enhanced, following the idea of L. Smolin
      \cite{0006137}. The
      multiplets now belong to $gl(1|32,R)$ super Lie algebra, and the 
      gauge symmetry for the large $N$ matrices is altered to
      $gl(N,R)$. We investigate\cite{virginal} 
      the possibility of this extended version of the
      cubic matrix model, again paying attention to the supersymmetry
      and the relationship with IKKT model.}
  \item{Sec 6. is devoted to the concluding remark and the outlook of
      our research.}
  \item{Appendix. A summarizes the notation of this paper, and
      introduces the knowledge of the properties of gamma matrix,
      supermatrices, $su(N)$ Lie algebra and the notion of the tensor
      product frequently used in the context of gauged theory.}
  \item{Appendix. B provides us with the miscellaneous calculation of this
      paper in full detail.}
  \item{Appendix. C. introduces the notion named the Wigner In{\"o}n{\"u} 
      contraction, which is needed in the discussion in Sec. 5.}
   \end{itemize}

\section{The brief review of Quantum Gravity in $D \leq 1$}
 We begin with a brief review of the old days - {\it the quantization
of gravity in} $D \leq 1$, in order to gain insight into the belief
that the constructive definition of superstring theory is described by 
matrix theory. This section is devoted to introducing a series of
works in the late 1980's, in which they succeeded in describing the
nonperturbative behavior of a noncritical string via matrix theory.

 \subsection{The quantization of $D \leq 1$ dimensional string theory}
  Distler and Kawai \cite{DDK} succeeded in the quantization of a
  non-critical string via conformal gauge. This subsection focuses on
  bosonic string, however their discussion readily extends to
  superstring theory with ease. The path integral of the bosonic
  Polyakov action is 
   \begin{eqnarray}
    Z = \int \frac{dXdg}{V_{diff}} \exp(-S_{M}) = \int
    dX d\phi dbdc \exp(-S_{M} - S_{bc}).  \label{2-1-}
   \end{eqnarray}
 \begin{itemize}
  \item{ $S_{M} = \frac{1}{4 \pi \alpha'} \int_{M} d^{2}z \sqrt{g} g^{ab}
      \partial_{a} X^{\mu} \partial_{b} X_{\mu} + \frac{\lambda}{4
      \pi} \int_{M} d^{2}z \sqrt{g} R$ is the Polyakov action of  bosonic 
      string theory. This action is, per se, classical, 
      and hence not subject to Weyl anomaly.}
  \item{$S_{bc} = \frac{1}{2 \pi} \int d^{2}z (b_{zz} \partial_{\bar
      z} c^{z} + b_{{\bar z} {\bar z}} \partial_{z} c^{\bar z})$ is the
      Faddeev-Popov ghost, which emerge as we gauge-fix the Polyakov
      action.}
  \item{We use a parameter $\phi$ for parameterizing a metric $g$
      according to a Weyl rescaling.} 
 \end{itemize}

  We consider this path integral in detail in order to gain insight
  into the quantum effect of the Weyl transformation of the
  non-critical string. The Weyl transformation is expressed by
    \begin{eqnarray}
     g_{ab} = {\hat g}_{ab} e^{\phi} \Rightarrow {\hat g}_{ab}.
     \label{2-2-}
    \end{eqnarray}

  Then, the Polyakov action, including the ghost effect, is subject to 
  this transformation in a quantum level. The quantum effect emerges
  when we consider the effective action by the above path integral
  (\ref{2-1-}). We define the effective action as
   \begin{eqnarray}
    \exp( -S_{eff}) = \int dX d\phi db dc \exp( -S_{M}-S_{bc}). \label{2-3-} 
   \end{eqnarray} 
  Note that this effective action is a {\it quantum object}, unlike
  the original Polyakov action. This action possesses Weyl anomaly, so 
  that this induces the Liouville action by Weyl transformation:
   \begin{eqnarray}
    S_{L} \stackrel{def}{=} S_{eff}(e^{\phi} {\hat g}) - S_{eff}({\hat 
    g}) = \frac{26-D}{48 \pi} \int d^{2}z \sqrt{\hat g}( \frac{1}{2}
    {\hat g}^{ab} \partial_{a} \phi \partial_{b} \phi + {\hat R} \phi
    + \mu (e^{\phi} -1) ), \label{2-4-}
   \end{eqnarray}
  where $D$ is a dimension of the spacetime and $\mu$ is an arbitrary
  integration constant. This is called the Liouville action, whose
  derivation we refer to \cite{Friedan}. For a critical string, which
  is realized when the dimension of the spacetime is $D=26$, the Weyl
  anomaly is cancelled between the matter and the ghost
  field. However, the same is no longer true of the noncritical
  string. This situation indicates that the Weyl parameter is not a
  gauge freedom but provides an additional interacting
  dimension\footnote{Note that this is why the noncritical string in
  $D$ dimensions is interpreted as a critical string in $d=D+1$
  dimensions. Here, we refer to the spacetime dimensions as that of
  the noncritical string.}. This is a fatal obstacle in considering
  this path integral. The cancer lies in the fact that {\it the
  measure of} $\phi$ {\it itself depends on} $\phi$. 

  The measures of the path integral are defined in terms of the norms 
  of the functional space:
   \begin{eqnarray}
     \parallel \delta \phi \parallel^{2}_{g} = \int d^{2}z \sqrt{g}
     (\delta \phi)^{2} = \int d^{2}z \sqrt{\hat g} e^{\phi(z)} (\delta 
     \phi)^{2}.  \label{2-5-}
   \end{eqnarray}
  This measure is too difficult to analyze because of the dependence
  of $\parallel \delta \phi \parallel$ itself on $\phi$. In order to
  remedy this situation, we transplant the cancer to the Jacobian, and 
  express the measure in terms of not $g$ but ${\hat g}$:
    \begin{eqnarray}
     [dX]_{g} [db]_{g} [dc]_{g} [d \phi]_{g} = [dX]_{\hat g}
     [db]_{\hat g} [dc]_{\hat g} [d \phi]_{\hat g} J. \label{2-6-}
    \end{eqnarray}
 $J$ is the Jacobian in question whose explicit form we do not
 know. However, we consider the Weyl transformation of $[dX]$, $[db]$
 and $[dc]$ in order to grasp the Jacobian $J$. We have already
 investigated the difference due to the Weyl transformation of the
 effective Polyakov action. Its path integral is expressed by
    \begin{eqnarray}
      Z = \int  [dX]_{g} [db]_{g} [dc]_{g} [d \phi]_{g} \exp( -S_{M} - 
      S_{bc} ). \label{2-7-}
    \end{eqnarray}
 As we have seen before, the classical Polyakov action and the ghost
 action $S_{M} + S_{bc}$ are, per se, Weyl invariant. However, when we 
 consider an effective action $S_{eff}$ with the quantum effect
 included in the path integral, $S_{eff}$ possesses Weyl anomaly. We
 express the dependence on the parameter of Weyl transformation {\it
 by the transformation of the measure.} This reflects the idea that
 the Weyl anomaly is due to the quantum effect, and hence that {\it
 the culprit is the measure of the path integral}. We depict this idea 
 by imposing the responsibility of the Weyl anomaly on the functional
 measure:
  \begin{eqnarray}
 & &  [dX]_{g} = [dX]_{\hat g} \exp (- \frac{-D}{48 \pi} \int d^{2}z
   \sqrt{g} ( \frac{1}{2} {\hat g}^{ab} \partial_{a} \phi \partial_{b} 
   \phi + {\hat R} \phi + \mu (e^{\phi} -1))), \label{2-8-} \\
 & & [dbdc]_{g} = [dbdc]_{\hat g} \exp (- \frac{26}{48 \pi} \int d^{2}z
   \sqrt{g} ( \frac{1}{2} {\hat g}^{ab} \partial_{a} \phi \partial_{b} 
   \phi + {\hat R} \phi + \mu (e^{\phi} -1))). \label{2-9-}
  \end{eqnarray}
 This never gives an explicit form of $J$, because of the difficulty
 in the path integral with respect to $\phi$. However, we can set the
 following ansatz, looking carefully at the formulae (\ref{2-8-}) and
 (\ref{2-9-}):
  \begin{eqnarray}
   J \stackrel{ansatz}{=} \frac{1}{8 \pi} \int d^{2}z ( \sqrt{\hat g}
   {\hat g}^{ab} \partial_{a} \phi \partial_{b} \phi - Q \sqrt{\hat g} 
   {\hat R} \phi + \mu_{1} \sqrt{\hat g} e^{\alpha \phi} ). \label{2-10-}
  \end{eqnarray}
 We emphasize that this is not a perfect answer but an assumption. We
 now determine the variables $Q$ and $\alpha$ according to the following two
 conditions\footnote{From now on, we insert the explicit quantity of
 Regge slope: We adopt the notation $\alpha'=2$.}. 
   \begin{itemize}
    \item{The partition function itself must not possess Weyl
        anomaly.}
    \item{The metric $g = e^{\alpha \phi} {\hat g}$ should be Weyl
        invariant.} 
   \end{itemize}
 We first consider the Weyl invariance of the partition function. This 
 condition is imposed because we would like to construct a consistent
 quantization of string, and the theory should be free from any kind
 of anomaly. The corresponding energy momentum tensor is
   \begin{eqnarray}
    T_{L} = - \frac{1}{2} : \partial \phi \partial \phi : -
    \frac{Q}{2} \partial^{2} \phi. \label{2-11-}
   \end{eqnarray}
 Taking the operator product expansion $T(z) T(w) \sim
 \frac{c_{L}}{2(z-w)^{4}} + \frac{2}{(z-w)^{2}} T(w) + \frac{1}{2}
 \partial T(w)$, we obtain a central charge $c_{L} =1 +
 3Q^{2}$. Therefore, in order for the partition function not to
 possess Weyl anomaly, the sum of the following charge is zero.
  \begin{itemize}
    \item{ $c_{L} = 1 + 3Q^{2}$ : This stems from the Liouville mode,
        where $Q$ is an unknown coefficient.}
    \item{ $c_{M} = D$ : This stems from the energy-momentum tensor of 
        the matter field $T_{M}(z) = - \frac{1}{\alpha'}: \partial
        X^{\mu} \partial X_{\mu} :$. This gives one central charge per 
        (spacetime) dimension.}
   \item{ $c_{bc} = -26$ : This is a contribution from the ghost
       energy momentum tensor.}
  \end{itemize}
 This gives a result $Q = \sqrt{ \frac{25-D}{3}}$.

  The next step is the evaluation of the unknown coefficient
  $\alpha$. This is determined by the latter condition: Weyl
  invariance of $g_{ab}$. This is equivalent to the statement that 
      \begin{eqnarray}
       ( \textrm{Conformal weight of } e^{\alpha \phi} ) = 1. \label{2-12-}
      \end{eqnarray}
  On the other hand, the OPE with the EM tensor gives a weight
  $-\frac{1}{2} \alpha ( Q + \alpha)$. Thus we obtain
   \begin{eqnarray}
    \alpha_{\pm} = -\frac{1}{2 \sqrt{3}} ( \sqrt{25-D} \mp \sqrt{1-D}
    ). \label{2-13-}
   \end{eqnarray}
  The agreement with the classical result ($D \to - \infty$) in
  achieved if we take the branch $\alpha = \alpha_{+} = - \frac{1}{2
  \sqrt{3}} ( \sqrt{25-D} - \sqrt{1-D} )$. Note that this answer is
  physically significant {\it only if} $D \leq 1$.
  \begin{itemize}
   \item{When $1 < D < 25$, we clearly give an inconsistent behavior
       of the theory, because $e^{\alpha \phi} = \exp (-  \frac{1}{2
       \sqrt{3}} \sqrt{25-D} + \frac{i}{2 \sqrt{3}} \sqrt{D-1} )$ and
       the imaginary part indicates the existence of tachyon
       vertex\footnote{Some attempts to surmount this problem and to
       extend this idea to the critical string theory using the idea
       of tachyon condensation are made in
       \cite{0004172}.}. Therefore, we must regard the vacuum of $\phi$ as
       unstable and this quantization cannot be applied to the case in 
       which $1<D<25$.}
  \item{When $D \geq 25$, both $Q$ and $\alpha$ are pure imaginary. It 
      seems that we can escape from the problem of tachyon vertex by
      the analytic continuation $\phi \to i \phi$. However, this
      changes the sign of the kinetic term of  
       (\ref{2-10-}). $\phi$ thus becomes a ghost field, and this
       procedure of the quantization cannot be applied to this case,
      either.}  
  \end{itemize}
 This quantization of gravity is valid only if the dimension of the
 spacetime is less than or equals 1 (hence $d \leq 2$, including the
 Liouville mode). Although this work only gives an answer to the
 quantization of gravity for a very low spacetime dimension, this
 result plays an essential role in providing the belief that string
 is expressed by matrix model. Another essential success of Distler
 and Kawai is the evaluation of the scaling law which remarkably
 agrees with the result of matrix theory, as we will explain in the
 subsequent section. We next consider the partition function as a
 function of the area of the world sheet. Let $A$ be the area of the
 worldsheet: $A = \int d^{2}z \sqrt{\hat g} e^{\alpha \phi}$. Then,
 \begin{eqnarray}
  Z = \int dX dg \exp(-S_{M}) \delta(\int d^{2}z e^{\alpha \phi}
  \sqrt{\hat g} - A), \label{2-14-} 
 \end{eqnarray}
 where we omitted the measure and the action of the ghost contribution 
 because these have nothing to do with the discussion of the scaling
 law. We expect the action to be invariant under the scaling of the
 parameter $\phi \to \phi + \frac{\rho}{\alpha}$. The shift of the
 Liouville action (\ref{2-10-}) and the delta function is
  \begin{eqnarray}
   & & J \to J - \frac{Q}{8 \pi} \int d^{2}z \sqrt{\hat g} {\hat R}
   \frac{\rho}{\alpha} = J - (1-h)\frac{Q \rho}{\alpha}, \label{2-15-} 
   \\
  & & \delta(\int d^{2}z e^{\alpha \phi} \sqrt{\hat g} - A) \to e^{-
   \rho} \delta( \int d^{2}z e^{\alpha \phi} \sqrt{\hat g} - e^{-
   \rho} A ), \label{2-16-}
  \end{eqnarray}
  where we have utilized the property of the Euler character: $\chi
  = \frac{1}{4 \pi} \int d^{2}z \sqrt{\hat g} {\hat R} = 2(1-h)$, and
  $h$ is the number of the genera of the world sheet. Therefore, the
  partition function  is  
   \begin{eqnarray}
    Z(A) = \exp ((\frac{Q(1-h)}{\alpha} - 1) \rho) Z(e^{-\rho} A) \to
    Z = K A^{\frac{Q}{\alpha}(1-h)-1}. \label{2-17-}
   \end{eqnarray}
  We define a quantity $\gamma$ named {\it string susceptibility}, as the
  exponent of the area of the world sheet:
   \begin{eqnarray}
    Z(A) \sim K A^{\gamma -3} \label{2-18-}
   \end{eqnarray}
  The string susceptibility $\gamma$ is thus 
   \begin{eqnarray}
    \gamma = \frac{1-h}{12}(D -25 - \sqrt{(25-D)(1-D)}) + 2.
    \label{2-19-} 
   \end{eqnarray}
  We have completed the evaluation of the string susceptibility, and
  this result is shown to match (for $D=0$) the analysis of 0
  dimensional QFT, which we explain in the subsequent section.
  The analysis of Distler and Kawai is applied to superstring theory
  \cite{DDK}. We omit in this review the extension of this discussion
  to superstring theory, but they have succeeded in the quantization
  of superstring theory only for $D \leq 1$. 

 \subsection{Random Triangulation}
  Although the time sequence is upside down, we next introduce the work 
  of F. David\cite{David} in 1985. Lattice gauge theory played an
  essential role in describing the nonperturbative region of QFT. The
  work of David inherits the idea of lattice gauge theory and
  attempted to construct the discritized version of string theory. His 
  idea was to divide the worldsheet into many polygons. For
  simplicity, we focus on 'random triangulation' but this idea
  applies to any polygon. We first emphasize that, although the
  quantization of string theory suggested by Distler and Kawai is
  extended to superstring theory, this 'random triangulation' cannot
  be applied to superstring theory. This is due to the difficulty in
  describing a chiral fermion in a discritized worldsheet. This is the 
  same kind of obstacle as is faced in lattice gauge
  theory. Therefore, we limit the following discussion only to the
  bosonic string theory.

  We consider $D=0$ dimensional string theory\footnote{The content of
  this section and the next section is based on the
  review\cite{9304011}.}. This is a pure theory of surfaces without any 
  coupling to matter degrees of freedom on the string worldsheet. The
  partition function of this string theory is
    \begin{eqnarray}
      Z = \sum_{h=0}^{\infty} \int dg \exp(-\beta A + \gamma
      \chi). \label{2-26-} 
    \end{eqnarray}
   \begin{itemize}
    \item{$h$ is a number of the genera of the worldsheet.}
    \item{$A$ is an area of the worldsheet, which is expressed by $A = 
        \int d^{D} x \sqrt{g}$. Polyakov action is

        $S_{M} = \frac{1}{4 
        \pi \alpha'} \int d^{2}z \sqrt{g} g^{ab} \partial_{a} X^{\mu}
        \partial_{b} X_{\mu}$, but this is the same as the area of the 
        worldsheet since we are now considering the $D=0$ theory.}
   \item{$\chi$ is the Euler character of the worldsheet: $\chi =
       \frac{1}{4 \pi} \int d^{2}z \sqrt{g}R = 2(1-h)$.}
   \item{$\beta$ and $\gamma$ are coefficients which do not play an
       important role in this context.}
   \end{itemize}
  This path integral is too difficult to solve explicitly, and we need 
  an approximation. We do not perform an integration for a continuous
  Riemann surface, but discritize the surface into many equilateral
  triangle. This is the well-known method 'random triangulation'.
   \begin{figure}[htbp]
   \begin{center}
    \scalebox{.5}{\includegraphics{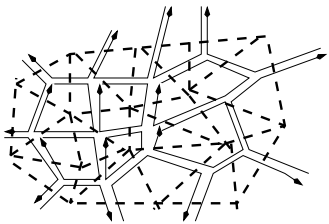} }
   \end{center}
   \caption{We discritize the worldsheet into many equilateral
     triangles in 'random triangulation'.}
  \label{randomtr}
  \end{figure}

  Then, the integration of the path integral is replaced with the sum
  of all random triangulation
    \begin{eqnarray}
     \sum_{h=0}^{\infty} \int dg \Rightarrow \sum_{\textrm{random
     triangulation}}.
    \end{eqnarray}
  This procedure does not make any change in Euler number. Look back
  on the terms in the path integral $\int \sqrt{g}R$. The sum
  $\sum_{\textrm{random triangulation}}$is regarded as the sum of the
  points at which the vertices of the triangles meet one
  another. Suppose there are $N_{i}$ incident equilateral triangle at
  the vertex $i$. The Ricci scalar at this point is $R_{i} = 2 \pi
  (\frac{6}{N_{i}} -1)$:
   \begin{eqnarray}
    \int \sqrt{g} R \Rightarrow \sum_{i} 4 \pi ( 1- \frac{N_{i}}{6} )
    = 4 \pi ( V - \frac{F}{2}) = 4 \pi (V-E+F) = 4 \pi \chi,
    \label{2-28-} 
   \end{eqnarray}
  where $V$, $E$ and $F$ are the number of vertices, edges and faces
  of the discritized worldsheet respectively, and we have utilized the 
  relationship $3F=2E$.

  This model is in fact described by 0 dimensional QFT of $\phi^{3}$
  theory. First, let us investigate the Feynman rule of the matrix
  theory whose action is
   \begin{eqnarray}
    S = \frac{1}{2} Tr M^{2}, \label{2-29-}
   \end{eqnarray}
  where $M$ si an hermitian $N \times N$ matrix and the trace $Tr$ is
  taken with respect to the $N \times N$ matrices. The partition
  function is now 
   \begin{eqnarray}
    Z = \int d^{N^{2}} M e^{-S} = \int d^{N^{2}} M \exp( -
    \frac{1}{2} TrM^{2}). \label{2-30-}
   \end{eqnarray}
  The propagator of this theory is given by
   \begin{eqnarray}
    \langle M_{ij} M_{kl} \rangle = \frac{1}{Z} \int d^{N^{2}} M
    M_{ij} M_{kl} \exp( - \frac{1}{2} Tr M^{2}) = \delta_{il}
    \delta_{jk}. \label{2-31-}
   \end{eqnarray}
  {\sf (Proof)
  We note that, due to the hermiticity of $M$, the trace is written as
  \begin{eqnarray}
   \frac{1}{2} Tr M^{2} = \frac{1}{2} \sum_{i,j=1}^{N} M_{ij} M_{ji}
  = \sum_{1 \leq i<j \leq N} M_{ij} M^{\star}_{ij} 
  + \frac{1}{2} \sum_{i=1}^{N} M_{ii} M_{ii}. 
  \end{eqnarray}
  Especially, we separate $M_{ij}$ into the real and the imaginary part as
  \begin{eqnarray}
    M_{ij} = \frac{X_{ij} + i Y_{ij}}{\sqrt{2}}
    (=M_{ji}^{\star}). 
  \end{eqnarray}
 Here, $X_{ij}$ and $Y_{ij}$ are real c-number.
 Then, the quadratic term  is written as
 \begin{eqnarray}
   \frac{1}{2} Tr M^{2} = \frac{1}{2} \sum_{i=1}^{N} M_{ii}
     + \frac{1}{2} \sum_{1\leq i<j \leq N} (X_{ij}^{2} + Y_{ij}^{2}).
 \end{eqnarray}
  The derivation of the propagator reduces to the simple Gaussian
  integral:
  \begin{eqnarray}
   \frac{1}{a} = \frac{\int^{+\infty}_{-\infty} dx x^{2}
   \exp(-\frac{ax^{2}}{2})}{\int^{+\infty}_{-\infty} dx
   \exp(-\frac{ax^{2}}{2})}. 
  \end{eqnarray}
  \begin{itemize}
   \item{$\langle M_{ii} M_{ll} \rangle$ survives only for $i=l$.}
   \item{For $\langle M_{ij} M_{kl} \rangle$ ($i \neq j$), we note the 
       following two results. Firstly, $\langle M_{ij} M_{ij}\rangle$
       is shown to vanish as 
    \begin{eqnarray}
      \langle M_{ij} M_{ij}\rangle = \frac{1}{2} \langle
          (\underbrace{X_{ij} X_{ij} - Y_{ij}
          Y_{ij}}_{\textrm{cancelled}} 
           + 2i \underbrace{X_{ij} Y_{ij}}_{(*)}) \rangle =
          \frac{1-1}{2}=0.
    \end{eqnarray}
     Here, (*) does not contribute ab initio, since this is a linear
          term of each $X_{ij}$ and $Y_{ij}$. 
     Secondly, we note that 
     \begin{eqnarray}
      \langle M_{ij} M_{ji} \rangle =
          \frac{1}{2} \langle 
          (X_{ij} X_{ij} + Y_{ij} Y_{ij}) \rangle = 1
     \end{eqnarray}
          survives (namely when $i=l, j=k$).
    }
  \end{itemize}
  This completes the proof of the Feynman rule
  (\ref{2-31-}). (Q.E.D.)
  }

   \begin{figure}[htbp]
   \begin{center}
    \scalebox{1}{\includegraphics{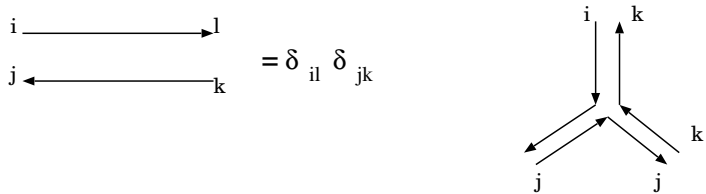} }
   \end{center}
   \caption{The Feynman rule of the $D=0$ dimensional matrix model. The 
     left describes the propagator while the right describes the vertex.}
  \label{zerodim}
  \end{figure}

 We next investigate the contribution of the vertex. For the
 $\phi^{3}$ theory, the action including the interaction is
   \begin{eqnarray}
    S_{int} = \frac{1}{2} TrM^{2} + \frac{\lambda}{3}
    TrM^{3}. \label{2-33-} 
   \end{eqnarray}
 The way to take the sum with respect to the indices is described by
 the Feynman rule, as shown in Fig. \ref{zerodim}. The vertex is
 ${\cal O}(\lambda)$. In this Feynman rule, the presence of the upper
 and lower matrix indices is represented by the double line, and this
 description inherits that of large $N$ QCD. This is a natural
 notation because $D=0$ dimensional QFT shares the structure of
 Feynman diagrams with large $N$ QCD.

  We next explain the relationship between the string theory described 
  by the 'random triangulation' and the $D=0$ dimensional
  QFT. Consider the partition function of the above large $N$ matrix
  model with $\phi^{3}$ interaction
   \begin{eqnarray}
    Z_{QFT} = \int d^{N^{2}} M \exp( -\frac{1}{2} TrM^{2} +
    \frac{g}{\sqrt{N}} TrM^{3} ), \label{2-34-}
   \end{eqnarray}
 where $M$ is a hermitian matrix, and $\lambda$ is now replaced with
 $\lambda \to - \frac{g}{\sqrt{N}}$. Note that this integral is
 defined by the analytic continuation in the coupling constant
 $g$. We perform a Taylor expansion of the interaction term $\exp
 ( \frac{g}{\sqrt{N}} TrM^{3})$. Then, this matrix integration is 
   \begin{eqnarray}
    Z_{QFT} = \sum_{n=0}^{\infty} \frac{1}{n!}
    (\frac{g}{\sqrt{N}})^{n} \int d^{N^{2}} M \exp( - \frac{1}{2}
    TrM^{2}) (TrM^{3})^{n}. \label{2-35-}
   \end{eqnarray}
 The $n$-th power describes the system in which there are $n$ 3-point
 vertices in the world sheet. And the above Feynman rule shows that
 this system is described by gluing $n$ equilateral triangles
 described by fig. \ref{randomtr}. Since each equilateral triangle
 possesses unit area, 
   \begin{eqnarray}
    A = (\textrm{area of the worldsheet}) = (\textrm{number of
    triangles}) = n. \label{2-36-}
   \end{eqnarray}
 We would like to identify this $D=0$ QFT with the partition function
 of the $D=0$ bosonic string theory. In this process, there are two
 identifications.
  \begin{itemize}
   \item{We can immediately discern the identification $e^{-\beta} =
       g$}
   \item{The other is to identify $e^{\gamma}$ with the size of the
       matrices $N$. However, this is less trivial than the
       previous identification, and we need some explanation. To
       discern the identification $N = e^{\gamma}$, let us rescale
       this matrix model by $M \to M \sqrt{N}$. The action is then
       described by
        \begin{eqnarray}
         - \frac{1}{2} TrM^{2} + \frac{g}{\sqrt{N}} TrM^{3} \to N(-
         \frac{1}{2} TrM^{2} + g TrM^{3}). \label{2-37-}
        \end{eqnarray}
       This rescale makes $N$ an overall factor, and the $N$
       dependence of the Feynman rule becomes transparent.
        \begin{itemize}
         \item{Vertex: This is clearly ${\cal O}(N)$, because of the
             interaction term $gN TrM^{3}$.}
         \item{Propagator: This is ${\cal O}(N^{-1})$, because of the 
             formula $\int_{-\infty}^{\infty} 
             x^{2} \exp( - \frac{ax^{2}}{2}) dx = \frac{1}{a}
             \int_{-\infty}^{\infty} \exp( - \frac{ax^{2}}{2} )dx$,
             with $a$ now being replaced by $N$.}
         \item{Loop: This is ${\cal O}(N)$. To comprehend this
             statement, let us have a look at a simple case. Consider
             the diagram of $\langle M_{ij} M_{kl} \rangle$ in
             Fig. \ref{zerodim}.  There are $N$ ways to connect the
             indices for each of $\delta_{i}^{l}$ and
             $\delta_{j}^{k}$. This induces $\sum_{i,j,k,l=1}^{N}
             \delta_{i}^{l} \delta_{j}^{k} = N^{2}$. The total
             contribution is therefore ${\cal O}(N^{-1} N^{2} ) =
             {\cal O}(N)$.}
        \end{itemize}
      These rules indicate that the total contribution of the power of
      $N$ is 
       \begin{eqnarray}
        {\cal O}(N^{V-E+F}) = {\cal O}(N^{\chi}), \label{2-38-}
       \end{eqnarray}
      where $\chi$ is the Euler character of the diagram, which is
      with ease identified with the Euler character of the worldsheet
      of the string theory. Therefore, the identification $N =
      e^{\gamma}$ is justified.}
  \end{itemize}
   This matrix model is formally identified with the string theory
   with on the discritized worldsheet by the identification of the
   quantities
   \begin{eqnarray}
    g \leftrightarrow e^{- \beta}, \hspace{3mm} N \leftrightarrow
    e^{\gamma}. \label{2-39-}
   \end{eqnarray}
  This is an evidence of the belief that string is described by a
  large $N$ reduced model\footnote{However, note that $D=0$
  dimensional QFT is not a reduced model of the 
  $\phi^{3}$ theory in QFT, because this $D=0$ dimensional model does
  not preserve the properties of the ${\phi}^{3}$ theory in
  QFT. Remember that the original proposal of large $N$ reduced
  model\cite{NEK} \cite{Das}. Their proposal, known as Eguchi-Kawai
  model, preserves the property of the original gauge theory in that
  the Eguchi-Kawai model reproduces the Schwinger-Dyson equation of
  the gauge theory.}. For $D \leq 1$, bosonic string theory is
  successfully quantized by Distler and Kawai\cite{DDK} (as is in fact 
  true of superstring theory). And as we will see later, the string
  susceptibility obtained by this matrix model matches the analysis of 
  Distler and Kawai. Owing to the work of Distler and Kawai, the
  matrix model may make a transition from being a mere 'toy model' to
  being a realistic nonperturbative description of string theory.
 An important remark is that this matrix model leaves only the leading 
 order of $N^{\chi} = N^{2 -2h}$ if we take a limit $N \to
 \infty$. This means that {\it only the effect of planar Riemann
 surface(without genus) survives}. We have seen a similar situation in 
 large $N$ QCD, in which only the effect of planar Feynman diagram
 survives in large $N$ limit.
 Although we have hitherto emphasized only the random triangulation,
 this analysis extends to any 'polygonulation' with ease. For example, 
 in order to consider the Riemann surfaces approximated by many
 equilateral squares, we have only to consider the matrix model
   \begin{eqnarray}
    Z_{QFT} = \int d^{N^{2}} M \exp( - \frac{1}{2} TrM^{2} +
    \frac{g}{N} TrM^{4} ) = \sum_{n=0}^{\infty} (\frac{g}{N})^{n} \int 
    d^{N^{2}} M \exp(-\frac{1}{2} TrM^{2}) (TrM^{4})^{n}. \label{2-40-}
   \end{eqnarray}
 The $n$-th power of this action describes the system in which there
 are $n$ 4-point vertices. This system is of grave importance in the
 analysis of pure gravity by Brezin and Kazakov in the subsequent
 section.

 \subsection{Orthogonal Polynomial Method}
  This section is devoted to introducing the method to analyze the
  above $D=0$ dimensional QFT in terms of large $N$ matrices. This
  method plays an essential role in solving this $D=0$ matrix model
  including the effect of higher genera, and thus makes the exact
  solution of this matrix model accessible. We start with the analysis 
  of the matrix model
   \begin{eqnarray}
    Z_{N}(g) = \int d^{N^{2}} M \exp( - V(M)). \label{2-41-}
   \end{eqnarray}
  Here, we focus on the pure gravity $V(M) = \frac{1}{2g}(TrM^{2} +
  \frac{1}{N} Tr M^{4})$. This model is analyzed in terms of the
  eigenvalues $\{ \lambda_{i} \}$ of the matrix $M$. The measure is
  known to be, whose proof we refer to Appendix. \ref{proofofmeasure},
  \begin{eqnarray}
      d^{N^{2}} M = \prod_{i} d \lambda_{i} \prod_{i<j} (\lambda_{i} -
     \lambda_{j})^{2} dU_{ij} = \prod_{i=1}^{N} d \lambda_{i}
     (detX)^{2} dU_{ij}, \nonumber \\
     \textrm{where } X = \left( \begin{array}{cccc}
     1 & 1 & \cdots & 1 \\
     \lambda_{1} & \lambda_{2} & \cdots & \lambda_{N} \\
     \lambda^{2}_{1} & \lambda^{2}_{2} & \cdots & \lambda^{2}_{N} \\
       \vdots & \vdots & \ddots & \vdots \\
     \lambda^{N-1}_{1} & \lambda^{N-1}_{2} & \cdots &
     \lambda^{N-1}_{N} \end{array} \right). \label{2-42-}
  \end{eqnarray}   
  Since the integrand depends only on the eigenvalues $\{ \lambda_{i}
  \}$, the effect of the integral $dU_{ij}$ is trivial. In order to
  solve this matrix model, we introduce a series of orthogonal
  polynomials $\{ P_{n}(\lambda) \}$ ($n=0, 1, \cdots $) which enjoy
  the following two properties.  
   \begin{itemize}
    \item{ $P_{n}(\lambda)$ is a polynomial of $n$-th degree, and the
        coefficient of the highest power of $\lambda$ is 1. There
        exists a series of coefficients $\{ a_{n,j} \}$ such that
        $P_{n} (\lambda) = \lambda^{n}  + \sum_{j=0}^{n-1} a_{n,j}
        \lambda^{j}$. }
   \item{ These polynomials are orthogonal with respect to the
       integral
     \begin{eqnarray}
       \int_{-\infty}^{\infty} d \lambda \exp(-V(\lambda))
       P_{n}(\lambda) P_{m}(\lambda) = h_{n} \delta_{mn}. \label{2-43-}
     \end{eqnarray}
    We call the quantity $h_{n}$ 'norm of $P_{n}(\lambda)$'. }
   \end{itemize}
  Vandermonde's determinant is rewritten as
   \begin{eqnarray}
    \det \left( \begin{array}{cccc}
     1 & 1 & \cdots & 1 \\
     \lambda_{1} & \lambda_{2} & \cdots & \lambda_{N} \\
     \lambda^{2}_{1} & \lambda^{2}_{2} & \cdots & \lambda^{2}_{N} \\
       \vdots & \vdots & \ddots & \vdots \\
     \lambda^{N-1}_{1} & \lambda^{N-1}_{2} & \cdots &
     \lambda^{N-1}_{N} \end{array} \right) = 
   \det \left( \begin{array}{cccc}
     P_{0}(\lambda_{1})  & P_{0}(\lambda_{2}) & \cdots &
     P_{0}(\lambda_{N})  \\
     P_{1}(\lambda_{1})  & P_{1}(\lambda_{2}) & \cdots &
     P_{1}(\lambda_{N})  \\
     P_{2}(\lambda_{1})  & P_{2}(\lambda_{2}) & \cdots &
     P_{2}(\lambda_{N})  \\
       \vdots & \vdots & \ddots & \vdots \\
     P_{N-1}(\lambda_{1}) & P_{N-1}(\lambda_{2}) & \cdots &
     P_{N-1}(\lambda_{N}) \end{array} \right). \label{2-44-}
   \end{eqnarray}
  Going back to the very definition of the determinant,
   \begin{eqnarray}
    \prod_{i,j=1}^{N} (\lambda_{i} - \lambda_{j})^{2} =
    \sum_{\sigma_{N}} (\prod_{i=1}^{N} P_{i} (\lambda_{\sigma_{N}(i)}
    ))^{2}, \label{2-45-}
   \end{eqnarray}
   where $\sigma_{N}(i)$ is the permutation, we readily obtain
    \begin{eqnarray}
      Z_{N}(g) = N! \prod_{i=0}^{N-1} h_{i} = N! h_{0}^{N}
      \prod_{i=1}^{N-1} f_{k}^{N-k}, \label{2-46-}
    \end{eqnarray}
  where $f_{k} = \frac{h_{k}}{h_{k-1}}$. We next seek the recursion
  formula of the orthogonal polynomials. The potential is here an even 
  function $V(\lambda) = \frac{1}{2g} (\lambda^{2} + \frac{1}{N}
  \lambda^{4})$. We consider the expansion $\lambda P_{n}(\lambda) =
  \sum_{i=0}^{n+1} c_{n,i} P_{i}(\lambda)$ with
  $c_{n,n+1}=1$. Likewise, the other coefficients are given by
  $c_{n,i} = h_{i}^{-1} \int_{-\infty}^{\infty} d \lambda
  e^{-V(\lambda)} P_{n}(\lambda) P_{i}(\lambda)$. Utilizing these
  properties and the orthogonality of the polynomials, we readily
  obtain the following relationships.
   \begin{itemize}
    \item{$c_{n,i}=0$ for $i=0,1,\cdots n-2$. This is trivial since
        $\lambda P_{i}(\lambda)$ can be expressed by the linear
        combination of $P_{0}(\lambda), P_{1}(\lambda), \cdots,
        P_{n-1}(\lambda)$.}
   \item{$c_{n,n-1} = h^{-1}_{n-1} \int_{-\infty}^{\infty} d \lambda
     e^{-V(\lambda)} P_{n}(\lambda) ( P_{n}(\lambda) +
     \sum_{j=0}^{n-1} c_{n-1,j} P_{j}(\lambda)) = h_{n} h^{-1}_{n-1} = 
     f_{n}$.}
   \item{$c_{n,n} = 0$ because the potential $V(\lambda)$ is an even
       function, whereas $\lambda (P_{n}(\lambda))^{2}$ is an odd
       function.}
   \end{itemize}
  Therefore, we have succeeded in deriving the recursion formula 
   \begin{eqnarray}
    \lambda P_{n}(\lambda) = P_{n+1}(\lambda) + f_{n}
    P_{n-1}(\lambda). \label{2-47-} 
   \end{eqnarray}
  Having obtained this relationship, we finally obtain the
  relationship amang the coefficients $\{ f_{i} \}$. The integration
  ${\cal I} = \int_{-\infty}^{\infty} e^{-V(\lambda)}
  \frac{dP_{n}(\lambda)}{d \lambda} \lambda P_{n}(\lambda)$ can be
  evaluated in two ways.
   \begin{itemize}
    \item{First, we consider the fact that 
      \begin{eqnarray} \lambda
        \frac{dP_{n}(\lambda)}{d \lambda} = \lambda(n \lambda^{n-1} +
        \sum_{j=1}^{n-1} a_{n,j} j \lambda^{j-1}) = n \lambda^{n} +
        \cdots. \nonumber
     \end{eqnarray}
     The above integral is readily given by ${\cal I} = n h_{n}$.}
   \item{The other way is to perform a partial integration. Here, we
       exploit an explicit form $V(\lambda) = \frac{1}{2g}(\lambda^{2} 
       + \frac{1}{N} \lambda^{4})$.  Then, the integral in question is
     \begin{eqnarray}
      {\cal I} = \int_{-\infty}^{\infty} d \lambda e^{-V(\lambda)}
      \frac{dP_{n}(\lambda)}{d \lambda} f_{n} P_{n-1} = f_{n}
      \int_{-\infty}^{\infty} e^{-V(\lambda)} \frac{d V(\lambda)}{d
      \lambda} P_{n}(\lambda) P_{n-1}(\lambda). \nonumber
     \end{eqnarray}
     Exploiting the fact that $\frac{dV(\lambda)}{d \lambda} =
     \frac{\lambda}{g} + \frac{2 \lambda^{3}}{gN}$, this integral is
     finally evaluated as
      \begin{eqnarray}
        {\cal I} = \frac{1}{g} h_{n} f_{n} + \frac{2}{gN} f_{n}
        \int_{-\infty}^{\infty} d \lambda P_{n}(\lambda) \lambda^{2}
        (P_{n}(\lambda) + f_{n-1} P_{n-1}(\lambda)) = \cdots =
        \frac{1}{g} f_{n} h_{n} + \frac{2}{gN} f_{n} h_{n} (f_{n-1} +
        f_{n} + f_{n+1}). \nonumber
      \end{eqnarray}}
   \end{itemize}
   Combining these two results, we finally obtain a recursion formula 
    \begin{eqnarray}
     gn = f_{n} + \frac{2}{N} f_{n}(f_{n-1} + f_{n} +
     f_{n+1}). \label{2-48-} 
    \end{eqnarray}

 \subsubsection{Analysis of the Planar Limit}
  Our next job is to solve the recursion formula (\ref{2-48-}). For
  simplicity, we first consider the planar ($N \to \infty$) limit. We
  approximate the series $\{ f_{k} \}$ by a continuous function 
   \begin{eqnarray}
    \frac{f_{n}}{N} \to f(\xi), \hspace{3mm} \frac{n \pm 1}{N} \to
    f(\xi \pm \epsilon), \label{2-49-}
   \end{eqnarray}
  where $\epsilon = \frac{1}{N}$ and $\xi = \frac{n}{N}$.  Since we
  take a limit $N \to \infty$, we discard the effect of $\epsilon =
  \frac{1}{N}$. This gives 
   \begin{eqnarray}
    g \xi = f(\xi) + 6f^{2}(\xi). \label{2-50-}
   \end{eqnarray}
 We define a new function $W(f)$ near its critical point $f \sim
 f_{c}$ as $W(f) = f + 6f^{2}$. We consider the saddle point of the
 function $W(f)$ : $\frac{dW(f)}{df} = 0$ at $f=f_{c}$. Defining
 $g_{c}$ as $g_{c} = W(f_{c})$, we obtain
   \begin{eqnarray}
    g \xi = g_{c} + \frac{1}{2} W''(f_{c}) (f(\xi) -
    f_{c})^{2}. \label{2-51-}
   \end{eqnarray}
 We compare this result with the quantized gravity. As we have
 explained before, the parameter named 'string susceptibility' is
 defined as (\ref{2-18-}). String susceptibility $\gamma$ is expressed 
 in the context of $D=0$ dimensional QFT as 
   \begin{eqnarray}
    f(\lambda) - f_{c} \sim (g_{c} - g \xi)^{-\gamma}. \label{2-52-}
   \end{eqnarray}
 The agreement of (\ref{2-52-}) with the very definition (\ref{2-18-}) 
 can be verified by reflecting the correspondence between the matrix
 model and the string theory on the discritized worldsheet. Here, we
 have approximated  the worldsheet by many equilateral square, because 
 we are now considering the potential $V(\lambda) = \frac{1}{2g}
 (\lambda^{2} + \frac{1}{N} \lambda^{4})$. The area of the worldsheet
 is obviously
   \begin{eqnarray}
    A = (\textrm{area of the worldsheet}) = (\textrm{number of
    square}) = n. \label{2-53-}
   \end{eqnarray} 
 On the other hand, the partition function is given by
 (\ref{2-46-}). Taking the logarithm, we can approximate (\ref{2-46-}) 
 by the continuous function
  \begin{eqnarray}
   \frac{1}{N^{2}} Z = \frac{1}{N} \sum_{k=0}^{N-1} ( 1- \frac{k}{N})
   \log f_{k} \sim \int_{0}^{1} d \xi (1 - \xi) \log
   f(\xi). \label{2-54-}
  \end{eqnarray}
 Performing the partial integration, we obtain a following remarkable 
 relationship:
  \begin{eqnarray}
   \frac{Z}{N^{2}} &\sim& \int_{0}^{1} d \xi (1-\xi) (f_{c} + (g_{c} - 
   g \xi)^{-\gamma}) \sim [(1-\xi)(g_{c} - g \xi)^{-\gamma+1}]_{0}^{1} 
   + \int_{0}^{1} d \xi (g_{c} - g \xi)^{- \gamma +1} \nonumber \\
   &\sim& (g_{c} - g \xi)^{- \gamma + 2} \sim \sum_{n=0}^{\infty}
   n^{\gamma -3} (\frac{g}{g_{c}})^{n} = \sum_{A=0}^{\infty} A^{\gamma 
   -3} (\frac{g}{g_{c}})^{A}. \label{2-55-}
  \end{eqnarray}
 Thus, in order to seek a string susceptibility, we have only to
 consider the scaling of the function $f(\xi)$. In this case, the
 behavior of the critical point has been obtained in
 (\ref{2-51-}). Since $r(\xi) - r_{c} \sim (g \xi -
 g_{c})^{\frac{1}{2}}$. The string susceptibility is thus
    \begin{eqnarray}
     \gamma = - \frac{1}{2}. \label{2-56-}
    \end{eqnarray}
 This is a result of planar ($N \times \infty$ and thus only the
 effect of $h=0$ survives) limit. Comparing this with the result
 in (\ref{2-19-}) \cite{DDK}, this corresponds to the result $D=0$ and 
 $h=0$.

 \subsubsection{Analysis of the Nonplanar Behavior}
 We next consider the effect of higher genera to solve this matrix
 model exactly. The essential difference from the previous analysis is 
 that we  do not discard the term $\epsilon = \frac{1}{N}$. We use the 
 function $W(r) = r + 6r^{2}$, and we obtain
  \begin{eqnarray}
  g \xi = g_{c} + \frac{1}{2} W''(r_{c}) (r(\xi) - r_{c})^{2} +
  2r(\xi) (r(\xi + \epsilon) + r(\xi - \epsilon) - 2r(\xi)) = g_{c} +
  \frac{1}{2} W''(r_{c})(r(\xi) - r_{c})^{2} + 2 \epsilon^{2}
  \frac{d^{2} r}{d \xi^{2}}. \label{2-57-}
  \end{eqnarray}
 Now, we take the double scaling limit $N \to \infty$ and $g \to
 g_{c}$. $g-g_{c}$ possesses dimension $[\textrm{length}]^{2}$, and it 
 is convenient to introduce a constant $a$ with dimension length. Then, 
 let 
  \begin{eqnarray} g - g_{c} = \kappa^{-\frac{4}{5}}
 a^{2}. \label{2-58-} \end{eqnarray}
 And we assume an ansatz
   \begin{eqnarray}
    \epsilon = \frac{1}{N} \stackrel{ansatz}{=} a^{\frac{5}{2}}. \nonumber
   \end{eqnarray}
 This is a maneuver to make the parameter $\kappa = (g-g_{c})^{-
 \frac{5}{4}} N^{-1}$ finite as we take the limit $N \to \infty$ and
 $g \to g_{c}$. In order to study this relationship, it is more
 convenient to change the variables as $g_{c} - g \xi = a^{2}z$. For
 this variables, we assume a following scaling ansatz
  \begin{eqnarray}
   f(\xi) - f_{c} \stackrel{ansatz}{=} a u(z). \label{2-59-} 
  \end{eqnarray}
 Then, we obtain a relationship $r(\xi + \epsilon) + r(\xi - \epsilon)
 - 2r(\xi) \sim \epsilon^{2} \frac{d^{2}r}{d \xi^{2}} \sim a^{2}
 \frac{d^{2} u}{dz^{2}}$, utilizing $\epsilon \frac{d}{d \xi} = -g
 a^{\frac{1}{2}} \frac{d}{dz}$. Then, we obtain a nonlinear
 differential equation for the function $u(z)$ called Painleve
 equation:
  \begin{eqnarray}
   z = u^{2} (z) + \frac{d^{2} u(z)}{dz^{2}}. \label{2-60-}
  \end{eqnarray}
 This equation provides us with a lot of information about the $D=0$
 dimensional QFT. First, we can obtain an exact form of the function
 $f(\xi)$ from the solution of Painleve equation. Noting the
 relationship of the relationship of the partition function
 (\ref{2-46-}), this gives an exact form of the partition function of
 the theory! The orthogonal polynomial method plays a splendid role in 
 solving the matrix model as $D=0$ dimensional QFT
 nonperturbatively. This is a great success of the matrix model in
 solving the string theory nonperturbatively, even though this
 discussion is limited to a very low spacetime dimension.

 Furthermore, the analysis of the string susceptibility  via
 orthogonal polynomial method agrees with the result of Distler and
 Kawai. In order to see this, let us consider the asymptotic solutions 
 of Painleve equation. First, we consider the solution of
 ({\ref{2-60-}) for $z \to \infty$. This corresponds to the planar
 behavior because we have taken a scaling $ \epsilon = \frac{1}{N} =
 a^{\frac{5}{2}}$ and $g_{c} - g \xi = a^{2}z$, and hence $a$ should
 be $a \to 0$. The asymptotic solution is 
   \begin{eqnarray}
    u(z) = \sqrt{z}, \hspace{2mm} (\textrm{as }z \to
    \infty). \label{2-61-} 
   \end{eqnarray}
 This makes sense because because substituting this answer into the
 Painleve equation, $z = (\sqrt{z})^{2} - \frac{1}{4} z^{-\frac{3}{2}}
 \sim z$ (as $z \to \infty$). This means that the string
 susceptibility of the string without any genus is $\gamma = -
 \frac{1}{2}$, from the property (\ref{2-52-}). We next add a
 contribution of the subleading term. Let the solution of Painleve
 equation be
  \begin{eqnarray}
   u(z) = \sqrt{z} + az^{b}. \label{2-62-}
  \end{eqnarray}
 Here, it is not the coefficient $a$ but the scaling $b$ that
 imports. Substituting this answer into Painleve equation, we obtain 
   \begin{eqnarray} 
    z = z + 2a z^{\frac{1}{2} + b} + a^{2} z^{2b} - \frac{1}{4}
    z^{-\frac{3}{2}} +ab(b-1) z^{b-2}. \label{2-63-}
   \end{eqnarray}
 In order for the equality of the above equation to hold, the scaling
 parameter $b$ should satisfy
  \begin{eqnarray}
   2b = b-2, \hspace{3mm} \frac{1}{2} + b = - \frac{3}{2}. \label{2-64-}
  \end{eqnarray}
 Both conditions are satisfied for $b=-2$. Them the solution including 
 the subleading term is
  \begin{eqnarray}
   u(z) = \sqrt{z} - \frac{1}{8} z^{-2}. \label{2-65-}
  \end{eqnarray}
 This indicates that the string susceptibility for $D=0$ and $h=1$ is
 $\gamma = 2$. Repeating the same procedure, we obtain an asymptotic
 solution 
  \begin{eqnarray}
   u(z) = \sqrt{z} (1 + \sum_{h=1}^{\infty} u_{h} z^{\frac{5h}{2}}),
   \label{2-66-} 
  \end{eqnarray}
 where $\{ u_{h} \}$ is a coefficient. This gives the string
 susceptibility for $D=0$ and all genera of the Riemann surface:
  \begin{eqnarray}
   \gamma = 2 - \frac{5}{2}(1-h). \label{2-67-}
  \end{eqnarray}
 This agrees with the analysis of Distler and Kawai
 (\ref{2-19-}). This striking agreement of the string susceptibility
 further solidifies the confidence that the nonperturbative behavior
 of string theory is described by matrix model.

 \subsection{Summary}
 We would like to conclude this section by summarizing the argument
 in this section.
  \begin{itemize}
   \item{Distler and Kawai succeeded in the quantization of both
       bosonic string and superstring for $D \leq 1$. And they
       computed a parameter 'string susceptibility' for all genera of
       the worldsheet.}
  \item{The proposal that string should be described by matrix model
      was first given by F. David. He discritized the worldsheet of
      bosonic string theory into many polygons, and related the
      discritized theory with a matrix model describing 0 dimensional
      QFT.}
  \item{Brezin and Kazakov solved the 0 dimensional QFT via orthogonal 
      polynomial method. Their answer describes the effects of higher
      genera of the worldsheet, and the string susceptibility for
      $D=0$ agrees with the results of Distler and Kawai.}
  \end{itemize}
  These series of works are of historical importance in that they are
  the canons of the belief that matrix theory is the Ariadne's thread
  to 'Theory of Everything'.
\section{The brief review of IKKT Model}
   In the late 1990's, several proposals for 'Theory of Everything' have
 been given, reflecting the progress of the research of superstring
 theory. The belief that the nonperturbative behavior of string should 
 be described by matrix theory has given three major proposals -
 BFSS model \cite{9610043}, IKKT model \cite{9612115} and matrix
 string theory \cite{9703030}. These proposals are the dimensional
 reduction of ${\cal N} =1$ SYM into 1, 0 and 2 dimensions
 respectively. Among them, the most successful proposal is IKKT model, 
 and this section is devoted to the review of IKKT model.

 \subsection{Definition and Symmetry of IKKT model}
  Inspired by the discovery of BFSS conjecture, Ishibashi, Kawai,
  Kitazawa and Tsuchiya proposed a matrix theory described by the 0
  dimensional reduction of ${\cal N}=1$ 10 dimensional SYM. This is
  IKKT model, whose action is
   \begin{eqnarray}
    S = - \frac{1}{g^{2}} Tr ( \frac{1}{4} \sum_{i,j=0}^{9} [A_{i},
    A_{j}] [A^{i}, A^{j}] + \frac{1}{2} \sum_{i,j=0}^{9} {\bar \psi}
    \Gamma^{i} [A_{i}, \psi ] ). \label{AZM31IKKT}
   \end{eqnarray}
  \begin{itemize}
   \item{$A_{i}$ are $N \times N$ Hermitian matrices, and these are 10 
       dimensional vectors.}
   \item{$\psi$ are also $N \times N$ Hermitian matrices, and these
       are 10 dimensional Majorana-Weyl 16 spinors.}
   \item{Throughout this paper, the
  indices $i,j, \cdots$ refer to 10 dimensions, while $\mu, \nu,
  \cdots $ refer to 11 dimensions. And the trace $Tr$ is for $N \times 
  N$ matrices in large $N$ reduced model.}
   \item{This matrix theory manifestly possesses $SO(9,1)$ Lorentz
       symmetry and $SU(N)$ gauge symmetry.}
   \item{This theory has no free parameter. The only parameter $g$ is
       absorbed into the fields by the rescaling $A_{i} \rightarrow
       g^{\frac{1}{2}} A_{i}$ and $\psi \rightarrow g^{\frac{3}{4}}
       \psi$.} 
  \end{itemize}

  This matrix theory has another name - {\it IIB matrix theory} -,
  because this theory is deeply related with type IIB
  superstring. There are several reasons to speculate that this model
  is a constructive definition of type IIB superstring, and one of
  these aspects is that this matrix model is the same as  the
  matrix reguralization of the Schild action of type IIB superstring.
  First, let us introduce Green-Schwarz action of type IIB
  superstring, whose derivation we refer to \cite{shino}. We introduce 
  a superspace for ${\cal N} =2$ superspace
   \begin{eqnarray}
     Z^{M} = (X^{i}, \theta^{1 \alpha}, \theta^{2 \alpha} ),
   \end{eqnarray}
  where $X^{i}$ are 10 dimensional vectors while two spinors
  $\theta^{1,2 \alpha}$ are each 16 Majorana Weyl spinors
  for 10 dimensional spacetime (therefore, in total we have 32
  spinors). For this superspace, the SUSY transformation is defined as 
   \begin{eqnarray}
     \delta_{S} X^{i} = i {\bar \epsilon^{1}} \Gamma^{i} \theta^{1} - i
     {\bar \epsilon^{2}} \Gamma^{i} \theta^{2} , \hspace{2mm}
     \delta_{S} \theta^{1,2} = \epsilon^{1,2}.
   \end{eqnarray}
    The following quantities are SUSY invariant :
    \begin{eqnarray}
     \Pi^{i}_{a} = \partial_{a} X^{i} - i {\bar \theta^{1}}
     \Gamma^{i} \partial_{a} \theta^{1} + i {\bar \theta^{2}} \Gamma^{i}
     \partial_{a} \theta^{2}, \hspace{2mm} \Pi^{1,2 \alpha}_{a} =
     \partial_{a} \theta^{1,2 \alpha}.
    \end{eqnarray}
   The trouble is that the degree of freedom for fermions are much
   larger that that of bosons, the former being $16 \times 2 = 32$,
   the latter being $11-2-1 + 8$. The maneuver to remedy this situation
   is to introduce Wess-Zumino term, whose explanation we owe to
   \cite{shino}. This maneuver is possible only for 
     \begin{eqnarray}
       D = \textrm{(dimensions of spacetime)} = 3,4,6,10.
     \end{eqnarray}  
  Then, the action should have another symmetry called $\kappa$ {\it
  symmetry}
    \begin{eqnarray}
     \delta_{\kappa} X^{i} = i {\bar \theta^{1}} \Gamma^{i} \alpha^{1} 
     - i {\bar \theta^{2}} \Gamma^{i} \alpha^{2}, \hspace{2mm}
     \delta_{\kappa} \theta^{1,2} = \alpha^{1,2},
    \end{eqnarray} 
  where 
   \begin{eqnarray}
    & &\alpha^{1} = (1 + {\tilde \Gamma}) \kappa_{1}, \hspace{2mm}
    \alpha_{2} = (1 - {\tilde \Gamma}) \kappa^{2}, \hspace{2mm} 
    {\tilde \Gamma} = \frac{1}{2! \sqrt{- M}} \epsilon^{ab}
    \Pi^{i}_{a} \Pi^{j}_{b} \Gamma_{ij}, \hspace{2mm} M = \det (\epsilon^{ab}
    \Pi^{i}_{a} \Pi^{j}_{b}), \\
    & &\epsilon_{01} = +1 \textrm{ (hence } \epsilon^{01} = -1 \textrm
    { ) }.
   \end{eqnarray}
  The action constructed to have these symmetry is the Green Schwarz
  form of the superstring possessing ${\cal N}=2$ symmetry: 
  \begin{eqnarray}
   S_{GS} = -T \int d^{2} \sigma [ \sqrt{-M} + i \epsilon^{ab}
   \partial_{a} X^{i} ( {\bar \theta^{1}} \Gamma_{i} \partial_{b}
   \theta_{1} + {\bar \theta^{2}} \Gamma_{i} \partial_{b} \theta_{2})
   + \epsilon^{ab} ({\bar \theta^{1}} \Gamma^{i} \partial_{a} \theta^{i})
   ({\bar \theta^{2}} \Gamma_{i} \partial_{b} \theta^{2}) ].
  \end{eqnarray}

  Type IIB superstring is defined so that the chirality of two spinors 
  are the same, and we set $\theta^{1} = \theta^{2} = \psi$. The
  action is rewritten as
   \begin{eqnarray}
    S_{GS} = -T \int d^{2} \sigma ( \sqrt{ -m} + 2i \epsilon^{ab}
    \partial_{a} X^{i} {\bar \psi} \Gamma_{i} \partial_{b} \psi ),
   \end{eqnarray}
  where $m= \det(m_{ab}) = \det(\partial_{a} X^{i} \partial_{b}
  X_{i})$. This
  action is ${\cal N} =2$ SUSY invariant provided we 
  redefine the SUSY by mixing the original SUSY with $\kappa$
  symmetry. The new SUSY is
   \begin{eqnarray}
    \delta \theta^{1,2} = \delta_{S} \theta^{1,2} + \delta_{\kappa}
    \theta^{1,2} \hspace{2mm} \delta X^{i} = \delta_{S} X^{i} +
    \delta_{\kappa} X^{i}, 
   \end{eqnarray}
 where we choose $\kappa$ symmetry to be $\kappa_{1} = \frac
 { -\epsilon^{1} + \epsilon^{2}}{2}, \kappa_{2} = \frac{\epsilon^{1} - 
 \epsilon^{2}}{2}$. Then, setting the SUSY parameter as
   \begin{eqnarray}
    \xi = \frac{\epsilon^{1} + \epsilon^{2}}{2}, \hspace{2mm} \epsilon 
    = \frac{\epsilon^{1} - \epsilon^{2}}{2}
   \end{eqnarray}
  we can express the ${\cal N}=2$ SUSY of the Green-Schwarz action as
  follows:
   \begin{eqnarray}
   & &  \delta^{(1)}_{\epsilon} \psi = \frac{1}{2} \sqrt{- m} m_{ij}
    \Gamma^{ij} \epsilon, \hspace{2mm} \delta^{(1)}_{\epsilon} X^{i} = 
    4 i {\bar \epsilon} \Gamma^{i} \psi, \label{AZM31SUSY1gs} \\
  & &  \delta^{(2)}_{\xi} \psi = \xi, \hspace{2mm} \delta^{(2)}_{\xi}
    X^{i} = 0,
   \end{eqnarray}
 where $m_{ij} = \epsilon^{ab} \partial_{a} X_{i} \partial_{b}
 X_{j}$.

 Our next job is to rewrite this Green-Schwarz action into Schild
 form. Defining $g_{ab}$ as the metric of the worldsheet, and the
 Poisson bracket as $\{X, Y \} = \frac{1}{\sqrt{g}} \epsilon^{ab}
 \partial_{a} X \partial_{b} Y$, this action is rewritten as 
  \begin{eqnarray}
    S_{Sh} = \int d^{2} \sigma [ \sqrt{g} \alpha (\frac{1}{4} 
    \{ X^{i}, Y^{i}  \}^{2} - \frac{i}{2} {\bar \psi} \Gamma^{i} 
    \{ X_{i}, \psi \} ) + \beta \sqrt{g} ].
  \end{eqnarray}
  This action is proven to be equivalent to the Green-Schwarz form of
  type IIB superstring. The ${\cal N}=2$ SUSY of this Schild action is
   \begin{eqnarray}
   & & \delta^{(1)}_{\epsilon} \psi = - \frac{1}{2} m_{ij} \Gamma^{ij}
    \epsilon , \hspace{2mm} \delta^{(1)} X^{i} = i {\bar \epsilon}
    \Gamma^{i} \psi, \label{AZM31SUSY1sh} \\
  & &  \delta^{(2)}_{\xi} \psi = \xi, \hspace{2mm} \delta^{(2)}_{\xi}
    X^{i} = 0.     
   \end{eqnarray}
  
  We perform a procedure called 'matrix regularization', whose
  detailed explanation we again owe to \cite{shino}. This procedure is,
  simply speaking, the mapping  from the Poisson bracket to the commutator
  of large $N$ matrices
   \begin{eqnarray}
    -i [ , ] \leftrightarrow  \{ , \} ,\hspace{2mm} Tr
    \leftrightarrow  \int d^{2} \sigma \sqrt{g}. 
   \end{eqnarray}
   The functions $X^{i}$ are now mapped into the $N \times N$ matrices 
  $A^{i}$. and we obtain the action similar to the original proposal
  of IKKT  model :
   \begin{eqnarray}
    S = -\alpha ( \frac{1}{4} Tr [A_{i}, A_{j}][A^{i}, A^{j}] +
    \frac{1}{2} Tr ({\bar \psi} \Gamma^{i} [A_{i}, \psi] ) ) + \beta
    Tr {\bf 1}.
   \end{eqnarray}
  By dropping the term $\beta Tr {\bf 1}$ and setting $\alpha =
  \frac{1}{g^{2}}$, we reproduce the action of IKKT model
  (\ref{AZM31IKKT}) . In this sence, it can be fairly said that IKKT
  model is a concept related to type IIB superstring theory. And we
  speculate that the matrix regularization of type IIB superstring
  theory is the constructive definition of type IIB superstring.

  Now, let us have a careful look at the SUSY of IKKT model. We have
  formulated the SUSY of Schild action of type IIB superstring. In
  IKKT model we consider that the matrix-regularized version of that
  SUSY is inherited. The ${\cal N}=2$ SUSY of IKKT model is then
  distinguished by
   \begin{itemize}
    \item{homogeneous: $\delta^{(1)}_{\epsilon} \psi = \frac{i}{2}
    [A_{i} , A_{j}] \Gamma^{ij} \epsilon, \hspace{2mm} 
    \delta^{(1)}_{\epsilon} A_{i} = i {\bar  \epsilon} \Gamma_{i} \psi$. \\
    The feature of the
    {\it 'homogeneous SUSY'} is that this SUSY transformation depends
    on the matter fields $A_{i}$ and $\psi$. And this SUSY
    transformation vanishes if there is no matter field.}
   \item{inhomogeneous: $\delta^{(2)}_{\xi} \psi = \xi, \hspace{2mm}
       \delta^{(2)}_{\xi}  A_{i} = 0$. \\
       The feature of the {\it
       'inhomogeneous SUSY'} is that the translation  survives without
       the matter fields.} 
   \end{itemize}
  The commutators of these SUSY's give the following important
  results:
   \begin{eqnarray}
    &(1)& [\delta^{(1)}_{\epsilon_{1}}, \delta^{(1)}_{\epsilon_{2}}]
    A_{i} = 0 ,\hspace{2mm} [\delta^{(1)}_{\epsilon_{1}},
    \delta^{(1)}_{\epsilon_{2}} ] \psi = 0,
    \label{AZM31SUSY1com} \\
    &(2)& [\delta^{(2)}_{\xi_{1}}, \delta^{(2)}_{\xi_{2}}] A_{i} =
    0, \hspace{2mm} [\delta^{(2)}_{\xi_{1}}, \delta^{(2)}_{\xi_{2}}] = 
    0, \label{AZM31SUSY2com} \\
    &(3)& [\delta^{(1)}_{\epsilon}, \delta^{(2)}_{\xi}] A_{i} = - i
    {\bar \epsilon} \Gamma_{i} \xi, \hspace{2mm}
    [\delta^{(1)}_{\epsilon}, \delta^{(2)}_{\xi}] \psi = 0.
    \label{AZM31SUSY3com} 
   \end{eqnarray}
 {\sf
 (Proof) These properties can be verified by taking the difference of
 the two SUSY transformations.
  \begin{enumerate}
   \item{This is the most complicated to compute. For the gauge field, 
       we should consider the following transformation
      \begin{eqnarray}
          A_{i} &\stackrel{\delta^{(1)}_{\epsilon_{2}}}{\rightarrow}& 
          A_{i} + i \epsilon_{2} \Gamma_{i} \psi
          \stackrel{\delta^{(1)}_{\epsilon_{1}}}{\rightarrow} A_{i} +
          i ({\bar \epsilon_{1}} + {\bar \epsilon_{2}}) \Gamma_{i}
          \psi - \frac{1}{2} {\bar \epsilon_{2}} \Gamma_{i} [A_{j},
          A_{k}] \Gamma^{jk} \epsilon_{1},   \\
          A_{i} &\stackrel{\delta^{(1)}_{\epsilon_{1}}}{\rightarrow}& 
          A_{i} + i \epsilon_{1} \Gamma_{i} \psi
          \stackrel{\delta^{(1)}_{\epsilon_{2}}}{\rightarrow} A_{i} +
          i ({\bar \epsilon_{1}} + {\bar \epsilon_{2}}) \Gamma_{i}
          \psi - \frac{1}{2} {\bar \epsilon_{1}} \Gamma_{i} [A_{j},
          A_{k}] \Gamma^{jk} \epsilon_{2}.
      \end{eqnarray}
      Then, the commutator is
     \begin{eqnarray}
       [\delta^{(1)}_{\epsilon_{1}}, \delta^{(1)}_{\epsilon_{2}}]
       A_{i} =  - \frac{1}{2} {\bar \epsilon_{2}} \Gamma_{i} [A_{j},
          A_{k}] \Gamma^{jk} \epsilon_{1}  + \frac{1}{2} {\bar
       \epsilon_{1}} \Gamma_{i} [A_{j}, A_{k}] \Gamma^{jk}
       \epsilon_{2}.         
     \end{eqnarray}
    Utilizing the formula $\Gamma^{i} \Gamma^{jk} = \Gamma^{ijk} +
    \eta^{ij} \Gamma^{k} - \eta^{ik} \Gamma^{j}$(for a general
    property, see Appendix. \ref{AZCproductgamma}) and the properties
    of the fermions in Appendix. \ref{AZCflipping} , we obtain
    \begin{eqnarray}
      [\delta^{(1)}_{\epsilon_{1}}, \delta^{(1)}_{\epsilon_{2}}]
    A_{i} = 2 {\bar \epsilon_{1}} \Gamma^{k} \epsilon_{2} [A_{i},
    A_{k}]. \label{AZM31SUSY11bos}
    \end{eqnarray}
   For the fermions, we have only to repeat the similar procedure:
    \begin{eqnarray}
     \psi &\stackrel{\delta^{(1)}_{\epsilon_{2}}}{\to}& \psi + \frac{i}{2}
     [A_{i}, A_{j}] \Gamma^{ij} \epsilon_{2}
     \stackrel{\delta^{(1)}_{\epsilon_{1}}}{\to} \psi + \frac{i}{2}
     [A_{i}, A_{j}] \Gamma^{ij} (\epsilon_{1} + \epsilon_{2}) -
     [A_{i}, {\bar \epsilon_{1}} \Gamma_{j} \psi ] \Gamma^{ij}
     \epsilon_{2}, \\
     \psi &\stackrel{\delta^{(1)}_{\epsilon_{1}}}{\to}& \psi + \frac{i}{2}
     [A_{i}, A_{j}] \Gamma^{ij} \epsilon_{1}
     \stackrel{\delta^{(1)}_{\epsilon_{2}}}{\to} \psi + \frac{i}{2}
     [A_{i}, A_{j}] \Gamma^{ij} (\epsilon_{1} + \epsilon_{2}) -
     [A_{i}, {\bar \epsilon_{2}} \Gamma_{j} \psi ] \Gamma^{ij}
     \epsilon_{1}.
    \end{eqnarray}    
    Using the formula of Fierz transformation, 
     \begin{eqnarray}
      {\bar \epsilon_{1}} \Gamma_{j} \psi \Gamma^{ij} \epsilon_{2} =
      {\bar \epsilon_{1}} \Gamma^{i} \epsilon_{2} \psi - \frac{7}{16}
      {\bar \epsilon_{1}} \Gamma^{k} \epsilon_{2} \Gamma_{k}
      \Gamma^{i} \psi - \frac{1}{16 \times 5!} {\bar \epsilon_{1}}
      \Gamma^{k_{1} \cdots k_{5}} \epsilon_{2} \Gamma_{k_{1} \cdots
      k_{5}} \Gamma^{i} \psi + (\textrm{rank 3 term}), \label{AZM31fierz}
     \end{eqnarray} 
  whose proof we present in Appendix \ref{AZCdecomposition},
  we verify that the commutator of SUSY transformation is 
    \begin{eqnarray}
      [\delta^{(1)}_{\epsilon_{1}},
    \delta^{(1)}_{\epsilon_{2}} ] \psi = 2 [\psi,   {\bar
    \epsilon_{1}} \Gamma^{i} \epsilon_{2} A_{i} ]. \label{AZM31aho}
    \label{AZM31SUSY11ferm} 
    \end{eqnarray}
  We explain some tips in arriving at the result (\ref{AZM31aho}).
  \begin{itemize}
   \item{The rank 3 term of (\ref{AZM31fierz}) does not contribute
   because the fermions $\epsilon_{1}, \epsilon_{2}$ are Majorana ones 
   and hence the difference ${\bar \epsilon}_{1} \Gamma^{ijk}
   \epsilon_{2} - {\bar \epsilon}_{2} \Gamma^{ijk} \epsilon_{1}$
   cancels due to the property presented in
   Appendix. \ref{AZCflipping}.}
  \item{We discern that the second and third terms of the Fierz
      transformation (\ref{AZM31fierz}) {\it with the help of the
      equation of motion of IKKT model}: id est, the SUSY
      transformation (\ref{AZM31SUSY1com}) hold true only on
      shell. We utilize the equation of motion 
       \begin{eqnarray}
     \frac{d S}{d {\bar \psi}} = - \frac{1}{g^{2}} \Gamma^{i} 
      [A_{i}, \psi] = 0.
       \end{eqnarray}
     The second and third terms of (\ref{AZM31fierz}) drops because
     this is proportional to the commutator $\Gamma^{i}[A_{i},
     \psi]$.}
  \end{itemize}

   Next, we note that the commutators of SUSY transformation
   (\ref{AZM31SUSY11bos}) and 
   (\ref{AZM31SUSY11ferm}) vanish up to the gauge transformation. The
   gauge transformation of IKKT model is to multiply the unitary
   matrix $\alpha \in U(N)$. The gauge transformation is expressed in
   the infinitesimal form as follows :
  \begin{eqnarray}
   A_{i}, \psi \rightarrow A_{i}+i[A_{i}, \alpha], \hspace{2mm} \psi
   + i [\psi, \alpha]. \label{AZM31gaugetr}
  \end{eqnarray}
   The SUSY transformation  (\ref{AZM31SUSY11bos}) and
   (\ref{AZM31SUSY11ferm}) can be gauged away by the gauge parameter
   $\alpha = 2 {\bar \epsilon_{1}} \Gamma^{k} \epsilon_{2} A_{k}$.  We 
   now complete the proof of (\ref{AZM31SUSY1com}) up to the gauge
   transformation. 
      }
   \item{This is trivial because the SUSY $\delta^{(2)}_{\xi}$
       involves only a constant.}
    \item{This can be proven by taking the difference of these two
        transformations  
       \begin{eqnarray}
        A_{i} &\stackrel{ \delta^{(2)}_{\xi}}{\to}&  A_{i}
       \stackrel{\delta^{(1)}_{\epsilon}}{\to} A_{i} +  i {\bar \epsilon}
      \Gamma_{i} \psi \textrm{, whereas }
       A_{i}\stackrel{\delta^{(1)}_{\epsilon}}{\to} A_{i} +  i
       {\bar \epsilon}  \Gamma_{i} \psi  \stackrel
       { \delta^{(2)}_{\xi}}{\to}  A_{i} +  i{\bar \epsilon}
       \Gamma_{i} (\psi + \xi),  \\ \nonumber 
      \psi  &\stackrel{ \delta^{(2)}_{\xi}}{\to}& \psi + \xi
       \stackrel{\delta^{(1)}_{\epsilon}}{\to}  \psi + \xi +  \frac{i}{2}
      \Gamma^{ij} [ A_{i} , A_{j} ] \epsilon \textrm{,
        whereas }\psi 
        \stackrel{\delta^{(1)}_{\epsilon}}{\to}   \psi +  \frac{i}{2}
      \Gamma^{ij} [ A_{i} , A_{j} ] \epsilon \stackrel
       { \delta^{(2)}_{\xi}}{\to} \psi + \xi +  \frac{i}{2}
      \Gamma^{ij} [ A_{i} , A_{j} ] \epsilon.
    \end{eqnarray}
  }
  \end{enumerate}
  This completes the proof of the commutation relation of the SUSY transformation.(Q.E.D.)
}\\

  We take a linear combination of the SUSY transformation to
  diagonalize the SUSY transformation
   \begin{eqnarray}
    {\tilde \delta}^{(1)} = \delta^{(1)} + \delta^{(2)} ,\hspace{2mm}
    {\tilde \delta}^{(2)} = i (\delta^{(1)} - \delta^{(2)}). 
   \end{eqnarray}
  We obtain a following ${\cal N} =2$ SUSY algebra 
   \begin{eqnarray}
   & & [{\tilde \delta}^{(1)}_{\epsilon}, {\tilde \delta}^{(1)}_{\xi}] 
        \psi =  [{\tilde \delta}^{(1)}_{\epsilon}, {\tilde
        \delta}^{(2)}_{\xi}] \psi =  [{\tilde
        \delta}^{(2)}_{\epsilon}, {\tilde \delta}^{(2)}_{\xi}] \psi =
        0, \label{AZM31SUSY4com} \\
  & &  [{\tilde \delta}^{(1)}_{\epsilon}, {\tilde \delta}^{(1)}_{\xi}] 
        A_{i} =  [{\tilde \delta}^{(2)}_{\epsilon}, {\tilde
        \delta}^{(2)}_{\xi}] A_{i} = -2i {\bar \epsilon} \Gamma_{i} \xi,
        \hspace{2mm}  [{\tilde \delta}^{(1)}_{\epsilon}, {\tilde
        \delta}^{(2)}_{\xi}] A_{i} = 0. \label{AZM31SUSY5com}
   \end{eqnarray} 
  We regard the SUSY ${\tilde \delta}^{(1)}$ and ${\tilde
  \delta}^{(2)}$ as a fundamental SUSY transformation. We now confirm
  that these fundamental SUSY transformations  actually
  satisfy Haag-Lopuszanski-Sohnius extension of Coleman-Mandula
  theorem, in which the supercharges must satisfy  
   \begin{eqnarray}
   \{ Q_{x}, Q_{x} \} = P_{i}, \hspace{3mm}  [P_{i}, Q_{x}] = 0 
    (\textrm{for } x=1,2), \label{AZ31hls}
   \end{eqnarray}
  where $P_{i}$ is an operator for the translation of the bosonic vector
  fields. We verify 
  these statements (\ref{AZ31hls}) one by one, for the ${\cal
  N}=2$ SUSY transformation of IKKT model.

 The former statement $\{ Q_{x}, Q_{x} \} = P_{i}$ is readily read off
 from the commutation relation (\ref{AZM31SUSY5com}). Let the
 supercharge $Q_{i}$ be the SUSY transformation $Q_{x}(\epsilon) = {\tilde
  \delta}^{(x)}_{\epsilon} $. Then, the relation
  (\ref{AZM31SUSY5com}) means 
   \begin{eqnarray}
    [ Q_{x}(\epsilon), Q_{x}(\xi) ] = -2i {\bar \epsilon} \Gamma_{i}
    \xi = \textrm{ (translation of the bosonic field)}. 
   \end{eqnarray} 
  The anti-commutator in (\ref{AZ31hls}) is replaced with the
  commutator because the SUSY parameters are Grassmann odd
  quantities. This indicates that the commutator of the supercharge
  translates the bosonic vector fields by $a_{i} = -2i {\bar \epsilon}
  \Gamma_{i} \xi$, and thus we have confirmed the first statement of
  Haag-Lopuszanski-Sohnius theorem. 

      The latter statement  $[P_{i}, Q] =0$ is trivial. Since the
      fundamental SUSY transformations are expressed by the linear
      combinations of $\delta^{(1)}$ and $\delta^{(2)}$, we have only to
      compute the commutation relation between the translation  $P_{i}$
      and the SUSY transformations $\delta^{(1)}, \delta^{(2)}$.
      \begin{itemize}
       \item{First, we verify the commutation relation $[P_{i},
           \delta^{(1)}] = 0$. This can be verified by taking the
           difference of the two paths for both matter fields and the
           fermionic fields.
           \begin{itemize}
           \item{$A_{i} \stackrel{P_{i}}{\to} A_{i} + a_{i}
               \stackrel{\delta^{(1)}_{\epsilon}}{\to} A_{i} + a_{i} + 
               i {\bar \epsilon} \Gamma_{i} \psi$, whereas $A_{i}
               \stackrel{\delta^{(1)}_{\epsilon}}{\to}  A_{i} + 
               i {\bar \epsilon} \Gamma_{i} \psi \stackrel{P_{i}}{\to}
               A_{i} + a_{i} +  i {\bar \epsilon} \Gamma_{i} \psi$.}
           \item{$\psi  \stackrel{P_{i}}{\to} \psi
               \stackrel{\delta^{(1)}_{\epsilon}}{\to} \psi$, whereas
               $\psi  \stackrel{\delta^{(1)}_{\epsilon}}{\to} \psi
               \stackrel{P_{i}}{\to} \psi$.}
           \end{itemize}
           These indicate that the commutator $[P_{i},\delta^{(1)}]$
               vanishes both for the bosons and the fermions.}
         \item{The second commutation relation $[P_{i},
           \delta^{(2)}] = 0$ is also verified in the same fashion. 
           \begin{itemize}
            \item{$A_{i} \stackrel{P_{i}}{\to} A_{i} + a_{i}
                \stackrel{\delta^{(2)}_{\epsilon}}{\to} A_{i} + a_{i}$,
                whereas $A_{i} \stackrel{\delta^{(2)}_{\epsilon}}{\to} A_{i}
                + a_{i}  \stackrel{P_{i}}{\to} A_{i} + a_{i}$.}
            \item{$\psi \stackrel{P_{i}}{\to} \psi
                \stackrel{\delta^{(2)}_{\epsilon}}{\to} \psi + \frac{1}{2}
                [A_{i}, A_{j}] \Gamma^{ij}$, whereas $\psi
                \stackrel{\delta^{(2)}_{\epsilon}}{\to} \psi + \frac{1}{2} 
                [A_{i}, A_{j}] \Gamma^{ij} \stackrel{P_{i}}{\to} \psi
                + \frac{1}{2} [A_{i}, A_{j}] \Gamma^{ij}$.}
        \end{itemize}}
       \end{itemize}
      This, although trivial,  completes the proof of the statement
      $[P_{i}, Q] =0$. 

  The important point is that the SUSY transformation
  (\ref{AZM31SUSY5com}) {\it triggers the interpretation of the
  spacetime as being a part of the degrees of freedom of the large $N$ 
  matrices}. IKKT model is a 0-dimensional quantum field theory, but
  the 10-dimensional spacetime emerges from the eigenvalues  of the
  bosonic matrices $A_{\mu}$. In other words, suppose that 
   \begin{eqnarray}
   (x^{(1)}_{\mu}, x^{(2)}_{\mu}, \cdots, x^{(N)}_{\mu})
  \end{eqnarray}
 are the eigenvalues of $A_{\mu}$. Then, the distribution of the points 
 $x^{(n)} = (x^{(n)}_{0}, x^{(n)}_{1}, \cdots, x^{(n)}_{9})$
 constitutes the 10-dimensional spacetime. Under this interpretation,
 the transformation (\ref{AZM31SUSY5com}) {\it generates the shift of
 this new spacetime} by $-2i {\bar \epsilon} \Gamma_{i} \xi$ in the
 $i$-th direction.
   \begin{figure}[htbp]
   \begin{center}
    \scalebox{.7}{\includegraphics{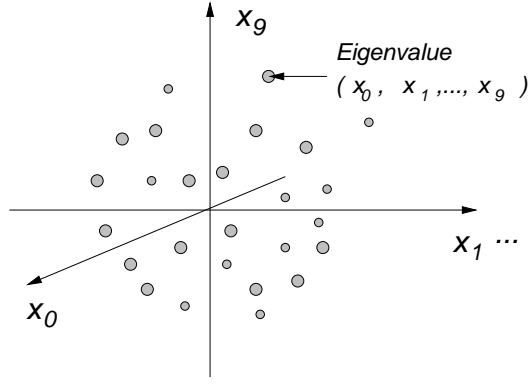} }
   \end{center}
   \caption{In IKKT model, the spacetimes are interpreted as the
     distribution of the eigenvalue of the bosonic matrices.}
  \label{spacetime}
  \end{figure}

  We have argued that IKKT model shares ${\cal  N} = 2$ SUSY with type 
  IIB superstring. This
  is a crucial property for the theory to contain gravity. If this
  theory includes massless spectrum, this theory must contain {\it
  spin 2 particles} - graviton. And this is the very reason why
  we had to reduce the {\it 10-dimensional, not the 4 or
  6-dimensional,} ${\cal N}=1$ SYM theory. Otherwise, this theory would 
  not have the maximal 32 SUSY\footnote{As we have explained in
  detail, the reduced model incorporates the {\it inhomogeneous SUSY}
  as well as the homogeneous SUSY which is the dimensional reduction
  of the original SYM theory, and
  thus the SUSY parameter doubles that of the original SYM
  theory.} and thus could not be a theory of gravity. 

  It is known that IKKT model induces
  IIB supergravity by one-loop effects. Computing the effective
  Lagrangian around the 10 dimensional background, we see the effects
  of graviton exchange, however we omit the computation in this
  review.

  \subsection{Description of Many-Body system}
  We now have a look at the aspects of IKKT model as a many-body
  system. We have argued that the matters are described by the bosonic 
  $N \times N$ matrices $A_{i}$. The amazing fact is that these large
  $N$ matrices can describe not only one-string effect but also
  multi-string effect. We concentrate on the simplest case -
  classical static D1-branes. We consider the classical equation of
  motion(EOM) of IKKT model, and hence we set the fermionic fields to be
  $\psi = 0$. Since the action (\ref{AZM31IKKT}) does not contain a
  kinetic term for $A_{i}$, the EOM is 
   \begin{eqnarray}
    [A_{i}, [A^{i}, A^{j}] ] = 0. \label{AZM32EOM}
   \end{eqnarray}
  Likewise, the EOM of Schild action of type IIB superstring is 
  $\{ X_{i}, \{ X^{i}, X^{j} \} \} = 0$. In terms of type IIB
  superstring, the solution of this EOM representing one D1-brane is
   \begin{eqnarray}
    X^{0} = T \tau ,\hspace{2mm} X^{1} = \frac{L}{2 \pi} \sigma,
    \hspace{2mm} X^{2} = \cdots X^{9} = 0, \label{AZM32D1brane1}
   \end{eqnarray}
  where $T$ and $L$ are the compactification radii of $X^{0}$ and
  $X^{1}$ directions respectively. The parameters $\tau$ and $\sigma$
  take values $0 \leq \tau \leq 1$ and $0 \leq \sigma \leq 2 \pi$.
   Therefore, the Poisson bracket is computed to be
   \begin{eqnarray}
    \{ X^{0}, X^{1} \} = \epsilon^{01} \partial_{0} X^{0} \partial_{1} 
    X^{1} = \epsilon^{01} \frac{TL}{2 \pi} = - \frac{TL}{2 \pi}.
   \end{eqnarray}
  Translating this relation into the language of large $N$ matrices,
  we want matrices which satisfy 
   \begin{eqnarray}
    -i [A_{0}, A_{1}] = \frac{TL}{2 \pi N} \Leftrightarrow \{ X_{0},
    X_{1} \} = (-1)  \{ X^{0}, X^{1} \} = \frac{TL}{2 \pi }.
   \end{eqnarray}
  Such a commutation relation is impossible if the size of the
  matrix $N$ is finite(this can be immediately seen by noting the
  cyclic property of trace $Tr$) , however taking the large $N$ limit
  we obtain a  following solution
   \begin{eqnarray}
    A_{0} = \frac{T}{\sqrt{2 \pi N}} q, \hspace{2mm} A_{1} =
    \frac{L}{\sqrt{2 \pi N}} p, \label{AZM31D1brane2}
   \end{eqnarray}
  where $q$ and $p$ are infinite size matrices satisfying the
  commutation relation $[q,p] = + i$. 

  Likewise, we express multi-string states by the matrix theory. we
  consider the following two cases.
   \begin{figure}[htbp]
   \begin{center}
    \scalebox{.5}{\includegraphics{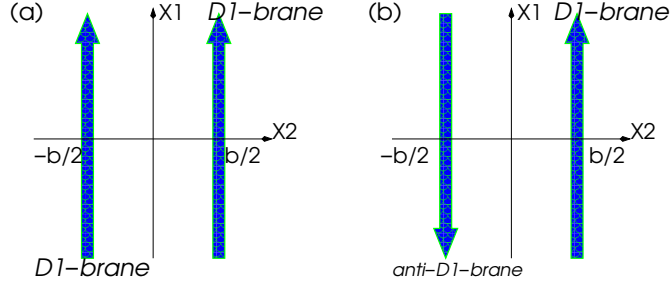} }
   \end{center}
   \caption{(a)$D1 - D1$ brane system. (b)$D1 - \overline{D1}$ brane system.}
  \label{D1brane}
  \end{figure}
  These systems are expressed in terms of the Schild type IIB
  superstring as
   \begin{eqnarray}
    (a) \left\{ \begin{array}{cc} 
     X^{(1)0} = T \tau^{(1)} & X^{(2)0} = T \tau^{(2)} \\
     X^{(1)1} = \frac{L}{2 \pi} \sigma^{(1)} & X^{(2)1} = \frac{L}{2
     \pi} \sigma^{(2)} \\
     X^{(1)2} = \frac{b}{2} & X^{(2)2} = - \frac{b}{2} \end{array}
     \right\}
     ,\hspace{5mm}
    (b)\left\{ \begin{array}{cc} 
     X^{(1)0} = T \tau^{(1)} & X^{(2)0} = T \tau^{(2)} \\
     X^{(1)1} = \frac{L}{2 \pi} \sigma^{(1)} & X^{(2)1} = - \frac{L}{2
     \pi} \sigma^{(2)} \\
     X^{(1)2} = \frac{b}{2} & X^{(2)2} = - \frac{b}{2} \end{array}
     \right\}.
   \end{eqnarray}
  We likewise translate these systems into the language of matrix
  theory by means of matrix regularization. Using two independent
  pairs of matrices satisfying canonical commutation relation
   \begin{eqnarray}
    [q,p] = +i, \hspace{2mm} [q',p'] = +i, \label{AZM32canonical2}
   \end{eqnarray}
  these two systems are described by block-diagonal matrices.
   \begin{eqnarray}
   (a) A_{0} = \left( \begin{array}{cc} \frac{T}{2 \pi n} q & 0 \\ 0 & 
   \frac{T}{2 \pi n} q' \end{array} \right) \equiv p_{0} ,\hspace{3mm} 
      A_{1} =  \left( \begin{array}{cc} \frac{T}{2 \pi n} p & 0 \\ 0 & 
   \frac{T}{2 \pi n} p' \end{array} \right) \equiv p_{1} ,\hspace{3mm}
      A_{2} =  \left( \begin{array}{cc} \frac{b}{2} & 0 \\ 0 & -
   \frac{b}{2} \end{array} \right) \equiv p_{2}, \label{AZM32D1D1} \\
   (b) A_{0} = \left( \begin{array}{cc} \frac{T}{2 \pi n} q & 0 \\ 0 & 
   \frac{T}{2 \pi n} q' \end{array} \right) \equiv p_{0} ,\hspace{3mm} 
      A_{1} =  \left( \begin{array}{cc} \frac{T}{2 \pi n} p & 0 \\ 0 & 
   - \frac{T}{2 \pi n} p' \end{array} \right) \equiv p_{1} ,\hspace{3mm}
      A_{2} =  \left( \begin{array}{cc} \frac{b}{2} & 0 \\ 0 & -
   \frac{b}{2} \end{array} \right) \equiv p_{2}, \label{AZM32D1antiD1}
  \end{eqnarray}
 where $n$ is a size of matrices $p,q,p',q'$ satisfying canonical
 commutation relation (\ref{AZM32canonical2}), hence $n$ should be 
 large enough. We have now scratched a beautiful aspect of IKKT model
 - the matrices $A_{i}$ in this matrix theory can describe many-body
 system by taking block-diagonal matrices like the above-mentioned
 example. Extending this idea, many-body systems can be embedded in one 
 large $N$ matrix. In this sense, IKKT model can be said to be a second
 quantization of superstring theory.

  We have now seen the description of many-body system, in which
  multi-string states can be described by a block-diagonal
  matrix. This system does not include any interaction of each object.
  Then, how can we consider the interaction in this matrix model? The
  answer to this question lies in the off-diagonal part of the
  matrices.
   \begin{figure}[htbp]
   \begin{center}
    \scalebox{.5}{\includegraphics{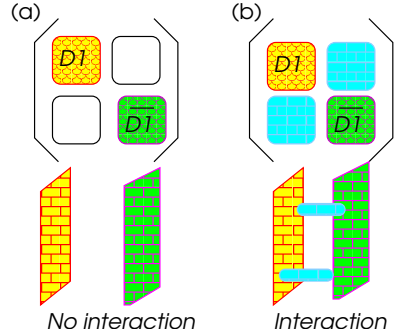} }
   \end{center}
   \caption{(a)The block diagonal matrices multi-string state
     including no interaction. (b)The off-diagonal part describes the
     interactions of multi-string states.} 
  \label{D1interaction}
  \end{figure}
 Here, we just sketch the idea of computing the interactions, and we
 leave the thorough explanation to \cite{9612115} or \cite{shino}.
 As depicted by Fig. \ref{D1interaction}, the off-diagonal
 part plays an essential  role in the interaction. We give an example 
 only for the simplest case - $D1-D1$ brane and $D1 - \overline{D1}$
 brane system. We separate the matrix $A_{i}$ and $\psi$ into the
 background and the quantum fluctuation
  \begin{eqnarray}
   A_{i} = p_{i} + a_{i}, \hspace{2mm} \psi = \chi + \phi,
  \end{eqnarray}
 $p_{i}$ and $\chi$ being the background, and $a_{i}$ and $\phi$ being 
 the quantum fluctuations. The action (\ref{AZM31IKKT}) is expanded up 
 to the second order of the quantum fluctuations,
  \begin{eqnarray}
   S_{2} &=& - Tr( \frac{1}{4}[p_{i}, p_{j}][p^{i}, p^{j}] + \frac{1}{2} 
   {\bar \chi} [p_{i}, \chi] - a_{j}([p_{i}, [p^{i}, p^{j}]] +
   \frac{1}{2} [ {\bar \chi} \Gamma^{i}, \chi]) + {\bar \psi}
   \Gamma^{i} [p_{i}, \phi] \nonumber \\
   &+& \frac{1}{2} [p_{i}, a_{j}][p^{i}, a^{j}] - \frac{1}{2} [p_{i},
   a^{i}][p_{j}, a^{j}] + [p_{i}, p_{j}[a^{i}, a^{j}] + \frac{1}{2}
   {\bar \phi} \Gamma^{i} [p_{i}, \phi] + {\bar \chi} \Gamma^{i}
   [a_{i}, \phi].
  \end{eqnarray}
 To fix the gauge invariance (\ref{AZM31gaugetr}), we must add a gauge 
 fixing term 
  \begin{eqnarray}
   S_{g.f.} = - Tr( \frac{1}{2}[p_{i}, a^{i}][p_{j}, a^{j}] + [p_{i},
   b][p^{i}, c] ). 
  \end{eqnarray}
  The one-loop effective potential is given by the following path
  integral
   \begin{eqnarray}
    W &=& - \log \int da d\phi db dc \exp( -(S_{2} + S_{g.f.})) \\
    \nonumber 
      &=& \frac{1}{2} Tr \log (P^{2}_{k} \delta_{ij} - 2i F_{ij}) -
    \frac{1}{4} Tr \log ((P^{2}_{k} + \frac{i}{2} F_{ij} \Gamma^{ij}
    )(\frac{1 + \Gamma^{\sharp}}{2}) - Tr \log(P^{2}_{k}) + i \theta,
   \end{eqnarray}
   where $P_{i} X = [ p_{i}, X]$, $F_{ij} X = [f_{ij}, X] = [ [p_{i},
   p_{j}], X]$ and $\theta$ is an anomaly term which vanishes in these
   simplest cases. We skip  the process to compute the interaction,
   and give only a result.
    \begin{itemize}
     \item{For $D1 - D1 $ brane system, this effective potential is $W 
         = 0$, id est, there is no interaction between these two $D1$
         branes. This is a natural result because the background $D1 - 
         D1$ system is supersymmetric, hence a stable system.}
     \item{For $D1 - \overline{D1}$ brane system, the leading term of
         the effective potential is $W \propto b^{-6}$. This result
         agrees with the interaction of $D1 - \overline{D1}$ brane in
         type IIB superstring theory.} 
    \end{itemize}
    This argument indicates that IKKT model succeeds in reproducing
    the interaction of $D1 - \overline{D1}$ brane system. This result
    solidifies the belief that IKKT model is a constructive definition 
    of type IIB superstring.

 \subsection{Noncommutative Yang-Mills(NCYM) in IKKT model}
  An important discovery in IKKT model is that this theory naturally
  induces noncommutative Yang-Mills theory by a simple mapping
  rule. The research of noncommutative Yang-Mills has recently become
  popular. We start with the idea of noncommutative
  space\cite{9908142}. A noncommutative space refers to a space
  defined by 
   \begin{eqnarray}
    [x^{i}, x^{j}] = i \theta^{ij}, \label{AZM33NCspace}
   \end{eqnarray}
  where $\theta^{ij}$ is an antisymmetric quantity called
  'noncommutative parameter'. What is the meaning of the relation
  $(\ref{AZM33NCspace})$ ? Note that the right-hand side is pure
  imaginary. This relation is reminiscent of the canonical
  commutation relation of the space coordinate and the
  momentum: $[q, p] = [q, -i \hbar \frac{\partial}{ \partial q} ] = +
  i \hbar$ (usually we adopt a God-given unit, and set $\hbar =
  1$). This canonical commutation relation causes the uncertainty
  relation between the space coordinate and its canonical
  momentum $(\Delta q) (\Delta p) \geq \frac{\hbar}{2}$ \footnote{This 
  relation is readily extended to any hermitian operators. We take 
  two hermitian operators ${\hat A}$ and ${\hat B}$ and a real number
  $\alpha$. Then we exploit the inequality $f(\alpha) = \int^{+
  \infty}_{-\infty} 
  d^{d} x | ({\hat A} - i \alpha {\hat B}) \psi |^{2}  = \langle ({\hat 
  A} + i \alpha {\hat B})( {\hat A} - i \alpha {\hat B}) \rangle =
  \alpha^{2} \langle {\hat B}^{2} \rangle - i \alpha \langle [{\hat
  A}, {\hat B}]  
  \rangle + \langle {\hat A}^{2} \rangle \geq 0$. This means that the
  equation of 2nd degree $f(\alpha) = 0$ never has two different real
  solutions. Taking the
  discriminant, we obtain $(\Delta {\hat A}) (\Delta {\hat B}) \geq
  \frac{1}{2} | \langle [{\hat A}, {\hat B}] \rangle | $. }
  The same idea is inherited in the noncommutative
  geometry. The commutation relation (\ref{AZM33NCspace})
  indicates the uncertainty relation {\it between the space
  coordinates}. This relation is a fundamental formula of '{\it the 
  quantization of a space}'. Utilizing the analogy of quantum
  mechanics, the physics of this space regards the coordinates as
  operators. 

  The noncommutativity of the space naturally emerges in superstring
  theory only by turning on B field \cite{9908142}. This discovery is
  attractive because the understanding of such a difficult spacetime
  can be achieved with ease through superstring theory. 
  According to the paper \cite{9908142}, for the string theory with NS 
  B field, the commutation relation (\ref{AZM33NCspace}) is satisfied
  with the noncommutative parameter 
   \begin{eqnarray}
    \theta^{ij} = 2 \pi \alpha' (\frac{1}{g + 2 \pi \alpha' B})_{A} =
    - (2 \pi \alpha ')^{2} (\frac{1}{g + 2 \pi \alpha' B}B \frac{1}{g
    - 2 \pi \alpha' B})^{ij},
   \end{eqnarray}
  which means that the coordinates possess the uncertainty 
    \begin{eqnarray}
     (\Delta x_{i}) (\Delta x_{j}) \geq \frac{1}{2} | \theta^{ij} |
     \sim {\cal O}(\alpha') = {\cal O}(\l_{s}^{2}).
    \end{eqnarray}
  The effect of the noncommutative space can be seen only if we
  consider the length scale as microscopic as the {\it string length}
  ${\cal O}(\l_{s})$ .
  This  noncommutative theory is described by replacing the naive
  product of  two operators with Moyal product 
   \begin{eqnarray}
    a(x) \star b(y) = \exp( \frac{i}{2} \theta^{ij}
    \frac{\partial^{2}}{\partial \xi^{i} \partial \eta^{\j}} ) a(x +
    \xi) b(y + \eta)|_{\xi = \eta = 0}. \label{AZM33Moyal}
   \end{eqnarray}
   This technique is attractive in that it enables us to deal with the 
   noncommutative field with the same ease as in the ordinary(commutative)
   fields. 

  Aoki, Ishibashi, Iso, Kawai, Kitazawa and Tada \cite{9908141}
  pointed out that IKKT model naturally includes the noncommutative
  Yang-Mills(NCYM) theory. This result is not surprising, because a
  matrix is by nature a noncommutative object. The classical EOM of
  IKKT model  (\ref{AZM31IKKT}) is expressed by (\ref{AZM32EOM}).  We
  pick up a set of classical solutions which satisfies  
   \begin{eqnarray}
    [{\hat p_{i}}, {\hat p_{j}}] = i B_{ij}, \label{AZM33classical1}
   \end{eqnarray}
  where $B_{ij}$ {\it are c-numbers} (therefore, it is trivial that
  this solution (\ref{AZM33classical1}) satisfies
  (\ref{AZM32EOM})). Of course this commutation relation cannot be
  satisfied for the finite size of the matrices $N$, however it is
  possible to approximate this solution by matrices whose size $N$ is
  large enough. Let ${\tilde d}$ be the rank of the matrices
  $B_{ij}$. And we separate the matrices $A_{i}$ between the classical 
  solution and its quantum fluctuation $A_{i} = {\hat p_{i}} + {\hat
  a_{i}}$. We perform Fourier transformation on the quantum
  fluctuation
   \begin{eqnarray}
    {\hat a}_{i} = \sum_{k} {\tilde a}_{i}(k) \exp (i C^{ij} k_{i} {\hat 
    p}_{j}) ,\hspace{2mm} {\hat \psi} = \sum_{k} {\tilde \psi}(k) \exp (i
    C^{ij} k_{i} {\hat p}_{j}), \label{AZM33fluc} 
   \end{eqnarray}
  where $C^{ij}$ is an inverse matrix of $B_{ij}$ (id est, $C^{ij}
  B_{jk} = \delta^{i}_{k}$). Because the matrices are hermitian, we
  require that ${\tilde a}^{\star}(k) = {\tilde a}(- k)$ and ${\tilde
  \psi}^{\star} (k) = {\tilde \psi} (- k)$. Note that there is no
  classical part of the fermion because $\psi =0$ in the classical
  solution. In order to gain insight into the correspondence between
  the matrix model and NCYM, we put together several properties of
  this Fourier transformation.

   Let us first consider the product of the Fourier transformation
  of two matrices ${\hat a}_{i} = \sum_{k} {\tilde a}_{i}(k) \exp (i
  C^{lm} k_{l} {\hat p}_{m})$ and ${\hat b}_{j} = \sum_{k} {\tilde
  b}_{j}(k) \exp (i C^{lm} k_{l} {\hat p}_{m})$. Utilizing
  Baker-Campbell-Hausdorff(BCH) formula $e^{A} e^{B} = \exp (A + B +
  \frac{1}{2} [A,B] \cdots )$, we obtain
   \begin{eqnarray}
   {\hat a_{i}} {\hat b_{j}} &=& \sum_{k,l} {\tilde a}_{i}(k) {\tilde
   b}_{j}(l) \exp (i C^{rs} k_{r} {\hat p}_{s}) \exp (i C^{tu} l_{t}
   {\hat p}_{u})
  = \sum_{k,l} {\tilde a}_{i}(k) {\tilde
   b}_{j}(l) \exp (i C^{rs} (k_{r}+ l_{r}) {\hat p}_{s} +
   \frac{i^{2}}{2} C^{rs} C^{tu} k_{r} l_{t} i B_{su}) 
   \nonumber \\
  &=&   \sum_{k,l} {\tilde a}_{i}(k) {\tilde
   b}_{j}(l) \exp (i C^{rs} (k_{r} + l_{r}) {\hat p}_{s} +
   \frac{-1}{2} C^{rs} C^{tu} k_{r} l_{t} i (-B_{us}) ) \nonumber \\  
  &=&  \sum_{k,l} {\tilde a}_{i}(k) {\tilde
   b}_{j}(l) \exp (i C^{rs} (k_{r} + l_{r}) {\hat p}_{s} + \frac{i}{2}
   C^{rs} k_{r} l_{s}). \label{AZM33product}
  \end{eqnarray}

 Next we consider the trace of this matrix theory in the simplest
 case : 2 dimensional system. In this system, the noncommutative
 parameter is $ B_{ij} \left( \begin{array}{cc} 0 & -B \\ B & 
   0  \end{array} \right) $ (hence $C^{ij} = \left( \begin{array}{cc}
 0 & B^{-1} \\ -B^{-1} &  0  \end{array} \right)$). And let the
 classical solution be the canonical pairs ${\hat
 p_{0}} = {\hat q}$ and ${\hat p_{1}} = {\hat p}$ such that $[{\hat
 q}, {\hat p}] = +i$. Then we obtain, utilizing BCH formula,   
  \begin{eqnarray}
   Tr \exp (i C^{ij} k_{i} {\hat p}_{j}) &=& Tr \exp( i B^{-1} (k_{0}
   {\hat p} - k_{1} {\hat q} ))
  = \int dq \langle q | \exp (i k_{0} {\hat p} B^{-1}) \exp ( -i
   k_{1} {\hat q} B^{-1}) \exp(ik_{0} k_{1} B^{-1}) |q \rangle
   \nonumber \\
   &=& 2 \pi   B \delta(k_{0}) \delta(k_{1}) = (2 \pi)^{2} \frac{1}{2
   \pi} \sqrt{ \det B} \delta(k_{0}) \delta(k_{1}). \label{AZM33trace}
  \end{eqnarray}
   
 Lastly let us see the effect of the adjoint operator $P_{i} {\hat o}
 \stackrel{def}{=} [ {\hat p_{i}} , {\hat o} ]$. This acts on ${\hat
 a_{i}}$ as  
  \begin{eqnarray}
   P_{i} {\hat a_{j}} = [{\hat p_{i}}, {\hat a_{j}}] 
  = \sum_{k} {\tilde a}_{j}(k) [{\hat p_{i}}, \exp( i C^{lm} k_{l}
  {\hat p}_{m} )] = \sum_{k} k_{i} {\tilde a}_{j} \exp(i C^{lm} k_{l}
  {\hat p}_{m} ).  \label{AZM33partial}
  \end{eqnarray}

 Having these results (\ref{AZM33product}), (\ref{AZM33trace}) and
 (\ref{AZM33partial}) in mind, we consider the mapping rule which
 transforms IKKT into NCYM. The key to this crucial relationship is
 very simple:
  \begin{eqnarray}
    {\hat a}_{i} \textrm{(IKKT)} \Rightarrow a_{i}(x) = \sum_{k}
    {\tilde a}_{i}(k) \exp( i k_{j} x^{j} ) \textrm{(NCYM)}.
    \label{AZM33map} 
 \end{eqnarray}
 \begin{itemize}
  \item{ This is a mapping from a $N \times N$ matrix to {\it a
        c-number function}. As we shall see later, this is a mapping
        into $U(1)$ NCYM theory.}
  \item{ The relationship (\ref{AZM33product}) indicates that the
      product of matrices is mapped into Moyal product in the language 
      of NCYM,
      \begin{eqnarray}
       {\hat a_{i}} {\hat b_{j}} \Rightarrow a_{i}(x) \star b_{j} (x)
       \equiv \exp (\frac{i C^{lm}}{2} \frac{\partial^{2}}{\partial
       \xi^{l} \partial \eta^{m} } ) a_{i} (x + 
       \xi) b_{j}(x + \eta)|_{\xi = \eta = 0}.
      \end{eqnarray}
      This relationship has a profound significance, in that the
  residual phase factor in BCH formula induces the noncommutativity
  of the space in the mapped world. This is the very reason why we
  regard the mapped world as the noncommutative spacetime.}
  \item{ The relationship (\ref{AZM33trace}) indicates that the trace
     of $N \times N$ matrices is translated into the integration over
     the spacetime,
     \begin{eqnarray}
     Tr ({\hat a_{i}}) \Rightarrow \sqrt{ \det B} (\frac{1}{2
     \pi})^{\frac{{\tilde d}}{2} } \int d^{\tilde d} x a(x),
     \end{eqnarray}
     where ${\tilde d}$ is {\it the rank of the matrix} $B_{ij}$,
     which equals to {\it the spacetime dimensions of the mapped
     world}. }
  \item{ The relationship (\ref{AZM33partial}) indicates that the
      adjoint operator $P_{i}$ is interpreted as {\it a differential
      operator} in the world of ${\tilde d}$ dimensional NCYM theory: 
       \begin{eqnarray}
        P_{i} {\hat a}_{j} (= [p_{i}, {\hat a_{j}}] ) \Rightarrow -i
        \partial_{i}.
       \end{eqnarray}
     Therefore, the commutator with the matrices of IKKT model $A_{i}$ 
     is translated into the {\it covariant derivative}:
       \begin{eqnarray}
         [A_{i}, {\hat o}] = [{\hat p}_{i} + {\hat a}_{i}, {\hat o}]
         \Rightarrow D_{i} o(x) \stackrel{def}{=} -i \partial_{i} o(x)
         + a_{i}(x) \star  o(x) - o(x) {\star} a_{i}(x). 
       \end{eqnarray}
     Especially, the commutator with two covariant derivative is a
     field strength
       \begin{eqnarray}
         [A_{i}, A_{j}] \Rightarrow -i (\partial_{i} a_{j}(x) -
         \partial_{j} a_{i}(x)) + [a_{i}(x), a_{j}(x)]_{\star} \equiv
         F_{ij}. \label{AZM33fieldstrength}
       \end{eqnarray} }
   \item{ This mapping rule induces the coordinate from the momentum
       in IKKT model. Having a careful look at the mapping rule
       (\ref{AZM33map}), the coordinate in NCYM is produced from the
       IKKT model:
        \begin{eqnarray}
          C^{ij} {\hat p}_{j} \Rightarrow x^{i}. \label{AZM33map2}
        \end{eqnarray}
       This correspondence possesses a profound meaning. We have
       before argued that the noncommutativity of the space 
       (\ref{AZM33NCspace}) is a quantization of the space, just as
       the commutator of the coordinate and the momentum is nonzero in 
       ordinary quantum mechanics. {\it That the coordinate is
       naturally induced from the momentum} strongly solidifies the
       correspondence between the ordinary quantum mechanics and the
       noncommutative geometry as the quantization of spacetime.}
 \end{itemize}

  These are the profound  features of the simple mapping rule
  (\ref{AZM33map}). Note that the long wavelength\footnote{hence 
  low energy,  noting the relationship $E \sim \frac{hc}{\textrm{wave
  length}}$} excitations  $|k|
  \ll \lambda$ ($\lambda$ refers to the spacing of the
  quanta) are commutative, again utilizing BCH formula, 
  \begin{eqnarray}
   [{\hat a}_{i}, {\hat a}_{j}] &=& \sum_{k,l} {\tilde a}_{i}(k) {\tilde
   b}_{j}(l) [ \exp( i C^{rs} k_{r}, {\hat p}_{s}), \exp(i C^{tu}
   l_{t} {\hat p}_{u} ) ]  \nonumber \\
  &=& \sum_{k,l} {\tilde a}_{i}(k)
   {\tilde a}_{j}(l) \exp(i C^{rs} (k_{r} 
   + l_{r} ) {\hat p}_{s}) (\exp( \frac{i^{2}}{2} C^{rs} C^{tu} k_{r} l_{s}
   i B_{su} ) - \exp(\frac{i^{2}}{2} C^{rs} C^{tu} k_{r} l_{s}
   i B_{us} ) ) \nonumber \\
  &\stackrel{B_{su} = - B_{us}}{=}& 2i \sum_{k,l}  {\tilde a}_{i}(k)
   {\tilde a}_{j}(l) \exp(i C^{rs} (k_{r} 
   + l_{r} ) {\hat p}_{s}) \sin (\frac{1}{2} C^{rs} k_{r} l_{s}).
   \label{AZM33lowenergy}
  \end{eqnarray}
  The low energy limit  $|k| \ll \lambda$ is regarded as the
  semiclassical limit of the space $x^{i} = C^{ij} {\hat
  p}_{j}$. Then, IKKT model is (when the rank of $B_{ij}$ is ${\tilde
  d}$), mapped into ${\tilde d}$ {\it dimensional NCYM theory} 
  \begin{eqnarray}
   & & - \frac{1}{g^{2}} Tr (\frac{1}{4} [A_{i}, A_{j}][A^{i}, A^{j}] +
   \frac{1}{2} {\bar \psi} \Gamma^{i} [A_{i}, \psi] ) \nonumber \\
  &\Rightarrow& \frac{{\tilde d} n B^{2}}{4 g^{2}} - \sqrt{\det B}
   (\frac{1}{2 \pi})^{\frac{\tilde d}{2} } \int d^{\tilde d} x
   \frac{1}{g^{2}} (\frac{1}{4} F_{ab} F^{ab} + \frac{1}{2} [D_{a},
   \alpha_{\rho} ][ D^{a}, \alpha^{\rho}] + \frac{1}{4}
   [\alpha_{\rho}, \alpha_{\chi}][\alpha^{\rho}, \alpha^{\chi}] 
   \nonumber \\ 
  &+&  \frac{1}{2} ( {\bar \psi}  \Gamma^{a} [D_{a}, \psi] +  {\bar
   \psi} \Gamma^{\rho} [\alpha_{\rho}, \psi] ) )_{\star}.
   \label{AZM33NCYM} 
  \end{eqnarray}
  \begin{itemize}
   \item{ The resulting NCYM theory possesses a gauge group $U(1)$
       because the matrix in IKKT model is mapped into functions. The
       Yang-Mills coupling is now $g_{YM} = g^{2} (\frac{2
       \pi}{B})^{\frac{\tilde d}{2}}$. }
   \item{ The indices $a,b,\cdots $ run over ${\tilde d}$ dimensional
       spacetime in the mapped NCYM theory. As we have before
       remarked, the commutator of the covariant derivative is a field 
       strength $F_{ab} \stackrel{def}{=} [D_{a}, D_{b}]_{\star} $.}
   \item{ The indices $\rho, \chi,\cdots$ here runs the transverse
       dimensions. In this residual dimension, there is no
       differentiation with respect to the field, and hence $D_{\rho}
       = \alpha_{\rho}$, where we have replaced $a_{\rho} \rightarrow
       \alpha_{\rho}$. }
  \end{itemize}
 
  In order to extend this result to NCYM theory with gauge group
  $U(m)$, we have only to map the matrices in IKKT model into $m
  \times m$ matrices. The argument is totally parallel to the $U(1)$
  case, and we replace element of ${\hat p}_{i}$ with 
    \begin{eqnarray}
      {\hat p}_{i} \rightarrow {\hat p}_{i} \otimes {\bf 1}_{m}. 
    \end{eqnarray} 
 The Fourier decomposition is similar to the $U(1)$ case:
    \begin{eqnarray}
    {\hat a}_{i} = \sum_{k} {{\tilde a}_{i}(k)}_{m \times m} \exp (i
    C^{ij} k_{i} {\hat  
    p}_{j}) ,\hspace{2mm} {\hat \psi} = \sum_{k} {{\tilde \psi}(k)}_{m 
    \times m} \exp (i
    C^{ij} k_{i} {\hat p}_{j}). \label{AZM33flucum} 
   \end{eqnarray}
 The difference is that ${\tilde a}_{i}(k)$ and ${\tilde \psi}(k)$ are 
 now $m \times m$ matrices (from now on we omit $_{m \times
 m}$). Therefore, the mapping rule is
   \begin{eqnarray}
       {\hat a}_{i} (\in M_{N \times N})  \Rightarrow a_{i}(x) = \sum_{k}
    {\tilde a}_{i}(k) \exp( i k_{j} x^{j} ) (\in M_{m \times m}),
   \end{eqnarray}
 where $M_{N \times N}$ is a set of $N \times N$ matrices. The
 resulting NCYM theory is similar to (\ref{AZM33NCYM}), except that
 the mapped theory  is with respect to $U(m)$ matrices, and hence the
 theory is non-abelian. 

%

\subsection{Summary}
  We have seen many beautiful properties of IKKT model.
  \begin{itemize}
  \item{IKKT model is defined as the 0 dimensional reduction of ${\cal 
        N}=1$ 10 dimensional SYM theory. And this theory is the same
        as the matrix regularization of the Schild form of type IIB
        superstring theory.}
  \item{IKKT model possesses no free parameter. The coupling constant
      can be absorbed into the field by the rescaling $A_{i}
      \rightarrow g^{\frac{1}{2}} A_{i}$ and $\psi \rightarrow
      g^{\frac{3}{4}} \psi$.}
  \item{ IKKT model possesses ${\cal N}=2$ supersymmetry, which is one 
      of the essential properties of type IIB superstring. This
      indicates that the theory should include spin 2 gravitons if
      this theory has massless particles.}
  \item{ IKKT model has an ability to describe many-body system only
      by one set of the matrices $A_{i}$. We have scratched the simplest
      case - how to describe $D1$-brane or anti-$D1$-brane. The
      computation of the interaction of $D1 - \overline{D1}$ brane
      based on this matrix theory beautifully reproduces the result of 
      type IIB superstring theory.}
  \item{ IKKT model naturally induces NCYM theory by a simple mapping
      from a momentum in IKKT to a coordinate NCYM, $x^{i} = C^{ij}
      {\hat p}_{j}$. This strongly serves to solidify the
      correspondence between 
      quantum mechanics and the noncommutative geometry as the
      quantization of spacetime.}  
  \end{itemize}
  There are other exciting properties of IKKT model. We list up some of the
  properties, but we omit the explanation.
   \begin{itemize}
    \item{Schwinger-Dyson equation of the action of IKKT model induces
        the light-cone string field  theory of type IIB superstring
        theory. }
    \item{ Utilizing the analogy of branched polymer, it is possible to 
        gain insight into how to induce our 4 dimensional world. We
        consider the branched polymer as the simplified model of the
        effective action of the spacetime points. The Hausdorff
        dimension of the branched polymer point is known to be 4.}
   \end{itemize}

  IKKT is a successful proposal for the constructive definition of
  superstring, and possesses many exciting properties, which
  solidifies the confidence that it is a constructive definition.

\section{$osp(1|32,R)$ Cubic Matrix Model}
  We have reviewed the successful aspects of the attempt to describe
  superstring theory in terms of matrix theory, and IKKT model
  indicates many promising aspects to be regarded as 'Theory of
  Everything'. Yet it is worth while to pursue a model exceeding IKKT
  model. L. Smolin presented a new approach to describe M-theory based 
  on a simple matrix model \cite{0002009}. This theory describes a
  dynamics of a  matrix which is built from the super Lie algebra
  $osp(1|32,R)$. The action of this model, suggested by L. Smolin is
  described by a simple cubic action, as we discuss later in detail.
  There  are several reasons we regard this model as attractive.
 
 The action suggested by L. Smolin is extremely simple.
           The dream of elementary particle physics is to pursue a
           'Theory of Everything' from which all the phenomena of the
           whole universe are derived. Once Einstein found that the
           mechanics including the effect of gravity is described by
           general relativity, and the Lagrangian of this theory $I =
           \int d^{d}x \sqrt{g} R$ is extremely simple. We have a
           belief that the 'mother of the whole physical theory' should 
           be described by a simple action. The proposal for the
           constructive definition of superstring theory was suggested 
           by Ishibashi, Kawai, Kitazawa and Tsuchiya. This is a
           dimensional reduction of ${\cal N} =1$ SYM theory, and this 
           proposal is described by a simple and beautiful action,
           even though this proposal was once criticized as {\it not
           as beautiful as general relativity.} That the
           theory exceeding IKKT model may be described by a
           simple action is an attractive proposal, worth pursuing its 
           validity and structure.

 $osp(1|32,R)$ has been known as the unique maximal simple
        super Lie algebra with 32 fermionic
        generators \cite{0003261}. This theory indicates a possibility
        that this may naturally include the existing matrix models, IKKT model
        or BFSS model. $osp(1|32,R)$ super Lie algebra, expressed in
        terms of 10 dimensional representation, includes two chiral
        spinors of both opposite chirality(IIA) and the same
        chirality(IIB). In this sense, we find $osp(1|32,R)$ super Lie
        algebra a natural framework for describing 'Theory of
        Everything', and we are inclined to speculate that
        L. Smolin's proposal is the clue to the ultimate
        theory.

 That the theory is expressed by a cubic action possesses a
        profound significance in two senses.
        One aspect is that the fundamental interaction of
        superstring theory is a three-point interaction, because four
        (or more) point interaction is identified with three-point
        interaction by conformal invariance of the Feynman graphs. Thus it
        is a quite natural idea that the 'Theory of Everything' which
        describes superstring theory comprehensively is a cubic matrix 
        theory.
      \begin{figure}[htbp]
        \begin{center}
        \scalebox{.5}{\includegraphics{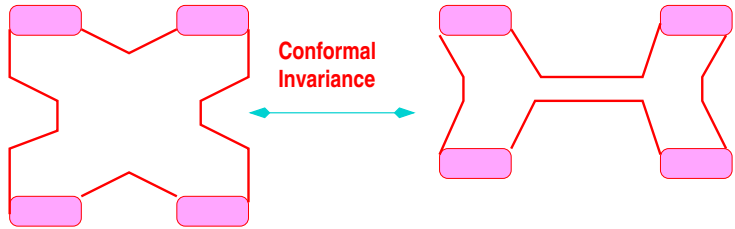} }
        \end{center}
        \caption{ The fundamental interaction of string theory is a
          cubic one.}
      \end{figure}
      The other respect is that  cubic action of string theory is
          identified with a 
          Chern-Simons theory if a due compactification is
          performed. Chern-Simons theory is known to be exactly
          solvable by means of Jones Polynomial in knot theory
          \cite{Jones}. If we find a correspondence between the cubic
          matrix model and Witten's technique of solving Chern-Simons
          theory, we may be able to solve exactly the behavior of
          superstring in nonperturbative region, just as Brezin and
          Kazakov succeeded in solving exactly the bosonic string in 0 
          spacetime dimension via orthogonal polynomial method.
       
 This theory deals with 11 dimensional spacetime, in terms
        of the 11 dimensional representation of $osp(1|32,R)$
        super Lie algebra. And this theory may be able to describe
        superstring theory in  such
        curved 10 dimensional spacetimes  as $S_{1} \times R_{9}$ or
        $AdS_{5} \times S^{5}$, as well as a flat 10 dimensional
        spacetime.

  This new proposal for describing superstring theory in terms of a
  supermatrix theory includes many interesting possibility, and the
  investigation of this cubic supermatrix theory is a fascinating
  issue. 

 \subsection{Definition of $osp(1|32,R)$ super Lie Algebra}
  Before entering the investigation of the superstring action, we
  settle the definition of a super Lie algebra. A super Lie algebra
  is an algebra of supermatrix in which both bosonic matrices and the
  fermionic matrices are embedded in one matrix.
  Supermatrices  possess many properties different from (ordinary)
  matrices, and these properties and the notations are summarized in
  detail in Appendix. \ref{AZMA091441}.  
  
 Let us start with the definition of $osp(1|32,R)$ super Lie
 algebra.
 \begin{eqnarray} 
  &\clubsuit& \textrm{ If } M \in osp(1|32,R),  \textrm{ then } {^{T}M }
  G + GM = 0 
  \textrm{ for }  G =  \left( \begin{array}{cc}  \Gamma^{0}  & 0  \\  0
  & i   \end{array} \right). \nonumber \\
  &\clubsuit& M \textrm{ is traceless with respect to 33 } \times
  \textrm{ 33 supermatrix.} \nonumber \\
  &\clubsuit& M \textrm{ is a real supermatrix in that } M^{\ast} =
  M. \nonumber 
 \end{eqnarray} 
   We confirm that, for the first condition,  $osp(1|32,R)$ forms a
  closed super Lie algebra. Suppose matrices $M_{1}$  
  and $M_{2}$ satisfy the condition 
  \begin{eqnarray} {^{T} M_{1}} G + G M_{1} = 0, \hspace{4mm}
    {^{T} M_{2}} G + G M_{2} = 0. \label{AZMspade1}\end{eqnarray}
  If $osp(1|32,R)$ is to be a closed super Lie algebra, we call for a
  following  condition
   \begin{eqnarray} {^{T}([M_{1} , M_{2}])} G + G [M_{1} , M_{2}]
  = 0. \label{AZMspade2} \end{eqnarray}
  Multiplying $G^{-1}$ on both (\ref{AZMspade1}) and (\ref{AZMspade2}) from 
  the left, they are respectively rewritten as 
   \begin{eqnarray} G^{-1} {^{T} M_{k}} G + M_{k} = 0,
  \label{AZMspade1+} \end{eqnarray}
  \begin{eqnarray} G^{-1} {^{T} ([M_{1} , M_{2}])} G +  [M_{1} , M_{2}] = 0,
  \label{AZMspade2+} \end{eqnarray}
 where $k=1,2$. The proof that $osp(1|32,R)$ is a closed super Lie
 algebra is equivalent to deriving (\ref{AZMspade2+}) utilizing
 (\ref{AZMspade1+}):
 \begin{eqnarray} 
  \textrm{(\ref{AZMspade2+})} = [ G^{-1} {^{T}M_{2}} G , G^{-1}
 {^{T}M_{1}} G] + [ M_{1} , M_{2}] \stackrel{(\ref{AZMspade1+})}{=} 
 [-M_{2} , -M_{1}] + [M_{1} , M_{2}] = 0.
   \end{eqnarray} 
  This statement, per se, can be satisfied whatever the matrix $G$ may
  be so long as $G$ has an inverse matrix. 

  Here comes one question:
 \begin{center}
 {\it Why do we define a metric $G$ as } $G
  = \left( \begin{array}{cc} \Gamma^{0} & 0 \\ 0 & i \end{array}
  \right)$ ?
 \end{center}
  This stems from the requirement that $M$ is a real
  matrix in that $M^{\ast} = ({^{T}M})^{\dagger} = M$. Let us
  consider the consistency between this reality condition and the
  very definition of $osp(1|32,R)$ super Lie algebra. Take a hermitian
  conjugate of the definition ${^{T}M} G + GM=0$. This gives 
   \begin{eqnarray}
    G^{\dagger} ({^{T}M})^{\dagger} + M^{\dagger} G^{\dagger} = 0.
   \end{eqnarray}  
  Utilizing the properties introduced in the
  Appendix. \ref{AZCaho}, and the reality condition, this is
  rewritten as 
   \begin{eqnarray}
   0 =  G^{\dagger} ({^{T}M})^{\dagger} + M^{\dagger} G^{\dagger}
   \stackrel{M^{\ast} = ({^{T}M})^{\dagger}}{=} G^{\dagger} M^{\ast} +
   M^{\dagger} G^{\dagger} \stackrel{M^{\dagger} = {^{T}(M^{\ast})}
   }{=} G^{\dagger} M^{\ast} + {^{T} (M^{\ast})} G^{\dagger}
   \stackrel{M=M^{\ast}}{=} G^{\dagger} M + {^{T}M} G^{\dagger}.
   \label{AZM3ccdef} \end{eqnarray}
  In order for the relationship (\ref{AZM3ccdef}) to be consistent
  with the very definition of $osp(1|32,R)$, we must call for a
  condition 
    \begin{eqnarray}
      G^{\dagger} = G.
    \end{eqnarray}
 And the starting point of $osp(1|32,R)$ is well-defined if we take a
 metric as $G = \left( \begin{array}{cc} \Gamma^{0} & 0 \\ 0 & i
 \end{array} \right)$.\\
  
 The next issue is to investigate the explicit form of this
 super Lie algebra. This is expressed by 
  \begin{eqnarray}
   \textrm{ If } M \in osp(1|32), \textrm{ then } M = \left
   ( \begin{array}{cc} m  & \psi \\   i {\bar \psi} & 0 \end{array}
   \right). \label{AZMsu1-16-16}
  \end{eqnarray}
 \begin{itemize}
  \item{ $m$ contains only the terms of rank 1,2,5. In other
      words,  $m$ is expressed as 
     \begin{eqnarray}
     m = u_{\mu_{1}} \Gamma^{\mu_{1}} + \frac{1}{2!} u_{\mu_{1} \mu_{2}}
      \Gamma^{\mu_{1} \mu_{2}} + \frac{1}{5!} u_{\mu_{1} \cdots
      \mu_{5} } \Gamma^{\mu_{1} \cdots \mu_{5} }.
     \end{eqnarray}
   } 
  \item{Here ${\bar \psi}$ denotes $\psi^{\dagger} 
      \Gamma^{0}$. However, this is equivalent to ${^{T} \psi}
      \Gamma^{0}$, because we are now considering a real super Lie algebra.}
  \end{itemize}
 {\sf (Proof)
 Let $M$ be the 
 element of $osp(1|32,R)$ and $M =  \left( \begin{array}{cc} m  & \psi
 \\ i {\bar \phi} & v  \end{array} \right)$ ,
 \begin{itemize}
  \item{ ${\bar \phi}$ is defined as ${\bar \phi} = \psi^{\dagger}
      \Gamma^{0} = {^{T} \phi} \Gamma^{0}$ .}
  \item{ Because $M$ is a real supermatrix, $m$, $v$ and $\psi$ are
      real, while $i {\bar \phi}$ is pure imaginary (and hence ${\bar
      \phi}$ is real).} 
 \end{itemize} 

 By substituting this formula into the very definition, we obtain 
  \begin{eqnarray}
   {^{T}M } G + GM =  \left( \begin{array}{cc} {^{T} m}  & - i {^{T}
 {\bar \phi}} \\ {^{T} \psi} & {^{T}v}  \end{array} \right) 
 \left( \begin{array}{cc}  \Gamma^{0}  & 0  \\  0  & i   \end{array}
 \right)  +  \left( \begin{array}{cc}  \Gamma^{0}  & 0  \\  0  & i
 \end{array}  \right) \left( \begin{array}{cc} m  & \psi
 \\ i {\bar \phi} & v  \end{array} \right) = \left( \begin{array}{cc}
 {^{T}m} \Gamma^{0} + \Gamma^{0} m  & {^{T}{\bar \phi}} + \Gamma^{0}
 \psi \\ {^{T} \psi} \Gamma^{0} - {\bar \phi} & 2iv  \end{array}
 \right) = 0. \nonumber 
  \end{eqnarray}
  We immediately obtain the relationship between two fermionic fields
  $\psi$ and $\phi$ from this definition: 
  \begin{eqnarray}
     {^{T}{\bar \phi}} + \Gamma^{0}\psi = {^{T} \Gamma^{0}} \phi +
     \Gamma^{0} \psi = - \Gamma^{0} \phi + \Gamma^{0} \psi = 0.
  \label{AZMdine}  \end{eqnarray}
 By multiplying $\Gamma^{0}$ on the both hand sides from the left , we 
 obtain the relationship between $\psi$ and $\phi$. We immediately note that
 ${^{T} \psi} \Gamma^{0} - {\bar \phi} = 0$ is equivalent to the
 equation (\ref{AZMdine}). 

 The constraint on $v$ is trivial,  and $v$ must vanish. On the other
 hand, the constraint on $m$ is worth a careful investigation.  $m$ is
 imposed on the constraint 
 \begin{eqnarray} 
 {^{T}m} \Gamma^{0} + \Gamma^{0} m = 0. \label{AZ41sp32}
 \end{eqnarray}
 This is the very definition of the  Lie algebra called
 $sp(32)$, and this statement indicates that the bosonic $32 \times
 32$ matrix of $osp(1|32,R)$ super Lie algebra must belong to $sp(32)$ Lie
 algebra. In analyzing
 this supermatrix theory, it is more convenient to decompose the
 bosonic part $m \in sp(32)$ in terms of the basis of arbitrary $32 \times 32$
 matrices \footnote{Actually this is a  $32 \times 32 = 1024$
 dimensional basis, because the dimension of this basis is
 $1 + {_{11}C_{1}} + \cdots {_{11}C_{5}} = 1 + 11 + 55 + 165 + 330 +
 462 = \frac{1}{2}(1+1)^{11} = 1024$. } ${\bf 1}_{32 \times 32}$,
 $\Gamma^{\mu}$, $\Gamma^{\mu_{1} \mu_{2}}$, $\Gamma^{\mu_{1} \mu_{2}
 \mu_{3}}$, $\Gamma^{\mu_{1} \cdots \mu_{4}}$ and $\Gamma^{\mu_{1}
 \cdots \mu_{5}}$
 , rather than to obey the expression in the paper \cite{0002009}.
 Our notation of the gamma matrices is introduced in
 Appendix. \ref{AZCgm} in detail.   
 The relationship (\ref{AZ41sp32}) determines what rank of the 11
 dimensional gamma matrices survive. 
  Suppose $m \in sp(32)$ are expressed in terms of the gamma matrices: 
   \begin{eqnarray}
    m = u {\bf 1} + u_{\mu} \Gamma^{\mu} + \frac{1}{2!} u_{\mu_{1}
    \mu_{2}} \Gamma^{\mu_{1} \mu_{2}} + \frac{1}{3!} u_{\mu_{1}
    \mu_{2} \mu_{3}} \Gamma^{\mu_{1} \mu_{2} \mu_{3}} + \frac{1}{4!}
    u_{\mu_{1} \cdots \mu_{4}} \Gamma^{\mu_{1} \cdots \mu_{4}} +
    \frac{1}{5!} u_{\mu_{1} \cdots \mu_{5}} \Gamma^{\mu_{1} \cdots
    \mu_{5}}. \label{AZ41assumption}
   \end{eqnarray}
 The  condition (\ref{AZ41sp32}) is rewritten as 
  \begin{eqnarray}
 m = - (\Gamma^{0})^{-1} ({^{T}m}) \Gamma^{0} = \Gamma^{0} ({^{T}m})
 \Gamma^{0}. \label{AZ41constm2}
  \end{eqnarray}
 Then, performing the following computation for $k=0, 1, \cdots, 5$, 
  \begin{eqnarray}
    \Gamma^{0} ({^{T}\Gamma^{\mu_{1} \cdots \mu_{k}}}) \Gamma^{0} &=&
    (-1)^{k-1} (\Gamma^{0} ({^{T} \Gamma^{\mu_{k}}})  \Gamma^{0} )
    \cdots (\Gamma^{0} ({^{T} \Gamma^{\mu_{1}}})  \Gamma^{0} ) =
    (-1)^{k-1} \Gamma^{\mu_{k} \cdots \mu_{1}} \nonumber \\
    &=& (-1)^{k-1}
    (-1)^{\frac{k(k-1)}{2}} \Gamma^{\mu_{1} \cdots \mu_{k}} =
    (-1)^{\frac{(k+2)(k-1)}{2}} \Gamma^{\mu_{1} \cdots \mu_{k}}. 
    \label{AZ41kth} 
  \end{eqnarray}
 This relationship is rewritten as, separating into two cases,
  \begin{itemize}
   \item{ For $k=1,2,5$, there is no sign in (\ref{AZ41kth}), so that
       $  \Gamma^{0} ({^{T}\Gamma^{\mu_{1} \cdots \mu_{k}}})
       \Gamma^{0} = \Gamma^{\mu_{1} \cdots \mu_{k}} $.}
   \item{ For $k=0,3,4$, the sign changes in (\ref{AZ41kth}), so that 
     $ \Gamma^{0} ({^{T}\Gamma^{\mu_{1} \cdots \mu_{k}}})
       \Gamma^{0} = - \Gamma^{\mu_{1} \cdots \mu_{k}} $, }
  \end{itemize} 
 where $k$ is the rank of the gamma matrices.
 Combining this result with the constraint of $sp(32)$ Lie algebra
 (\ref{AZ41constm2}), we can discern that only the gamma matrices of
 rank 1, 2 and 5 survive.  We are thus finished with verifying the
 explicit form of $osp(1|32,R)$ super Lie algebra. (Q.E.D.)
}

 \subsection{Action of the Cubic Matrix Theory}
  In considering the action of a large $N$ reduced model, we promote the 
  component of the super Lie algebra to an $N \times N$ matrix. L. Smolin
  suggested a following action of this matrix theory:
  \begin{eqnarray}
   I &=& \frac{i}{g^{2}} Tr_{N \times N} \sum_{Q,R=1}^{33}
  (( \sum_{p=1}^{32} {M_{p}}^{Q} [ {M_{Q}}^{R},
  {M_{R}}^{p}] ) - {M_{33}}^{Q} [ {M_{Q}}^{R}, {M_{R}}^{33} ] )
  \nonumber \\
    &=& \frac{i}{g^{2}} \sum_{a,b,c=1}^{N^{2}} Str_{33 \times 33}
  ( M^{a}  M^{b} M^{c} ) Tr_{N \times N}(T^{a} [T^{b}, T^{c}])
  \label{AZ42action}  \\
   &=& - \frac{f_{abc}}{2g^{2}} Str( M^{a} M^{b} M^{c}
  ). \label{AZM42cubic2}   
 \end{eqnarray}
  We explain the meaning of this action in the following remarks.
  \begin{itemize}
  \item{$M$ is a supermatrix belonging to $osp(1|32,R)$ super Lie
      algebra\footnote{Each c-number component is promoted to a
      hermitian $N \times N$ matrix, as we will explain later.}.}
  \item{$P, Q, R, \cdots$ are indices running from $P, Q, R, \cdots  =
     1, 2, \cdots 33$, whereas $p, q, r, \cdots$ runs from $p, q, r,
     \cdots = 1, 2, \cdots 32$.  
     These indices  represent the $33 \times 33$ elements of the
     supermatrices.} 
  \item{In this matrix model, each c-number component of the $osp(1|32,R)$
      super Lie algebra is promoted to the elements of $su(N)$ Lie
      algebra, whose basic properties we refer to
      Appendix. \ref{AZCsun}. This is the same idea as IKKT model, and 
      we depict the way these hermitian $N \times N$
      are embedded in the matrix model in Fig. \ref{33times33}. 
       \begin{figure}[htbp]
        \begin{center}
        \scalebox{.5}{\includegraphics{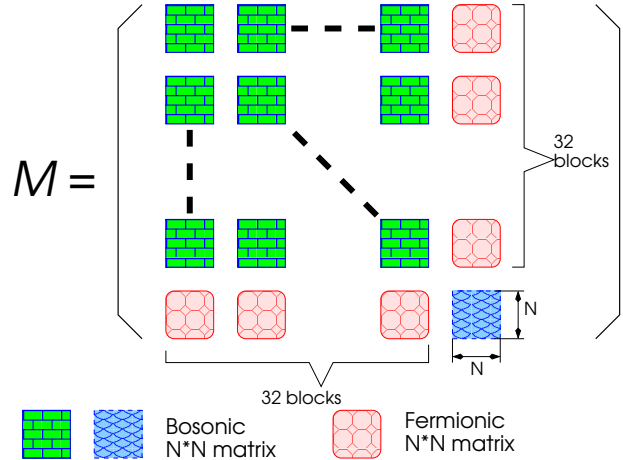} }
        \end{center}
        \caption{ The way large $N$ matrices are embedded in this
          matrix model.}
         \label{33times33}
       \end{figure} }
   \item{The commutator in (\ref{AZ42action}) is taken with
      respect to the $N \times N$ matrices, rather than the $33 \times 33$
      supermatrices.}
  \item{ The traces with respect to $32 \times 32$ matrices and $N \times 
      N$ large matrices are confusing. Throughout in this paper we use 
        \begin{itemize}
         \item{ $Tr$ as a trace of $N \times N$ large matrices.}
         \item{ $tr$ as a trace of $32 \times 32$ matrices, and $Str$
             as a supertrace of $33 \times 33$ matrices.}
        \end{itemize}
      We do not necessarily write explicitly the size of the
      matrices.}
  \item{It is often convenient to
      utilize the representation of color indices of $su(N)$ Lie
      algebra. Each component ${M_{P}}^{Q}$, now promoted to a $N
      \times N$ matrix, is expanded by the basis of 
      the generators of $SU(N)$ Lie group as follows:
        \begin{eqnarray}
         {M_{P}}^{Q} = \sum_{a=1}^{N^{2}} {(M^{a})_{P}}^{Q} T^{a},
        \end{eqnarray}
      where ${(M^{a})_{P}}^{Q}$ are c-number coefficients of the
      expansion with respect to the generators $\{ T^{a} \}$. Since
      the commutator is taken with respect to $N \times N$ hermitian
      matrices, it is easy to see that the action is rewritten as
      (\ref{AZ42action}) by the representation of the color indices.
      And the expression in terms of the structure constant is
      (\ref{AZM42cubic2}), noting that $Tr(T^{a} [ T^{b}, T^{c}]) =
      \frac{i}{2} f^{abc}$.
  }  
  \item{ This matrix model possesses no free parameter. This property
      is easier to see than in IKKT model. The parameter $g$ is
      absorbed in the supermatrices $M$ only by the scale redefinition 
      $M \rightarrow g^{\frac{2}{3}} M$. }
  \item{ It is indispensable to multiply the 
      overall $i$ so that the action should be a real quantity,
      because $Tr(T^{a} [ T^{b}, T^{c}]) =
      \frac{i}{2} f^{abc}$, where the structure constant is a real
      number. We require the action to be real because of the analog of
 quantum field theory. In considering the quantum field theory in
 Minkowski space, we usually consider the action to be real so that
 the Hamiltonian of the theory is  an hermitian object.}
  \end{itemize}

  This model apparently possesses the following pathological
  properties \cite{0002009}. The first is that this theory is not
  bounded from above or below. This pathology stems from the fact that 
  the action is {\it cubic}, and can be seen by a naive
  field redefinition $M \rightarrow M \lambda$. The action is
  rewritten as  $I = \frac{i \lambda^{3}}{g^{2}} Tr_{N \times N}
  ( {M_{p}}^{Q}  [ {M_{Q}}^{R},{M_{R}}^{p} ] - {M_{33}}^{Q}
  [ {M_{Q}}^{R}, {M_{R}}^{33} ] )$, 
  and we can set this action to be $I \to \pm \infty$ by setting the
  parameter to be $\lambda \to \pm \infty$. This means that the path
  integral of the theory $S = \int e^{-I}$ does not converge. However, 
  note that this pathology is shared by general relativity. 
  This pathological property of general relativity can be seen by Weyl
  transformation
  \begin{eqnarray}
   I_{einstein} = \int d^{d} x \sqrt{g} R = \int d^{d} x e^{(\frac{d}{2} -1)
   \omega} \sqrt{g} ( R - (d-1) \nabla^{2} \omega -
   \frac{(d-2)(d-1)}{4} \partial_{\mu} \omega \partial^{\mu} \omega).
  \nonumber \end{eqnarray}
  Although the pathology of Smolin's proposal is not a happy aspect,
  this may be regarded as a good news because this may indicate that
   Smolin's proposals may include general relativity by taking a due
   limit. The second problem is that this theory possesses no explicit 
   time coordinate. This is again a pathology shared by general
   relativity. We can introduce a time coordinate by expanding the
   theory around a certain background. Once a time coordinate is
   introduced, we can construct a Hamiltonian of this theory. We will
   investigate the compactification later.

  \subsubsection{Supertrace}
  We have seen in (\ref{AZ42action}) that this cubic action is
  described by the supertrace. Let us give a brief explanation of the
  notion of supertrace for general supermatrices\footnote{This
  discussion is not limited to $osp(1|32,R)$ super Lie algebra.}.
      A supertrace is defined as 
       \begin{eqnarray}
         Str_{33 \times 33} M \stackrel{def}{=} (\sum_{p=1}^{32} {M_{p}}^{p}) -
         {M_{33}}^{33}. 
       \end{eqnarray}
      Note that {\it the last bosonic part is subtracted, rather than
      summed}. In other words, when the supermatrix is expressed by $M
      = \left( \begin{array}{cc} m & \psi \\ i {\bar \phi} & v \end{array}
      \right)$, the supertrace is 
        \begin{eqnarray}
          Str M = (tr_{32 \times 32} m) - v.
        \end{eqnarray}

    The supertrace of the supermatrices guarantees the cyclic
    rule. This property  can be verified by the following argument. We
    consider two arbitrary matrices: 
  \begin{eqnarray} Z_{1} =  \left( \begin{array}{cc} A_{11} &
  \alpha_{12} \\ \alpha^{T}_{21} & a_{22} \end{array} \right) ,
  \hspace{3mm} Z_{2} = \left( \begin{array}{cc} B_{11} & \beta_{12} \\ 
  \beta^{T}_{21} & b_{22} \end{array} \right), \end{eqnarray}
   where $A_{11}$ and $B_{11}$ are 32 $\times$ 32 bosonic matrices,
  $\alpha_{12}$ , $\alpha_{21}$ , $\beta_{12}$ and $\beta_{21}$ are
  32-component fermionic vectors, and $a_{22}$ and $b_{22}$ are
  c-numbers\footnote{ This argument holds true even if we promote the
  components into large $N$ matrices.}. We consider the supertrace of
  $Z_{1} Z_{2}$ and $Z_{2}   Z_{1}$: 
  \begin{eqnarray} Str(Z_{1} Z_{2}) = tr(A_{11} B_{11} +
  \alpha_{12} \beta^{T}_{21} ) - (\alpha^{T}_{21} \beta_{12} + a_{22}
  b_{22}), \end{eqnarray} 
  \begin{eqnarray} Str(Z_{2} Z_{1}) = tr(A_{11} B_{11} +
  \beta_{12} \alpha^{T}_{21} ) - (\beta^{T}_{21} \alpha_{12} + a_{22}
  b_{22}). \end{eqnarray} 
  Since $\alpha$ and $\beta$ are fermionic quantities, the sign changes
  if we change the order. Therefore, we establish that it is {\it not an
  ordinary trace but a supertrace} that the cyclic rule  $Str(Z_{1}
  Z_{2}) = Str(Z_{2} Z_{1})$ holds true of the supermatrix.

 \subsubsection{Gauge Symmetry}
  Let us investigate the gauge symmetry of this cubic action. We have
  reviewed in the previous section that the gauge group of IKKT model
  is $SO(9,1) \times SU(N)$. The same is true of the promotion of this
  cubic matrix model from the c-number elements of $osp(1|32,R)$
  multiplets to hermitian $N \times N$ matrices. The gauge group of
  this cubic matrix model is $OSp(1|32,R) \times SU(N)$. This means
  that the generator ${\cal X}$ of this gauge group is:
   \begin{eqnarray}
     {\cal X} = (X \otimes {\bf 1}_{N \times N}) + ( {\bf 1}_{33
     \times 33} \otimes u), \textrm{ with } X \in osp(1|32,R) \textrm
     { and } u \in su(N),
   \end{eqnarray}
  with the tensor product $\otimes$ taken with respect to two
  matrices\footnote{Therefore, this is a normal usage of the tensor
  product $\otimes$.}. Therefore, like IKKT model, the gauge
  transformation with respect to  the  $osp(1|32,R)$ Lie algebra and
  $su(N)$ Lie  algebra is taken {\it independently}. The gauge
  transformation with  respect to $su(N)$ Lie algebra  is the same as
  the conventional proposal for large $N$ reduced models, and we do
  not repeat it. On the other hand, the gauge invariance of $su(N)$
  transformation teaches us the physical significance of taking the
  supertrace with respect to $33 \times 33$ supermatrices. The
  infinitesimal gauge transformation is of course 
    \begin{eqnarray}
     \textrm{For an arbitrary element of } u \in osp(1|32,R), 
     \textrm{ the gauge transformation is } M \rightarrow M + [u, M]. 
    \end{eqnarray}
  This indicates that the gauge invariance becomes possible only if
  the action possesses the cyclic symmetry. As we have investigated
  before, the cyclic symmetry for the supermatrix is guaranteed not
  for the ordinary trace but the supertrace. Therefore, it is
  indispensable to take the supertrace with respect to the supermatrix 
  if we are to construct a physically consistent theory.
  
\subsubsection{Explicit Form of the Action}
  When we analyze this matrix theory, it is convenient to express this 
  action in terms of the components of $osp(1|32,R)$ group $M = 
  \left( \begin{array}{cc} m & \psi \\ i {\bar \psi} & 0 \end{array}
  \right)$. The computation is easier to perform if we regard the
  components of $32 \times 32$ matrices as a c-number utilizing 
  (\ref{AZM42cubic2}). 
    \begin{eqnarray}
 & &  Str(M^{a} M^{b} M^{c})  \nonumber \\ 
 &=& tr({(m^{a})_{p}}^{q} {(m^{b})_{q}}^{r} {(m^{c})_{r}}^{p}) + i
    tr({(m^{a})_{p}}^{q}  (\psi^{b})_{q} {(\bar \psi^{c})}^{p}) + i
    tr((\psi^{a})_{p} {(\bar \psi^{b})}^{q} {(m^{c})_{q}}^{p})
  - i {(\bar \psi^{a})}^{q} {(m^{b})_{q}}^{r} (\psi^{c})_{r}
 \nonumber \\
 &\stackrel{\ast}{=}& tr({(m^{a})_{p}}^{q} {(m^{b})_{q}}^{r}
 {(m^{c})_{r}}^{p})  - i tr( {(\bar \psi^{c})}^{p}
   {(m^{a})_{p}}^{q} (\psi^{b})_{q} )
   - i tr({(\bar \psi^{b})}^{q} {(m^{c})_{q}}^{p} (\psi^{a})_{p}) - i
   {(\bar \psi^{a}) }^{q}
   {(m^{b})_{q}}^{r} (\psi^{c})_{r} \nonumber \\
 &=& tr(m^{a} m^{b} m^{c}) - 3i  {\bar \psi^{a}} m^{b} \psi^{c}.
   \label{AZ42mpp} 
   \end{eqnarray}
 In the equality $\stackrel{\ast}{=}$, we utilized the cyclic symmetry
      of the $32 
      \times 32$ trace. However there is a important point in treating 
      fermionic quantities. Since {\it a fermion has jumped over
      another fermions} in the cyclic procedure, the sign must change
      in the cyclic rule. Therefore, the action of this matrix model
      is rewritten as  
  \begin{eqnarray}
   I = - \frac{f_{abc}}{2g^{2}} ( tr(m^{a} m^{b} m^{c}) - 3i {\bar
     \psi^{a}} m^{b} \psi^{c})  
     = \frac{i}{g^{2}} Tr( {m_{p}}^{q} [ {m_{q}}^{r} , {m_{r}}^{p}
     ] ) - 3i {\bar \psi} [m, \psi] ).  
   \label{AZ42actioncomp} 
  \end{eqnarray}
  The former is equivalent to the latter, 
  the former written in terms of the color index and the latter
  adopting the large $N$ matrix representation.  From now on, we express
  the bosonic $32 \times 32$ matrices in terms 
  of the basis of 11 dimensional gamma matrices:
  \begin{eqnarray}
    m = u_{\mu_{1}} \Gamma^{\mu_{1}} + \frac{1}{2}
    u_{\mu_{1} \mu_{2}} \Gamma^{\mu_{1} \mu_{2}} + \frac{1}{5!}
    u_{\mu_{1} \cdots \mu_{5}} \Gamma^{\mu_{1} \cdots \mu_{5}}.
  \end{eqnarray}
 The computation of the bosonic cubic terms in terms of this
  representation is a bit tedious, but it is worth while to overcome
  this obstacle because the physics described by this matrix model can 
  be understood more transparently if we consider the theory in terms
  of 11 dimensional framework. Originally, $osp(1|32,R)$ super Lie
  algebra is known as
  an 'ultimate symmetry of M-theory', and the 11 dimensional
  representation of $osp(1|32,R)$ super Lie algebra is the
  best description to investigate the physics of unified superstring
  theory. 

  Our problem in this paper is the correspondence between our cubic
  matrix model and the existing 10 dimensional matrix theories, such as
  IKKT model. For this purpose, it is more convenient to express the
  induces for 10 dimensions, and we introduce the following new
  variables. 
   \begin{eqnarray}
 & &    Z = u, \hspace{3mm}
    W = u_{\sharp}, \hspace{3mm}
    A^{(\pm)}_{i} = u_{i} \pm  u_{i \sharp}, \hspace{3mm}
    C_{i_{1} i_{2}} = u_{i_{1} i_{2}}, \hspace{3mm}
    D_{i_{1} i_{2}} = u_{i_{1} i_{2} \sharp}, \hspace{3mm} \nonumber
    \\
 & & E^{(\pm)}_{i_{1} i_{2} i_{3}} = u_{i_{1} i_{2} i_{3}} \pm
    u_{i_{1} i_{2} i_{3} \sharp}, \hspace{3mm}
    G_{i_{1} \cdots i_{4}} = u_{i_{1} \cdots i_{4}}, \hspace{3mm}
    H_{i_{1} \cdots i_{4}} = u_{i_{1} \cdots i_{4} \sharp}, \hspace{3mm}
    I^{(\pm)}_{i_{1} \cdots i_{5}} = \frac{1}{2} ( u_{i_{1} \cdots
    i_{5}} \pm {\tilde u}_{i_{1} \cdots i_{5}} ). \label{AZ43fieldredef}
   \end{eqnarray}
  \begin{itemize}
   \item{ The indices $i_{1}, i_{2}, \cdots$ runs $0, 1, \cdots, 9$,
       excluding the $x^{\sharp}$ direction. Throughout this paper,
       $\sharp$ denotes the 10th direction.}
   \item{ On the other hand, the conventional induces $\mu, \nu,
       \cdots$ runs $0, 1, \cdots, 9, \sharp$.}
  \item{The quantity ${\tilde u}_{i_{1} \cdots i_{5}}$ denotes the
      dual of $u_{i_{1} \cdots i_{5}}$: 
   \begin{eqnarray}
    {\tilde u}_{i_{1} \cdots i_{5}} \stackrel{def}{=} \frac{-1}{5!}
    u_{i_{6} \cdots i_{10}} \epsilon^{i_{1} \cdots i_{10} \sharp}.
   \end{eqnarray}
   The quantities $I^{(+)}_{i_{1} \cdots i_{5}}$ and $I^{(-)}_{i_{1}
   \cdots i_{5}}$ are defined to be self-dual and anti-self-dual,
   respectively:
     \begin{eqnarray}
        I^{(+)}_{i_{1} \cdots i_{5}} = {\tilde I}^{(+)}_{i_{1} \cdots
        i_{5}}, \hspace{3mm}
        I^{(-)}_{i_{1} \cdots i_{5}} = - {\tilde I}^{(-)}_{i_{1} \cdots
        i_{5}}.
     \end{eqnarray}
    }
  \item{ Since we are considering $osp(1|32,R)$ matrix model, only the 
      variables $W$, $A^{(\pm)}_{i}$, $C_{i_{1} i_{2}}$, $H_{i_{1}
      \cdots i_{4}}$ and $I^{(\pm)}_{i_{1} \cdots i_{5}}$ concerns
      the discussion in this section. However, we introduce the
      variables of other ranks for future reference.}
  \end{itemize}
  These new variables see other spacetime directions than the 10th
  direction. We give the direction $x^{\sharp}$ a special treatment
  because we would like to preserve the Lorentz symmetry of the 10
  dimensional description. Utilizing these variables, and adopting the
 formulation of $N \times N$ matrices, this
 action is computed to be   

 \begin{eqnarray}
  I_{b} &=& \frac{i}{g^{2}} Tr_{N \times N} ( -96[A^{(+)}_{i_{1}} ,
  A^{(-)}_{i_{2}} ] C^{i_{1} i_{2}} - 96 W [A^{(+) i} , A^{(-)}_{i}] + 
  \frac{4}{5} W [ I^{(+)}_{i_{1} \cdots i_{5}} , I^{(-) i_{1} \cdots
  i_{5}}] \nonumber \\
  &-& 4 H_{i_{1} \cdots i_{4}} ( [ A^{(+)}_{i_{5}} , I^{(-) i_{1}
  \cdots i_{5}}] - [ A^{(-)}_{i_{5}} , I^{(+) i_{1} \cdots i_{5}}]  )
  - 8 C_{i_{1} i_{2}} [ {I^{(+)i_{1}}}_{i_{3} \cdots i_{6}} , I^{(-)
  i_{2} \cdots i_{6}}]  \nonumber \\
  &+&  \frac{8}{3} {H^{jk}}_{i_{1} i_{2}} ( [{I^{(+)}}_{jk  i_{3}
  i_{4} i_{5}} ,  I^{(-) i_{1} \cdots i_{5}}] - 
  [{I^{(-)}}_{jk i_{3} i_{4} i_{5}} ,  I^{(+) i_{1} \cdots
  i_{5}}]  ) \nonumber \\
  &+& 32 [ {C^{i_{1}}}_{i_{2}} , C_{i_{1} i_{3}}] C^{i_{2} i_{3}} - 16 
  C_{i_{1} i_{2}} [ {H^{i_{1}}}_{i_{3} i_{4} i_{5}} , H^{i_{2} \cdots
  i_{5}}] 
  + \frac{1}{27} H_{i_{1} \cdots i_{4}} [ {H^{j}}_{i_{5} \cdots
  i_{7}} , H_{j i_{8} i_{9} i_{10}}] \epsilon^{i_{1} \cdots i_{10}
  \sharp} ), \label{AZM52actionpmb} \\
 I_{f} &=& \frac{i}{g^{2}} Tr_{N \times N} ( 
   -3i (- {\bar \psi_{L}}  [ W , \psi_{R}] +  {\bar \psi_{R}} [ W ,
  \psi_{L}] ) 
  - 3i  ({\bar \psi_{L}} \Gamma^{i} [ A^{(+)}_{i} , \psi_{L}] 
     + {\bar \psi_{R}} \Gamma^{i} [ A^{(-)}_{i} , \psi_{R}] ) \nonumber \\ 
   &-&\frac{3i}{2!} ( {\bar \psi_{L}} \Gamma^{i_{1} i_{2}} [ C_{i_{1} i_{2}} ,
  \psi_{R}] + {\bar \psi_{R}} \Gamma^{i_{1} i_{2}} [ C_{i_{1} i_{2}} ,
  \psi_{L}] )
   -\frac{3i}{4!} ( - {\bar \psi_{L}} \Gamma^{i_{1} i_{2} i_{3} i_{4}}
  [ H_{i_{1} i_{2} i_{3} i_{4}} ,   \psi_{R}] +  {\bar \psi_{R}}
  \Gamma^{i_{1} i_{2} i_{3} i_{4}} [ H_{i_{1} i_{2} i_{3} i_{4}} ,
  \psi_{L}] ) \nonumber \\ 
   &-& \frac{3i}{5!} ( 2 {\bar \psi_{L}} \Gamma^{i_{1} i_{2} i_{3}
  i_{4} i_{5}}  
 [ {I^{(+)}}_{i_{1} i_{2} i_{3} i_{4} i_{5}} , \psi_{L}] 
    + 2 {\bar \psi_{R}} \Gamma^{i_{1} i_{2} i_{3} i_{4} i_{5}} 
 [ {I^{(-)}}_{i_{1} i_{2} i_{3} i_{4} i_{5}} ,\psi_{R}] ) ),
  \label{AZM52actionpmf} 
 \end{eqnarray}
 where $I_{b}$ and $I_{f}$ are the bosonic and fermionic parts of the
 action $I$, respectively. The whole action is $I = I_{b} + I_{f}$. 
 The computation of this action is lengthy, and we refer the proof to
 Appendix. \ref{AZCospactionres}. 
\subsection{Identification of SUSY with IKKT Model}
  We next investigate the SUSY of this cubic matrix model. The SUSY of 
  the theory is an essential property because this symmetry gives a
  wealth of information about the theory. As we have seen in the
  previous section, IKKT model shares ${\cal N}=2$ SUSY with type IIB
  superstring theory, and this SUSY teaches us a lot about the
  properties of IKKT model. For example, the ${\cal N} =2$ SUSY indicates
  the existence of massless graviton, which is an essential property
  for a theory including gravity.
  
  The investigation of the SUSY of this cubic matrix model is an
  interesting issue to be acquainted with the properties of this
  theory. Azuma, Iso, Kawai and Ohwashi \cite{virginal} pointed out
  that this cubic matrix model possesses ${\cal N} = 2$ SUSY to be
  identified with that of IKKT model.
  The discovery of the existence of ${\cal N} =2$ is of significance
  in that they  discovered the property shared by the existing matrix
  model.  

  One of the motivation of tackling with  this cubic matrix is to
  investigate a matrix model which naturally contains the existing
  proposal for the constructive definition of superstring, such as
  IKKT or BFSS model. That the $osp(1|32,R)$ cubic model may
  contain these models is conjectured from the group theoretical
  consideration of $osp(1|32,R)$ super Lie algebra. Performing the
  dimensional reduction on the 11 dimensional representation into the
  10 dimensional representation, we obtain a symmetry of both type IIA 
  and type IIB \cite{0003261}. The work \cite{virginal} solidified
  the belief that this cubic matrix model naturally includes IKKT
  model by the identification of the ${\cal N} = 2$ SUSY with that of
  IKKT model  (\ref{AZM31SUSY4com}) and
  (\ref{AZM31SUSY5com}). 
  
  \subsubsection{Definition of Supercharge}
  First, let us consider what is the SUSY of this cubic matrix
  model. The SUSY is a symmetry relating the fermions and the bosons : 
  a SUSY transformation turns a bosonic state into a fermionic state
  and vice versa. Let $Q$ be an operator which generates such
  transformations $Q$ is called a supercharge, and this satisfies 
   \begin{eqnarray}
    Q| \textrm{Boson} \rangle = |\textrm{Fermion} \rangle,
    \hspace{3mm} Q | \textrm{Fermion} \rangle = | \textrm{Boson}
    \rangle. \label{AZ43defsusy}
   \end{eqnarray}
  A supercharge must be therefore a fermionic quantity, and there is
  no room for the fermionic fields to enter the supercharge. Then, we
  define a supercharge of this cubic matrix theory as follows:
   \begin{eqnarray}
    Q = \left( \begin{array}{cc} 0 & \chi \\ i {\bar \chi} & 0
    \end{array} \right). \label{AZ43defsusy2}
   \end{eqnarray}
  Originally, the supercharge must satisfy the
  Haag-Lopuszanski-Sohnius extension of Coleman-Mandula
  theorem, as we have mentioned in the previous chapter
  (\ref{AZ31hls}). However, we have yet to understand the meaning of
  the translation of the bosonic fields in this cubic matrix theory. And it
  is impossible to justify here that the definition
  (\ref{AZ43defsusy2}) is an eligible supercharge. We defer this
  argument later, and we consider the justification of this
  SUSY transformation by comparing this 'SUSY' transformation with
  that of IKKT model.
  
 \subsubsection{${\cal N} =2$ SUSY transformation}
  We consider the 'SUSY' transformations by the above
  supercharge (\ref{AZ43defsusy2}). The SUSY transformation of this model is
  expressed by the infinitesimal transformation by this
  supercharge. For a matter field of this supermatrix theory $M =
  \left( \begin{array}{cc} m & \psi \\ i {\bar \psi} & 0 \end{array}
  \right)$, the SUSY  transformation $\delta_{\chi} M$ is the
  commutator with the supercharge: 
   \begin{eqnarray}
    \delta_{\chi} M = [Q, M] 
   = [\left( \begin{array}{cc} 0 & \chi \\ i {\bar \chi} & 0
   \end{array} \right) , 
      \left( \begin{array}{cc} m & \psi \\ i {\bar \psi} & 0 \end{array}
   \right) ]
   = \left( \begin{array}{cc} i (\chi {\bar \psi} - \psi {\bar
   \chi}) & - m \chi \\ i {\bar \chi} m & 0 \end{array} \right).
   \end{eqnarray}
  There is another kind of supersymmetry in this cubic
  matrix theory. Note that the action of this matrix theory
  (\ref{AZ42action}) includes the commutators. This means that this
  cubic matrix theory is invariant under a naive 
  translation of the fermionic field $\psi \rightarrow \psi +
  \epsilon$. This is 
  the same situation as emerged in the case of IKKT model, where the
  fermionic field emerged only as a commutator with the matter
  field. 

  The notion of {\it 'homogeneous'} and {\it 'inhomogeneous'} SUSY
  transformations  is used in the same sense as in IKKT model.
  \begin{itemize}
  \item{  
  The former SUSY, the infinitesimal transformation by the supercharge,
  is called {\it homogeneous} SUSY, and we use
  $\delta^{(1)}_{\chi}$ to represent this homogeneous SUSY by the
  supercharge $Q_{\chi} = \left( \begin{array}{cc}  0 & \chi
  \\ i {\bar \chi} & 0 \end{array} \right)$.   }
  \item{  The latter SUSY is called  {\it inhomogeneous}
   SUSY. $\delta^{(2)}_{\epsilon}$ is defined as the inhomogeneous SUSY
   with the translation of the fermion $\psi \rightarrow \psi + \epsilon$.}
 \end{itemize}

 We have discerned that this cubic matrix model possesses ${\cal N}
 =2$ SUSY in the same sense as IKKT model. This stems from the fact
 that the action is written by the commutator of the
 matrices.  The SUSY of the $osp(1|32,R)$ cubic matrix model is
 summarized as
   \begin{itemize}
    \item{ homogeneous: $\delta^{(1)}_{\chi} m = i ( \chi {\bar 
          \psi} - \psi {\bar \chi} )$ , 
          $\delta^{(2)}_{\epsilon} \psi = - m \chi$. }
    \item{ inhomogeneous : $\delta^{(2)}_{\epsilon} m = 0$,
        $\delta^{(2)}_{\epsilon} \psi = \epsilon $. } 
   \end{itemize}

  We have seen that this cubic model possesses ${\cal N} =2$ SUSY, and 
  the origin of the homogeneous and the inhomogeneous SUSY is similar
  to the case of the IKKT model. However, there are two questions in
  order to solidify the correspondence of the SUSY of the cubic matrix 
  model with that of IKKT model. 
        One question stems from the discrepancy of the number of the
        SUSY parameters between the cubic model and IKKT model. 
        This cubic matrix model possesses twice as many SUSY
        parameters as IKKT model. The SUSY parameter of the homogeneous SUSY 
        in IKKT model is 10 dimensional Majorana-Weyl spinors, and the 
        number of the spinors are 16. On the other hand, The SUSY
        parameter of the inhomogeneous SUSY, id est the naive translation,
        is also 16. Therefore, IKKT model possesses in total 32 SUSY
        parameter. On the other hand, our cubic model possesses 32
        homogeneous SUSY parameters $\chi$ and 32 inhomogeneous SUSY
        parameters $\epsilon$, in total 64 SUSY parameters. In order
        to identify the SUSY of the cubic matrix model with that of
        IKKT model, it seem to be necessary to divide the SUSY
        parameter into two groups each of which consists of 32
        fermions. 
        Another issue to tackle with is the identification of the
        fields of the cubic matrix model with those of IKKT model. We
        investigate this issue through the SUSY transformation on the
        fields of the matter fields. This issue is interesting in the
        light of not only the identification of the supersymmetry but
        also in that this question provides us with the clues of what
        fields we should regard as the fundamental fields.

  The clue to the first question lies in the chiral decomposition of
  the fermionic fields. We separate 64 fermionic SUSY parameters into
  the group of left and right chirality, and thus divide these 64 SUSY 
  parameters into two groups.  We have introduced the definition of the
  chiral projection in Appendix. \ref{AZCfermch}.  
  We define a homogeneous SUSY transformation
   $\delta^{(1)}_{\chi_{L}}$ and  $\delta^{(1)}_{\chi_{R}} $ as the
  transformation by the  supercharge $Q_{\chi_{L}}$ and $Q_{\chi_{R}}$
  respectively:
   \begin{eqnarray}
    Q_{\chi_{L}} = \left( \begin{array}{cc} 0 & \chi_{L} \\ i
    {\bar \chi_{L}} & 0 \end{array} \right), \hspace{3mm}
    Q_{\chi_{R}} = \left( \begin{array}{cc} 0 & \chi_{R} \\ i
    {\bar \chi_{R}} & 0 \end{array} \right).
   \end{eqnarray}
  Likewise, the inhomogeneous SUSY transformation
  $\delta^{(2)}_{\epsilon_{L,R}}$ as the translation of the
  fermions\footnote{The abbreviation $\delta^{(2)}_{\epsilon_{L.R}}
  \psi = \psi + \epsilon_{L.R}$ means that the transformation
  $\delta^{(2)}_{\epsilon_{L}}$ corresponds to the translation by
  $\epsilon_{L}$, and that $ \delta^{(2)}_{\epsilon_{R}}$ corresponds
  to the translation by $\epsilon_{R}$.}:
    \begin{eqnarray}
    \delta^{(2)}_{\epsilon_{L.R}} \psi =  \epsilon_{L.R}.
   \end{eqnarray}  

   We next pursue the problem of  how each of the components of the
 matter fields 
 \begin{eqnarray}
    m &=& W \Gamma^{\sharp} 
 + \frac{1}{2} A^{(+)}_{i} \Gamma^{i}( 1 + \Gamma^{\sharp} )
 + \frac{1}{2} A^{(-)}_{i} \Gamma^{i}( 1 - \Gamma^{\sharp} )
 + \frac{1}{2!} C_{i_{1} i_{2}} \Gamma^{i_{1} i_{2}}
 + \frac{1}{4!} H_{i_{1} \cdots i_{4}} \Gamma^{i_{1} \cdots i_{4}}
 \nonumber \\
 &+& \frac{1}{5!} I^{(+)}_{i_{1} \cdots i_{5}} \Gamma^{i_{1} \cdots
 i_{5}} (1 + \Gamma^{\sharp} )
  + \frac{1}{5!} I^{(-)}_{i_{1} \cdots i_{5}} \Gamma^{i_{1} \cdots
 i_{5}} (1 - \Gamma^{\sharp} ) \label{AZ43m}
 \end{eqnarray} 
 are transformed under
 this SUSY transformation. We have verified that the commutator of 
 the homogeneous and inhomogeneous SUSY transformation gives a
 nonvanishing contribution. In extracting the SUSY transformation
 of each of $W$, $A^{(\pm)}_{i}$, $C_{i_{1} i_{2}}$, $H_{i_{1} \cdots
 i_{4}}$ and $I^{(\pm)}_{i_{1} \cdots i_{5}}$, we use the technique
 introduced in  Appendix. \ref{AZCdecomposition}. 
 We find it a natural interpretation to regard the fields of rank 1 as 
 identified with the fields of IKKT model. 

  The homogeneous SUSY transformation is computed as follows. We would 
  like to extract from the SUSY transformation $\delta^{(1)}_{\chi} m
  = i (\chi {\bar \psi} - \psi {\bar \chi} )$ the transformations of
  the fields $A^{(+)}_{i}$ and $B^{(+)}_{i}$. We utilize the formulae
  in Appendix. \ref{AZCdecomposition} to extract the SUSY
  transformation for these fields. Utilizing the formulae
  (\ref{AZMAdec7777}), we obtain\footnote{Note that $A^{(\pm)}_{i}$ is 
  defined to be $A^{(\pm)}_{i} = u_{i} \pm u_{i \sharp}$. $u_{i
  \sharp}$ is a quantity of rank 2 with respect to the 11 dimensional
  indices, and thus we need a minus in the formulae (\ref{AZMAdec7777}).}
   \begin{eqnarray}
   \delta^{(1)}_{\chi} A^{(+)}_{i} &=& \frac{1}{32} tr( (\delta^{(1)}_{\chi} 
   m ) \Gamma_{i} ) + \frac{-1}{32} tr( (\delta^{(1)}_{\chi}
   m)\Gamma_{i \sharp} ) 
  =  \frac{i}{32} tr((\chi {\bar \psi} - \psi {\bar
   \chi})\Gamma_{i} (1 - \Gamma_{\sharp} ) \nonumber \\
  &=& \frac{i}{32}( - {\bar \psi}
   \Gamma_{i} (1 - \Gamma_{\sharp}) \chi + {\bar \chi} \Gamma_{i}(1-
   \Gamma_{\sharp}) \psi) 
  = \frac{i}{16} ({\bar \chi} \Gamma_{i} ( 1 - \Gamma_{\sharp})
   \psi) =  \frac{i}{8} {\bar \chi_{R}}
   \Gamma_{i} \psi_{R}, \label{AZ43homoa+} \\
  \delta^{(1)}_{\chi} A^{(-)}_{i} &=& \frac{1}{32} tr( (\delta^{(1)}_{\chi} 
   m ) \Gamma_{i} ) - \frac{-1}{32} tr( (\delta^{(1)}_{\chi}
   m)\Gamma_{i \sharp} ) 
  =   \frac{i}{8} {\bar \chi_{L}} \Gamma_{i} \psi_{L},  
   \label{AZ43homoa-} \\
  \delta^{(1)}_{\chi} \psi &=&  - m \chi.  \label{AZ43homopsi}
  \end{eqnarray}
 $ \delta^{(1)}_{\chi} A^{(-)}_{i}$ is computed in the same manner as
 $ \delta^{(1)}_{\chi} A^{(+)}_{i}$.
 These transformation laws of the fields possesses the same structure
 as the homogeneous SUSY transformation of IKKT model. Now, we have
 discerned the correspondence between the bosonic fields and the fermionic
 fields. 
  \begin{center} \begin{tabular}{|c|c|} \hline
      bosons $A_{i}$ (IKKT) & fermions $\psi$(IKKT)  \\ \hline \hline
      $A^{(+)}_{i}$ & $\psi_{R} = \frac{1 -
    \Gamma^{\sharp}}{2} \psi$ \\ \hline 
      $A^{(-)}_{i}$ & $\psi_{L} = \frac{1 +
    \Gamma^{\sharp}}{2} \psi$ \\ \hline
  \end{tabular} \end{center} 
 
 \subsubsection{Commutation relation of SUSY transformation}
  We next consider the commutation relation of the SUSY
  transformations. In IKKT model, we have seen the      
  commutators of the supersymmetries (\ref{AZM31SUSY1com}),
  (\ref{AZM31SUSY2com}) and (\ref{AZM31SUSY3com}). We wish to find
  the same structure in the cubic matrix model. We investigate the
  commutators of the SUSY transformations for each chirality
  \begin{eqnarray}
  (1) [\delta^{(1)}_{\chi}, \delta^{(1)}_{\epsilon} ] A^{(\pm)}_{i},
  \hspace{3mm} 
  (2) [\delta^{(2)}_{\chi}, \delta^{(2)}_{\epsilon} ] A^{(\pm)}_{i},
  \hspace{3mm} 
  (3)  [\delta^{(1)}_{\chi}, \delta^{(2)}_{\epsilon} ] A^{(\pm)}_{i}.
  \label{AZ43namco} \end{eqnarray}
  As we have reviewed in the previous chapter, these commutation
  relations reveal what is the translation of the bosonic 
  vector fields, and served to examine that the ${\cal N} = 2$ SUSY
  transformations satisfy Haag-Lopuszanski-Sohnius theorem 
  $\{ Q, Q \} = P_{i} =$ (translation of bosonic fields). The investigation
  of the commutators is an important problem in this cubic matrix
  model, too. And we investigate these commutators one by one.
  It is a trivial matter that the commutator (2)
  vanishes because the inhomogeneous SUSY transformation is just a
  naive translation of the fermions, and we investigate the rest of the SUSY 
  transformations.

  \paragraph{ SUSY transformation $(3)  [\delta^{(1)}_{\chi},
  \delta^{(2)}_{\epsilon} ] A^{(\pm)}_{i}$} .\\
   Even though the order is upside down, let us investigate the SUSY
   of this type.  In IKKT model, this is the only nonvanishing
   commutator, and this commutation relation would give us a big
   information about the correspondence of the SUSY between the cubic
   model and IKKT model. In our cubic model, this commutation
   relation is obtained by 
   \begin{eqnarray}
    [ \delta^{(1)}_{\chi}, \delta^{(2)}_{\epsilon} ] m = -i 
    ( \chi {\bar \epsilon} - \epsilon {\bar \chi} ), \hspace{3mm}
    [ \delta^{(1)}_{\chi}, \delta^{(2)}_{\epsilon} ] \psi = 0.
  \label{AZ43SUSYcom69}
   \end{eqnarray}
 {\sf (Proof)
 We perform the similar procedure, as in the case of IKKT model.
 In considering the commutator of $[ \delta^{(1)}_{\chi},
      \delta^{(2)}_{\epsilon} ]$, it is convenient to compare the two
      paths of the SUSY transformations.
      \begin{itemize}
       \item{$m \stackrel{\delta^{(2)}_{\epsilon}}{\to} m 
         \stackrel{\delta^{(1)}_{\chi}}{\to} m + i (\chi {\bar \psi} -
         \psi {\bar \chi} )$, whereas $m
         \stackrel{\delta^{(1)}_{\chi}}{\to} m + i (\chi {\bar
         \psi} - \psi {\bar \chi} )  
         \stackrel{\delta^{(2)}_{\epsilon}}{\to} m + i \chi 
         ({\bar \psi} + {\bar \epsilon})  - i (\psi + \epsilon) {\bar
         \chi} $.}  
       \item{$\psi \stackrel{\delta^{(2)}_{\epsilon}}{\to} \psi +
           \epsilon  \stackrel{\delta^{(1)}_{\chi}}{\to} \psi +
           \epsilon - m \chi$, whereas $\psi
           \stackrel{\delta^{(1)}_{\chi}}{\to} \psi - m \chi
           \stackrel{\delta^{(2)}_{\epsilon}}{\to}  \psi +
           \epsilon - m \chi$.}
      \end{itemize}
     Taking the difference of these two paths for the bosonic fields
     and the fermions respectively, we obtain the above commutation
     relation. (Q.E.D.) }
 
  We extract the commutators of the SUSY transformation with respect
  to the fields of rank 1: $A^{(\pm)}_{i}$. For this purpose, we make
  use of the formulae in Appendix. \ref{AZCdecomposition}, in the
  same fashion as in the homogeneous SUSY transformation itself:
   \begin{eqnarray}
  & &  [\delta^{(1)}_{\chi}, \delta^{(2)}_{\epsilon}] A^{(+)}_{i} =
    \frac{1}{32} tr(( [\delta^{(1)}_{\chi}, \delta^{(2)}_{\epsilon}] m)
    \Gamma_{i}) + \frac{-1}{32} tr(( [\delta^{(1)}_{\chi},
    \delta^{(2)}_{\epsilon}] m) \Gamma_{i \sharp}) =
    \frac{i}{32}tr((\epsilon {\bar \chi} - \chi {\bar \epsilon})
    \Gamma_{i} (1 - \Gamma_{\sharp}) ) \nonumber \\
  & & \hspace{22mm} = \frac{i}{32} ( - {\bar \chi} \Gamma_{i} (1 -
    \Gamma_{\sharp}) 
    \epsilon + {\bar \epsilon} \Gamma_{i} ( 1- \Gamma_{\sharp})
    \chi) = \frac{i}{16} {\bar \epsilon} \Gamma_{i} ( 1 -
    \Gamma_{\sharp}) \chi = \frac{i}{8} {\bar \epsilon}_{R} \Gamma_{i} 
    \chi_{R}, \label{AZM43SUSYcom12+} \\
  & &  [\delta^{(1)}_{\chi}, \delta^{(2)}_{\epsilon}] A^{(-)}_{i} =
    \frac{1}{32} tr (( [\delta^{(1)}_{\chi}, \delta^{(2)}_{\epsilon}]m)
    \Gamma_{i}) - \frac{-1}{32} tr(( [\delta^{(1)}_{\chi},
    \delta^{(2)}_{\epsilon}] m) \Gamma_{i \sharp}) =
    \frac{i}{8} {\bar \epsilon}_{L} \Gamma_{i} \chi_{L}.
    \label{AZM43SUSYcom12-} 
   \end{eqnarray}

  Extracting the specific chirality of the SUSY parameters, the
  nonvanishing commutators are now clear:
   \begin{eqnarray}
   & & [ \delta^{(1)}_{\chi_{L}}, \delta^{(2)}_{\epsilon_{L}} ]
   A^{(+)}_{i} = 0 , \hspace{15mm}
     [ \delta^{(1)}_{\chi_{R}}, \delta^{(2)}_{\epsilon_{R}} ]
   A^{(+)}_{i} = \frac{i}{8} {\bar \epsilon}_{R} \Gamma_{i} \chi_{R},
 \nonumber \\
  & &  [\delta^{(1)}_{\chi_{L}}, \delta^{(2)}_{\epsilon_{L}} ]
   A^{(-)}_{i} = \frac{i}{8} {\bar \epsilon}_{L} \Gamma_{i} \chi_{L},
   \hspace{3mm} [\delta^{(1)}_{\chi_{R}}, \delta^{(2)}_{\epsilon_{R}} ]
   A^{(-)}_{i} = 0.  \label{AZ43SUSYmainres}
   \end{eqnarray}
  The commutators with different chirality of the two SUSY parameters
  clearly vanishes:
   \begin{eqnarray}
       [\delta^{(1)}_{\chi_{L}}, \delta^{(2)}_{\epsilon_{R}} ]
       A^{(\pm)}_{i} =  [\delta^{(1)}_{\chi_{R}},
       \delta^{(2)}_{\epsilon_{L}} ]  A^{(\pm)}_{i} = 0.
   \end{eqnarray}

  These combinations of SUSY transformations clearly resemble the
  structure of the SUSY transformations of IKKT model. And the
  correspondence between the vector fields and the chirality of the
  fermionic fields matches that obtained by the correspondence of the
  homogeneous transformation itself. 

  \paragraph{ SUSY transformation $(1) [\delta^{(1)}_{\chi},
  \delta^{(1)}_{\epsilon} ] A^{(\pm)}_{i} $} .\\
  In the case of IKKT model, this SUSY transformation vanishes up to
  the gauge transformation, as we have seen in
  (\ref{AZM31SUSY1com}). We want to find the same vanishing in our cubic
  model, if we are to identify the SUSY transformation. 
 However, it turns out that this commutator does not vanish.
 In investigating this commutation relation, it is easier
 to utilize the  following identity:
        \begin{eqnarray}
         [ \delta^{(1)}_{\chi}, \delta^{(1)}_{\epsilon} ] M =
         [ [Q_{\chi}, Q_{\epsilon} ], M]. \label{AZ43comm1964}
        \end{eqnarray}
       This can be verified by the explicit computation of the both
       hand sides:
        \begin{eqnarray}
    & & [ \delta^{(1)}_{\chi}, \delta^{(1)}_{\epsilon} ]
       M = [ Q_{\chi}, [Q_{\epsilon} , M]] -  [ Q_{\epsilon},
       [Q_{\chi} , M]] \nonumber \\
     & & \hspace{10mm} =   Q_{\chi} Q_{\epsilon} M - 
       Q_{\chi} M Q_{\epsilon} - Q_{\epsilon} M Q_{\chi} + M
       Q_{\epsilon} Q_{\chi} - Q_{\epsilon} Q_{\chi} M + Q_{\chi} M
       Q_{\epsilon} + Q_{\epsilon} M Q_{\chi} - M Q_{\chi}
       Q_{\epsilon} = [ [Q_{\chi}, Q_{\epsilon}], M]. \nonumber
    \end{eqnarray}
     This commutator of the
       SUSY transformation is thus obtained by
        \begin{eqnarray}
          [ \delta^{(1)}_{\chi}, \delta^{(1)}_{\epsilon} ] M =
          [ \left( \begin{array}{cc} i (\chi {\bar \epsilon} -
          \epsilon {\bar \chi} ) & 0 \\ 0 & 0 \end{array} \right),
          \left( \begin{array}{cc} m & \psi \\ i {\bar \psi} & 0
          \end{array} \right) ] = \left( \begin{array}{cc} [i(\chi
          {\bar \epsilon} - \epsilon {\bar \chi}), m] & i (\chi
          {\bar \epsilon} - \epsilon {\bar \chi}) \psi \\ -i {\bar
          \psi} (\chi {\bar \epsilon} - \epsilon {\bar \chi}) &  0
          \end{array} \right). \label{AZM43SUSY1coma}
        \end{eqnarray}      
    This relation reveals that the commutators of the two
    homogeneous transformation are
     \begin{eqnarray}
    & &  [ \delta^{(1)}_{\chi}, \delta^{(1)}_{\epsilon} ] m = i [(\chi
          {\bar \epsilon} - \epsilon {\bar \chi}) ,m ], \label{AZ43hhm} \\
    & &   [ \delta^{(1)}_{\chi}, \delta^{(1)}_{\epsilon} ] \psi = i (\chi
          {\bar \epsilon} - \epsilon {\bar \chi}) \psi. \label{AZ43hhpsi}
     \end{eqnarray}
  We have seen in
   (\ref{AZ43SUSYmainres}) which chiralities of the SUSY parameters
   correspond to the vector fields $A^{(\pm)}_{i}$. According to this
   correspondence, we examine the commutation relation with respect 
   to the following cases. 

    We first investigate the commutation relation  
    $ [ \delta^{(1)}_{\chi_{L}}, \delta^{(1)}_{\epsilon_{L}}
          ]$. We examine
          this commutation relation of this SUSY transformation:
          \begin{eqnarray}
        & &  [ \delta^{(1)}_{\chi_{L}}, \delta^{(1)}_{\epsilon_{L}}
          ] A^{(-)}_{i} = \frac{1}{32} tr( 
          ( [ \delta^{(1)}_{\chi_{L}}, \delta^{(1)}_{\epsilon_{L}}] m) 
          \Gamma_{i} )- \frac{-1}{32} tr( ( [ \delta^{(1)}_{\chi_{L}},
          \delta^{(1)}_{\epsilon_{L}} ] m) \Gamma_{i \sharp} )
          \nonumber \\
        & & \hspace{10mm} = \frac{i}{32} tr( (\chi_{L} {\bar \epsilon_{L}}
         - \epsilon_{L} {\bar \chi_{L}} ) m \Gamma_{i}(1
         + \Gamma^{\sharp} ) - m (\chi_{L} {\bar \epsilon_{L}} - \epsilon_{L} 
         {\bar \chi_{L}} ) \Gamma_{i} (1 + \Gamma^{\sharp}) ) \nonumber \\
        & & \hspace{10mm} = \frac{i}{16} tr(  (\chi_{L} {\bar \epsilon_{L}}
         - \epsilon_{L} {\bar \chi_{L}} ) m \Gamma^{i} -  m (\chi_{L}
          {\bar \epsilon_{L}} - \epsilon_{L} {\bar \chi_{L}} ) \Gamma_{i}
          ) = \frac{i}{16} tr(  (\chi_{L}
          {\bar \epsilon_{L}} - \epsilon_{L} {\bar \chi_{L}} ) [ m,
          \Gamma_{i}] ) \nonumber \\
        & & \hspace{10mm} = \frac{i}{16} ( - {\bar \epsilon_{L}}  [ m,
          \Gamma_{i}] \chi_{L} +  {\bar \chi_{L}} [ m,
          \Gamma_{i}] \epsilon_{L} ) = \frac{i}{8} ( {\bar \chi_{L}}
          [m , \Gamma_{i}] \epsilon_{L} ),  \label{AZM43SUSY11a-}
        \end{eqnarray}
        where we have utilized the fact that $[m,\Gamma_{i}] \in
        sp(32)$ and the flipping property of the
        fermions in the last equality.
        Unfortunately, this commutation relation does
        not vanish exactly. And we examine which fields of $m$ in the
        commutator of (\ref{AZM43SUSY11a-}) survive. Now, $m$ is
        expanded as in (\ref{AZ43m}). The commutation relation of
        $[m,\Gamma^{i}]$ and the relations
        (\ref{AZMA21fermvanish}) clarify which components of $m$
        survive or perish. 
        \begin{itemize}
         \item{ ${\bar \chi_{L}} [ W \Gamma_{\sharp}, \Gamma_{i} ]
             \epsilon_{L} = {\bar \chi_{L}} W \Gamma_{\sharp} \Gamma_{i} 
             \epsilon_{L} (\neq 0)$, }
        \item{${\bar \chi_{L}} [A^{(+)}_{j} \Gamma^{j}( 1 +
              \Gamma^{\sharp} ), \Gamma_{i}] \epsilon_{L} =  
               {\bar \chi_{L}} A^{(+)}_{j} ({\Gamma^{j}}_{i} +
              {\eta^{j}}_{i} (1  
              + 2 \Gamma^{\sharp})) \epsilon_{L} = 0$, }
        \item{${\bar \chi_{L}} [A^{(-)}_{j} \Gamma^{j}( 1 -
              \Gamma^{\sharp} ), \Gamma_{i}] \epsilon_{L} =  
               {\bar \chi_{L}} A^{(-)}_{j} ({\Gamma^{j}}_{i} +
              {\eta^{j}}_{i} (1 - 2 \Gamma_{\sharp})) \epsilon_{L} = 0$. \\ 
               The fields $A^{(\pm)}_{i}$ is shown to vanish in the
              SUSY transformation (\ref{AZM43SUSY11a-}). }
        \item{ Computing the commutators likewise, the fields
           $C_{i_{1} i_{2}} $ and 
           $H_{i_{1} \cdots i_{4}}$ survive, while $I^{(\pm)}_{i_{1} 
           \cdots i_{5}}$ vanish.}
        \end{itemize}
       These commutation relations reveal that the fields of even
       rank $W$, $C_{i_{1} i_{2}}$ and $H_{i_{1} \cdots i_{4}}$
       survive in the commutator (\ref{AZM43SUSY11a-}). This is the
       complicated structure of the SUSY transformation of the cubic
       matrix model. The difference from IKKT model lies in the
       homogeneous SUSY transformation of fermionic fields. In our
       cubic matrix model, the homogeneous SUSY transformation of
       fermionic field is represented by (\ref{AZ43homopsi}). 
       Since we are involved in the mixing of the fields
       of other rank than 1, the structure of this commutator is not
       so simple as IKKT model.

       However this can be interpreted as a good news in that we have
       succeeded in 
       excluding the very fields to be identified with those of IKKT
       model: $A^{(\pm)}_{i}$.  In
       this sense, we can regard the commutation relation
       (\ref{AZM43SUSY11a-}) as identical to that of IKKT model. 

      We next investigate the commutation relation of other
      chirality.  $ [ \delta^{(1)}_{\chi_{R}},
      \delta^{(1)}_{\epsilon_{R}}]$. The corresponding field is now
      $A^{(+)}_{i}$, and the commutation relation is now 
         \begin{eqnarray}
          [ \delta^{(1)}_{\chi_{R}}, \delta^{(1)}_{\epsilon_{R}}
          ] A^{(+)}_{i} = \frac{1}{32} tr( 
          ( [ \delta^{(1)}_{\chi_{R}}, \delta^{(1)}_{\epsilon_{R}}] m) 
          \Gamma_{i} ) + \frac{-1}{32} tr( ( [ \delta^{(1)}_{\chi_{R}},
          \delta^{(1)}_{\epsilon_{R}} ] m) \Gamma_{i \sharp} ) = 
          \frac{i}{8} ( {\bar \chi_{R}} 
          [m , \Gamma_{i}] \epsilon_{R} ),  \label{AZM43SUSY11a+}
        \end{eqnarray}
     by completely computing in the same fashion as in
     (\ref{AZM43SUSY11a-}). And totally the same problem emerges, and
     we do not repeat this case. 
 
     An important point is that the commutator
     $[\delta^{(1)}_{\chi_{L}}, \delta^{(1)}_{\epsilon_{R}}]$ does not
     trivially vanish. Computing these commutation relations in the 
     same fashion as before, we obtain 
      \begin{eqnarray}
  & & [\delta^{(1)}_{\chi_{L}}, \delta^{(1)}_{\epsilon_{R}}] A^{(+)}_{i} = 
      \frac{1}{32} tr(([\delta^{(1)}_{\chi_{L}},
      \delta^{(1)}_{\epsilon_{R}}] m) \Gamma_{i} ) + \frac{-1}{32} tr(
      ( [\delta^{(1)}_{\chi_{L}}, \delta^{(1)}_{\epsilon_{R}}] m)
      \Gamma_{i \sharp} )   \label{AZ43SUSY11lra+} \\ 
  & & \hspace{10mm} = \frac{i}{32} tr( (\chi_{L} {\bar \epsilon_{R}} - 
      \epsilon_{R} {\bar \chi}_{L}) m \Gamma_{i} (1 - \Gamma^{\sharp}) 
      - m (\chi_{L} {\bar \epsilon_{R}} - \epsilon_{R} {\bar \chi_{L}} 
      ) \Gamma_{i} (1 - \Gamma^{\sharp}) )  =  \frac{i}{16} ({\bar
      \chi}_{L} m \Gamma_{i} 
      \epsilon_{R} + {\bar \epsilon}_{R} \Gamma_{i} m \chi_{L} ),
      \nonumber \\ 
  & & [\delta^{(1)}_{\chi_{L}}, \delta^{(1)}_{\epsilon_{R}}]
      A^{(-)}_{i} =   \frac{1}{32} tr( ([\delta^{(1)}_{\chi_{L}},
      \delta^{(1)}_{\epsilon_{R}}] m ) \Gamma_{i} ) - \frac{-1}{32}
      tr(( [\delta^{(1)}_{\chi_{L}}, \delta^{(1)}_{\epsilon_{R}}] m)
      \Gamma_{i \sharp} ) \label{AZ43SUSY11lra-} \\
  & & \hspace{10mm} = \frac{i}{32} tr( (\chi_{L} {\bar \epsilon_{R}} - 
      \epsilon_{R} {\bar \chi}_{L}) m \Gamma_{i} (1 + \Gamma^{\sharp}) 
      - m (\chi_{L} {\bar \epsilon_{R}} - \epsilon_{R} {\bar \chi_{L}} 
      ) \Gamma_{i} (1 + \Gamma^{\sharp}) )   = -  \frac{i}{16} ({\bar
      \epsilon}_{R} m \Gamma_{i} 
      \chi_{L} + {\bar \chi}_{L} \Gamma_{i} m \epsilon_{R}
      ). \nonumber
    \end{eqnarray}
  The disastrous fact is that these commutators include the fields
  $A^{(+)}_{i}$ and $A^{(-)}_{i}$. 
  \begin{itemize}
   \item{For the former commutation relation, we extract from $m$
       the fields of rank 1 : $m \to A^{(+)}_{i} \Gamma^{i}
       \frac{1 + \Gamma^{\sharp}}{2} + A^{(-)}_{i} \Gamma^{i} \frac{1
       - \Gamma^{\sharp}}{2}$. Then the commutator is 
    \begin{eqnarray}
  & &  [\delta^{(1)}_{\chi_{L}}, \delta^{(1)}_{\epsilon_{R}}] A^{(+)}_{i}
    \to \frac{i}{16} ({\bar \chi}_{L} ( A^{(+)}_{j} \Gamma^{j}
       \frac{1 + \Gamma^{\sharp}}{2} + A^{(-)}_{j} \Gamma^{j} \frac{1
       - \Gamma^{\sharp}}{2}) \Gamma_{i} \epsilon_{R}
       - {\bar \chi}_{L} \Gamma_{i} (A^{(+)}_{j} \Gamma^{j}
       \frac{1 + \Gamma^{\sharp}}{2} + A^{(-)}_{j} \Gamma^{j} \frac{1
       - \Gamma^{\sharp}}{2}) \epsilon_{R} ) \nonumber \\
   & & \hspace{20mm} = - \frac{i}{8} {\bar \chi}_{L} 
    A^{(+)}_{j} {\Gamma_{i}}^{j} \epsilon_{R}. \label{AZ43lrmix+} 
    \end{eqnarray} }
  \item{ For the latter, we are faced with the same problem as in the
      previous case:
    \begin{eqnarray}
 & & [\delta^{(1)}_{\chi_{L}}, \delta^{(1)}_{\epsilon_{R}}]
    A^{(-)}_{i} \to - \frac{i}{16} ({\bar \epsilon}_{R}  ( A^{(+)}_{j}
    \Gamma^{j} 
       \frac{1 + \Gamma^{\sharp}}{2} + A^{(-)}_{j} \Gamma^{j} \frac{1
       - \Gamma^{\sharp}}{2}) \Gamma_{i} \chi_{L} 
      - {\bar \epsilon}_{R} \Gamma_{i}  ( A^{(+)}_{j} \Gamma^{j}
       \frac{1 + \Gamma^{\sharp}}{2} + A^{(-)}_{j} \Gamma^{j} \frac{1
       - \Gamma^{\sharp}}{2}) \chi_{L}) \nonumber \\
  & & \hspace{20mm} = - \frac{i}{8} {\bar \chi}_{L} 
     A^{(-)}_{j} {\Gamma_{i}}^{j} \epsilon_{R}.
    \label{AZ43lrmix-} 
    \end{eqnarray} }
  \end{itemize}
  These commutation relations reveal that the two-fold SUSY's are
  not independent of each other, but are connected by not the impurity 
  $W, C_{i_{1} i_{2}}$ and $H_{i_{1} \cdots i_{4}}$, but the fields
  $A^{(\pm)}_{i}$. This is an  unfavorable situation in the analysis
  of the SUSY transformation of this cubic model. 

  \paragraph{Structure of ${\cal N}=2$ SUSY transformation}. \\
  The investigation of the commutation relations of this cubic
  matrix model possess two significances. First is that the
  relation (\ref{AZ43SUSYmainres}) clarified the correspondence
  between the fields of rank 1 $A^{(\pm)}_{i}$ and the chirality of
  the SUSY parameters. This relation clarified that the
  correspondence of the fields $A^{(\pm)}_{i}$ and its chirality is
  the same as the fermionic fields $\psi_{L.R}$. The correspondence of 
  the fermionic fields is 
   \begin{itemize}
    \item{For $A^{(+)}_{i}$, the SUSY parameters $\epsilon_{R}$ and
        $\chi_{R}$ corresponds to the SUSY parameters of IKKT model.}
     \item{For $A^{(-)}_{i}$, the SUSY parameters $\epsilon_{L}$ and
        $\chi_{L}$ corresponds to the SUSY parameters of IKKT model.}
   \end{itemize}
  In both cases, the number of the corresponding fermionic SUSY
  parameters is 16+16=32, and this agrees with IKKT model.

   The next significance is that the commutation relation serves
   to clarify whether the 
   supercharge $Q = \left( \begin{array}{cc} 0 & \chi \\ i {\bar \chi} & 
   0 \end{array} \right)$, and the translation of the fermions really
   constitutes an  ${\cal N} =2 $ SUSY transformation. We would like
   to confirm that these SUSY transformations truly satisfy
   Haag-Lopuszanski-Sohnius theorem (\ref{AZ31hls}) if we are to
   assert that these SUSY transformations are actually eligible SUSY 
   transformations.

   We first consider the validity of the former statement $\{Q, Q \} = 
   P_{i}$ by considering the commutation relations of the SUSY
   transformations. We have identified the set of the fields and 
   the supercharges $(A^{(+)}_{i}, \psi_{R}, \chi_{R}, \epsilon_{R})$ and
   $(A^{(-)}_{i}, \psi_{L}, \chi_{L}, \epsilon_{L})$ with those of
   IKKT model, each 
   of which possesses just 32 supercharges. We have completely
   reproduced the commutation relations\footnote{Here, we frequently
   use such  abbreviations as $ [\delta^{(1)}_{\chi_{R,L}} ,
   \delta^{(2)}_{\epsilon_{R,L}} ]$. This means the correspondence of
   the chirality $ [\delta^{(1)}_{\chi_{R}} ,
   \delta^{(2)}_{\epsilon_{R}} ]$ and  $ [\delta^{(1)}_{\chi_{L}} ,
   \delta^{(2)}_{\epsilon_{L}} ]$.}
   $[\delta^{(1)}_{\chi_{R,L}} , \delta^{(2)}_{\epsilon_{R,L}} ]$ and
   $[\delta^{(2)}_{\chi_{R,L}} , \delta^{(2)}_{\epsilon_{R,L}} ]$.
  However, we have failed to reproduce the commutation relation 
   $[\delta^{(1)}_{\chi_{R,L}} , \delta^{(1)}_{\epsilon_{R,L}} ]$. This
   commutation relation includes the unnecessary fields $W,
   C_{i_{1} i_{2}}$ and  $H_{i_{1} \cdots i_{4}}$. This means that a 'noise'
   is involved in the commutation relation of the supercharge. The 
   fundamental supercharge of ${\cal N} =2$ SUSY is now\footnote{Here, 
   we again use the abbreviation $ Q^{fund(1)}_{\epsilon_{R,L}} =
   \delta^{(1)}_{\epsilon_{R,L}} +
   \delta^{(2)}_{\epsilon_{R,L}}$. This means $Q^{fund(1)}_{\epsilon_{R}} =
   \delta^{(1)}_{\epsilon_{R}} + \delta^{(2)}_{\epsilon_{R}}$ and that 
   $Q^{fund(1)}_{\epsilon_{L}} =
   \delta^{(1)}_{\epsilon_{L}} + \delta^{(2)}_{\epsilon_{L}}$.}
    \begin{eqnarray}
     Q^{fund(1)}_{\epsilon_{R,L}} = \delta^{(1)}_{\epsilon_{R,L}} +
     \delta^{(2)}_{\epsilon_{R,L}}, \hspace{3mm}
     Q^{fund(2)}_{\epsilon_{R,L}} = i ( \delta^{(1)}_{\epsilon_{R,L}} -
     \delta^{(2)}_{\epsilon_{R,L}} ).
    \end{eqnarray}
   The noise $W$, $C_{i_{1} i_{2}}$ and $H_{i_{1} \cdots i_{4}}$
   emerging in the commutation relation
   $[\delta^{(1)}_{\chi_{R,L}} , \delta^{(1)}_{\epsilon_{R,L}} ]$ reveals
   that\footnote{Again, note that the anti-commutator in the original
   Haag-Lopuszanski-Sohnius theorem is replaced with the commutator,
   because the SUSY parameters are Grassmann odd.}
    \begin{eqnarray}
     [Q^{fund(x)}_{R,L}, Q^{fund(x)}_{R,L}] = P_{iR,L}
     \textrm{(translation of bosons)} +
     ( \textrm{Other fields } W, C, H ) \textrm{ (with }x=1,2 \textrm{)}
    \end{eqnarray}
   The operator $P_{iR.L}$ is the translation of the bosonic
   fields. The commutation relations (\ref{AZ43SUSYmainres}) reveal
   the way $P_{iR,L}$ translates the bosonic fields of rank 1.
    \begin{itemize}
     \item{ $P_{i,R}$ translates only the bosonic fields $A^{(+)}_{i}$ by
         $a^{(+)}_{i} = \frac{i}{8} {\bar \epsilon}_{R} \Gamma_{i}
         \chi_{R}$.}
     \item{ $P_{i,L}$ translates only the bosonic fields $A^{(-)}_{i}$ by
         $a^{(-)}_{i} = \frac{i}{8} {\bar \epsilon}_{L} \Gamma_{i}
         \chi_{L}$.}
    \end{itemize}
   The first statement of Haag-Lopuszanski-Sohnius theorem is thus shown to 
   hold true of this matrix model up to the impurity $W,C,H$. 
 
    The latter statement  $[P_{i}, Q]=0$ is easily verified 
    by computing the commutators of the translation of the matter fields and 
    the SUSY transformations $\delta^{(1)}_{\chi_{R,L}}$ and
    $\delta^{(2)}_{\epsilon_{R,L}}$. This argument does not rely on the
    commutation relations (\ref{AZ43namco}).
    \begin{itemize}
   \item{The commutation relation $[ P_{iR,L},
       \delta^{(1)}_{\chi_{R,L}}] =0 $  is trivial. In the following
       argument, note again that the transformation of right chilarity
       corresponds to the bosonic fields $A^{(+)}_{i}$ and that the
       transformation of left chirality corresponds to $A^{(-)}_{i}$.
      \begin{itemize}
       \item{$A^{(\pm)}_{i} \stackrel{P_{iR,L}}{\to} A^{(\pm)}_{i} +
           a^{(\pm)}_{i} \stackrel{\delta^{(1)}_{\chi_{R,L}}}{\to}
           A^{(\pm)}_{i} + a^{(\pm)}_{i} +   
           \frac{i}{8} {\bar \chi_{R,L}} \Gamma_{i} \psi_{R,L}$,
           whereas \\ $A^{(\pm)}_{i}
           \stackrel{\delta^{(1)}_{\chi_{R,L}}}{\to} A^{(\pm)}_{i} +
           \frac{i}{8} {\bar \chi_{R,L}} \Gamma_{i} \psi_{R,L}
           \stackrel{P_{iR,L}}{\to}  A^{(\pm)}_{i} + a^{(\pm)}_{i} + 
           \frac{i}{8} {\bar \chi_{R,L}} \Gamma_{i} \psi_{R,L}$.} 
      \item{$\psi_{R,L} \stackrel{P_{iR,L}}{\to}  \psi_{R,L} 
           \stackrel{\delta^{(1)}_{\chi_{R,L}}}{\to} \psi_{R,L} - (m
           \chi)_{R,L}$, whereas $\psi_{R,L}
           \stackrel{\delta^{(1)}_{\chi_{R,L}}}{\to} \psi_{R,L} - (m 
           \chi)_{R,L}  \stackrel{P_{iR,L}}{\to}
           \psi_{R,L} - (m \chi)_{R,L}$.}
      \end{itemize}}
     \item{
    The commutation relation $[P_{iR,L}, \delta^{(2)}_{\epsilon_{R,L}}]
    =0 $ is also trivial, because $P_{iR,L}$ moves only the bosons while
    $\delta^{(2)}_{\epsilon_{R,L}}$ moves only the fermions. 
     \begin{itemize}
       \item{$A^{(\pm)}_{i} \stackrel{P_{iR,L}}{\to} A^{(\pm)}_{i} +
           a^{(\pm)}_{i}  \stackrel{\delta^{(2)}_{\epsilon_{R,L}}}{\to}
           A^{(\pm)}_{i} + a^{(\pm)}_{i}$, whereas 
           $ A^{(\pm)}_{i} \stackrel{\delta^{(2)}_{\epsilon_{R,L}}}{\to}
           A^{(\pm)}_{i} \stackrel{P_{iR,L}}{\to} A^{(\pm)}_{i} +
           a^{(\pm)}_{i}$.}
      \item{$\psi_{R,L} \stackrel{P_{iR,L}}{\to} \psi_{R,L}
          \stackrel{\delta^{(2)}_{\epsilon_{R,L}}}{\to} \psi_{R,L} +
          \epsilon_{R,L}$, whereas $\psi_{R,L}
          \stackrel{\delta^{(2)}_{\epsilon_{R,L}}}{\to} \psi_{R,L} +
          \epsilon_{R,L} \stackrel{P_{iR,L}}{\to}   \psi_{R,L} +
          \epsilon_{R,L}$.}
     \end{itemize} }
  \end{itemize}    
   Therefore, the commutation relation $[P_{i}, Q]=0$ is trivial, and 
   this does not involve a noise $W, C, H$.

  This completes the justification of the SUSY, which we 'defined' in
   (\ref{AZ43defsusy}), and is consistent with
   Haag-Lopuszanski-Sohnius theorem  apart from the fields $W, C, H$.

   And the serious difficulty is that the relation
   (\ref{AZ43lrmix+}) and (\ref{AZ43lrmix-}) indicates that these
   two-fold  SUSY  transformations $Q^{fund(x)}_{\chi_{L}}$  and
   $Q^{fund(x)}_{\chi_{R}}$  are not independent of each other. 
   These two-fold IKKT-like SUSY's are related by the fields
   $A^{(\pm)}_{i}$ themselves, and we have yet to succeed in
   interpreting  this impurity  physically.

  \subsubsection{Summary of the results}
  The homogeneous SUSY transformation (\ref{AZ43homoa+}) and
   (\ref{AZ43homoa-}) and the commutation relation of the SUSY
   transformations 
   $(\ref{AZ43SUSYmainres})$ for the vector fields $A^{(\pm)}_{i}$
   answer  both of the questions.
   First, these results give us an answer of the  
  identification of the fields of the cubic model with those of IKKT
  model. The answer is now straightforward. There are two choices :
  the fields of IKKT model $A^{IKKT}_{i}$ should be identified with
  either $A^{(+)}_{i}$ or $A^{(-)}_{i}$. 

  The other problem with the SUSY transformation of this cubic model
  was how to cope with the fermionic SUSY parameters twice as many as
  those of IKKT. The answer to this question is readily read off from
  the final results (\ref{AZ43SUSYmainres}).
  We have succeeded in separating the 64 fermionic parameters in the
  cubic matrix theory into the two 
  groups of 32 fermionic parameters by noting the chirality. These
  supersymmetry asserts that there are two distinct worlds each of
  which possesses a SUSY of IKKT model.
   \begin{itemize}
     \item{For the fields $A^{(+)}_{i}$, the fermionic fields to be 
         identified with that of IKKT model are {\it those of right
         chirality} : Therefore, the fermions of the 
         right chirality $\psi_{R}$ should correspond to the fermionic 
         fields of IKKT model, if we regard $A^{(+)}_{i}$ as the
         fundamental vector field. 
         And we regard the SUSY parameters of right chirality
         $\chi_{R}$ and $\epsilon_{R}$ as corresponding to those of
         IKKT model. Each of the fermionic
         parameters are just half of the whole parameter $\chi$ and
         $\epsilon$, and each of $\chi_{R}$ and $\epsilon_{R}$
         possesses 16 fermions. 
         This is just what it should be in IKKT model. 
         And we have justified the decomposition of the SUSY
         parameters by noting the commutation relation
         (\ref{AZ43SUSYmainres}). 
        }
    \item{For the fields $A^{(-)}_{i}$, the fermionic fields
        $\psi_{L}$ should be identified with IKKT model. Therefore,
        the SUSY parameter corresponding to IKKT model is $\chi_{L}$
        and $\epsilon_{L}$. In this case, we have succeeded in
        reducing the parameters of the SUSY transformation to 16 + 16
        = 32, too.} 
   \end{itemize}
  \begin{figure}[htbp]
   \begin{center}
    \scalebox{.5}{\includegraphics{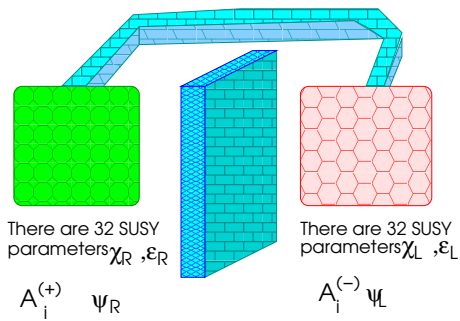} }
   \end{center}
   \caption{The two-fold SUSY structures of this cubic matrix
     model. These two worlds are not independent of each other.}
   \label{susyident}
  \end{figure}

  However, the relations (\ref{AZ43lrmix+}) and (\ref{AZ43lrmix-}) 
  indicate that these two-fold SUSY structures are not independent of
  each other, but are related with the fields $A^{(\pm)}_{i}$. This
  is not a beautiful structure, and the physical interpretation of
  this mixing is not clear yet.

  We have seen the SUSY transformation of this cubic matrix model,
  and this cubic model possesses ${\cal N} = 2$ supersymmetry. And
  analyzing this ${\cal N} = 2$ SUSY more deeply, we have clarified the way
  this cubic model embeds the symmetry of IKKT model : there are
  two-fold  symmetries of 32 fermionic fields, each of which have
  shown  agreement with the SUSY of IKKT model. The
  investigation of supersymmetry is a powerful tool to analyze the
  structure of the model. 

  \subsection{The Induction of IKKT model}
  We have seen the correspondence of this cubic matrix model with
  IKKT model in terms of the SUSY transformation, and we have shown
  that this cubic matrix model embeds two supersymmetries of IKKT
  model, by dividing the chirality of the fermionic parameters. 
  Then here comes a next question : 
  \begin{center}
   {\it How can we induce IKKT model from this cubic model?}
  \end{center}
  In this section, we consider the answer to this question, however
  this is a difficult problem, and we give just a hand-waving argument
  to this issue.

  We have seen the explicit form of the action of this cubic matrix
  model (\ref{AZM52actionpmb}) and (\ref{AZM52actionpmf}).  As we have
  seen in Sec. 3, IKKT model can be interpreted as 
  non-commutative Yang-Mills theory if this theory is expanded around
  a certain classical solution to the equation of motion
  \cite{9908141}. Likewise, the starting point to analyze the behavior 
  is to expand the theory around a classical solution.

  This cubic model possesses no time or space derivative from the
  beginning, as is true of IKKT model. The classical equations of
  motion should be given by Euler-Lagrange equation of the action
  (\ref{AZM52actionpmb}) and (\ref{AZM52actionpmf})
   \begin{eqnarray}
    \partial_{\mu} \frac{ \partial I}{ \partial (\partial_{\mu} X)} -
    \frac{\partial I}{\partial X} = 0,
   \end{eqnarray}
 where $X$ is an arbitrary field of this theory. Since we are 
 interested in the classical solution, we do not consider the
 fermionic fields in the classical equations of motion. These
 equations of motion are explicitly given by
     \begin{eqnarray}
 \frac{\partial I}{\partial W} &=& - 96
   [A^{(+)}_{i_{1}} , A^{(-) i_{1}}] + \frac{4}{5}[ I^{(+)}_{i_{1}
   \cdots i_{5}} , I^{(-) i_{1} \cdots i_{5}}] = 0, \label{AZ44W} \\
 \frac{\partial I}{\partial A^{(+)}_{i_{1}}}  &=& - 96
   [A^{(-)}_{i_{2}} , C^{i_{1} i_{2}}] + 96 [ W , A^{(-)i_{1}}] + 4
   [ H_{i_{2} \cdots i_{5}} , I^{(-)i_{1} \cdots i_{5}}] = 0,
   \nonumber \\
 \frac{\partial I}{\partial A^{(-)}_{i_{1}}} &=&  - 96 
   [ A^{(+)}_{i_{2}} , C^{i_{1} i_{2}}] - 96 [ W, A^{(+) i_{1}}] - 4
   [ H_{i_{2} \cdots i_{5}} , I^{(+)i_{1} \cdots i_{5}}] = 0,
   \label{AZ44A-} \\ 
 \frac{\partial I}{\partial C_{i_{1} i_{2}}} &=& -  96
    [ A^{(+) [i_{1}} , A^{(-) i_{2}]} ] - 8 [ {I^{(+) 
    [i_{1}}}_{i_{3}  \cdots i_{6}} , I^{(-) i_{2}] \cdots i_{6}}] + 96
    [ {C^{ [i_{1}}}_{\rho}  , C^{i_{2}] \rho}] \nonumber \\ 
    &-& 16 [ {H^{[i_{1}}}_{i_{3} \cdots i_{5}} , H^{i_{2}] \cdots
   i_{5}}]  = 0,  \label{AZ44C} \\
 \frac{\partial I}{\partial H_{i_{1} \cdots i_{4}}} &=& 
    -4 [ A^{(+)}_{i_{5}} , I^{(-) i_{1} \cdots i_{5}}] + 4
    [ A^{(-)}_{i_{5}} , I^{(+) i_{1} \cdots i_{5}}] +
    \frac{8}{3} [ {I^{(+) [i_{1} i_{2}}}_{i_{5} i_{6} i_{7}} , I^{(-) 
    i_{3} i_{4}] i_{5} i_{6} i_{7}} ] \nonumber \\
    &-& \frac{8}{3}  [ {I^{(-) [i_{1} i_{2}}}_{i_{5} i_{6}
    i_{7}} , I^{(+) i_{3} i_{4}] i_{5} i_{6} i_{7}} ] - 32
   [ {C_{\rho}}^{[i_{1}} , 
    H^{\rho i_{2} i_{3} i_{4}] }] + \frac{1}{54} [ {I^{\rho
    \sigma}}_{i_{5} i_{6} i_{7}} , I_{\rho \sigma i_{8} i_{9} i_{10}}] 
    \epsilon^{i_{1} \cdots i_{10} \sharp} \nonumber \\
    &+& \frac{1}{27} [ {H^{\rho}}_{i_{5} i_{6} i_{7}} , H_{\rho i_{8}
    i_{9} i_{10}} ] \epsilon^{i_{1} \cdots i_{10} \sharp} +
    \frac{2}{27} [ {H^{i_{1}}}_{i_{5} i_{6} i_{7}} , H_{i_{8} \cdots
    i_{11}}] \epsilon^{i_{2} \cdots i_{11} \sharp} = 0, \label{AZ44H} \\
  \frac{\partial I}{\partial I^{(+)}_{i_{1} \cdots i_{5}} } &=&
   \frac{-4}{5} [ W, I^{(-) i_{1} \cdots i_{5} } ] - 4
   [ A^{(-) [ i_{1} }, H^{i_{2} \cdots i_{5}] }] -8 [ C^{k [i_{1}},
   {{I^{(-)}}_{k}}^{i_{2} \cdots i_{5}]} ]  - \frac{16}{3} [ {H_{k
   l}}^{[ i_{1} i_{2}}, I^{(-) k l i_{3} i_{4} i_{5}] } ] = 0,
    \nonumber \\
  \frac{\partial I}{\partial I^{(-)}_{i_{1} \cdots i_{5}}} &=&
  \frac{4}{5} [W, I^{(+) i_{1} \cdots i_{5} }] + 4 [A^{(+) [ i_{1}},
   H^{i_{2} \cdots i_{5}] } ] -8 [C^{k [i_{1}}, {{I^{(+)}}_{k}}^{i_{2}
   \cdots i_{5}]} + \frac{16}{3} [ {H_{kl}}^{[ i_{1} i_{2}}, I^{(+) kl
   i_{3} i_{4} i_{5}]  }] = 0. \label{AZ44I-}
 \end{eqnarray}
 Now, we consider the classical solution of the above equations of
 motion. It is almost impossible to solve the above equations of
 motions exactly, but we can find a classical solution which possesses 
 a physical consequence. As we have seen in the case of IKKT model,
 the canonical pairs play an essential role in introducing  kinetic
 terms of the theory:
   \begin{eqnarray}
    A^{(+)}_{0} = p_{1}, A^{(+)}_{1} = q_{1}, \cdots A^{(+)}_{2d-2} =
    p_{d}, A^{(+)}_{2d-1} = q_{d}, \textrm{ (all the other fields)}=0, 
   \label{AZ44cs}
   \end{eqnarray}
 where $p_{k}, q_{l}$ are $d$ independent pairs of canonical pairs satisfying
 $[q_{k}, p_{l}] =  i \delta_{kl}$.  $2d$ is a number of the dimension 
 of the spacetime produced by this classical solution, and we now
 focus on the case in which $2d = 10$.  If we expand the theory around
 this classical 
 solution, we can introduce the spacetime derivative in the same
 fashion as in IKKT model \cite{9908141}. We separate the field
 $A^{(+)}_{i}$ between the classical solution and the
 fluctuation:
  \begin{eqnarray}
   A^{(+)}_{i} = {\hat p}_{i} + a^{(+)}_{i},
  \end{eqnarray}
 where ${\hat p}_{i}$  denotes the above classical solution\footnote
 {Be careful not to be confused with the canonical pairs $p_{k},
 q_{l}$.}. We can then introduce a spacetime derivative by mapping the 
 commutators with the $A^{(+)}_{i}$ fields with the covariant
 derivative, where $X$ is an arbitrary field,
  \begin{eqnarray}
    [ A^{(+)}_{i}, X ] = - i \partial_{i} X + [a^{(+)}_{i}, X].
  \end{eqnarray}
  Applying this mapping rule to the action (\ref{AZM52actionpmb}) and
  (\ref{AZM52actionpmf}), we can introduce a kinetic term in this
  action, and thus we obtain the following action:
  \begin{eqnarray}
     I_{b} &=& \frac{1}{g^{2}} Tr_{N \times N} ( -96(\partial_{i_{1}}
     A^{(-)}_{i_{2}} ) C^{i_{1} i_{2}} + 96 ( \partial_{i} W )
     A^{(-) i} + 4 (\partial_{i_{1}} H_{i_{2} \cdots i_{5}} ) I^{(-)
     i_{1} \cdots i_{5}} ) \label{AZ44partial} \\
 &+&
  \frac{i}{g^{2}} Tr_{N \times N} ( -96[a^{(+)}_{i_{1}} ,
  A^{(-)}_{i_{2}} ] C^{i_{1} i_{2}} - 96 W [a^{(+) i} , A^{(-)}_{i}] + 
  \frac{4}{5} W [ I^{(+)}_{i_{1} \cdots i_{5}} , I^{(-) i_{1} \cdots
  i_{5}}] \nonumber \\
  &+& 4 ( [ a^{(+)}_{i_{1}} , H_{i_{2} \cdots i_{5}}] I^{(-) i_{1}
  \cdots i_{5}} - [ A^{(-)}_{i_{1}} , H_{i_{2} \cdots i_{5}}] I^{(+)
  i_{1} \cdots i_{5}}  ) 
    - 8 C_{i_{1} i_{2}} [ {I^{(+)i_{1}}}_{i_{3} \cdots i_{6}} , I^{(-)
  i_{2} \cdots i_{6}}]  \nonumber \\
  &+&  \frac{8}{3} {H^{kl}}_{i_{1} i_{2}} ( [{I^{(+)}}_{k l i_{3}
     i_{4} i_{5}} ,  I^{(-) i_{1} \cdots i_{5}}] - 
  [{I^{(-)}}_{k l  i_{3} i_{4} i_{5}} ,  I^{(+) i_{1} \cdots
  i_{5}}]  ) \nonumber \\
  &+& 32 [ {C^{i_{1}}}_{i_{2}} , C_{i_{1} i_{3}}] C^{i_{2} i_{3}} - 16 
  C_{i_{1} i_{2}} [ {H^{i_{1}}}_{i_{3} i_{4} i_{5}} , H^{i_{2} \cdots
  i_{5}}] 
  + \frac{1}{27} H_{i_{1} \cdots i_{4}} [ {H^{k}}_{i_{5} \cdots
  i_{7}} , H_{k i_{8} i_{9} i_{10}}] \epsilon^{i_{1} \cdots i_{10}
  \sharp} ), \label{AZ44others} \\
  I_{f} &=& \frac{1}{g^{2}} Tr_{N \times N} ( - 3i  {\bar
     \psi_{L}} \Gamma^{i} \partial_{i} \psi_{L} ) \label{AZ44fermpartial} \\
   &+& \frac{i}{g^{2}} Tr_{N \times N} (-3i ( - {\bar \psi_{L}} [ W ,
     \psi_{R}] +  {\bar  \psi_{R}} [ W , \psi_{L}] )  
   - 3i ( {\bar \psi_{L}} \Gamma^{i} [ a^{(+)}_{i} , \psi_{L}] 
     + {\bar \psi_{R}} \Gamma^{i} [ A^{(-)}_{i} , \psi_{R}] ) \nonumber \\ 
   &-&\frac{3i}{2!} ( {\bar \psi_{L}} \Gamma^{i_{1} i_{2}} [ C_{i_{1} i_{2}} ,
  \psi_{R}] + {\bar \psi_{R}} \Gamma^{i_{1} i_{2}} [ C_{i_{1} i_{2}} ,
  \psi_{L}] )
   -\frac{3i}{4!} ( - {\bar \psi_{L}} \Gamma^{i_{1} i_{2} i_{3} i_{4}}
  [ H_{i_{1} i_{2} i_{3} i_{4}} ,   \psi_{R}] +  {\bar \psi_{R}}
  \Gamma^{i_{1} i_{2} i_{3} i_{4}} [ H_{i_{1} i_{2} i_{3} i_{4}} ,
  \psi_{L}] ) \nonumber \\ 
   &-& \frac{3i}{5!} ( 2 {\bar \psi_{L}} \Gamma^{i_{1} i_{2} i_{3}
     i_{4} i_{5}}  
 [ {I^{(+)}}_{i_{1} i_{2} i_{3} i_{4} i_{5}} , \psi_{L}] 
   +  2 {\bar \psi_{R}} \Gamma^{i_{1} i_{2} i_{3} i_{4} i_{5}} 
 [ {I^{(-)}}_{i_{1} i_{2} i_{3} i_{4} i_{5}} ,\psi_{R}] ) ).
     \label{AZ44fermother} 
  \end{eqnarray}

  The cubic action, per se, does not include a quartic term of vector
  fields in the action. However, we can interpret that the bosonic
  term is induced by the fermionic term of the IKKT model. The idea
  that the theory consisting of  fermionic fields and
  a Dirac operator induces Einstein gravity and Yang-Mills
  theory has long been suggested. The proposal of Connes and
  Chamseddine is one of these 
  suggestions of induced gravity \cite{9606001}. Based on these
  suggestions, we find it a natural idea that 
  the bosonic term of IKKT model is induced from its fermionic
  term. And we hypothesize that the whole IKKT model should be induced
  only by the fermionic field. 

  Our goal is thus to find the fermionic terms in this cubic matrix model
  to be identified with that of IKKT model. 
  We have seen in the previous section the correspondence between the
  vector fields and the fermionic fields according to the
  identification of ${\cal N}=2$  SUSY with that of IKKT model. And
  the fermionic terms to be identified with that if IKKT model is
    \begin{eqnarray}
     {\bar \psi_{R}} \Gamma^{i} A^{(+)}_{i} \psi_{R}
     \stackrel{or}{\Leftrightarrow} 
     {\bar \psi_{L}} \Gamma^{i} A^{(-)}_{i} \psi_{L}. \label{AZ44goal} 
    \end{eqnarray}
  If either of these terms exists in the action, this can be
  identified with the fermionic term of IKKT model.  However, the
  kinetic term of the fermionic fields 
  (\ref{AZ44fermpartial}), per se, does not include such terms as
  (\ref{AZ44goal}), and does not serve to induce IKKT
  model due to the discrepancy of the correspondence of the vector
  fields and the chirality of the fermions.
  In order to remedy this situation, we consider inducing the terms
  (\ref{AZ44goal}) by multi-loop effect. The action (\ref{AZ44partial})
  $\sim$ (\ref{AZ44fermother}) just tells us that there is no terms to 
  be identified with IKKT model {\it at a tree level}. Since the idea
  of 'induced IKKT model' is to construct the bosonic term by one-loop
  effect of fermionic Lagrangian, the idea of 'induced theory' is, per
  se, a notion based on multi-loop effect of the perturbative
  theory. In this sense, there is no problem if the fermionic term is
  induced by the multi-loop effect.

  In order to consider the multi-loop effect of the cubic theory, we
  start from considering the Feynman diagram of this theory. The
  kinetic terms contributes to the propagators, and the cubic terms
  contributes to the vertices of this theory. These Feynman rules can
  be read off from the action (\ref{AZ44partial}) $\sim$
  (\ref{AZ44fermother}).
  \begin{itemize}
   \item{Propagator : The propagators of this cubic matrix theory
       stems from the kinetic terms (\ref{AZ44partial}) and
       (\ref{AZ44fermpartial}). 
   \begin{figure}[htbp]
   \begin{center}
    \scalebox{.5}{\includegraphics{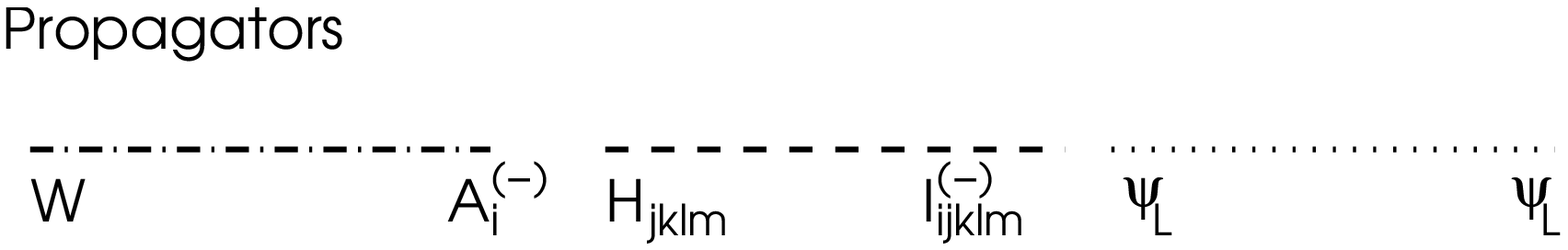} }
   \end{center}
    \caption{Propagators of this cubic matrix theory.}
   \label{feynprop}
  \end{figure}
  Apparently, the action (\ref{AZ44partial}) $\sim$
  (\ref{AZ44fermother}) seems to include the propagators $\langle W
  A^{(-) i} \rangle$, $\langle A^{(-)}_{i} C^{ij} \rangle$,
  $\langle H_{jklm} I^{(-) ijklm} \rangle$ and $\langle \psi_{L}
  \psi_{L} \rangle$. However, a careful investigation of this theory
  reveals that the propagator $\langle A^{(-)}_{i} C^{ij} \rangle$ is
  prohibited from existing, while the other propagators $\langle W
  A^{(-)}_{i} \rangle$,  $\langle H_{jklm} I^{(-) ijklm} \rangle$
  and $\langle \psi_{L} \psi_{L} \rangle$ do exist in this theory. We
  leave the discussion of the existence of these propagators to
  Appendix. \ref{AZCkillac}. 
 }
  \item{Vertex: This is also read off from the action of this theory.
   \begin{figure}[htbp]
   \begin{center}
    \scalebox{.45}{\includegraphics{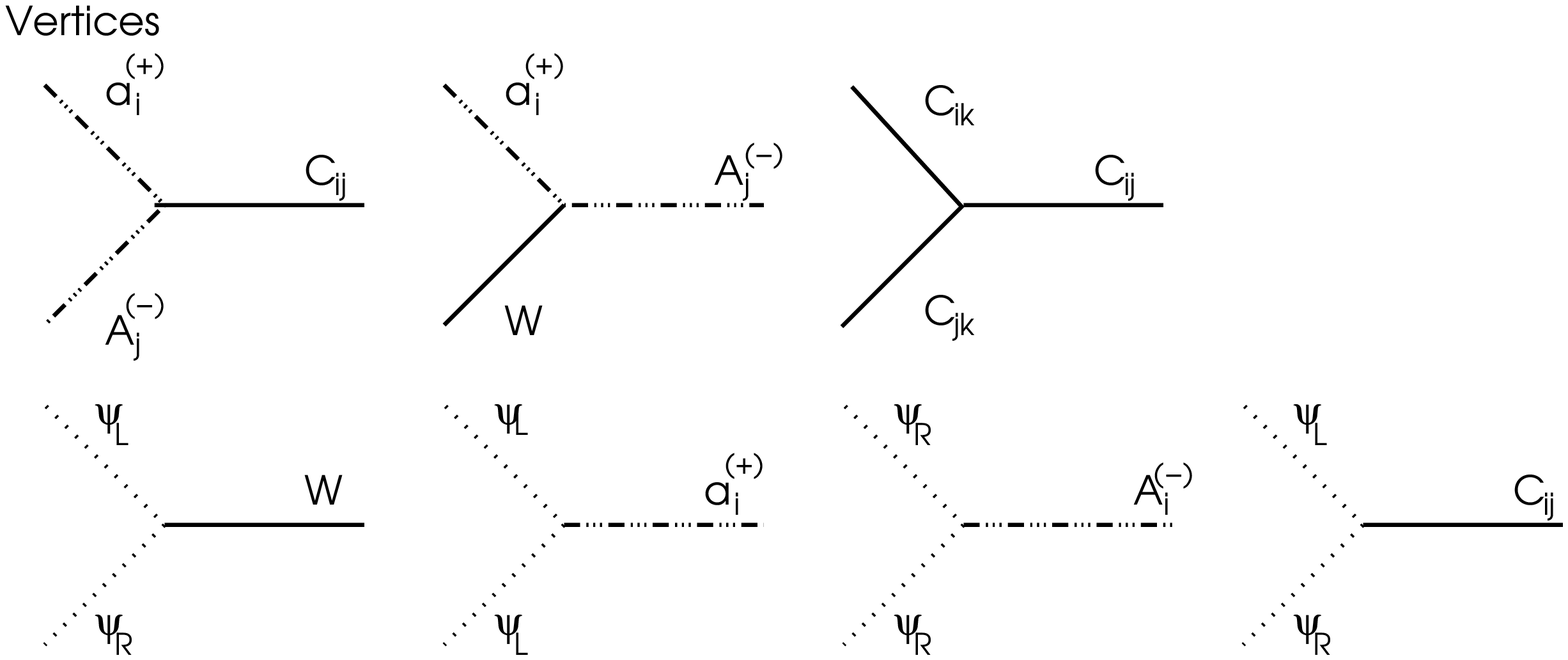} }
   \end{center}
    \caption{Vertices of this cubic matrix theory.}
   \label{feynvertex}
  \end{figure}
  Of course, the vertices for the fields of higher rank (such as
  $H_{i_{1} \cdots i_{4}}$ and $I^{(\pm)}_{i_{1} \cdots i_{5}}$) 
  exist, however we omit drawing their Feynman diagrams because these 
  diagrams do not contribute to our discussion of inducing IKKT
  model. }
  \end{itemize}

  Our goal is to build a fermion vertex (\ref{AZ44goal}) by means of
  the multi-loop effect. In order to build such a term, the above
  Feynman diagrams do not suffice. Especially we are lacking the
  propagators of the fields. It is thus necessary to build an induced
  propagators by means of the multi-loop effect of the existing
  Feynman rule. These induced propagators are constructed by the
  following loop effect.
   \begin{figure}[htbp]
   \begin{center}
    \scalebox{.45}{\includegraphics{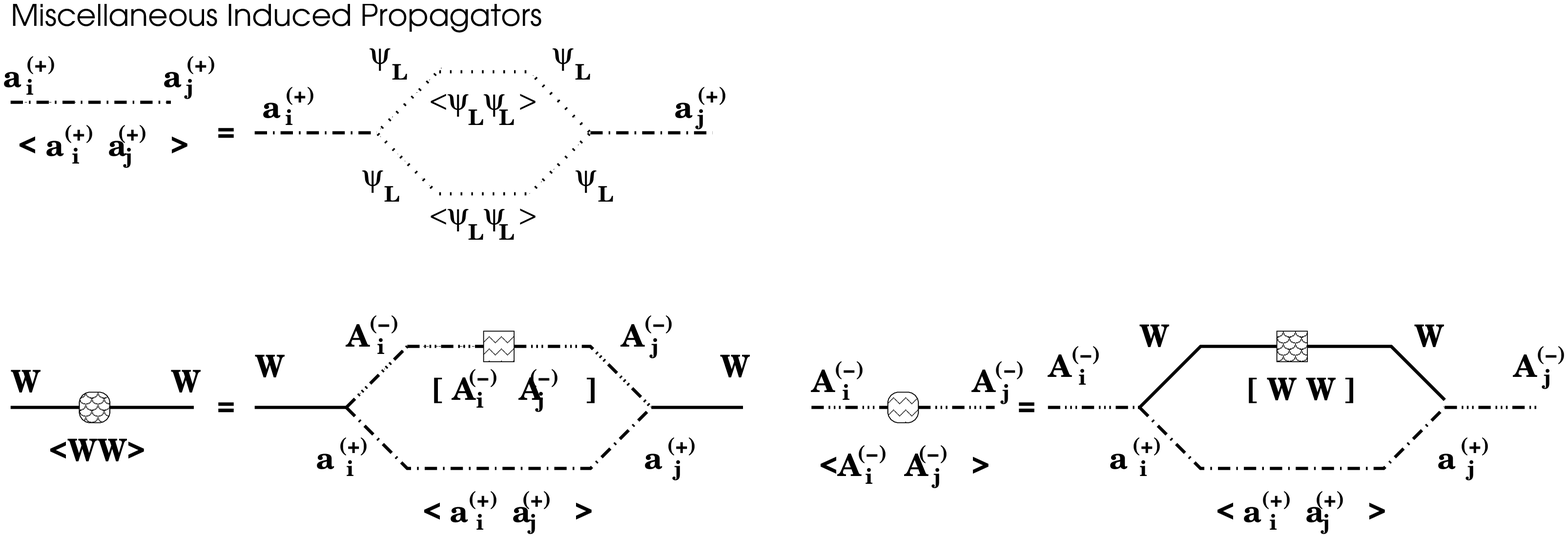} }
   \end{center}
    \caption{The induced propagators of the bosonic fields}
   \label{propwwaa}
  \end{figure}
 \begin{itemize}
  \item{$\langle a^{(+)}_{i} a^{(+)}_{j} \rangle$ : It is easy to construct
      this induced propagator. All we have to do is to connect the
      fermions using the existing fermion propagators $\langle
      \psi_{L} \psi_{L} \rangle$. }
  \item{$\langle WW \rangle$ and $\langle A^{(-)}_{i} A^{(-)}_{j}
      \rangle$ : The construction of these propagators is a difficult
      problem. Unfortunately, it is impossible to construct these
      propagators perturbatively, and we disprove their existence in
      Appendix. \ref{AZCkillwwaa}. However, this is not the end of the 
      story. Even if we fail to induce these propagators
      perturbatively, we have a choice to induce these propagators by
      means of the nonperturbative effect. These propagators may be
      induced by the following recursive structure. 
  \begin{figure}[htbp]
   \begin{center}
    \scalebox{.45}{\includegraphics{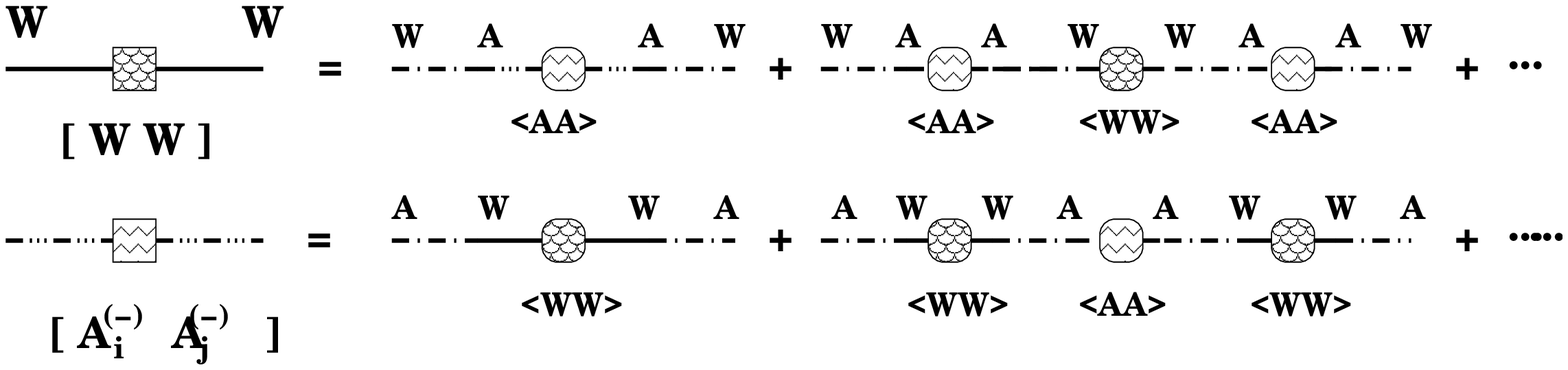} }
   \end{center}
    \caption{The induced propagators of the bosonic fields}
   \label{inducedwwaa}
  \end{figure}

      This structure is reminiscent of the self-consistency condition
      of Nambu-Jona-Lasino model. And when we consider the
      nonperturbative\footnote{Here, we mean the word 'nonperturbative' 
      by the effect not stemming from the multi-loop effect of
      Feynman diagram.} effect, there is no particular conservation
      law which prohibits the existence of the propagators $\langle WW
      \rangle$ or $\langle A^{(-)}_{i} A^{(-)}_{j} \rangle$ 
      \footnote{ The argument of the charge in Appendix \ref{AZCkillwwaa}
      is not a conservation law applicable to the non-perturbative
      framework, because this argument is based on the perturbative
      multi-loop context.}.  Throughout our discussion, we assume the
      existence of these propagators. }
 \end{itemize}
 
  Now we are ready to construct the induced propagators of this cubic
  model. In constructing the vertices of the fermionic terms
  (\ref{AZ44goal}).  The answer is now easy, and the IKKT-like vertex
  is constructed by the following procedure.
   \begin{figure}[htbp]
   \begin{center}
    \scalebox{.45}{\includegraphics{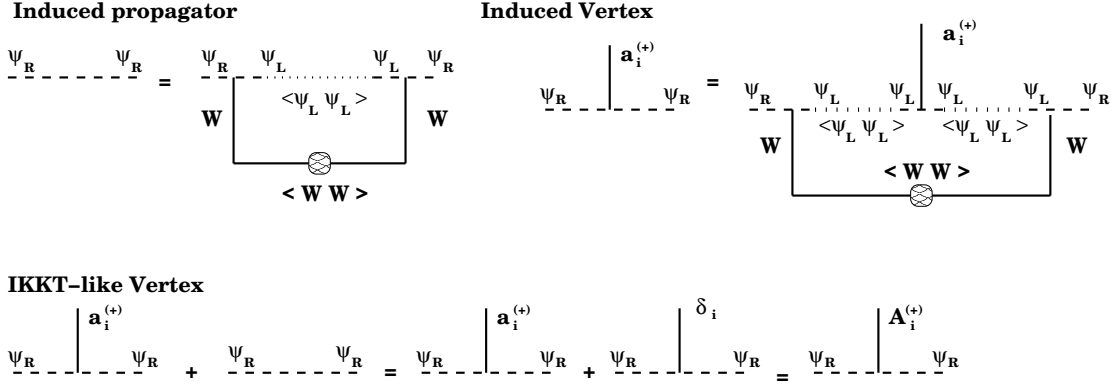} }
   \end{center}
    \caption{The induced IKKT-like vertex ${\bar \psi} \Gamma^{i}
      A^{(+)}_{i} \psi$}
   \label{truefinalanswer}
  \end{figure}
  \begin{itemize}
   \item{To construct the induced vertex corresponding to
       (\ref{AZ44goal}), we must first construct two objects. One is
       the propagator $\langle \psi_{R} \psi_{R} \rangle$. This is
       easily constructed once we admit the existence of the
       propagator $\langle WW \rangle$. }
   \item{ Another object is the vertex ${\bar \psi_{R}} \Gamma^{i}
       a^{(+)}_{i} \psi_{R}$. It is also easy to construct this vertex 
       utilizing the induced propagator $\langle WW \rangle$.}
   \item{ These two objects serve to induce our desired term ${\bar
         \psi_{R}} \Gamma^{i} A^{(+)}_{i} \psi_{R}$. Note that this induced
       propagator indicates that the kinetic term ${\bar \psi_{R}}
       \Gamma^{i} \partial_{i} \psi_{R}$. This is diagrammatically
       regarded as the vertex of $\psi_{R}$, $\psi_{R}$ and
       $\partial_{i} \sim {\hat p}_{i}$, where ${\hat p}_{i}$ is the
       classical solution (\ref{AZ44cs}), around which we have
       expanded the theory. Therefore, the sum of these two objects is 
         regarded as 
       \begin{eqnarray}
       \langle \psi_{R} \psi_{R} \rangle + {\bar \psi_{R}} \Gamma^{i}
       a^{(+)}_{i} \psi_{R} = -i {\bar \psi_{R}} \partial_{i} \psi_{R} 
       + {\bar \psi_{R}} \Gamma^{i} a^{(+)}_{i} \psi_{R} =
       {\bar \psi_{R}} \Gamma^{i} A^{(+)}_{i} \psi_{R}.
       \end{eqnarray} }
  \end{itemize}

  Now, we have completed constructing the fermionic term to be
  identified with that of IKKT model, considering the correspondence
  of ${\cal N}=2$ SUSY transformation:
   \begin{eqnarray}
    I \sim {\bar \psi_{R}} \Gamma^{i} A^{(+)}_{i} \psi_{R}.
   \end{eqnarray}

  Now, here comes two objections of the prosecutor to our argument.
   \begin{enumerate}
    \item{ One objection is that we have yet to succeed in
        decoupling the fermions. Even though we have constructed the
        fermionic term ${\bar \psi_{R}} \Gamma^{i} A^{(+)}_{i} \psi_{R}$,
        the term ${\bar \psi_{L}} \Gamma^{i} A^{(+)}_{i} \psi_{L}$ is
        still mixed in this action, and we have no system to decouple
        the  'impurity' ${\bar \psi_{L}} \Gamma^{i} A^{(+)}_{i}
        \psi_{L}$. Then, it may be artificial to extract the desired
        term  ${\bar \psi_{R}} \Gamma^{i} A^{(+)}_{i} \psi_{R}$ by
        hand, and it cannot be said that the fermionic term of IKKT
        model is {\it naturally} induced.}
    \item{ The other objection is that we have assumed the existence
        of the propagator $\langle WW \rangle$ and $\langle
        A^{(-)}_{i} A^{(-)}_{j} \rangle$. The analogue of
        Nambu-Jona-Lasino model can only assert that, {\it if} such 
        propagators {\it exist}, these are due to the nonperturbative
        effect. Like Nambu-Jona-Lasino model, we are
        required to investigate the {\it existence} of the solution
        of self-consistency condition. We have not succeeded in
        verifying the existence of these propagators.}
   \end{enumerate} 

  These are fatal objections to our discussion, however the pleader
  refutes these objections as follows. 
   \begin{enumerate}
    \item{ For the first problem, we have a viewpoint that the fermion 
        $\psi_{L}$ is integrated out in constructing the vertex
        operator ${\bar \psi_{R}} \Gamma^{i} A^{(+)}_{i}
        \psi_{R}$. Since we have performed a loop integration with
        respect to the fermionic propagator $\langle \psi_{L} \psi_{L} 
        \rangle$, it is  a natural interpretation that the $\psi_{L}$
        has been integrated out and no longer exists in the effective
        theory. We reckon this integration as the system to ostracize
        the impurity  ${\bar \psi_{L}} \Gamma^{i} A^{(+)}_{i}
        \psi_{L}$  from the theory.}
    \item{ The latter is a tough objection, and we have yet to find a
        definite answer. However, there is no physical conservation
        law to prohibit the existence of these propagators, and we
        consider the existence of these propagators to be a decent
        hypothesis.}
   \end{enumerate}

  The action with only the fermionic term of the theory is
  known to induce the bosonic part of IKKT model. The bosonic part
  stems from the one-loop effect of this fermionic vertex, as shown in 
  Fig. \ref{inducedikkt}.   These induced terms
  $(A^{(\pm)}_{i})^{2}$ are conjectured to emerge as a commutator
  $[A^{(\pm)}_{i}, A^{(\pm)}_{j} ]$. Therefore, we can speculate that
  this cubic matrix model may induce IKKT model by the multi-loop
  effect: 
   \begin{eqnarray}
    I_{IKKT} = - \frac{1}{g^{2}_{IKKT}} ( \frac{1}{4} [A_{IKKTi},
    A_{IKKTj}][A_{IKKT}^{i} , A_{IKKT}^{j}] + \frac{1}{2} {\bar
    \psi}_{IKKT} \Gamma_{i} [A_{IKKT}^{i} , \psi_{IKKT}].
   \end{eqnarray}
    \begin{figure}[htbp]
   \begin{center}
    \scalebox{.45}{\includegraphics{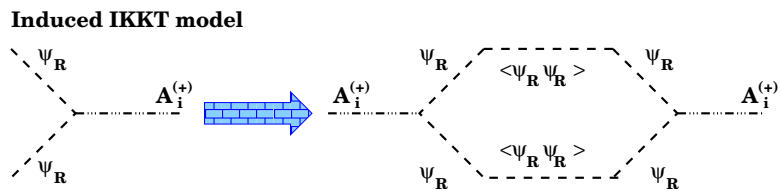} }
   \end{center}
   \caption{Induced IKKT model from the fermionic term. The bosonic
     term will be induced by the one-loop effect.}
   \label{inducedikkt}
  \end{figure} 

  Although we have developed just a hand-waving argument, it may
  at least indicate a sign that this $osp(1|32,R)$  cubic model
  has a scenario to induce IKKT model, with the induced terms agreeing
  with the correspondence of the identification of ${\cal N} =2 $
  SUSY.  

 \subsection{Summary}
  Let us conclude this section by summarizing the discussion concerning 
  the $osp(1|32,R)$ cubic matrix model. 
   \begin{itemize}
    \item{ L. Smolin suggested a new type of cubic matrix theory:
     \begin{eqnarray}
   I &=& \frac{i}{g^{2}} Tr_{N \times N} \sum_{Q,R=1}^{33}
  (( \sum_{p=1}^{32} {M_{p}}^{Q} [ {M_{Q}}^{R},
  {M_{R}}^{p}] ) - {M_{33}}^{Q} [ {M_{Q}}^{R}, {M_{R}}^{33} ] )
  \nonumber \\
    &=& \frac{i}{g^{2}} \sum_{a,b,c=1}^{N^{2}} Str( M^{a}  M^{b} M^{c} )
  Tr(T^{a} [T^{b}, T^{c}]).  \nonumber
     \end{eqnarray}
   The biggest novelty of this cubic matrix model lies in the fact
   that both  bosons and fermions are embedded in one multiplet of
   super Lie algebra $osp(1|32,R)$.  } 
   \item{This cubic model possesses two-fold ${\cal N} =2$ SUSY
       structures, each for 32 fermionic SUSY parameters. Each of
       these two-fold  SUSY structures is identified with the ${\cal
       N} =2$ SUSY of IKKT model. The correspondence of the fields of
       the cubic model and those of IKKT model is as follows.\\
  \begin{center} \begin{tabular}{|c||c|c|} \hline
    &  bosons $A_{i}$ (IKKT) & fermions $\psi$(IKKT)  \\ \hline \hline
    SUSY I   &  $A^{(+)}_{i}$ & $\psi_{R} = \frac{1 -
    \Gamma^{\sharp}}{2} \psi$ \\ \hline 
    SUSY II  &  $A^{(-)}_{i}$ & $\psi_{L} = \frac{1 +
    \Gamma^{\sharp}}{2} \psi$ \\ \hline
  \end{tabular} \end{center} 
  However, these two-fold SUSY's are not independent of each other, but
  are connected by the fields $A^{(\pm)}_{i}$, the very fields  to be
  identified with those of IKKT model.} 
 \item{We speculate by a hand-waving argument that this cubic model
     induces IKKT model by a multi-loop effect of the fermionic
     term. And we constructed the
     fermionic terms with the correspondence with the vector fields
     imposed by the identification of ${\cal N}=2$ SUSY. This can be
     regarded as the induced IKKT model,  with the bosonic part
     induced by the fermionic part.} 
  \end{itemize}

  Our discussion in this section solidifies the conjecture that
  Smolin's proposal may naturally include IKKT model, as is
  predictable from the symmetry of $osp(1|32,R)$ super Lie
  algebra. However, our scenario to derive IKKT model from Smolin's
  proposal is hand-waving, and it is an issue of interest and
  importance to pursue a more solid scenario. If we succeed in proving 
  the embedding of IKKT model in a more natural way, we will be more
  convinced that Smolin's proposal truly exceeds IKKT
  model.

\section{$gl(1|32,R) \otimes gl(N,R)$ Gauged Cubic Matrix Model}
   We have seen a new proposal for the constructive definition of
  superstring theory in the previous chapter, and we have considered
  the cubic matrix model with the multiplet belonging to
  $osp(1|32,R)$ super Lie algebra. However, it is an interesting
  problem to consider the  extended version of this model.  
  The action we investigate here is named 'gauged' action in terms of
  Smolin's proposal \cite{0002009} \cite{0006137}. The action is
  described by the following formula:
   \begin{eqnarray}
        I = \frac{1}{g^{2}} Tr_{N \times N} \sum_{Q,R=1}^{33}
        ((\sum_{p=1}^{32}  {M_{p}}^{Q} {M_{Q}}^{R}
        {M_{R}}^{p})  -{M_{33}}^{Q} {M_{Q}}^{R}
        {M_{R}}^{33} ) = \frac{1}{g^{2}} Tr_{N \times N}( Str_{33
        \times 33} M^{3}). \label{AZ51action}
   \end{eqnarray}
  This action is characterized by the fact that we have promoted the
  tensor  product of the two gauge group from that of {\it Lie groups}
  ($\times$) to  that of {\it Lie algebra} ($\otimes$). 
  The difference of these two notions has been treated in
  Appendix. \ref{AZCtensorpr}.  This action is in this sense  
  named {\it 'gauged action'} \footnote{ In contrast to the gauged
  action, the action proposed in the previous chapter is named {\it
  nongauged action}. } Although this 
  action is deprived of the translation symmetry, the symmetry of
  this action possesses far larger gauge symmetry because the direct
  product is promoted to the tensor product of two matrices. 

  Originally L. Smolin proposed this new version as a complexification 
  of $osp(1|32,R)$ matrix model, and the super Lie algebra is taken to 
  be $u(1|16,16)$. The complexification of $osp(1|32,R)$ matrix model has
  enriched the symmetry of the theory\footnote{The meaning of the
  enhancement of the gauge symmetry is explained in
  Appendix. \ref{AZCtensorpr}.}, and Smolin conjectures that
  this gauged theory may include loop quantum gravity
  \cite{0006137}. 

  We investigate a {\it real version} of Smolin's proposal, rather
  than the original $u(1|16,16)$ multiplets. The
  matrix model we pursue can be obtained by the 'analytic continuation'
  of Smolin's proposal in two respects.
   \begin{itemize}
    \item{ We adopt $gl(1|32,R)$ super Lie algebra as a symmetry of
        this action. This is a super Lie algebra composed of real
        supermatrix. }
    \item{ We take the gauge symmetry of large $N$ matrices as a real Lie 
        algebra $(N)$, rather than $su(N)$.}
   \end{itemize} 
   Throughout this chapter, we focus on the
        $gl(1|32,R) \otimes gl(N,R)$ gauge symmetry. In this section, we
        sometimes refer to the  
        relationship between our argument and the features of
        $u(1|16,16)$ gauge symmetry, which is a cousin of 
        our  $gl(1|32,R) \otimes gl(N,R)$ gauge group.

 \subsection{Definition of $gl(1|32,R)$ Super Lie Algebra}
  This section is devoted to introducing the super Lie algebra
  $gl(1|32,R)$, noting the difference from the cousin $u(1|16,16)$
  super Lie algebra, which is introduced as a complexification of
  $osp(1|32,R)$ super Lie algebra. Here we make a bit of excursion of
  the complex $u(1|16,16)$ super Lie algebra, and then we define
  $gl(1|32,R)$ super Lie algebra and compare these two super Lie algebras.
 
 \subsubsection{Excursion of $u(1|16,16)$ Super Lie Algebra}
 The very definition of this super Lie algebra is that 
 \begin{eqnarray} 
  \textrm{ If } M \in u(1|16,16),  \textrm{ then } M^{\dagger} G + GM = 0
  \textrm{ for }  G =  \left( \begin{array}{cc}  \Gamma^{0}  & 0  \\  0
  & i   \end{array} \right).
 \end{eqnarray} 
  That this is a complexification of $osp(1|32,R)$ can be seen from
  the following aspect. Unlike the $osp(1|32,R)$ super Lie algebra, we
  do not restrict $M$ to be a real supermatrix, where the reality of
  the supermatrix is defined as $M = M^{\ast} = ({^{T}
  M})^{\dagger}$. Therefore, we must replace the {\it transpose} by
  the {\it hermitian conjugate}. Note that in the real version the
  complex conjugate is equivalent to the transpose according to the
  property in Appendix. \ref{AZMA091441}. 

  We can confirm that the super Lie algebra $u(1|16,16)$ actually
   closes in totally the same fashion as in $osp(1|32,R)$. The
   legitimacy of the metric also stems from the same logic as in
   $osp(1|32,R)$.  
   We specify the element of $u(1|16,16)$ group according to the
 above definition.  The result is that 
  \begin{eqnarray}
   \textrm{ If } M \in u(1|16,16) , \textrm{ then } M = \left
   ( \begin{array}{cc} m  & \psi \\   i {\bar \psi} & v \end{array}
   \right). \label{AZ51su11616}
  \end{eqnarray}
 \begin{itemize}
  \item{ $v$ is restricted to be a pure imaginary number.} 
  \item{ $u_{\mu_{1}}$ ,$u_{\mu_{1} \mu_{2}}$ and
      $u_{\mu_{1} \cdots \mu_{5}}$ are real numbers, while $u$, $u_{\mu_{1}
      \mu_{2} \mu_{3}}$ and $u_{\mu_{1} \cdots \mu_{4}}$ are
      pure imaginary numbers.} 
  \item{Here ${\bar \psi}$ denotes $\psi^{\dagger} \Gamma^{0}$.}
  \end{itemize}
{\sf (Proof)
 This result is derived from the very definition of the super Lie
 algebra $u(1|16,16)$. The complex conjugate of supermatrices is defined
 in Appendix. \ref{AZMA091441}:
  \begin{eqnarray} M^{\dagger} G + GM =  \left( \begin{array}{cc}
 m^{\dagger}   & \phi \\   \psi^{\dagger} & v^{\dagger} \end{array}
 \right)  \left( \begin{array}{cc} \Gamma^{0}   & 0 \\ 0 & i
 \end{array} \right) +  \left( \begin{array}{cc} \Gamma^{0}   & 0 \\ 0 & i
 \end{array} \right)  \left( \begin{array}{cc} m   & \psi \\
 \phi^{\dagger} & v \end{array} \right) = \left( \begin{array}{cc}
 m^{\dagger} \Gamma^{0} + \Gamma^{0} m  & i \phi + \Gamma^{0} \psi \\
 \psi^{\dagger} \Gamma^{0} + i \phi^{\dagger}  & i( v + v^{\dagger})
 \end{array} \right) = 0.
 \end{eqnarray}
  \begin{itemize}
  \item{ It is a trivial matter to understand that $v$ is a pure
      imaginary number from $v + v^{\dagger} = 0$. }
  \item{We first investigate the constraint of the bosonic matrix
      $m$. These are decomposed, as we have done in $osp(1|32,R)$
      matrix model, in terms of the basis of gamma matrices:
      \begin{eqnarray}
       m &=& u {\bf 1} + u_{\mu_{1}} \Gamma^{\mu_{1}} + \frac{1}{2!}
       u_{\mu_{1} \mu_{2}} \Gamma^{\mu_{1} \mu_{2}} + \frac{1}{3!}
       u_{\mu_{1} \mu_{2} \mu_{3}} \Gamma^{\mu_{1} \mu_{2} \mu_{3}} +
       \frac{1}{4!} u_{\mu_{1} \cdots \mu_{4}} \Gamma^{\mu_{1} \cdots
       \mu_{4}} + \frac{1}{5!} u_{\mu_{1} \cdots \mu_{5}}
       \Gamma^{\mu_{1} \cdots \mu_{5}}. \nonumber
      \end{eqnarray}
     We utilize the relationship of the gamma matrices
     ${^{T} \Gamma^{\mu_{1} \cdots \mu_{k}} } \Gamma^{0} = \pm
     \Gamma^{0} \Gamma^{\mu_{1} \cdots \mu_{k}}$ with the sign $+$ for 
     $k=1,2,5$ and the sign $-$ for $k=0,3,4$. The reveals that the
     coefficients must satisfy the following results.
      \begin{itemize}
        \item{ $(u_{\mu_{1} \cdots \mu_{k}})^{\ast} = u_{\mu_{1}
              \cdots \mu_{k}}$ for $k=1,2,5$. These are thus
              restricted to be real numbers.}
        \item{ $(u_{\mu_{1} \cdots \mu_{k}})^{\ast} = - u_{\mu_{1}
              \cdots \mu_{k}}$ for $k=0,3,4$ These are thus restricted 
              to be pure imaginary numbers.} 
     \end{itemize}  }
  \item{We investigate the relationship between two fermions $\psi$ and
  $\phi^{\dagger}$ utilizing the result  $ i \phi + \Gamma^{0} \psi=0$: 
 \begin{eqnarray}
  \phi^{\dagger} = (i \Gamma^{0} \psi)^{\dagger} = (-i)
  \psi^{\dagger} ({^{T} \Gamma^{0}})  = (-i) \psi^{\dagger}
  (-\Gamma^{0})  = i {\bar \psi}.
 \end{eqnarray}
 We can verify that this is consistent with the condition
 ${\psi}^{\dagger} \Gamma^{0} + i {\phi}^{\dagger} = 0 $.  }
  \end{itemize}
 We are thus finished with the determination of the elements of
 $U(1|16,16)$ super Lie algebra. (Q.E.D.)
 }

  The important property of $u(1|16,16)$ super Lie algebra is
  that these can be uniquely decomposed into the direct sum of two
  different representations of $osp(1|32,R)$. We introduce two
  different representations of $osp(1|32,R)$ super Lie algebra
   \begin{eqnarray}
   &\clubsuit& {\cal H} \stackrel{def}{=} \{ M = 
   \left( \begin{array}{cc} m_{h} & \psi_{h} \\ i {\bar \psi_{h}} & 0
   \end{array} \right) | m_{h} = u_{\mu_{1}} \Gamma^{\mu_{1}} +
   \frac{1}{2!} u_{\mu_{1} \mu_{2}} \Gamma^{\mu_{1} \mu_{2}} +
   \frac{1}{5!} u_{\mu_{1} \cdots \mu_{5}} \Gamma^{\mu_{1} \cdots
   \mu_{5}}, \nonumber \\
  & &  \hspace{45mm} u_{\mu_{1}}, u_{\mu_{1} \mu_{2}}, u_{\mu_{1}
   \cdots \mu_{5}}, \psi_{h} \in {\cal R}  \}, \nonumber \\
  &\clubsuit&  {\cal A}' \stackrel{def}{=} \{ M = 
   \left( \begin{array}{cc} m_{a} & i \psi_{a} \\ {\bar \psi_{a}}
   & iv \end{array} \right) | m_{a} = u + \frac{1}{3!} u_{\mu_{1}
   \mu_{2} \mu_{3}} \Gamma^{\mu_{1} \mu_{2} \mu_{3}} + \frac{1}{4!}
   u_{\mu_{1} \cdots \mu_{4}} \Gamma^{\mu_{1} \cdots
   \mu_{4}}, \nonumber \\
  & & \hspace{45mm} u, u_{\mu_{1} \mu_{2} \mu_{3}}, u_{\mu_{1} \cdots
   \mu_{4}}, i \psi_{a}, iv \in (\textrm{pure imaginary}) \}. \nonumber 
   \end{eqnarray}

  And let $H$ and $A'$ be the element of ${\cal H}$ and ${\cal A}'$
  respectively. Because these are real (pure imaginary), these
  elements  respectively satisfy $H^{\dagger} = {^{T} H'}$ and
  $A'^{\dagger} = - {^{T} A'}$. Therefore, these elements satisfy the
  following property
   \begin{eqnarray}
     {^{T} H} G + G H = 0, \textrm{ for } H \in {\cal H}, \hspace{3mm} 
     {^{T} A}' G - G A' = 0, \textrm{ for } A' \in {\cal A}'.
     \label{AZ51hasu}
   \end{eqnarray}
  The set ${\cal H}$ is, by definition, $osp(1|32,R)$ super Lie
  algebra itself.
  We investigate an important property of the subset of these two
  subalgebras. The commutation and anti-commutation relations are
  properties of grave importance in getting  acquainted with the
  Algebras of these groups.
    \begin{eqnarray}
    & & (1) [H_{1} , H_{2}  ] \in {\cal H}, \hspace{3mm}
        (2) [H     , A'     ] \in {\cal A}',  \hspace{3mm}
        (3) [A'_{1}, A'_{2} ] \in {\cal H}, \nonumber \\
    & & (4)\{H_{1} , H_{2} \} \in {\cal A}', \hspace{3mm}
        (5)\{H     , A'    \} \in {\cal H}, \hspace{3mm}
        (6)\{A'_{1}, A'_{2}\} \in {\cal A}'. \label{AZ51comhasu}
   \end{eqnarray} 
   where $H, H_{1}, H_{2} \in {\cal H}$ and $A', A'_{1}, A'_{2} \in {\cal
   A'}$. \\
    {\sf (Proof)
 These properties can be verified by noting the properties
 (\ref{AZ51hasu}). 
  \begin{enumerate}
   \item{${^{T}[H_{1}, H_{2}]} G 
   = {^{T}H_{2}} {^{T}H_{1}} G -{^{T}H_{1}} {^{T}H_{2}} G
   = {^{T} H_{2}}(- G H_{1}) - {^{T}H_{1}} (-G H_{2})
   = G H_{2} H_{1} - G H_{1} H_{2} = -  G [H_{1}, H_{2}]$, }
    \item{${^{T}[H, A']} G
   = {^{T}A'} {^{T}H} G - {^{T}H} {^{T}A'} G   = {^{T}A'} ( -G H) -
   {^{T}H} (G A') = - G A' H + G H A' = G [H, A']$,} 
    \item{${^{T} [A'_{1}, A'_{2}]} G
   = {^{T}A'_{2}} {^{T}A'_{1}} G - {^{T}A'_{1}} {^{T}A'_{2}} G
   = {^{T}A'_{2}} ( G A'_{1} ) - {^{T}A'_{1}} (G A'_{2})
   = G A'_{2} A'_{1} - G A'_{1} A'_{2} = - G [A'_{1}, A'_{2}]$,}
    \item{${^{T}\{H_{1}, H_{2} \}} G 
   = {^{T}H_{2}} {^{T}H_{1}} G + {^{T}H_{1}} {^{T}H_{2}} G 
   = {^{T}H_{2}} (-G H_{1} ) + {^{T}H_{1}} (-G H_{2})
   = G H_{2} H_{1} + G H_{1} H_{2} = G \{ H_{1}, H_{2} \}$,}
   \item{${^{T}\{ H_{1}, A'_{1} \}} G 
   = {^{T}A'_{1}} {^{T}H_{1}} G + {^{T}H_{1}} {^{T}A'_{1}} G 
   = {^{T}A'_{1}} (-G H_{1}) + {^{T}H_{1}} (G A'_{1})
   = - G A'_{1} H_{1} -  G H_{1} A'_{1} = - G \{ H_{1}, A'_{1} \}$,}
  \item{${^{T} \{ A'_{1}, A'_{2} \}} G
   = {^{T}A'_{2}} {^{T}A'_{1}} G + {^{T} A'_{1}} {^{T}A'_{2}} G
   = {^{T}A'_{2}} G A'_{1} + {^{T}A'_{1}} G A'_{2}
   = G A'_{2} A'_{1} + G A'_{1} A'_{2} = G \{ A'_{1}, A'_{2} \}$.}
  \end{enumerate} 
  This completes the proof of the above properties. (Q.E.D.)
 }

  Utilizing these relations, we can discern that
  ${\cal A}'$, as well as ${\cal H}'$ is the representations of $osp(1|32,R)$
  and also  that  the algebra ${\cal A}'$ is a
         representation of $osp(1|32,R)$ super Lie algebra by the
         commutation relation  $(2) [H, A' ] \in {\cal
         A}'$ for $H \in {\cal H}$ and $A' \in {\cal A}'$. This
         commutation relation states that $A'$ remain in the super 
         Lie algebra ${\cal A}'$ after the 
         infinitesimal translation by the elements $H \in {\cal
         H}$. In this sense, we can understand that ${\cal A}'$ is
         another representation of $osp(1|32,R)$.

  The introduction of these two representations of $osp(1|32,R)$
  teaches us the relationship of $osp(1|32,R)$ and $u(1|16,16)$
  super Lie algebras. ${\cal H} ( = osp(1|32,R))$ is a real part of
  $u(1|16,16)$ Lie algebra, while ${\cal A}'$ is its imaginary
  part. It is clear that the elements of $u(1|16,16)$ can be
  uniquely decomposed into the direct sum of ${\cal H}$ and ${\cal
  A}'$.
  \begin{eqnarray}
    u(1|16,16) \equiv {\cal H} \oplus {\cal A}',
  \end{eqnarray}
 where $\oplus$ denotes the direct sum of two sets.
  
 \subsubsection{Definition of $gl(1|32,R)$ Super Lie Algebra}
  The definition of $gl(1|32,R)$ super Lie algebra is, per se, simple:
   \begin{eqnarray}
     &\clubsuit&  \textrm{If } M \in gl(1|32,R), \hspace{3mm} \textrm
     { then } M = \left( \begin{array}{cc} m & \psi \\ i {\bar \phi} & 
     v \end{array} \right).
   \end{eqnarray}
  \begin{itemize}
   \item{ $m$ is an element of the Lie algebra $gl(32,R)$, id est, $m$
       is allowed to be {\it an arbitrary $32 \times 32$ bosonic
       matrix}. Decomposing this by the gamma matrices, this can be
       expressed by 
        \begin{eqnarray}
         m = u {\bf 1} + u_{\mu_{1}} \Gamma^{\mu_{1}} + \frac{1}{2!}
         u_{\mu_{1} \mu_{2}} \Gamma^{\mu_{1} \mu_{2}} + \frac{1}{3!}
         u_{\mu_{1} \mu_{2} \mu_{3}} \Gamma^{\mu_{1} \mu_{2} \mu_{3}}
         + \frac{1}{4!} u_{\mu_{1} \cdots \mu_{4}} \Gamma^{\mu_{1}
         \cdots \mu_{4}} + \frac{1}{5!} u_{\mu_{1} \cdots \mu_{5}}
         \Gamma^{\mu_{1} \cdots \mu_{5}}, 
        \end{eqnarray} 
       where the coefficients $u_{\mu_{1} \cdots }$ are all real numbers.}
   \item{ $\psi$ and $\phi$ are independent fermionic vectors. Each of 
       them possesses 32 components, and the components are fermionic
       real number.}
   \item{ $v$ is also a real number.}
  \end{itemize}
  The definition of $gl(1|32,R)$ states nothing. This definition just
  states that an arbitrary real $33 \times 33$ supermatrix is an eligible
  member of the super Lie algebra $gl(1|32,R)$. Although this
  definition does not give any restriction to the elements, the
  correspondence with the complex group $u(1|16,16)$ is an
  interesting aspect of $gl(1|32,R)$ super Lie algebra. Since $\psi$
  and $\phi$ are independent fermionic vectors, these can be rewritten 
  as 
   \begin{eqnarray}
    \psi = \psi_{1} + \psi_{2}, \hspace{2mm} \phi = \psi_{1} - \psi_{2}.
   \end{eqnarray}
  And the bosonic $32 \times 32$ matrices are separated by $m = m_{1}
  + m_{2}$, where
   \begin{eqnarray}
    & & m_{1} = u_{\mu_{1}} \Gamma^{\mu_{1}} + \frac{1}{2!} u_{\mu_{1} 
    \mu_{2}} \Gamma^{\mu_{1} \mu_{2}} + \frac{1}{5!} u_{\mu_{1} \cdots 
    \mu_{5}} \Gamma^{\mu_{1} \cdots \mu_{5}}, \textrm{ where }
    u_{\mu_{1}}, u_{\mu_{1} \mu_{2}}, u_{\mu_{1} \cdots \mu_{5}}
    \textrm{ are real numbers.} \nonumber \\
    & & m_{2} = u {\bf 1} + \frac{1}{3!} u_{\mu_{1} \mu_{2} \mu_{3}}
    \Gamma^{\mu_{1} \mu_{2} \mu_{3}} + \frac{1}{4!} u_{\mu_{1} \cdots
    \mu_{4}} \Gamma^{\mu_{1} \cdots \mu_{4}}, \textrm{ where } u,
    u_{\mu_{1} \mu_{2} \mu_{3}}, u_{\mu_{1} \cdots \mu_{4}} \textrm
    { are real numbers.} \nonumber
   \end{eqnarray}
  Then, we define the sets ${\cal H}$ and ${\cal A}$ as follows.
  \begin{eqnarray}
   &\clubsuit& {\cal H} \stackrel{def}{=} \{ M = 
   \left( \begin{array}{cc} m_{1} & \psi_{1} \\ i {\bar \psi_{1}} & 0
   \end{array} \right) | m_{1} = u_{\mu_{1}} \Gamma^{\mu_{1}} +
   \frac{1}{2!} u_{\mu_{1} \mu_{2}} \Gamma^{\mu_{1} \mu_{2}} +
   \frac{1}{5!} u_{\mu_{1} \cdots \mu_{5}} \Gamma^{\mu_{1} \cdots
   \mu_{5}}, \nonumber \\
  & &  \hspace{45mm} u_{\mu_{1}}, u_{\mu_{1} \mu_{2}}, u_{\mu_{1}
   \cdots \mu_{5}}, \psi_{1} \in {\cal R}  \},
   \nonumber \\
  &\clubsuit&  {\cal A} \stackrel{def}{=} \{ M = 
   \left( \begin{array}{cc} m_{2} & \psi_{2} \\  -i {\bar \psi_{2}}
   & v \end{array} \right) | m_{2} = u + \frac{1}{3!} u_{\mu_{1}
   \mu_{2} \mu_{3}} \Gamma^{\mu_{1} \mu_{2} \mu_{3}} + \frac{1}{4!}
   u_{\mu_{1} \cdots \mu_{4}} \Gamma^{\mu_{1} \cdots
   \mu_{4}}, \nonumber \\
  & &  \hspace{45mm} u, u_{\mu_{1} \mu_{2} \mu_{3}}, u_{\mu_{1} \cdots 
   \mu_{4}}, \psi_{2}, v \in {\cal R} \}.
   \nonumber 
  \end{eqnarray}
  The super Lie algebra $gl(1|32,R)$ is clearly the direct sum of
  these two super Lie algebras 
   \begin{eqnarray}
    gl(1|32,R) = {\cal H} \oplus {\cal A}.
   \end{eqnarray}

  These two subalgebras are also the two different representations of
  $osp(1|32,R)$ super Lie algebra. ${\cal H}$ is $osp(1|32,R)$ itself, 
  and the same super Lie algebra as in introduced in $u(1|16,16)$.
  On the other hand, the  subalgebra ${\cal A}$ is ${\cal
  A}' = i {\cal A}$.\footnote{ This means that, if $A \in {\cal A}$,
  then $i A \in {\cal A}'$.} 
  And the elements of these subalgebras readily satisfy
   \begin{eqnarray}
   {^{T} H} G + G H = 0, \textrm{ for } H \in {\cal H},
   \hspace{3mm} {^{T} A} G - GA = 0, \textrm{ for } A \in {\cal A}.   
   \label{AZ51ha}
   \end{eqnarray}
  And it is clear that these two subalgebras obey
  totally the same commutation relations as those of  ${\cal H}$ and ${\cal 
  A}'$: 
   \begin{eqnarray}
    & & (1) [H_{1}, H_{2} ] \in {\cal H}, \hspace{3mm}
        (2) [H    , A     ] \in {\cal A},  \hspace{3mm}
        (3) [A_{1}, A_{2} ] \in {\cal H}, \nonumber \\
    & & (4)\{H_{1}, H_{2}\} \in {\cal A}, \hspace{3mm}
        (5)\{H    , A    \} \in {\cal H}, \hspace{3mm}
        (6)\{A_{1}, A_{2}\} \in {\cal A}, \label{AZB1comha}
   \end{eqnarray} 
  where $H, H_{1}, H_{2} \in {\cal H}$ and $A, A_{1}, A_{2} \in {\cal
  A}$. The proof is completely the same as that of
  (\ref{AZ51comhasu}), and we  do not repeat it.  
  The commutation relation $[{\cal H}, {\cal A}] \in {\cal A}$
  indicates that ${\cal A}$ is a representation of $osp(1|32,R)$ super 
  Lie algebra. \\

  Now, the relationship of the three super Lie algebras $osp(1|32,R)$,
  $u(1|16,16)$ and $gl(1|32,R)$ is clear. We have seen that both
  $u(1|16,16)$ and $gl(1|32,R)$ are represented by the direct sum
  of two different representations of $osp(1|32,R)$:
   \begin{eqnarray}
    u(1|16,16) = {\cal H} \oplus {\cal A}' ,\hspace{3mm} 
    gl(1|32,R)   = {\cal H} \oplus {\cal A}. \label{AZ51lovetriangle}
   \end{eqnarray}
  The relationship between ${\cal A}$ and ${\cal A}'$ is 
   \begin{eqnarray}
    {\cal A}' = i {\cal A} \Rightarrow \textrm{ If } A \in {\cal A},
    \textrm{ then } i A \in {\cal A}'.
   \end{eqnarray}
  In this sense, we can regard $gl(1|32,R)$ super Lie algebra as {\it
  the analytic continuation} of $u(1|16,16)$. Although we adopt a
  matrix theory with the gauge symmetry $gl(1|32,R)$ unlike Smolin's
  original proposal \cite{0006137}, we note that $gl(1|32,R)$ is a
  cousin of the original $u(1|16,16)$.

  \subsection{Promotion of the elements to large $N$ matrices.}
   The grave difference from the non-gauged case emerges when we
   promote the elements of $gl(1|32,R)$ super Lie algebra to large $N$
   matrices. In the non-gauged case, the multiplets of the theory are
   the generators of the gauge group $OSp(1|32,R) \times SU(N)$.  
   And the same is true of IKKT model, with the gauge group being
   $SO(9,1) \times  SU(N)$. However, {\it this no longer holds true of
   the proposal of gauged theory} \cite{0006137}. The meaning of {\it
   gauged matrix theory} is that  
   the gauge symmetry is enhanced from the Lie algebra of the two gauge
   groups to the tensor product of the two Lie algebras of the gauge
   symmetry. The notions of the tensor product are defined in
   Appendix. \ref{AZCtensorpr}.   

   This property drastically changes the meaning of closed
   algebra. The Lie algebra of the gauge symmetry must close  
   with respect to the {\it commutator of the tensor
   products}: 
    \begin{eqnarray}
     [A \otimes B, C \otimes D] = \frac{1}{2} (\{ A ,C \} \otimes [B,D]) +
     \frac{1}{2} ([A,C] \otimes \{ B, D \}). \label{AZ53truecomm}
    \end{eqnarray}
   This clarifies the reason for the choice of the gauge group both
   for Smolin's original version and our $gl(1|32,R) \otimes gl(N,R)$
   gauge symmetry. We investigate the gauged theory one by one. We have
   seen the commutation relations of two representations of
   $osp(1|32,R)$ super Lie algebra in (\ref{AZ51comhasu}) ( the
   version for its analytic  continuation is the same, and this is listed in
   (\ref{AZB1comha}). )
   We need the commutation relations of $N \times N$ hermitian and
   anti-hermitian matrices, and these relations are listed in
   Appendix. \ref{AZCsun}, including their proof. But we repeat the
   result because these play an essential role in the analysis of the
   gauged Lie algebra:
  \begin{eqnarray}
  & & (1) [h_{1}, h_{2} ] \in {\bf A}, \hspace{3mm}
      (2) [h    , a     ] \in {\bf H},    \hspace{3mm}
      (3) [a_{1}, a_{2} ] \in {\bf A}, \nonumber \\
  & & (4)\{h_{1}, h_{2}\} \in {\bf H},    \hspace{3mm}
      (5)\{h    , a    \} \in {\bf A}, \hspace{3mm}
      (6)\{a_{1}, a_{2}\} \in {\bf H}. \label{AZ52suncomm}
 \end{eqnarray}
  \begin{itemize}
       \item{Hermitian matrices: ${\bf H} = \{ M \in M_{N \times N}({\bf
            C}) | M^{\dagger} = M \} $. $h, h_{1} h_{2}$ belong to
            ${\bf H}$. }
    \item{Anti-hermitian matrices : ${\bf A} = \{ M \in M_{N \times N}
          ({\bf C}) | M^{\dagger} = -M \} $. $a, a_{1}, a_{2}$ belong
          to ${\bf A}$. }
    \item{In this context, we do not use the elements
            of the super Lie algebra ${\cal H}$ and ${\cal A}$. Do not 
            confuse this ${\bf H}$ with $H \in {\cal H}$ and so on.}
   \end{itemize}

    \subsubsection{Gauged version of $osp(1|32,R)$ matrix model}
    First we clarify the reason why we must complexify  $osp(1|32,R)$
    super Lie algebra in considering its gauged theory.  The naive
    alteration of the product from the Lie algebra of $OSp(1|32,R)
    \times SU(N)$ gauge group to $osp(1|32,R) \otimes {\bf H}$ never
    constitutes a closed set. It is due to the commutation relation
    (\ref{AZ53truecomm}) \footnote{ We discuss where the commutators
    or the anti-commutators of the group belong, and we are sloppy
    about the coefficient $\frac{1}{2}$. }. Utilizing the commutation
    relations (\ref{AZ51comhasu}) and (\ref{AZ52suncomm}), we
    obtain \footnote{ Since we are now considering the complex
    version, we consider the group ${\cal H}$ and ${\cal A}'$. }
     \begin{eqnarray}
      [ ({\cal H} \otimes {\bf H}) , ({\cal H} \otimes {\bf H}) ] = 
     (\{ {\cal H}, {\cal H} \} \otimes [{\bf H},{\bf H}]) \oplus 
     ( [ {\cal H}, {\cal H}] \otimes \{ {\bf H}, {\bf H} \})      =
    ({\cal A}' \otimes {\bf A}) \oplus ({\cal H} \otimes {\bf H}).
     \label{AZ52naivefail}
     \end{eqnarray} 
    This commutation relation teaches us that the set  ${\cal H}
    \otimes {\bf H} = osp(1|32,R) \otimes su(N)$ does not close, and thus is
    not an eligible gauge symmetry.

    In order to remedy this situation, we enlarge the gauge symmetry into 
    $ ({\cal H} \otimes {\bf H}) \oplus ({\cal A}' \otimes {\bf A})$. This is
    verified to be a closed Lie algebra by noting the commutation
    relation (\ref{AZ52naivefail}) and 
    \begin{eqnarray}
     [ ( {\cal A}' \otimes {\bf A}) , ({\cal A}' \otimes {\bf A}) ] = 
     ( \{ {\cal A}', {\cal A}' \} \otimes [{\bf A},{\bf A}] ) \oplus 
     ( [ {\cal A}', {\cal A}'] \otimes \{ {\bf A}, {\bf A} \} )      =
     ( {\cal A}' \otimes {\bf A} ) \oplus ( {\cal H} \otimes {\bf H} ).
     \label{AZ52close111} 
   \end{eqnarray}

   This indicates that $({\cal H} \otimes {\bf H}) \oplus ({\cal A}' \otimes
   {\bf A})$ is actually an eligible gauge symmetry.
   We have clarified that $osp(1|32,R) \check{\otimes} su(N)$, which
   is the smallest closed Lie algebras including $osp(1|32,R) \otimes
   su(N)$, is not  $osp(1|32,R) \otimes su(N)$ itself but 
    \begin{eqnarray}
      osp(1|32,R) \check{\otimes} su(N) = ({\cal H} \otimes {\bf H})
      \oplus ({\cal A}' \otimes{\bf A}). 
    \end{eqnarray}
    At the same time,
   this discussion teaches us how to promote the elements of c-number
   $U(1|16,16)$ to large $N$ matrices.  
    \begin{itemize}
     \item{ The elements of ${\cal H} = osp(1|32,R)$, the real
         part of $u(1|16,16)$ should be promoted to hermitian
         matrices.}
     \item{ The elements of ${\cal A}$, the imaginary part of
         $u(1|16,16)$ should be promoted to anti-hermitian matrices.}
    \end{itemize}
 
  \subsubsection{Gauged version of $gl(1|32,R)$ matrix model}
  We likewise investigate what gauge group of $N \times N$ matrices
  is appropriate. First, let us consider the conventional gauge group 
  $SU(N)$. Noting that $gl(1|32,R)$ super Lie algebra is a direct sum
  $gl(1|32,R) = {\cal H} \oplus {\cal  A}$. We first consider the
  following tensor product as a candidate of the gauge symmetry: 
    \begin{eqnarray}
       gl(1,32|R) \otimes su(N) = ({\cal H} \oplus {\cal A}) \otimes
       {\bf H}. 
    \end{eqnarray}
 It turns out that this set does not close with respect to the
  commutator (\ref{AZ53truecomm}):
    \begin{eqnarray}
   & &   [ ({\cal H} \oplus {\cal A}) \otimes {\bf H}, ({\cal H}
     \oplus {\cal A}) \otimes {\bf H}] =  
     ( \{ ({\cal H} \oplus {\cal A}), ({\cal H} \oplus {\cal A}) \}
      \otimes [{\bf H},{\bf H}] ) \oplus ( [ ({\cal H} \oplus {\cal A}),
      ({\cal H} \oplus {\cal A}) ] \otimes \{ {\bf H}, {\bf H} \}
      \nonumber \\ 
   &=& (({\cal H} \oplus {\cal A}) \otimes {\bf A} ) \oplus (({\cal H} \oplus
      {\cal A}) \otimes {\bf H} ) = ({\cal H} \oplus {\cal A}) \otimes
      ( {\bf H} \oplus {\bf A} ).
    \end{eqnarray}
  This indicates that the $gl(1|32,R)$ gauged cubic matrix theory
  closes if we enlarge the algebra of $N \times N$ matrices to
  {\it the set of all $M_{N}({\bf C})$ matrices}. Id est, the set
   $gl(1,32|R) \otimes gl(N,R)$ is no longer an eligible gauge group but
    \begin{eqnarray}
      gl(1|32,R) \check{\otimes} su(N) = ({\cal H} \oplus {\cal A}) \otimes
      ( {\bf H} \oplus {\bf A} ) = gl(1|32,R) \otimes M_{N}({\bf C}).
    \end{eqnarray}

  However, this gauge symmetry is too large. And we consider another
  choice of closed gauge group. If we take the gauge group to be
  $M_{N} ({\bf R}) = gl(N,R)$, instead of $M_{N} ({\bf C}) = {\bf H}
  \oplus {\bf A}$, the tensor product 
  \begin{eqnarray}
   gl(1|32,R) \otimes gl(N,R)
  \end{eqnarray}
  trivially constitutes a closed Lie algebra. In other words,
  $gl(1|32,R) \check{\otimes} gl(N,R)$, which is the smallest closed Lie 
  algebra including $gl(1|32,R) \otimes gl(N,R)$, is 
   \begin{eqnarray}
     gl(1|32,R) \check{\otimes} gl(N,R) = gl(1|32,R) \otimes gl(N,R).
   \end{eqnarray}
 
  Therefore, we choose to enlarge the gauge symmetry to $gl(N,R)$ Lie
  algebra. This is an unhappy result, because we must abandon the
  virtues of conventional $SU(N)$ gauge group. We investigate the
  physics constituted by $GL(N,R)$ gauge group in the next section.

  \subsection{Novelty of $GL(N,R)$ Gauge Group}
  This section is devoted to the discussion of how decent a physics
  $GL(N,R)$ gauge group constitutes. We usually consider such gauge
  group as $SU(N)$, and IKKT model or Smolin's original proposal is no 
  exception. The gauge group $SU(N)$ possesses many virtues, such as
 the ability to construct a canonical pair $[q,p]=+i$, and the
  compactness of the group. However, as we have seen in the preceding 
  section, we have no choice but to pursue the unfamiliar gauge
  group $GL(N,R)$, instead of $SU(N)$ if we are to consider real
  $gl(1|32,R)$ gauge group.  Then it is necessary to gain
  insight into the physics of $GL(N,R)$ gauge group, and we are
  interested in the decency of the world of $GL(N,R)$ gauge theory. 

  \subsubsection{Generators and Structure Constants}
    We consider a matrix theory with real gauge group $GL(N,R)$. We
  explain this group by comparing the ordinary $SU(N)$ gauge
  theory. For simplicity we compare a toy model $GL(2,R)$ and $SU(2)$
  gauge theory. The generators of both Lie algebras are \footnote{ We
  attach the subscript $^{G}$ to the generators of $GL(N,R)$ group here,
  just in order to distinguish them from those of $SU(N)$. After this
  section we omit this awkward subscript.} 
   \begin{itemize}
    \item{ $GL(2,R)$: $T^{1G} =\frac{1}{2} \left( \begin{array}{cc} 0 & 1 \\ 
            1 & 0 \end{array} \right)$, 
           $T^{2G} = \frac{1}{2} \left( \begin{array}{cc} 0 & -1 \\ 1
            & 0 \end{array} \right)$, 
           $T^{3G} = \frac{1}{2}\left( \begin{array}{cc} 1 & 0 \\ 0 & -1
            \end{array} \right) $.}
    \item{ $SU(2,C)$ : $T^{1} = \frac{1}{2}\left( \begin{array}{cc} 0 & 1 \\
            1 & 0 \end{array} \right)$, 
           $T^{2} = \frac{1}{2}  \left( \begin{array}{cc} 0 & -i \\ i
            & 0 \end{array} \right) $ , 
           $T^{3} = \frac{1}{2} \left( \begin{array}{cc} 1 & 0 \\ 0 & -1
            \end{array} \right) $.}
   \end{itemize}
  The generators of $SU(N)$ Lie group are composed of only
  hermitian matrices by definition. However, this is not the case with 
  the generators of $GL(N,R)$ Lie group. As we see from the above
  trivial example, the generators of $GL(N,R)$ includes anti-hermitian
  (and hence anti-symmetric, because this group is real)
  matrices. These generators are related with those of $SU(N)$ by
  analytic continuation. In the above case,
   \begin{eqnarray}
    T^{2G} = \frac{1}{2} \left( \begin{array}{cc} 0 & -1 \\ 1 & 0
    \end{array} \right) \leftrightarrow T^{2} = \frac{1}{2}
     \left( \begin{array}{cc} 0 & -i \\ i & 0 \end{array}  \right). 
   \end{eqnarray}
 
 Let us have a look at the properties of the structure constant of
 $gl(N,R)$ Lie algebra.  We define the structure 
 constant $f'^{G}_{abc}$  and $d'^{G}_{abc}$ as a real number satisfying 
  \begin{eqnarray}
    [ T^{aG}, T^{bG}] = f'^{G}_{abc} T^{cG}, \hspace{2mm} \{ T^{aG},
    T^{bG} \} = d'^{G}_{abc} T^{cG}.
  \end{eqnarray}
  There are two major differences from the $su(N)$ Lie algebra.
  First note that we do not have to multiply $i$. Since the generators 
  are real matrix, their product is also a real matrix. This is
  different from the case of $su(N)$ Lie algebra, in which the commutator of 
  hermitian matrices are {\it anti-hermitian}. 
 
  The second difference from $su(N)$ is that the cyclic symmetry of
  the structure constant no longer holds true of $gl(N,R)$ Lie
  algebra\footnote{For example,  $f'^{G}_{123} = 2$, $f'^{G}_{231} =
  2$, and  $f'^{G}_{312} = -2$ for the above basis of $gl(2,R)$.}. 
  However, this difference does not matter because the frequently used
  quantities  
  \begin{eqnarray}
   Tr(T^{aG} [T^{bG}, T^{cG}]) = \frac{f^{G}_{abc}}{2}, \hspace{2mm} 
   Tr(T^{aG} \{ T^{bG}, T^{cG} \}) = \frac{d^{G}_{abc}}{2}
   \label{AZ52structureconst}
  \end{eqnarray}
 do preserve the cyclic symmetry, which is trivial from the cyclic
 symmetry of the trace. It goes without saying that the quantities
 $f^{G}_{abc}$ and $d^{G}_{abc}$ are real numbers. 

 \subsubsection{Canonical Pairs in $gl(N,R)$ Lie algebra}
 We next investigate the canonical commutation relation of the Lie
 algebra $gl(N,R)$. As we have seen in IKKT model or the
 $osp(1|32,R)$ matrix  model, canonical pairs play an essential role
 in introducing kinetic  terms in the theory, and thus making a
 mapping rule into  Yang-Mills theory. The existence of the canonical
 pairs is essential in considering such matrix model as IKKT model or
 Smolin's proposals. 

 Originally we would like to seek the elements of $gl(N,R)$ which 
 satisfies $[p,q] = - i$, where $p,q \in gl(N,R)$.  However, it is
 clearly impossible to generate such a pair of matrices  
 from a real algebra $gl(N,R)$. Therefore we instead consider a following
 correspondence 
  \begin{eqnarray}
   [p,iq] = +1 \textrm{ , where } p, iq \in gl(N,R).
  \end{eqnarray}
 This commutation relation can be achieved by using real
 matrices. Of course it is impossible to give such matrices of finite
 size which satisfies the canonical relation, and we need to
 approximate them by large enough matrices. The key to generate such
 matrices is the analogy from the creation-annihilation operators. We
 define two real operators,
  \begin{eqnarray}
   a = \frac{1}{\sqrt{2}} (p-iq), \hspace{2mm} a^{\dagger} =
   \frac{1}{\sqrt{2}} (p+iq) \textrm{ ,where } a, a^{\dagger} \in
   gl(N,R).
  \end{eqnarray}
 These operators clearly satisfies a commutation relation as a
 creation-annihilation operators $[a,a^{\dagger}] = 1$. The basis of
 this large enough dimensional vector space is the excited state of
 the harmonic oscillator $| n \rangle$ such that 
   \begin{eqnarray}
    a | n \rangle = \sqrt{n} | n-1 \rangle ,\hspace{2mm} a^{\dagger} | 
    n \rangle = \sqrt{n+1} | n+1 \rangle.
   \end{eqnarray}
 The infinite dimensional matrix representations of these
 creation-annihilation operators are
  \begin{eqnarray}
   a = \left( \begin{array}{ccccc} 0 & 1 & 0& 0 &0 \\ 0  & 0 & \sqrt{2} &
  0 & 0 \\ 0 & 0 & \ddots & \sqrt{3} & 0 \\ 0 & 0 &0  & \ddots &
  \ddots \\ 0 & 0 & 0  & 0 & 0 \end{array} \right), \hspace{4mm} a^{\dagger} = 
   \left( \begin{array}{ccccc} 0 &  0 & 0 & 0 & 0 \\ 1 & 0 & 0  & 0
  & 0 \\ 0 & \sqrt{2} &  \ddots & 0  & 0 \\ 0  & 0 & \sqrt{3} &  \ddots & 0
  \\ 0 & 0 & 0 & \ddots & 0
   \end{array} \right).
  \end{eqnarray}
 If we are to consider a background of D-branes, we need to consider
 the backgrounds such that 
   \begin{eqnarray}
    u_{1} = p_{1}, \hspace{2mm} u_{2} = iq_{1},\hspace{2mm} u_{3} =
    p_{2}, \hspace{2mm} u_{4} = iq_{2}.
   \end{eqnarray}
 Then, the noncommutativity of the background is different from the
 ordinary ($SU(N)$ gauge) theory. 
   \begin{eqnarray}
    [ u_{\mu}, u_{\nu}] = B_{\mu \nu} \textrm{ , where } B_{\mu \nu}
    \in {\bf R}.
   \end{eqnarray}
  The crucial difference is that the noncommutative parameter is {\it
  not pure imaginary but real}. We consider the correspondence between 
  the matrix theory and NCYM. As we have already seen, the mapping
  rule from the matrix model to NCYM is 
   \begin{eqnarray}
    x^{\mu} = C^{\mu \nu} u_{\nu},
   \end{eqnarray}
  where $C_{\mu \nu}$ is an inverse matrix of $B_{\mu \nu}$ such that
  $B^{\mu \nu} C_{\nu \rho} = \delta^{\mu}_{\rho}$. The
  noncommutativity is then
  \begin{eqnarray}
    [x^{\mu}, x^{\nu}] = C^{\mu \nu}. 
  \end{eqnarray}
 Note that the noncommutativity parameter is no longer pure imaginary, 
 and hence the noncommutative contribution is now no longer a mere
 phase factor. The new Moyal bracket should be 
   \begin{eqnarray}
    a(x) \star b(x) = \exp ( \frac{1}{2} C^{\mu \nu}
    \frac{\partial^{2}}{\partial \xi^{\mu} \partial \eta^{\nu}}) a(x +
    \xi) b(x + \eta)|_{\xi = \eta = 0}.
   \end{eqnarray}

 \subsubsection{Pathology of Non-compactness} 
  Next, we consider the action of Yang-Mills theory in order to gain
  an insight into an important  pathology of $GL(N,R)$ gauge group.
  For simplicity, let us have a look at the action of the $GL(2,R)$
  gauge theory:
   \begin{eqnarray}
  \sum_{k=1}^{3} tr_{2 \times 2} [ \partial_{\mu} + A_{\mu}^{k}
    T^{kG} , \partial_{\nu}  + A_{\nu}^{k} T^{kG} ]^{2} = \frac{1}{2}
    \{ (\partial_{\mu} A_{\nu}^{1} - \partial_{\nu} A_{\mu}^{1})^{2} -
    (\partial_{\mu} A_{\nu}^{2} - 
    \partial_{\nu} A_{\mu}^{2})^{2} +
    (\partial_{\mu} A_{\nu}^{3} - \partial_{\nu} A_{\mu}^{3})^{2}  \}. 
   \end{eqnarray}
  Look at the sign of $ (\partial_{\mu} A_{\nu}^{2} - \partial_{\nu}
  A_{\mu}^{2})^{2}$. The coefficient is {\it not 1 but -1}, which means 
  that the theory is no longer positive definite! Why does such a
  disease emerge? The answer lies in the fact that $tr
  (T^{2G})^{2} = - \frac{1}{2}$ while $tr (T^{1G})^{2} = tr
  (T^{3G})^{2} = \frac{1}{2}$. In order to remedy this situation, we
  should perform an 
  analytic continuation with respect to $- (\partial_{\mu}
  A_{\nu}^{2} - \partial_{\nu} A_{\mu}^{2})^{2}$ by replacing
  $A_{\mu}^{2} \rightarrow i A_{\mu}^{2}$.

  This example of Yang-Mills theory teaches us in a pedagogical
  way that $GL(N,R)$ gauge theory reduces to an ordinary $SU(N)$ gauge
  theory by analytic continuation. This is the physical meaning of
  $GL(N,R)$ gauge theory, with which we are not familiar. These
  arguments reveal the physical interpretation of $GL(N,R)$ gauge
  theory, and have given a  confidence that this gauge group is not so
  malignant a pathology.  

 \subsection{Action of $gl(1|32,R) \otimes gl(N,R)$ Gauged Theory}
  The next job is to investigate the action of this theory. The basic
  idea is similar to $osp(1|32,R)$ non-gauged cubic matrix model, and
  we proceed rather quickly. The action is 
  \begin{eqnarray}
   I &=& \frac{1}{g^{2}} Tr_{N \times N} \sum_{Q,R=1}^{33}
        ((\sum_{p=1}^{32}  {M_{p}}^{Q} {M_{Q}}^{R}
        {M_{R}}^{p})  -{M_{33}}^{Q} {M_{Q}}^{R}
        {M_{R}}^{33} ) = \frac{1}{g^{2}} Tr_{N \times N}( Str_{33
        \times 33} M^{3}) \nonumber \\
    &=& \frac{1}{g^{2}} \sum_{a,b,c=1}^{N^{2}} Str(M^{a} M^{b} M^{c}
     ) Tr(T^{a} T^{b} T^{c}). \label{AZ54action}
  \end{eqnarray}
   \begin{itemize}
    \item{$M$ is now a multiplet of $gl(1|32,R)$ super Lie algebra,
        with each component promoted to the element of $gl(N,R)$ Lie
        algebra.}
    \item{As we have explained in the previous section, the indices
        $P, Q, R, \cdots$ runs $P, Q, R, \cdots = 1, \cdots, 33$,
        while $p, q, r, \cdots = 1, \cdots, 32$. }
    \item{Of course, this matrix model possesses no free parameter,
        totally in the same sense as $osp(1|32,R)$ non-gauged cubic
        matrix model.}
    \item{We do not need to multiply $i$ in order to make this action
        real,  because the $gl(N,R)$ Lie algebra is real, so
        that the reality of this action is trivial.}
  \item{We have promoted the $33 \times 33$ matrix $M$ to a large
        $33N \times 33N$ matrices. However, the structure of the promotion is
      completely different from $osp(1|32,R)$ model. The gauge
      transformation is with respect to not the separate $gl(1|32,R)$
      and $gl(N,R)$, but the tensor product of the Lie algebra
        $gl(1|32,R) \otimes gl(N,R)$. This 
      drastically enhances the gauge symmetry of the theory, in the
        sense explained in Appendix. \ref{AZCtensorpr}. 
      The gauge symmetry is thus 
    \begin{eqnarray}
     \textrm{For an arbitrary element of } u \in gl(1|32,R) \otimes
     gl(N,R) , \textrm
     { the gauge transformation is } M \Rightarrow M + [u, M] \nonumber
     \end{eqnarray}
     }
  \item{  It is possible to rewrite this action in terms of the structure
   constant. The matrices $M$ are rewritten using the basis of $gl(N,R)$
   gauge group:
    \begin{eqnarray}
      {M_{P}}^{Q} = \sum_{a=1}^{N^{2}} {(M^{a})_{P}}^{Q} T^{a}.
    \end{eqnarray}
  $M^{a}$ are $33 \times 33$ supermatrices with each component being
  {\it real c-numbers}, not large $N$  matrices.
  $\{ T^{a} \}$ are now the basis of $gl(N,R)$ Lie algebra. The
  action is then  rewritten as in (\ref{AZ54action}).}
   \end{itemize}

  Note that the trace of the generators in (\ref{AZ54action}) can be
  written using commutators and anti-commutators. This can be easily
  performed using the structure constants introduced in
  (\ref{AZ52structureconst}):
   \begin{eqnarray}
    Tr(T^{a} T^{b} T^{c}) = \frac{1}{2} Tr( T^{a} [T^{b}, T^{c}]) +
    \frac{1}{2} Tr( T^{a} \{ T^{b}, T^{c} \})
    = \frac{1}{4} (f_{abc} + d_{abc}).
   \end{eqnarray}
 We can now switch the large $N$ matrix
  representation and the representation in terms of the color indices
  as follows:
   \begin{eqnarray}
 & &   I = \frac{1}{g^{2}} Tr_{N \times N} \sum_{Q,R=1}^{33} 
   ( ( \sum_{p=1}^{32} {M_{p}}^{Q} {M_{Q}}^{R}
   {M_{R}}^{p}) - {M_{33}}^{Q} {M_{Q}}^{R}{M_{R}}^{33} ) =
   \frac{1}{g^{2}} Tr_{N \times N} ( Str_{33 \times 33} M^{3})
    \nonumber \\
 &\Leftrightarrow& \frac{1}{4 g^{2}} (f_{abc} + d_{abc} ) Str(M^{a}
   M^{b} M^{c} ). \label{AZ53colormatrix}  
   \end{eqnarray}
  
  In the case of $gl(1|32,R) \otimes gl(N,R)$ cubic matrix model, it is
  convenient to analyze the action in terms of the color indices $M^{a} = 
  \left( \begin{array}{cc} m^{a} & \psi^{a} \\ i {\bar \phi}^{a} &
  v^{a} \end{array} \right)$. Computing the third power of this
  supermatrix, we obtain 
  \begin{eqnarray}
 & &   Str(M^{a} M^{b} M^{c} ) \nonumber  \\
 &=& tr( m^{a} m^{b} m^{c} + i m^{a} \psi^{b} {\bar \phi}^{c} + i
 \psi^{a} {\bar \phi}^{b} m^{c} + i \psi^{a} v^{b} {\bar \phi}^{c} ) -
 ( i {\bar \phi}^{a} m^{b} \psi^{c} + i {\bar \phi}^{a} \psi^{b} v^{c}
 + i v^{a} {\bar \phi}^{b} \psi^{c} + v^{a} v^{b} v^{c}) \nonumber \\ 
 &=& tr(m^{a} m^{b} m^{c}) - 3i {\bar \phi}^{a} m^{b} \psi^{c} - 3i
 {\bar \phi}^{a} \psi^{b} v^{c} - v^{a} v^{b} v^{c}. 
  \end{eqnarray}
 
  Therefore, this action is expressed by, using the components,
   \begin{eqnarray}
 & & I = \frac{1}{4g^{2}} (f_{abc} + d_{abc}) ( tr(m^{a} m^{b} m^{c}) -
 3i {\bar \phi}^{a} m^{b} \psi^{c} - 3i 
 {\bar \phi}^{a} \psi^{b} v^{c} - v^{a} v^{b} v^{c} ), \nonumber \\
 &\Leftrightarrow&
 I = \frac{1}{g^{2}} Tr ( tr(m^{3}) - 3i {\bar \phi} m \psi - 3i {\bar 
 \phi} \psi v - v^{3}). \label{AZ53action}
   \end{eqnarray}
  In this case again, the former formulation utilizing the color
  indices and the latter $N \times N$ matrix formulation is
  equivalent. In the following
  analysis, we utilize the large $N$ matrix description and 
  again express the bosonic matrix $m$ in terms of the
  basis of 11 dimensional gamma matrices  
   \begin{eqnarray}
    m &=& Z {\bf 1} + W \Gamma^{\sharp} + \frac{1}{2} ( A^{(+)}_{i}
     \Gamma^{i} ( 1 + \Gamma^{\sharp}) + A^{(-)}_{i} \Gamma^{i} ( 1 -
     \Gamma^{\sharp}) ) + 
     + \frac{1}{2} C_{i_{1} i_{2}} \Gamma^{i_{1}
    i_{2}} + \frac{1}{2} D_{i_{1} i_{2}} \Gamma^{i_{1} i_{2} \sharp}
    \nonumber \\
   &+&  \frac{1}{2 \times 3!} ( E^{(+)}_{i_{1} i_{2} i_{3}}
     \Gamma^{i_{1} i_{2} i_{3}} ( 1 + \Gamma^{\sharp} ) +
     E^{(-)}_{i_{1} i_{2} i_{3}} \Gamma^{i_{1} i_{2} i_{3}} (1 -
     \Gamma^{\sharp} ) 
   + \frac{1}{4!} G_{i_{1} \cdots i_{4}} \Gamma^{i_{1} \cdots
    i_{4}} + \frac{1}{4!} H_{i_{1} \cdots i_{4}} \Gamma^{i_{1} \cdots
    i_{4} \sharp} \nonumber \\
  &+& \frac{1}{5!} ( I^{(+)}_{i_{1} \cdots i_{5}} \Gamma^{i_{1}
    \cdots i_{5}} ( 1 + \Gamma^{\sharp} ) + I^{(-)}_{i_{1} \cdots
     i_{5}} ( 1 - \Gamma^{\sharp}) ).
   \end{eqnarray}
  However, we do not rely on the complicated expansion of the trace
  $tr(m^{3})$ in our discussion. The explicit expansion is explained
  in full detail in Appendix.  \ref{AZCgaugedactionres}, however the
  computation or techniques therein does not concern the following
  discussion at all.

  \subsection{Structure of ${\cal N} =2$ SUSY}
  We have investigated the SUSY transformation of $osp(1|32,R)$
  non-gauged cubic matrix model. It turned out the this matrix model
  possesses two-fold ${\cal N}=2$ SUSY transformations each of which is 
  to be identified with that of IKKT model. We investigate the
  structure of SUSY transformation for the $gl(1|32,R) \otimes gl(N,R)$
  gauged theory.  

 \subsubsection{Effective Action}
  Before the investigation of the SUSY structure, let us compare this
  $gl(1|32,R) \otimes gl(N,R)$ gauged model with $osp(1|32,R)$ non-gauged
  model. The common feature is that both theories possess 64 fermionic
  SUSY parameters \footnote{ The supercharge of this gauged theory is 
  introduced later.}, twice as many as IKKT model. And it is a natural 
  speculation  that this gauged matrix model may possess two-fold
  ${\cal N} =2$ SUSY structures in the same sense as in $osp(1|32,R)$
  non-gauged model. The grave difference is that it is impossible to
  introduce a SUSY transformation by a naive translation of the fermionic
  fields. Because we have excluded the commutator from the
  action, the naive translation is no longer a supersymmetry of the
  theory. And we require a maneuver to introduce an inhomogeneous SUSY 
  transformation. 

  In order to remedy this situation, we consider the physics in ${\bf
  R}^{9,1} AdS$ space\footnote{In \cite{virginal}, the Wigner
  In\"{o}n\"{u} contraction on a ten dimensional sphere is
  investigated, but this 
  discrepancy causes no essential difference from \cite{virginal} at
  all.},  in which the radius of the hyperboloid is 
  extremely large. This is called the 'Wigner-In{\"o}n{\"u} contraction,
  whose brief review is given in 
  Appendix. \ref{AZCwicont}. Just as we perceive the earth as a flat
  2 dimensional space because the earth is much bigger than we, we
  consider the physics in apparently '10 dimensional' space because of 
  the large radius of the hyperboloid. In order to consider this
  situation, we alter the action a bit, and consider the following
  action. From now on, we set the coupling constant $g$ to 1, because
  this quantity does not play an essential role in the following
  discussion:
   \begin{eqnarray}
    I = \frac{1}{3} Tr_{N \times N} Str( M_{t}^{3}) - R^{2} Tr_{N \times
    N} Str M_{t}. \label{AZ54veryaction}
   \end{eqnarray}
  We add the second term so that we can consider the Wigner
  In{\"o}n{\"u} contraction. In other words, we altered the action in
  order for the theory to possess a classical solution $\langle m
  \rangle = R \Gamma^{\sharp}$. We can verify that this is a classical
  solution of (\ref{AZ54veryaction}) without solving the equations of
  motion for the complicated explicit form in Appendix.
  \ref{AZCgaugedactionres}. The equation of motion with respect to
  the matrix $M_{t}$ is 
   \begin{eqnarray}
    \frac{\partial I}{\partial M_{t}} = M_{t}^{2} - R^{2} {\bf 1}_{33 \times
    33} = 0.
   \end{eqnarray}
  One of the classical solution is 
  \begin{eqnarray}
  \langle M \rangle = \left( \begin{array}{cc} R \Gamma^{\sharp}
  \otimes {\bf 1}_{N \times N} & 0 \\ 0 & R \otimes {\bf 1}_{N \times N}
  \end{array} \right), 
  \end{eqnarray}
   and the investigation of the theory in terms of this classical solution is
  equivalent to the Wigner In{\"o}n{\"u} contraction \footnote{This
  classical solution is impossible in the original $u(1|16,16)$
  gauged theory, because the (33,33) component $v$ is restricted to
  be an anti-hermitian matrix. If we are to consider the
  Wigner-In{\"o}n{\"u} contraction, one way is to consider the quintic 
  action $I_{u(1|16,16)} = \frac{1}{5} Str(M^{5}_{t}) - R^{4} Str
  M_{t}$. Then, the classical solution $\langle M \rangle = 
  \left( \begin{array}{cc} R  \Gamma^{\sharp} \otimes {\bf 1}_{N
  \times N} & 0 \\ 0 & i R \otimes {\bf 1}_{N \times N} \end{array}
  \right)$ is possible because $ i R \otimes {\bf 1}_{N \times N}$ is now an
  anti-hermitian matrix. Another caution is that the gauge group must
  be not $u(1|16,16) \otimes SU(N,C)$ but $u(1|16,16) \otimes
  U(N,C)$. If the gauge group is $u(1|16,16) \otimes SU(N,C)$, the
  linear term in (\ref{AZ54veryaction}) vanishes because the
  generators are $Tr(T^{a})=0$, and the Wigner In{\"o}n{\"u}
  contraction is impossible from the beginning.}. Then, we separate
  the original matrix    between the classical solution and the
  fluctuation  and the classical solution as 
   \begin{eqnarray}
    M_{t} = \langle M \rangle + M = \left( \begin{array}{cc} R
    \Gamma^{\sharp} & 0 \\ 0 & R \end{array} \right) + \left
    ( \begin{array}{cc} m & \psi \\ i {\bar \phi} & v \end{array}
    \right).
   \end{eqnarray}
  Then, the action is 
   \begin{eqnarray}
    I = \frac{1}{3} tr( (m + R \Gamma^{\sharp})^{3}) - i ( {\bar \phi} 
    m \psi + {\bar \phi} \psi v + R {\bar \phi} (1 + \Gamma^{\sharp} ) 
    \psi ) - \frac{(v+R)^{3}}{3} - R^{2} ( tr(m + R \Gamma^{\sharp}) - 
    v).
   \end{eqnarray}
    We ignore the terms of ${\cal O}(R^{3})$, because this is just a constant.
    And we ignore the terms of ${\cal O}(R^{2})$, because this is a
    linear term with respect to the fluctuation. Then, the action is
    expressed as follows: 
   \begin{eqnarray}
    I = R( tr(m^{2} \Gamma^{\sharp}) - v^{2} - i {\bar \phi} (1 +
    \Gamma^{\sharp}) \psi)  + \frac{1}{3} tr(m^{3}) -
    \frac{v^{3}}{3} -i ({\bar  \phi} m {\psi} + v {\bar \phi} \psi ). 
   \end{eqnarray}
  In investigating this action, we do not rely on the disastrously
  complicated explicit form in Appendix. \ref{AZCgaugedactionres}, but 
  we distinguish the fluctuation around the classical solution as
  follows.
  \begin{itemize}
   \item{ The bosonic matter fields are distinguished into the
       following two terms $m = m_{e} + m_{o}$
      \begin{itemize}
        \item{$m_{e}$ consists of the components of the even
            rank. From now on, we use the word 'rank' of the gamma
            matrices with respect to 10 dimensional indices: 
            \begin{eqnarray}
             m_{e} = Z {\bf 1} + W \Gamma^{\sharp} + \frac{1}{2}
             ( C_{i_{1} i_{2}} \Gamma^{i_{1} i_{2}} + D_{i_{1} i_{2}}
             \Gamma^{i_{1} i_{2} \sharp} ) + \frac{1}{4!} ( G_{i_{1}
             \cdots i_{4}} \Gamma^{i_{1} \cdots i_{4}} +
             H_{i_{1} \cdots i_{4}} \Gamma^{i_{1} \cdots i_{4}
             \sharp}).
            \end{eqnarray} }
       \item{$m_{o}$ consists of the components of the odd rank
          \begin{eqnarray}
           m_{o} &=& \frac{1}{2} ( A^{(+)}_{i} \Gamma^{i} ( 1 +
           \Gamma^{\sharp}) + A^{(-)}_{i} \Gamma^{i} (1-
           \Gamma^{\sharp}) ) + 
           \frac{1}{2 \times 3!} (E^{(+)}_{i_{1} i_{2} i_{3}}
           \Gamma^{i_{1} i_{2} i_{3}} (1 + \Gamma^{\sharp} ) +
           E^{(-)}_{i_{1} i_{2} i_{3}} \Gamma^{i_{1} i_{2} i_{3}} ( 1
           - \Gamma^{\sharp} ) ) \nonumber \\
           &+& \frac{1}{5!} (I^{(+)}_{i_{1} \cdots i_{5}}
           \Gamma^{i_{1} \cdots i_{5}} (1 + \Gamma^{\sharp}) +
           I^{(-)}_{i_{1} \cdots i_{5}} \Gamma^{i_{1} \cdots i_{5}} (1
           - \Gamma^{\sharp}) ).
          \end{eqnarray} }
      \end{itemize} }
   \item{Fermionic fields are decomposed according to their
       chiralities.}
  \end{itemize}
  The action is then written as follows. The proof is given in
  Appendix. \ref{AZCcocoro}:
   \begin{eqnarray}
    I &=& R( tr(m^{2}_{e} \Gamma^{\sharp}) - v^{2} - 2i {\bar \phi}_{R}
    \psi_{L} ) + tr (\frac{1}{3} m^{3}_{e} + m_{e} m^{2}_{o} )
    \nonumber \\
    &-& i ( {\bar \phi}_{R} (m_{e} + v) \psi_{L}  + {\bar \phi}_{L}
    (m_{e} + v) \psi_{R} + {\bar \phi}_{L} m_{o} \psi_{L} + {\bar
    \phi}_{R} m_{o} \psi_{R} ) - \frac{1}{3} v^{3}. \label{AZ55cocoro} 
   \end{eqnarray}

  We now would like to integrate out the fields of order ${\cal
  O}(R)$ and consider the effective theory, however, the obstacle is
  the cubic term $tr(m_{e}^{3})$. In order to exclude such a nuisance, 
  we consider the following rescaling:
   \begin{eqnarray}
  & &   m_{t} = R \Gamma^{\sharp} + m = R \Gamma^{\sharp} +
    R^{-\frac{1}{2}} m'_{e} + R^{\frac{1}{4}} m'_{o}, \hspace{3mm}
    v_{t} = R + v = R + R^{-\frac{1}{2}} v', \nonumber \\
  & & \psi = \psi_{L} + \psi_{R} = R^{-\frac{1}{2}} \psi'_{L} +
    R^{\frac{1}{4}} \psi'_{R} , \hspace{3mm} {\bar \phi} = {\bar
    \phi}_{L} + {\bar \phi}_{R} = 
    R^{\frac{1}{4}} {\bar \phi}'_{L} + R^{-\frac{1}{2}} {\bar
    \phi}'_{R}.
   \end{eqnarray}
  Following this rescaling, $tr(m'^{3}_{e})$ and ${\bar \phi}'_{R}
  (m'_{e} + v') \psi_{L}$ are excluded, because this is rescaled as
  ${\cal O}(R^{- \frac{3}{2}})$. This theory is thus rescaled to be
   \begin{eqnarray}
    I = ( tr(m'^{2}_{e} \Gamma^{\sharp}) - v'^{2} + tr(m'_{e}
    m'^{2}_{o}) ) -i ( 2 {\bar \phi}'_{R} \psi'_{L} + {\bar \phi}'_{L}
    (m'_{e} + v') \psi'_{R} + {\bar \psi}'_{L} m'_{o} \psi'_{L} +
    {\bar \phi}'_{R} m'_{o} \psi'_{R} ).
   \end{eqnarray}
  We integrate out the fields $m'_{e}$, $\psi'_{L}$ and ${\bar
  \phi}'_{R}$ by Gaussian integration. Completing this action square,
  the action is 
   \begin{eqnarray}
   I &=& tr(  \{ m'_{e} + \frac{1}{2} ( m'^{2}_{o}
   \Gamma^{\sharp} + i (\psi'_{R} {\bar \phi}'_{L} ) \Gamma^{\sharp})
   \}^{2} \Gamma^{\sharp} ) - \frac{1}{4} tr( \{ m'^{2}_{o} +
  i (\psi'_{R} {\bar \phi}'_{L} ) \}^{2} \Gamma^{\sharp} ) \nonumber \\
    &-& (v' + \frac{i}{2} ({\bar \phi}_{L} \psi_{R}) )^{2} -
   \frac{1}{4} ({\bar \phi}'_{L} \psi_{R})^{2} - 2i ({\bar \phi}'_{R}
   + \frac{1}{2} {\bar \phi}'_{L} m_{o} )( \psi'_{L} + \frac{1}{2}
   m'_{o} \psi'_{R} ) + \frac{i}{2} ( {\bar \phi}'_{L} m'^{2}_{o}
   \psi'_{R} ), \label{AZ55compsq}
   \end{eqnarray}
 where we have utilized the fact that $(\Gamma^{\sharp})^{2} = {\bf
 1}_{32 \times 32}$, and thus $\Gamma^{\sharp}$ possesses an inverse
 matrix. 
 The effective action is given by the following path
 integral\footnote{ The analogy of this path integral is a 
 following toy model. (i)For the action $S=ax^{2}+bx+c$, the path
 integral is $e^{-W} = \int dx \exp( -(ax^{2}+bx+c)) =
 \int dx  \exp(-a(x+\frac{b}{2a})^{2} ) \exp( -c+ \frac{b^{2}}{4a})
 \propto  \exp(-c+ \frac{b^{2}}{4a})$. And thus the effective action
 is $W = \frac{-b^{2} + 4ac}{4a}$. (ii)The second example is the
 following action, $S=axy+bx+cy$, with $x,y$ being auxiliary
 fields. This integration is performed by $e^{-W} = \int dx dy \exp (
 a(x+ \frac{b}{a})(y+\frac{c}{a}) + \frac{bc}{a} )$, and thus the
 effective action is $W = - \frac{bc}{a}$. This holds even if $x,y$
 are Grassmann odd quantities.}. However, this effective action 
 vanish:
   \begin{eqnarray}
 & &    e^{-W} = \int dm'_{e} d \psi'_{L} d {\bar \phi}'_{R} dv e^{-I} 
 \nonumber \\
 &\Rightarrow& W = - \frac{1}{4}  tr(  \{ m'^{2}_{o} + i  (\psi'_{R}
 {\bar \phi}'_{L} ) \}^{2} \Gamma^{\sharp} ) - \frac{1}{4}  ( {\bar 
 \phi}'_{L} \psi_{R})^{2}  + \frac{i}{2} ( {\bar \phi}'_{L} m'^{2}_{o}
   \psi'_{R} ) \nonumber \\
 & & \hspace{5mm} =  -  \frac{1}{4} tr( m'^{4}_{o} \Gamma^{\sharp} ) +
 \frac{i}{2} ( tr( - m'^{2}_{o} (\psi'_{R} {\bar \phi}'_{L}) 
 \Gamma^{\sharp} ) + {\bar \phi}'_{L} m'^{2}_{o} \psi'_{R} ) +
 \frac{1}{4} ( tr( (\psi'_{R} {\bar \phi}'_{L} )^{2} \Gamma^{\sharp} ) - 
 ({\bar \phi}'_{L} \psi'_{R})^{2} ) = 0. \label{AZ55effectiveac}
   \end{eqnarray}
  That the first term vanishes is discerned from  the
  anti-commutativity of the matrices $m'_{o}$ and $\Gamma^{\sharp}$:
  $m_{o} \Gamma^{\sharp}= - \Gamma^{\sharp} m'_{o}$. Noting this fact,
  this term is rewritten as 
       \begin{eqnarray}
         -  \frac{1}{4} tr( m'^{4}_{o} \Gamma^{\sharp} ) 
         \stackrel{\{ \Gamma_{\sharp}, m'_{o} \} = 0}{=} \frac{1}{4}
         tr( m'^{3}_{o} \Gamma^{\sharp} m'_{o} ) \stackrel{cyclic}{=}
         \frac{1}{4} tr( m'^{4}_{o} \Gamma^{\sharp} )  = 0. 
       \end{eqnarray}
 We will consider the physical meaning of this empty action later in this 
 section. 
 
 \subsubsection{Symmetry of the Effective Theory}
  We analyze the effective theory (\ref{AZ55effectiveac}) by
  investigating the symmetry of the fields. There are two kinds of
  transformations in the theory.

  First is the supersymmetry of this theory. As we have explained in
  the previous chapter, the SUSY is a transformation from fermions to
  the bosons and vice versa, and the SUSY parameter is restricted to
  be fermionic. We 'define' the supercharge of this theory
  to be an arbitrary fermionic matrix:
   \begin{eqnarray}
    Q_{\chi \epsilon} \stackrel{?}{=} \left( \begin{array}{cc} 0 &
    \chi \\ i {\bar \epsilon} & 0 \end{array} \right). \label{AZ55charge?}
   \end{eqnarray}
  If  this is to be a bona fide  supercharge, this must satisfy
  Haag-Lopuszanski Sohnius theorem (\ref{AZ31hls}). Here, we can see
  the miracle of the Wigner-In{\"o}n{\"u} contraction on the hyperboloid
  in ${\bf R}^{9,1} AdS$ spacetime. Unlike the case of the non-gauged theory,
  the translation is trivially introduced. As we have seen in
  Appendix. \ref{AZCwicont}, the translation in the direction of $x^{i}$ is
  introduced by  the rotation in $x^{i} x^{\sharp}$ plane. The
  generator of the translation is  $P_{i} = \frac{1}{R} \Gamma_{i
  \sharp}$. Using this fact, both of the statements of
  Haag-Lopuszanski-Sohnius are easily verified. 
   \begin{itemize}
    \item{In considering the former statement, we consider the
        commutation relation of the two 'supercharges' 
        $Q_{\chi \epsilon} = \left( \begin{array}{cc} 0 & \chi \\ i
        {\bar \epsilon} & 0 \end{array} \right)$ and $Q_{\rho \eta} =
        \left( \begin{array}{cc} 0 & \rho \\ i {\bar \eta} & 0
        \end{array} \right) $. The commutator of these charges is 
       \begin{eqnarray}
        [Q_{\chi \epsilon} , Q_{\rho \eta}] = \left( \begin{array}{cc} 
        i (\chi {\bar \eta} - \rho {\bar \epsilon}) & 0 \\ 0 & i
        ({\bar \epsilon} \rho  - {\bar \eta} \chi) \end{array} \right).
       \end{eqnarray}  
        The bosonic $32 \times 32$ part is now obtained by $i (\chi
        {\bar \eta} - \rho {\bar \epsilon}) $. In order for these
        supercharge to be well-defined ones, this bosonic part must
        include the translation $P_{i} = \frac{1}{R} \Gamma_{i \sharp}$. We
        extract the coefficient of the gamma matrix $\Gamma_{i
        \sharp}$ by the prescription in
        Appendix. \ref{AZCdecomposition}: 
     \begin{eqnarray}
      \textrm{(coefficients of }  \Gamma_{i \sharp} \textrm{)} = 
      \frac{i}{32} ( {\bar \eta} \Gamma_{i \sharp} \chi - {\bar
      \epsilon} \Gamma_{i \sharp} \rho).   
     \end{eqnarray}
     In order for this term to survive, the chirality of the following 
     pairs $(\chi \leftrightarrow  \eta)$ and $(\epsilon
     \leftrightarrow \rho)$ must possess the same chirality
     {\it respectively}. Therefore, it is possible to leave this
     coefficients non-zero whatever the chiralities of $\chi$ and $\epsilon$ 
     are. An arbitrary fermionic translation thus satisfies the first
     statement of Haag-Lopuszanski-Sohnius theorem.}
   \item{The second condition is trivially verified by noting the fact 
       the the radius of the hyperboloid is large:
      \begin{eqnarray}
        [P_{i}, Q] = \lim_{R \to \infty} \frac{1}{R} [\Gamma_{i
        \sharp}, Q] = 0.
      \end{eqnarray}}
   \end{itemize}
  Therefore, the supercharge (\ref{AZ55charge?}) is verified to
  represent a well-defined SUSY transformation:  
    \begin{eqnarray}
      \textrm{(supercharge)} = Q =  \left( \begin{array}{cc} 0 &
    \chi \\ i {\bar \epsilon} & 0 \end{array} \right). \label{AZ55charge}
    \end{eqnarray}

  The next transformation is the bosonic effect. For example, we must 
  consider the translation of the bosonic vector fields $P_{i}$ and the
  Lorentz transformations. The former is, of course, generated by $
  \theta_{i} \Gamma_{i \sharp}$ where $\theta_{i}$ is a translation
  parameter. The latter is represented by the rotation on $x^{i}
  x^{j}$ plane, $\Gamma^{ij}$. These transformations, including
  the effects of other ranks, is represented by the charge $
  \left( \begin{array}{cc} a & 0 \\  0 & b \end{array} \right)$.\\

  The charge of both fermionic and bosonic transformation is expressed 
  by 
  \begin{eqnarray}
  A = \left( \begin{array}{cc} a & \chi \\ i {\bar \epsilon} & b
  \end{array} \right).
  \end{eqnarray}
  The transformation of the matter field with respect to this charge is
  now given by 
   \begin{eqnarray}
   [A, M] = \left( \begin{array}{cc} [a, m + R \Gamma^{\sharp}] + i
   ( \chi {\bar \phi} - \psi {\bar \epsilon}) & - (m + R
   \Gamma^{\sharp}) \chi + a \psi - b \psi + \chi v \\ i {\bar
   \epsilon} (m + R \Gamma^{\sharp}) - i ( {\bar \phi} a + v {\bar
   \epsilon} ) + i b {\bar \phi} & i ({\bar \epsilon} \psi - {\bar
   \phi} \chi ) + [b, v] \end{array} \right). \label{AZ55subotr}
   \end{eqnarray}

  These parameters are decomposed in the same manner as the matter fields, 
  and consider the following rescaling:
   \begin{eqnarray}
    A = \left( \begin{array}{cc} a'_{e} + R^{-\frac{3}{4}} a'_{o} &
    \chi'_{L} + R^{-\frac{3}{4}} \chi'_{R} \\ i ( R^{-\frac{3}{4}} {\bar 
    \epsilon}'_{L} + {\bar \epsilon}'_{R}) & b' \end{array} \right).
    \label{AZ55chres} 
   \end{eqnarray}
  We next compute the transformation of each field, paying attention to
  the scaling law. The conclusion is
   \begin{eqnarray}
      \delta m'_{e} &=&  [a_{e}, m'_{e}] + [a_{o}, m'_{o}] + i
     ( \chi'_{L} {\bar \epsilon}'_{R} + \chi'_{R} {\bar \epsilon'_{L}}
     ) - i ( \psi'_{L} {\bar \epsilon}'_{R} + \chi'_{R} {\bar
     \epsilon}'_{L} ), \label{AZ55tr1964me} \\
       \delta m'_{o} &=& [ a_{o}, \Gamma^{\sharp}] + [a'_{e}, m'_{o}]
     + i ( \chi'_{L} {\bar \phi}'_{L} - \psi'_{R} {\bar \epsilon}'_{R}
     ), \label{AZ55tr1964mo} \\
      \delta \psi'_{L} &=& - ( m'_{o} \chi'_{R} + m'_{e} \chi_{L}) +
       (a_{o} \psi'_{R} + a_{e} \psi'_{L}) - b' \psi'_{L} + v'
       \chi'_{L}, \label{AZ55tr1964psil} \\
      \delta \psi'_{R} &=& 2 \chi'_{R} + ( a'_{e} \psi'_{R}  - b
       \psi'_{R} - (m'_{o} \chi'_{L}) ), \label{AZ55tr1964psir} \\
      \delta {\bar \phi}'_{L} &=& - 2 {\bar \epsilon}'_{L} + ( ({\bar
         \epsilon}'_{R} m'_{o}) + b' {\bar \phi}'_{L} - {\bar
         \phi}'_{L} a'_{e}  ), \label{AZ55tr1964phil} \\
   \delta {\bar \phi}'_{R} &=& ( {\bar \epsilon}'_{R} m'_{e} +
      {\bar \epsilon}'_{L} m'_{o} ) + b' {\bar \phi}'_{R} - ({\bar
      \phi}'_{L} a'_{o} + {\bar \phi}'_{R} a'_{e} ) - v' {\bar
      \epsilon}'_{R}, \label{AZ55tr1964phir} \\
    \delta v' &=& i ( {\bar \epsilon}'_{R} \psi'_{L} + {\bar
         \epsilon}'_{L} m'_{o} ) - i ({\bar \phi}'_{R} \chi'_{L} +
         {\bar \phi}'_{L} \chi'_{R} ) + [b', v']. \label{AZ55tr1964v}
   \end{eqnarray}
  We give the idea to derive the above transformations. However, we
  explain only the first two transformations, because the idea to
  derive the rest is therein included. 
  \begin{itemize}
   \item{$\delta m_{e}$: This is given by $ \delta m_{e} = [a, m + R
       \Gamma^{\sharp}] + i ( \chi {\bar \phi} - \psi {\bar
       \epsilon})$. In picking up these terms, we note the following
       two points.
        \begin{itemize}
         \item{The formula of the product of the gamma matrices
             (\ref{AZproduct}) indicates that only the product $\Gamma^{e}
             \Gamma^{e}$ or $\Gamma^{o} \Gamma^{o}$ produces the
             gamma matrices of even rank.}
        \item{The property in (\ref{AZMA21fermvanish}) and the
            decomposition in Appendix. \ref{AZCdecomposition} reveals
            that the fermions of {\it different chirality} produces
            the terms of even rank.}
        \end{itemize} 
        Therefore, the surviving transformation is 
     \begin{eqnarray}
     R^{-\frac{1}{2}} \delta m'_{e} = R [ a_{e}, \Gamma^{\sharp}] +
     R^{-\frac{1}{2}} ([a_{e}, m'_{e}] + [a_{o}, m'_{o}] + i
     ( \chi'_{L} {\bar \epsilon}'_{R} + \chi'_{R} {\bar \epsilon'_{L}}
     ) - i ( \psi'_{L} {\bar \epsilon}'_{R} + \chi'_{R} {\bar
     \epsilon}'_{L} ) ). \nonumber
     \end{eqnarray}
  Note that the commutator $[\Gamma^{\sharp}, \Gamma^{e}]$
  vanishes. Then taking the limit $R \to \infty$, we obtain the
  result. In this case, there is no term excluded by this
  rescaling. However, in the following analysis, we must exclude the
  subleading term. Then, we obtain the result (\ref{AZ55tr1964me}). }
   \item{ $\delta m'_{o}$: This transformation is read off by the
      similar logic:
     \begin{eqnarray}
      R^{\frac{1}{4}} \delta m'_{o} = R^{\frac{1}{4}} [ a_{o},
      \Gamma^{\sharp}] + R^{-\frac{5}{4}} [a_{o}, m'_{e}] +
      R^{\frac{1}{4}} [a'_{e}, m'_{o}]
     + i (R^{\frac{1}{4}} \chi'_{L} {\bar \phi}'_{L} +
      R^{-\frac{5}{4}} \psi'_{R} {\bar \phi}'_{R} )  
     - i (R^{-\frac{5}{4}} \psi'_{L}  {\bar \phi}'_{L} +
      R^{\frac{1}{4}} \psi'_{R} {\bar \epsilon}'_{R} ). \nonumber 
     \end{eqnarray}
   Taking the limit $R \to \infty$, the terms of ${\cal
   O}(R^{-\frac{5}{4}}) $ is neglected, and the transformation
   (\ref{AZ55tr1964mo}) survives. This is of grave significance
   because this enables us to exclude the variables to be integrated
   out from the transformation of the remaining variables. }
  \item{
   Performing the similar analysis, and distinguishing the fields with 
   respect to the chiralities, we obtain the transformations of other 
   fields (\ref{AZ55tr1964psil}) $\sim$ (\ref{AZ55tr1964v}). }
  \end{itemize}   

  Let us summarize the features of these transformations.
  \begin{enumerate}
   \item{ In the effective theory, we integrate out $m'_{e}$,
       $\psi'_{L}$, ${\bar \phi}'_{R}$ and $v'$, while we leave
       $m'_{o}$, $\psi'_{R}$ and ${\bar \phi}'_{L}$ unerased.}
   \item{ The transformation of the remaining fields  $m'_{o}$,
       $\psi'_{R}$ and ${\bar \phi}'_{L}$ always include the
       inhomogeneous translations, which do not depend on the matter
       fields. This is a virtue of the Wigner-In{\"o}n{\"u} contraction,
       and the expectation value $\langle M \rangle$ is the source of
       these translations.}
   \item{ The transformation of the remaining fields never involves the
       fields to be integrated out. This is because we have excluded
       the terms of ${\cal O}(R^{-\frac{5}{4}} )$ by taking the limit
       $R \to \infty$. This is also a virtue of the Wigner-In{\"o}n{\"u}
       contraction on a very large hyperboloid in $AdS$ space.} 
  \end{enumerate}
  
  These results possess two major significances. First is that this
  reconfirms that the effective action (\ref{AZ55effectiveac})
  vanish. Because of the second and third features, the effective
  action is invariant under the transformation of the fields $m'_{o}$,
  $\psi'_{R}$ and ${\bar \phi}'_{L}$. Therefore, the effective action
  makes no difference if we translate these fermions
  arbitrarily. Taking the fields $m'_{o}$, $\psi'_{R}$ and ${\bar
  \phi}'_{L}$ to be all zero, the effective action is $W =
  0$. Therefore, the effective action remains $W=0$ even if we translate
  these fields into non-zero values. 

  The second significance is that this clarifies the structure of the
  SUSY of the effective theory. This gauged theory originally
  possesses no trivial translation like $osp(1|32,R)$ non-gauged 
  model, because this theory is deprived of the symmetry of the
  commutator. However, the miracle of the Wigner-In{\"o}n{\"u} contraction 
  introduced an inhomogeneous translation of the fields by considering the
  translation in the direction of $x^{i}$. The following table summarizes
  the way the SUSY parameters involve the SUSY transformation of the
  remaining fields. 
  \begin{center} \begin{tabular}{|c||c|c|c|c|} \hline
     & $\chi_{L}$ & ${\bar \epsilon}_{L}$ & $\chi_{R}$ & ${\bar 
     \epsilon}_{R}$ \\ \hline \hline
  $\psi'_{R}$ & $H$ & $\times$ & $I$ & $\times$ \\ \hline
  ${\bar \phi}'_{L}$ & $\times$ & $I$ & $\times$ & $H$ \\ \hline
  $m'_{o}$ & $H$ & $\times$ & $\times$ & $H$ \\ \hline
 \end{tabular}  \end{center}
  \begin{itemize}
   \item{$H$ means that the SUSY parameters are included in the
       transformation in a homogeneous way, id est, the contribution of
       the parameter vanishes without the matter fields.}
   \item{$I$ means that the SUSY parameters are included in the
       inhomogeneous way. This is a  translation independent of the
       matter fields. }
   \item{$\times$ means that the SUSY parameters are not included in
       the transformation.}
   \item{The inhomogeneous translation of the matter fields is supplied by
       $[\Gamma^{\sharp}, a_{o}]$, the translation in the direction of
       $x^{i}$. }
  \end{itemize}
  
  The SUSY transformation of these remaining fields is obtained by
  extracting only the transformation with respect to the fermionic
  parameters from (\ref{AZ55tr1964mo}), (\ref{AZ55tr1964psir}) and
  (\ref{AZ55tr1964phil}):
   \begin{eqnarray}
       \delta m'_{o} =  i ( \chi'_{L} {\bar \phi}'_{L} - \psi'_{R}
      {\bar \epsilon}'_{R}), \hspace{4mm} 
      \delta \psi'_{R} = 2 \chi'_{R}  - (m'_{o} \chi'_{L}),
      \hspace{4mm}
      \delta {\bar \phi}'_{L} = - 2 {\bar \epsilon}'_{L} + ( {\bar
         \epsilon}'_{R} m'_{o} ). \label{AZ55susy1964} 
   \end{eqnarray}
   We speculate that this structure is identical to the two-fold ${\cal 
   N}=2$ SUSY transformations, and consider the following correspondence
   between the fields of rank 1 and the vector fields of IKKT model.
  
  \paragraph{Correspondence between $A'^{(-)}_{i}$ and ($\chi_{L}$,
   ${\bar \epsilon_{L}}$)} .\\
  We first consider the transformation of the vector fields by these SUSY 
  parameters. We define a SUSY transformation $\delta^{(1)}_{\chi_{L}}$ and
  $\delta^{(2)}_{\epsilon_{L}}$ as follows:
   \begin{eqnarray}
    \delta^{(1)}_{\chi_{L}} = [Q_{\chi_{L}}, \hspace{1mm} \bullet
     \hspace{1mm} ] = [ 
     \left( \begin{array}{cc}  0 & \chi_{L} \\ 0 & 0 \end{array}
     \right), \hspace{1mm} \bullet \hspace{1mm} ], \hspace{3mm} 
    \delta^{(2)}_{\epsilon_{L}} = [Q_{\epsilon_{L}}, \hspace{1mm}
     \bullet \hspace{1mm} ] = [ 
     \left( \begin{array}{cc} 0 & 0 \\ i {\bar \epsilon}_{L} & 0
     \end{array} \right) , \hspace{1mm} \bullet \hspace{1mm} ]. 
   \end{eqnarray}
  The former is regarded as the homogeneous transformation, and the
  latter is the inhomogeneous translation, form the correspondence in
  the above table. We first consider the transformation of the matter
  field. As we have seen, the contribution of the SUSY transformation of
  left chirality is $\delta m'_{o} \to i \chi_{L} {\bar \phi}'_{L}$,
  Using the prescription in Appendix. \ref{AZCdecomposition}, we
  obtain 
   \begin{eqnarray}
 \delta^{(1)}_{\chi_{L}} A'^{(+)}_{i} &=& \frac{1}{32} tr(i \chi_{L}
    {\bar \phi}'_{L} \Gamma_{i}) + \frac{-1}{32} tr(i \chi_{L} {\bar
    \phi}'_{L} \Gamma_{i \sharp}) = - \frac{i}{32} {\bar \phi}'_{L}
    \Gamma_{i} (1 - \Gamma_{\sharp}) \chi_{L} = 0, \nonumber \\
 \delta^{(1)}_{\chi_{L}} A'^{(-)}_{i} &=& \frac{1}{32} tr(i \chi_{L}
    {\bar \phi}'_{L} \Gamma_{i}) - \frac{-1}{32} tr(i \chi_{L} {\bar
    \phi}'_{L} \Gamma_{i \sharp}) = - \frac{i}{32} {\bar \phi}'_{L}
    \Gamma_{i} (1 + \Gamma_{\sharp}) \chi_{L} = - \frac{i}{16} {\bar
    \phi}'_{L} \Gamma_{i} \chi_{L}.     
   \end{eqnarray}
 The inhomogeneous translation of course do not affect the transformation
 of the vector fields. This clarifies the correspondence between the
 fields $A'^{(-)}_{i}$ and the fermions ${\bar \phi}'_{L}$. \\

  We next consider the commutator of these transformations. We have seen
  a complicated structure of the commutator of the SUSY
  transformation in $osp(1|32,R)$ non-gauged theory. Especially, the
  commutator of the two homogeneous 
  transformations was a grave nuisance, and this caused many unfavorable
  structures. However, this is not the case with the SUSY
  structure of the gauged theory. The correspondence of the SUSY
  transformations is now crystal 
  clear. The commutators of the two SUSY  transformations of the same
  kind trivially vanish:
   \begin{eqnarray}
   & & [\delta^{(1)}_{\chi_{L}}, \delta^{(1)}_{\rho_{L}}] = 
 [ [ \left( \begin{array}{cc} 0 & \chi_{L} \\ 0 & 0 \end{array}
  \right),   \left( \begin{array}{cc} 0 & \rho_{L} \\ 0 & 0
  \end{array} \right)], \hspace{1mm} \bullet \hspace{1mm} ]  = 0,
 \label{AZ55SUSYcom1} \\ 
   & &  [\delta^{(2)}_{\epsilon_{L}}, \delta^{(2)}_{\eta_{L}}] =
  [ [ \left( \begin{array}{cc} 0 & 0 \\ i {\bar \epsilon}_{L} & 0
  \end{array} \right), \left( \begin{array}{cc} 0 & 0 \\ i {\bar
  \eta}_{L} & 0 \end{array} \right)] , \hspace{1mm} \bullet
  \hspace{1mm} ]= 0, \label{AZ55SUSYcom2} 
   \end{eqnarray}
 where we have utilized the relations (\ref{AZ43comm1964}). 
 These relations hold without any room for the impurities to be
 involved. We next investigate the commutator between the homogeneous
 and inhomogeneous transformation. Since the 
 commutator of the charge is 
  \begin{eqnarray}
 [Q_{\chi_{L}}, Q_{\epsilon_{L}}] = [ \left( \begin{array}{cc} 0 &
  \chi_{L} \\ 0 & 0 \end{array} 
  \right), \left( \begin{array}{cc} 0 & 0 \\ i {\bar \epsilon}_{L} & 0
  \end{array} \right) ] =  \left( \begin{array}{cc} i \chi_{L} {\bar
  \epsilon}_{L}  & 0  \\ 0  & 0 \end{array} \right), 
  \end{eqnarray}
 the commutator of these SUSY transformations with respect to the
 bosonic fields is
  \begin{eqnarray}
   [ \delta^{(1)}_{\chi_{L}}, \delta^{(2)}_{\epsilon_{L}}] (m'_{o} + R 
   \Gamma^{\sharp}) = i [ ( \chi_{L} {\bar \epsilon}_{L} ), m'_{o} + R
   \Gamma^{\sharp} ]. \label{AZ55susycommat}
  \end{eqnarray}
  It is trivial that this commutator with respect  to the fermionic
  fields vanish. We would like to extract the SUSY transformation with
  respect to the fields $A'^{(\pm)}_{i}$: 
   \begin{eqnarray}
   & & [ \delta^{(1)}_{\chi_{L}}, \delta^{(2)}_{\epsilon_{L}}]
   A'^{(+)}_{i} = 0, \nonumber \\
   & & [ \delta^{(1)}_{\chi_{L}}, \delta^{(2)}_{\epsilon_{L}}]
   A'^{(-)}_{i} = \frac{1}{32} tr( i [ ( \chi_{L} {\bar \epsilon}_{L}
   ), m'_{o} + R \Gamma^{\sharp} ] \Gamma_{i} ) - \frac{-1}{32} tr( i
   [ ( \chi_{L} {\bar \epsilon}_{L} ), m'_{o} + R \Gamma^{\sharp} ]
   \Gamma_{i \sharp} ) \nonumber \\ 
  & & \hspace{23mm} = \frac{iR}{16} {\bar \epsilon}_{L}
   \Gamma_{i \sharp}  \chi_{L} - \frac{i}{16} {\bar \epsilon}_{L}
   [m'_{o}, \Gamma_{i}] \chi_{L} =  \frac{iR}{16} {\bar \epsilon}_{L}
   \Gamma_{i}  \chi_{L}.  \label{AZ55finalresult1}
    \end{eqnarray}
  The term ${\bar \epsilon}_{L} [m'_{o}, \Gamma_{i}] \chi_{L}$
  is understood to vanish, because the commutator $[m'_{o},
  \Gamma_{i}]$ produces only the gamma matrices of even rank from the
  formula (\ref{AZproduct}). And considering the relations
  (\ref{AZMA21fermvanish}), this term is prohibited from existing
  in this contribution. 

 \paragraph{Correspondence between $A'^{(+)}_{i}$ and ($\chi_{R}$,
   ${\bar \epsilon_{R}}$)} .\\
 The analysis goes in the same way as in the previous case. The above
 table indicates the homogeneous and inhomogeneous SUSY transformation is
   \begin{eqnarray}
   \delta^{(1)}_{\epsilon_{R}} = [Q_{\epsilon_{R}}, \hspace{1mm}
   \bullet \hspace{1mm} ] = [ \left( \begin{array}{cc} 0 & 0 \\ i {\bar 
   \epsilon}_{R} & 0 \end{array} \right), \hspace{1mm} \bullet
   \hspace{1mm} ], \hspace{3mm} \delta^{(2)}_{\chi_{R}} =
   [Q_{\chi_{R}}, \hspace{1mm}  \bullet \hspace{1mm} ] =
   [ \left( \begin{array}{cc} 0 & \chi_{R} \\ 0 & 0 \end{array}
   \right), \hspace{1mm} \bullet \hspace{1mm} ].
   \end{eqnarray}
 Since the transformation of the bosonic matter field with respect to the
 SUSY transformation of right chirality is $\delta m'_{o} \to -i
 \psi'_{R} {\bar \epsilon}_{R}$ , the transformation of the fields is as 
 follows. 
  \begin{eqnarray}
   \delta^{(1)}_{\epsilon_{R}} A'^{(+)}_{i} &=& \frac{1}{32} tr( -i
   \psi'_{R} {\bar \epsilon}_{R} \Gamma_{i}) + \frac{-1}{32} tr( -i
   \psi'_{R} {\bar \epsilon}_{R} \Gamma_{i \sharp}) = 
   \frac{i}{32} {\bar \epsilon}_{R} \Gamma_{i} ( 1 - \Gamma_{\sharp})
   \psi'_{R} = \frac{i}{16} {\bar \epsilon}_{R} \Gamma_{i} \psi'_{R},
   \label{AZ55trmata+} \\
    \delta^{(1)}_{\epsilon_{R}} A'^{(-)}_{i} &=& \frac{1}{32} tr( -i
   \psi'_{R} {\bar \epsilon}_{R} \Gamma_{i}) - \frac{-1}{32} tr( -i
   \psi'_{R} {\bar \epsilon}_{R} \Gamma_{i \sharp}) = 
   \frac{i}{32} {\bar \epsilon}_{R} \Gamma_{i} ( 1 + \Gamma_{\sharp})
   \psi'_{R} = 0. \nonumber 
  \end{eqnarray}
  Likewise, the inhomogeneous translation does not affect the bosonic
  fields. This clarifies the correspondence between the vector fields
  $A'^{(+)}_{i}$ and the fermions $\psi_{R}$. 

  It is again clear that the commutators of the SUSY
  transformations of the same kind vanish:
   \begin{eqnarray}
   & & [\delta^{(1)}_{\epsilon_{R}}, \delta^{(1)}_{\eta_{R}} ] = [ [ 
   \left( \begin{array}{cc} 0 & 0 \\ i {\bar \epsilon}_{R} & 0
   \end{array} \right),  \left( \begin{array}{cc} 0 & 0 \\ i {\bar
   \eta}_{R} & 0 \end{array} \right) ] , \hspace{1mm} \bullet
   \hspace{1mm} ] = 0, \label{AZ55SUSYcom3} \\
   & & [\delta^{(2)}_{\chi_{R}}, \delta^{(2)}_{\rho_{R}} ] = [ [ 
   \left( \begin{array}{cc} 0 & \chi_{R} \\ 0 & 0 \end{array} \right),
   \left( \begin{array}{cc} 0 & \rho_{R} \\ 0 & 0 \end{array} \right)
   ] , \hspace{1mm} \bullet \hspace{1mm} ] = 0. \label{AZ55SUSYcom4}  
   \end{eqnarray}
  Utilizing the commutation relation $
  [\delta^{(1)}_{\epsilon_{R}}, \delta^{(1)}_{\chi_{R}} ] = [  
  \left( \begin{array}{cc} 0 & 0 \\ i {\bar \epsilon}_{R} & 0
  \end{array} \right) ,  \left( \begin{array}{cc} 0 & \chi_{R} \\ 0 &
  0 \end{array} \right) =  
   \left( \begin{array}{cc} -i \chi_{R} {\bar \epsilon}_{R}  & 0 \\ 0
   & 0 \end{array} \right)$, we obtain
  \begin{eqnarray}
   [ \delta^{(1)}_{\epsilon_{R}}, \delta^{(2)}_{\chi_{R}}] (m'_{o} + R 
   \Gamma^{\sharp}) = - i [ ( \chi_{R} {\bar \epsilon}_{R} ), m'_{o} + R
   \Gamma^{\sharp} ]. \label{AZ55susycomma-}
  \end{eqnarray}
  Therefore, the commutator of these SUSY transformations with respect
  to the fields of rank 1 is 
   \begin{eqnarray}
    & & [\delta^{(1)}_{\epsilon_{R}}, \delta^{(2)}_{\chi_{R}}]
    A'^{(+)}_{i} = \frac{1}{32} tr( - i [ ( \chi_{R} {\bar
    \epsilon}_{R} ), m'_{o} + R \Gamma^{\sharp} ] \Gamma_{i} ) +
    \frac{-1}{32}  tr( - i [ ( \chi_{R} {\bar
    \epsilon}_{R} ), m'_{o} + R \Gamma^{\sharp} ] \Gamma_{i \sharp} )
   \nonumber \\
   & & \hspace{23mm} = \frac{iR}{16} {\bar \epsilon}_{R} \Gamma_{i} \chi_{R} +
    \frac{i}{16} {\bar \epsilon}_{R} [m'_{o}, \Gamma_{i}] \chi_{R} =
    \frac{iR}{16} {\bar \epsilon}_{R} \Gamma_{i} \chi_{R}, 
   \label{AZ55finalresult2} \\
    & & [\delta^{(1)}_{\epsilon_{R}}, \delta^{(2)}_{\chi_{R}}]
    A'^{(-)}_{i} = 0, \nonumber 
   \end{eqnarray}
  where ${\bar \epsilon}_{R} [m'_{o}, \Gamma_{i}] \chi_{R}$ vanishes
  because of the same reasoning as before.

 \paragraph{Structure of ${\cal N}=2$ SUSY transformation} . \\
  The transformation of the fields $A^{(\pm)}_{i}$ and the commutator of
  the SUSY transformation clarifies the structure of the SUSY
  transformation. First, we have clarified the correspondence between
  the fields $A'^{(\pm)}_{i}$ and the chirality of the fermions. The
  correspondence is the same as the $osp(1|32,R)$ non-gauged theory. 
  \begin{center} \begin{tabular}{|c||c|c|c|} \hline
    &  bosons $A'_{i}$ (IKKT) & fermions $\psi$(IKKT) & SUSY parameters 
    \\ \hline \hline
    SUSY I   &  $A'^{(+)}_{i}$ & $\psi'_{R}$ & $\chi_{R}$,
    $\epsilon_{R}$  \\ \hline  
    SUSY II  &  $A'^{(-)}_{i}$ & ${\bar \phi}'_{L}$ & $\chi_{L}$,
    $\epsilon_{L}$  \\ \hline 
  \end{tabular} \end{center} 
 And the corresponding SUSY parameter is of the same chilarity as the
 corresponding fermionic fields. In the same sense as in $osp(1|32,R)$ 
 model, we have constructed two ${\cal N}=2$ SUSY structures by
 deriving the commutation relations of the SUSY transformations
 (\ref{AZ55finalresult1}) and (\ref{AZ55finalresult2}), to be
 identified with IKKT  model. The original 64 SUSY parameters are 
 happily  separated into two groups each of which comprises 32 SUSY
 parameters. 

  The gauged theory analyzed by means
 of the Wigner-In{\"o}n{\"u} contraction is much more beautiful than in
 $osp(1|32,R)$ non-gauged theory. In the previous section, we have seen
 that $osp(1|32,R)$ non-gauged theory does not completely reproduce
 the SUSY structures of the IKKT model. The commutator of the
 supercharge produces not only the translation of the vector fields but also
 included the impurities $W$, $C_{i_{1} i_{2}}$ and $H_{i_{1} \cdots
 i_{4}}$\footnote{These are distinguished as $m_{e}$ in the contemporary
 context.}. And furthermore the two-fold ${\cal N}=2$ SUSY structures
 were not independent of each other. However, the analysis by means of
 The Wigner-In{\"o}n{\"u} contraction solves these problems splendidly.
 In the gauged theory, these two diseases are completely cured, and
 the two SUSY strictures are now regarded as independent. This happy
 result is predictable from the fact that the gauged theory possesses
 a gauge symmetry much richer than that of the non-gauged $osp(1|32,R)$
 model.

 However, there are two problems with this effective theory. First is
 that the matter fermions subject to the homogeneous SUSY transformation is
 {\it different from} the fermions which is translated by the
 inhomogeneous SUSY transformation. This discrepancy of the fermions
 is discerned from the table given above. For example, with respect to 
 the SUSY ($A'^{(-)}_{i}$, $\chi_{L}$, $\epsilon_{L}$), the fermion
 $\psi'_{R}$ receives the homogeneous transformation, whereas ${\bar
 \phi}'_{L}$ receives the inhomogeneous transformation. We have yet to 
 gain insight into the physical interpretation of this discrepancy, and it is 
 unclear how we should cope with this problem. 

  The second problem is that the effective action obtained by
  integrating out the fields $m'_{e}$, $\psi'_{L}$, ${\bar \phi}'_{R}$ 
  and $v'$ vanish completely : $W=0$.  Even if we have identified the SUSY
  structures of this effective theory, the effective action  possesses 
  only an empty structure. This is a serious problem, however there
  are two interpretations of this situation. First is to interpret
  that this effective action becomes zero just to the
  order\footnote{Note that the order in the following 
  discussion  is for the theory before the rescaling.} ${\cal
  O}(R^{-1})$. In other
  words, our analysis just says $ W = 0 + {\cal O}(R^{-2})$.
  We cannot abandon the possibility that there may emerge a non-zero
  contribution in the lower order. If we succeed in the analysis of
  this order, the structure of the action to be identified with
  IKKT model may lie in the world of the lower order. The second
  interpretation is that this situation may be related to the
  suggestion of topological matrix model \cite{9708039}. Hirano and
  Kato made a bold proposal that IKKT model is induced from aught,
  though it deviates from the proverb 'Nothing comes out of 
  nothing'.  Even though their proposal may sound perverse, this is related to
  the topological symmetries of IKKT model. Their proposal may be
  related to the scenario to derive IKKT model from this  gauged cubic
  matrix model.

  Even though the scenario to induce IKKT model from this model is
  tougher than in $osp(1|32,R)$ cubic matrix model, this model has a
  great advantage that the two-fold ${\cal N}=2$ SUSY structures are
  realized much more beautifully than in $osp(1|32,R)$ model, because
  of the vast symmetry. 

 \subsection{Summary}
  We have considered the gauged cubic matrix model as an extension of
  the conventional suggestion of the non-gauged theory.
   \begin{itemize}
    \item{We have investigated the gauged cubic matrix model with the
        gauge group $gl(1|32,R) \otimes gl(N,R)$,which is the analytic 
        continuation of the  conventional Smolin's proposal
        $u(1|16,16)$.  Both $gl(1|32,R)$ and $u(1|16,16)$ are
        constructed by the direct sum of the two different
        representations of $osp(1|32,R)$ super Lie algebra. Although
        this model is deprived of a symmetry of the commutator, this
        model possesses a gauge symmetry much richer than that of the
        non-gauged theory.}
    \item{We have investigated this theory by means of the
        Wigner-In{\"o}n{\"u} contraction. 
        \begin{itemize}
         \item{This gauged theory also possesses the two-fold ${\cal
               N}=2$ SUSY structure.}
         \item{ The effective action happened to vanish : $W=0$.} 
        \end{itemize}}
   \end{itemize} 
  The scenario to derive IKKT model from the gauged action, which is
  the first step for these model to exceed IKKT model, is still
  tougher than in the non-gauged IKKT model, however the vast symmetry 
  of this theory is a splendid aspect of this gauged theory. The
  investigation of this model, paying attention to the conventional
  matrix model, is an interesting issue to pursue.

\section{Concluding Remark}
 We have hitherto investigated the possibility that the matrix
 model describes superstring theory. This idea stems from the attempt
 to describe string theory in terms of large $N$ matrix in the late
 1980's. These works just succeeded in describing bosonic string
 theory in a very low spacetime dimension. The extension to this idea
 to superstring theory was not successful because of the same
 difficulty that we are faced  in describing chiral fermions in
 lattice gauge theory. These works were far from describing the real
 superstring theory residing in 10 dimensional spacetime. However these
 works gave a way to describe the nonperturbative behavior of string 
 theory, and the exact solution is obtained by a non-linear
 differential equation named Painleve equation. Although this is a
 mere toy model, these works gave us a confidence that the
 nonperturbative behavior of superstring theory may be described by
 matrix model.

 In the late 1990's, the description of the nonperturbative
 superstring theory has become a more fascinating issue because the
 discovery of D-brane has clarified that different kinds of
 superstring theories are in fact related with each other by
 duality. At that time, there has emerged a belief that the
 constructive definition of superstring theory is 'Theory of
 Everything', which unifies all the interactions in the universe. The
 most powerful existing proposal for the nonperturbative description
 of superstring theory is IKKT model. 

 In this paper, we have attempted to construct a matrix model
 exceeding IKKT model. The clue to this challenging issue is the
 proposal of L. Smolin. He conjectured that the cubic matrix model
 described  by $OSp(1|32,R) \times SU(N)$ gauge group and its
 gauged model may be the clue to the new theory truly
 exceeding IKKT model. We have investigated these model, especially
 paying attention to the structure of supersymmetry and  the way IKKT
 model is embedded.

  We have discovered that both the non-gauged $osp(1|32,R)$ model and
  the gauged $gl(1|32,R) \otimes gl(N,R)$ model 
  possess two-fold structures of ${\cal N}=2$ SUSY of IKKT model. This
  is a predictable result, noting that these theories possesses 64
  fermionic SUSY parameters. This may indicate that the world
  described by this model may include two worlds of IKKT model. 

  However, it was a tough problem to induce the action of IKKT model
  from these cubic matrix model.  We have yet to obtain a definite
  scenario to induce IKKT model from these cubic actions. This issue
  is the first step to assert that these cubic models truly describe
  superstring theory. 

  These matrix models suggest many other interesting issues than we
  have treated in this paper. 
   \begin{itemize}
    \item{ These two-fold SUSY structures are reminiscent of the brane world
  scenario, in which there are two 4-dimensional worlds described by
  D3-branes in the extra dimensional spacetime. Investigating the
  two worlds of IKKT model, we may derive a relationship with the
  conventional brane world scenario.}
    \item{These models are formulated in 11 dimensional spacetime, and 
       may enable us to treat the curved 10 dimensional spacetime. This
       is an impossible problem in IKKT model, because the theory is
       described in 10 dimensional flat Minkowski space. These models
       have a possibility to describe such curved space as $S_{1}
       \times R^{9}$ or $AdS_{5} \times S^{5}$ spacetime. The
       description of $AdS_{5} \times S^{5}$ space is an interesting
       problem, in connection with AdS/CFT correspondence
       \cite{9711200} . The direct 
       test of AdS/CFT correspondence has been attempted by
       investigating the strong-coupling region of ${\cal N}=4$ SYM
       theory \cite{0003055} \cite{0010274}. These models may play
       an essential role in  describing the whole region of the coupling
       constant of the string theory.}
    \item{ As we have mentioned at the beginning of Sec. 4, this model 
        is related to Chern-Simons theory by a due
        compactification. This indicates the possibility to describe
        the nonperturbative behavior of superstring theory exactly by
        means of Jones Polynomial \cite{Jones}. If you succeeded in
        this issue, you can be Brezin and Kazakov in superstring
        theory. }
   \end{itemize}

  Mankind  has yet to grasp what is the true 'Theory of Everything'
  These cubic models described by $osp(1|32,R)$ (or its
  extension), which is the ultimate symmetry group of M-theory, may be 
  an answer to this ultimate and most difficult question of elementary
  particle physics.

  \paragraph{Acknowledgment} \hspace{0mm} \\
  It is my great pleasure to acknowledge the collaboration and the advice to
  complete the present work. I would like to express my sincere
  gratitude to Prof. H. Kawai for his ardent guidance. The
  discussion with him was literally an eye-opener, and I find his
  perusal of this manuscript quite an asset for me. And I would
  like to express my thanks to Prof. S. Iso and Dr. Y. Ohwashi for the 
  collaboration. 

\appendix
\section{Notation}
  \subsection{Definitions of the Gamma Matrices} \label{AZCgm}
  We follow the conventions of \cite{Pol}. 
 The gamma matrices obey the Clifford algebra 
    \begin{eqnarray} \{ \Gamma^{\mu} , \Gamma^{\nu} \} = 2 \eta^{\mu
  \nu}, \hspace{2mm} \textrm{where} \hspace{2mm} \eta^{\mu \nu} =
  diag(-1, +1, \cdots, +1). \label{AZclifford} \end{eqnarray}
    The explicit forms of the gamma matrices are as follows:
    \begin{eqnarray}  \Gamma^{i} = \left( \begin{array}{cc} \gamma^{i}
  & 0  \\ 0 & - \gamma^{i} \end{array} \right) ,
 \hspace{2mm} \Gamma^{10} = \Gamma^{\sharp} = \left( \begin{array}{cc}
 0  & I  \\ I & 0  \end{array} \right),
 \hspace{2mm} \Gamma^{0}
 = \left( \begin{array}{cc} 0  & -I \\ I & 0 
 \end{array} \right),  \label{AZgammaexplicit}  \end{eqnarray}
  where $I$ is a $16 \times 16$ unit matrix, and $\gamma^{i}$ are $16
  \times 16$ matrices following the Clifford 
  algebra $ \{ \gamma^{i} , \gamma^{j} \} = 2 \delta^{ij}$, so that
  these matrices obeys 
   \begin{eqnarray} 
    (\Gamma^{0})^{2} = -1, \hspace{2mm}  (\Gamma^{\mu})^{2} =
    1 (\mu = 1,2, \cdots \sharp). 
   \end{eqnarray}
   The transpose of the gamma matrices is, as is clear from the
   explicit form (\ref{AZgammaexplicit}), 
    \begin{eqnarray}
      {^{T} \Gamma^{0} } = - \Gamma^{0}, \hspace{3mm}
      {^{T} \Gamma^{\sharp}} = \Gamma^{\sharp}. 
    \end{eqnarray}
   The gamma matrix $\Gamma^{\sharp}$ is originally defined as 
   \begin{eqnarray} \Gamma^{\sharp} = \Gamma^{0} \Gamma^{1} \Gamma^{2} 
     \cdots \Gamma^{9}, \end{eqnarray}
  so that the following property holds: 
   \begin{eqnarray} ( \Gamma^{\sharp})^{2} = \Gamma^{0}
  \Gamma^{1} \cdots \Gamma^{9} \Gamma^{\sharp} = (-1)^{9+8+ \cdots +1} 
  (\Gamma^{0})^{2} \cdots (\Gamma^{9})^{2} = (-1)^{46} 1 = 1.
  \end{eqnarray}

 \subsection{Miscellaneous properties of the Gamma Matrices}
  \label{AZCgammaproperty} 
   \subsubsection{Chirality of Fermions} \label{AZCfermch}
  We have defined the gamma matrices $\Gamma^{\sharp}$ in the previous 
  section. This notion is deeply related to the chirality of the
  fermions.   The fermions with left and right chirality is in this
  paper defined  as follows: \\
  \shadowbox{ \parbox{16cm}{
  \begin{eqnarray}
    \psi_{L} \stackrel{def}{=} \frac{1 + \Gamma_{\sharp}}{2} \psi,
    \hspace{3mm} 
    \psi_{R} \stackrel{def}{=} \frac{1 - \Gamma_{\sharp}}{2} \psi.
  \end{eqnarray} }} \\
    We next introduce some relationships of the chiral fermion we
    frequently utilize in the analysis. Let the gamma matrices
    $\Gamma^{o}$ and $\Gamma^{e}$ be of odd or even rank with respect
    to 10 dimensional indices respectively. \footnote{For example,
    $\Gamma^{i \sharp}$ or $\Gamma^{ijk}$ is abstractly expressed by
    $\Gamma^{o}$. On the other hand, $\Gamma^{\sharp}$ or
    $\Gamma^{ij}$ belongs to the family $\Gamma^{e}$. }
   Then, the following relationships stand: \\
   \shadowbox{ \parbox{16cm}{
    \begin{eqnarray}
    (1) {\bar \chi}_{L} \Gamma^{e} \epsilon_{L} = {\bar \chi}_{R}
    \Gamma^{e} \epsilon_{R} = 0, \hspace{3mm}
    (2) {\bar \chi}_{L} \Gamma^{o} \epsilon_{R} = {\bar \chi}_{R}
    \Gamma^{o} \epsilon_{L} = 0. \label{AZMA21fermvanish}
    \end{eqnarray} }}

  {\sf (Proof)
   These relationships can be verified by writing the contributions of 
   the fermions explicitly. 
  \begin{enumerate}
   \item{ The former is shown to vanish by the following computation.
    Let $k$ be a positive  integer, and thus $2k$ be an even number. The
    indices $i_{1}, i_{2}, \cdots i_{2k}$ runs $0,1, \cdots, 9$:
     \begin{eqnarray}
      {\bar \chi}_{L} \Gamma^{i_{1} \cdots i_{2k}} \epsilon_{L} &=& 
       {^{T} \chi} {^{T} (\frac{1 + \Gamma^{\sharp}}{2})} \Gamma^{0} 
       \Gamma^{i_{1} \cdots i_{2k}} \frac{1 + \Gamma^{\sharp}}{2}
       \epsilon = {^{T} \chi} \frac{1 + \Gamma^{\sharp}}{2} \frac{1 +
       (-1)^{2k+1} \Gamma^{\sharp}}{2} \Gamma^{0} \Gamma^{i_{1} \cdots
       i_{2k}} \epsilon \nonumber \\
    &=&  {^{T} \chi} \frac{1 + \Gamma^{\sharp}}{2} \frac{1 -
       \Gamma^{\sharp}}{2} \Gamma^{0} \Gamma^{i_{1} \cdots i_{2k}}
       \epsilon = 0, \label{AZMA21lol} \\
    {\bar \chi}_{R} \Gamma^{i_{1} \cdots i_{2k}} \epsilon_{R} &=&
       {^{T} \chi} {^{T} (\frac{1 - \Gamma^{\sharp}}{2})} \Gamma^{0}  
       \Gamma^{i_{1} \cdots i_{2k}} \frac{1 - \Gamma^{\sharp}}{2}
       \epsilon = {^{T} \chi} \frac{1 - \Gamma^{\sharp}}{2} \frac{1 -
       (-1)^{2k+1} \Gamma^{\sharp}}{2} \Gamma^{0} \Gamma^{i_{1} \cdots
       i_{2k}} \epsilon \nonumber \\
    &=&  {^{T} \chi} \frac{1 - \Gamma^{\sharp}}{2} \frac{1 +
       \Gamma^{\sharp}}{2} \Gamma^{0} \Gamma^{i_{1} \cdots i_{2k}}
       \epsilon = 0. \label{AZMA21ror} 
     \end{eqnarray}}
   \item{The latter properties can be shown in the similar fashion:
    \begin{eqnarray}
      {\bar \chi}_{L} \Gamma^{i_{1} \cdots i_{2k-1}} \epsilon_{R} &=& 
       {^{T} \chi} {^{T} (\frac{1 + \Gamma^{\sharp}}{2})} \Gamma^{0} 
       \Gamma^{i_{1} \cdots i_{2k-1}} \frac{1 - \Gamma^{\sharp}}{2}
       \epsilon = {^{T} \chi} \frac{1 + \Gamma^{\sharp}}{2} \frac{1 -
       (-1)^{2k} \Gamma^{\sharp}}{2} \Gamma^{0} \Gamma^{i_{1} \cdots
       i_{2k-1}} \epsilon \nonumber \\
    &=&  {^{T} \chi} \frac{1 + \Gamma^{\sharp}}{2} \frac{1 -
       \Gamma^{\sharp}}{2} \Gamma^{0} \Gamma^{i_{1} \cdots i_{2k-1}}
       \epsilon = 0, \label{AZMA21ler} \\
    {\bar \chi}_{R} \Gamma^{i_{1} \cdots i_{2k-1}} \epsilon_{L} &=&
       {^{T} \chi} {^{T} (\frac{1 - \Gamma^{\sharp}}{2})} \Gamma^{0}  
       \Gamma^{i_{1} \cdots i_{2k-1}} \frac{1 + \Gamma^{\sharp}}{2}
       \epsilon = {^{T} \chi} \frac{1 - \Gamma^{\sharp}}{2} \frac{1 +
       (-1)^{2k} \Gamma^{\sharp}}{2} \Gamma^{0} \Gamma^{i_{1} \cdots
       i_{2k-1}} \epsilon \nonumber \\
    &=&  {^{T} \chi} \frac{1 - \Gamma^{\sharp}}{2} \frac{1 +
       \Gamma^{\sharp}}{2} \Gamma^{0} \Gamma^{i_{1} \cdots i_{2k-1}}
       \epsilon = 0. \label{AZMA21rel} 
    \end{eqnarray} }
  \end{enumerate}
  Thus, we have verified the relationships
  (\ref{AZMA21fermvanish}). Note that the proof makes no difference
  even if the indices of $\Gamma^{e}$ or $\Gamma^{o}$ include $\sharp$
  (for example, $\Gamma^{i \sharp}$ or $\Gamma^{ijkl \sharp}$),
  because $\Gamma^{\sharp}$ serve only to flip the sign of $\chi_{R}$, 
  id est, $\Gamma^{\sharp} \chi_{L} = \chi_{L}$ whereas
  $\Gamma^{\sharp} \chi_{R} = - \chi_{R}$.  (Q.E.D.)  } 

   \subsubsection{Duality}
    We exhibit the way to describe the product of $d$ gamma
    matrices in term of the product of $(11-d)$ gamma matrices.
    Before that, let us settle the conventions: \\
   \shadowbox { \parbox{16cm}{ 
    \begin{itemize}
     \item{Epsilon Tensor : $\epsilon_{0123 \cdots 9 \sharp}=1$, so
   that $\epsilon^{0123 \cdots \sharp} = -1$. }
     \item{Antisymmetrized Gamma Matrices : $\Gamma^{\mu_{1} \cdots
   \mu_{k}} = \frac{1}{k!} \sum_{\sigma \in S_{n}} sgn(\sigma)
   \Gamma^{\mu_{\sigma(1)}} \Gamma^{\mu_{\sigma(2)}} \cdots
   \Gamma^{\mu_{\sigma(k)}}$. }
  \end{itemize}
   }} \\
     Under this convention, we obtain an important property: \\
    \shadowbox{ \parbox{16cm}{
    \begin{eqnarray} \Gamma^{\mu_{0} \mu_{1} \cdots \mu_{k}} =
     \frac{(-1)^{\frac{k(k+1)}{2}}}{(10-k)!} \epsilon^{\mu_{0} \mu_{1} 
     \cdots \mu_{9} \mu_{\sharp} } \Gamma_{\mu_{k+1} \mu_{k+2}
     \cdots \mu_{\sharp} }. \label{AZgammaduality}
    \end{eqnarray} }} \\
  This property can be understood by comparing the sign of
     $\Gamma^{012 \cdots k}$ and $\frac{1}{(10-k)!} \epsilon^{012
     \cdots \sharp} 
     \Gamma_{(k+1) \cdots \sharp} = - \Gamma^{(k+1) \cdots
     \sharp}$. Suppose that $\Gamma^{012 \cdots k} = -a \Gamma^{(k+1)
     \cdots \sharp}$ .Of course, $\Gamma^{012 \cdots k} = \Gamma^{0}
     \Gamma^{1} \cdots \Gamma^{k}$.  Then multiplying $\Gamma^{012
     \cdots k }$ from  the left on the both hand sides, we obtain 
    \begin{eqnarray} (\Gamma^{012 \cdots k})^{2} = (-1)^{k+ \cdots +1} 
     (\Gamma^{0})^{2} \cdots (\Gamma^{k})^{2} =
     (-1)^{\frac{k(k+1)}{2}+1} 1 = -a \Gamma^{0} \cdots
     \Gamma^{\sharp} = -a 1.
    \end{eqnarray}
     Therefore, we understand that the relative sign is $ a =
     (-1)^{\frac{k(k+1)}{2}}$. \\

  \subsubsection{Multiplication law of the Gamma Matrices}
  \label{AZCproductgamma}
    Next we exhibit another frequently used property of the gamma
    matrices. The products of $\Gamma^{\mu_{1} \mu_{2} \cdots
    \mu_{m}}$ and $\Gamma^{\nu_{1} \nu_{2} \cdots \nu_{n}}$ is known
    to be \\
      \shadowbox{ \parbox{16cm}{
     \begin{eqnarray} \Gamma^{\mu_{1} \cdots \mu_{m}} \Gamma^{\nu_{1}
    \cdots \nu_{n}} &=& \Gamma^{\mu_{1} \cdots \mu_{m} \nu_{1} \cdots
    \nu_{n}} + {(-1)^{m-1}} {_{m}C_{1}}{_{n}C_{1}}
    {\eta^{[\mu_{1}}}^{[\nu_{1}} {\Gamma^{\mu_{2} \cdots
    \mu_{m}]}}^{\nu_{2} \cdots \nu_{n}]} \nonumber \\
   &+& {(-1)^{(m-1)+(m-2)}} {_{m}C_{2}}{ _{n}C_{2}} 2!
    {\eta^{[\mu_{1}}}^{[\nu_{1}} {\eta^{\mu_{2}}}^{\nu_{2}}
    {\Gamma^{\mu_{3} \cdots \mu_{m}]}}^{\nu_{3} \cdots \nu_{n}]}
    \label{AZproduct} \\
  &+& {(-1)^{(m-1)+(m-2)+(m-3)}} {_{m}C_{3}}{ _{n}C_{3}} 3!
    {\eta^{[\mu_{1}}}^{[\nu_{1}} {\eta^{\mu_{2}}}^{\nu_{2}}
    {\eta^{\mu_{3}}}^{\nu_{3}} {\Gamma^{\mu_{4} \cdots
    \mu_{m}]}}^{\nu_{4} \cdots \nu_{n}]} + \cdots,  \nonumber
     \end{eqnarray} }} \\
      where the indices $\mu_{1}, \cdots, \mu_{n}$ and $\nu_{1} ,\cdots
   \nu_{n}$ in the right hand are antisymmetrized. For clarity we give 
   a simple example of this notation:
    \begin{eqnarray}
     \Gamma^{i} \Gamma^{jk} = \Gamma^{ijk} + 2 (-1)^{1-1} \eta^{i [j}
     \Gamma^{k]} = \Gamma^{ijk} + \eta^{ij} \Gamma^{k} - \eta^{ik}
     \Gamma^{j}. 
    \end{eqnarray}
 
 \subsubsection{Flipping Property of the Fermions} \label{AZCflipping} 
  We next investigate the properties for the fermionic field. This
  section is devoted to the proof of the following properties \\
  \shadowbox{ \parbox{16cm}{
   \begin{eqnarray}
    & &(1)\textrm{For } k=1,2,5 \Rightarrow {\bar \chi}
    \Gamma^{\mu_{1} \cdots \mu_{k}} \eta = - {\bar \eta}
    \Gamma^{\mu_{1} \cdots \mu_{k}} \chi, \label{AZMA2flip125} \\
    & &(2)\textrm{For } k=0,3,4 \Rightarrow {\bar \chi} 
    \Gamma^{\mu_{1} \cdots  \mu_{k}} \eta = + {\bar \eta}
    \Gamma^{\mu_{1} \cdots \mu_{k}} \chi,  \label{AZMA2flip034} 
   \end{eqnarray} }} \\
  where  $k$ is a rank of the gamma matrix and $\chi$ and $\eta$ are
  both Majorana spinors. \\
  {\sf (Proof)
 First we prepare some basic properties frequently used
  in this proof:
   \begin{eqnarray} 
    (a)(\Gamma^{0})^{2} = - {\bf 1},\hspace{2mm} 
    (b)({^{T} \Gamma^{0}}) = - \Gamma^{0} , \hspace{2mm} 
    (c)({^{T} \Gamma^{i}}) = \Gamma^{i} ,\hspace{2mm} 
    (d)({^{T} \Gamma^{\sharp}}) = \Gamma^{\sharp},
  \end{eqnarray}
  which is immediately understood from the definition of the gamma
  matrices. We utilize these properties in computing the flipping
  properties. From the above properties, it is easy to verify that 
  \begin{eqnarray}
   \Gamma^{0}  ({^{T}\Gamma^{\mu}}) \Gamma^{0} = \Gamma^{\mu}.
   \label{AZMA21st}
   \end{eqnarray}
  \begin{itemize}
  \item{ For $\mu=0$, $\Gamma^{0} ({^{T}\Gamma^{0}}) \Gamma^{0}
      \stackrel{(b)}{=} - (\Gamma^{0})^{3} \stackrel{(a)}{=} 
      \Gamma^{0}$.}
  \item{ For $\mu= 1, \cdots 9, \sharp$, the reasoning is a bit
      different from the previous case. 
      $\Gamma^{0}  ({^{T}\Gamma^{\mu}}) \Gamma^{0} 
       \stackrel{(c)(d)}{=} \Gamma^{0} \Gamma^{\mu} \Gamma^{0} 
       \stackrel{(\ref{AZclifford})}{=} - (\Gamma^{0})^{2} \Gamma^{\mu} 
       \stackrel{(a)}{=} \Gamma^{\mu}$.}
  \end{itemize}

  Utilizing this property, we obtain a following relationship which
  plays an  essential role in the proof of the flipping property,
  \begin{eqnarray}
     \Gamma^{0} ({^{T}\Gamma^{\mu_{1} \cdots \mu_{k}}}) \Gamma^{0}
    &\stackrel{\ast}{=}&
    (-1)^{k-1} (\Gamma^{0} ({^{T} \Gamma^{\mu_{k}}})  \Gamma^{0} )
    \cdots (\Gamma^{0} ({^{T} \Gamma^{\mu_{1}}})  \Gamma^{0} ) 
    \stackrel{\ast \ast}{=}
    (-1)^{k-1} \Gamma^{\mu_{k} \cdots \mu_{1}} \nonumber \\
    &\stackrel{\ast \ast \ast}{=}& (-1)^{k-1}
    (-1)^{\frac{k(k-1)}{2}} \Gamma^{\mu_{1} \cdots \mu_{k}} =
    (-1)^{\frac{(k+2)(k-1)}{2}} \Gamma^{\mu_{1} \cdots \mu_{k}}.
    \label{AZMkth}  
  \end{eqnarray}
  \begin{itemize}
   \item{ $\ast$ : We first rewrote the gamma matrix
        ${^{T}\Gamma^{\mu_{1} \cdots \mu_{k}}} = ({^{T}
        \Gamma^{\mu_{k} }})({^{T} \Gamma^{\mu_{k-1}}}) \cdots ({^{T}
        \Gamma^{\mu_{1}}})$. And then  we inserted $(\Gamma^{0})^{2}
        = -1 $ for each of the $(k-1)$ intervals between ${^{T}
        \Gamma^{\mu_{l} }}$ and ${^{T} \Gamma^{\mu_{l-1} }}$ with
        $l=2, 3, \cdots, k$.}
   \item{ $\ast \ast$ : We have utilized the property
       (\ref{AZMA21st}) for each of $ (\Gamma^{0} ({^{T}
       \Gamma^{\mu_{l}}})  \Gamma^{0} )$ with $l=1,2,\cdots k$.}
   \item{$\ast \ast \ast$ : We have changed the order of the
       indices of $\Gamma^{k \cdots 1}$. Since  $\Gamma^{k \cdots 1}$
       is anti-symmetric with respect to the exchange of the indices,
       we gain another factor of $(-1)^{\frac{k(k-1)}{2}}$.}
  \end{itemize} 
  Having this property of the gamma matrices in mind, we enter the
  proof of the flipping properties:
 \begin{eqnarray}
    {\bar \chi} \Gamma^{\mu_{1} \cdots \mu_{k}} \eta
  &\stackrel{\ast}{=}& - ({^{T} \chi}) (\Gamma^{0} \Gamma^{\mu_{1}
   \cdots \mu_{k}} \Gamma^{0}) \Gamma^{0} \eta 
   \stackrel{(\ref{AZMkth}) }{=} - (-1)^{\frac{(k+2)(k-1)}{2}} ({^{T}
   \chi}) ({^{T} \Gamma^{\mu_{1} \cdots \mu_{k}}} ) \Gamma^{0} \eta 
  \nonumber \\
  &\stackrel{\ast \ast}{=}&  +
   (-1)^{\frac{(k+2)(k-1)}{2}}({^{T} \eta}) ({^{T} \Gamma^{0}})
   (\Gamma^{\mu_{1} \cdots \mu_{k}}) \chi 
   \stackrel{(b)} = - (-1)^{\frac{(k+2)(k-1)}{2}} {\bar \eta}
   \Gamma^{\mu_{1} \cdots \mu_{k}} \chi.
   \end{eqnarray}
  \begin{itemize}
   \item{ $\ast$ : We have rewritten the barred quantity as ${\bar
         \chi} = \chi^{\dagger} \Gamma^{0} = {^{T}\chi} \Gamma^{0}$
         and inserted $(\Gamma^{0})^{2}  = - {\bf 1}$.} 
   \item{ $\ast \ast$ : We have taken an over all transpose. Since
       this quantity is not a $32 \times 32$ matrix but a c-number,
       the overall transpose in itself gives the same
       quantity. However, be cautious of the fact that {\it we have
       flipped the two fermionic numbers} in the transpose. That is
       the reason we must multiply (-1).} 
  \end{itemize}
  Let us note the power of $(-1)$.
   On one hand, $- (-1)^{\frac{(k+2)(k-1)}{2}} = - 1$ for $k=1,2,5$. 
   On the other hand, $-(-1)^{\frac{(k+2)(k-1)}{2}} = 1$  for
   $k=0,3,4$. This completes the proof of the properties
   (\ref{AZMA2flip125}) and (\ref{AZMA2flip034}). (Q.E.D.)
  }

 \subsubsection{The Trace of Gamma Matrices} \label{AZCvanish}
    \shadowbox{ \parbox{16cm}{
  {\sf [Prop] All the Gamma matrices $\Gamma^{\mu}, 
  \Gamma^{\mu \nu}, \Gamma^{\mu \nu \lambda}, \Gamma^{\mu \nu \lambda
  \rho}$ and $\Gamma^{\mu \nu \lambda \rho \sigma}$ are traceless.}
  }}\\

  {\sf [Proof] We verify the tracelessness of the matrices one by one.
  \begin{itemize}
   \item{The tracelessness of $\Gamma^{\mu}$ are trivial from the
       explicit form (\ref{AZgammaexplicit}).The tracelessness of
       $\Gamma^{\mu \nu \lambda}$ and 
      $\Gamma^{\mu \nu \lambda \rho \sigma}$ is clear, because these
      are composed of the product of an odd number of traceless
      matrices.}
  \item{For $\Gamma^{\mu \nu}$, the tracelessness readily follows
      from the cyclic rule of the trace. That is, $tr(\Gamma^{\mu \nu} )
      = \frac{1}{2} tr(\Gamma^{\mu} \Gamma^{\nu} - \Gamma^{\nu}
      \Gamma^{\mu}) = 0$. }
    \item{The tracelessness of $ \Gamma^{\mu \nu \lambda  \rho}$ can be 
       understood in the similar fashion to to second case. 
       $tr(\Gamma^{\mu} \Gamma^{\nu} \Gamma^{\lambda} \Gamma^{\rho})=
       tr(\Gamma^{\nu} \Gamma^{\lambda} \Gamma^{\rho} \Gamma^{\mu})$
       holds because of the cyclic symmetry of the 
        trace. However ,  $\Gamma^{\mu \nu \lambda \rho}$ is defined in
       the Sec.1. $sgn(1234)=+1$ while $sgn(2341)=-1$ , and it follows 
       that  $tr(\Gamma^{\mu \nu \lambda \rho})=0$. (Q.E.D.)}
  \end{itemize} }

 \subsubsection{The Decomposition with respect to the Gamma Matrices.}
       \label{AZCdecomposition}
  We introduce a technique of decomposing an arbitrary $32 \times 32$
  matrix ${\cal W}$ in terms of the basis ${\bf 1}, \Gamma^{\mu},
  \cdots,  \Gamma^{\mu_{1} \cdots \mu_{5}}$. For simplicity, let $X, Y,
  \cdots $ denote in general all the indices $\emptyset, \mu_{1},
  \cdots, \mu_{1} \mu_{2} \mu_{3} \mu_{4} \mu_{5}$( $\emptyset$ denotes 
  ${\bf 1}_{32 \times 32}$). The properties
  introduced in Appendix. \ref{AZCproductgamma} and
   \ref{AZCvanish} gives
   \begin{eqnarray}
       \frac{1}{32}tr({\bf 1}{\bf 1}) &=& 
       \frac{1}{32} tr( \Gamma_{\mu} \Gamma^{\mu} ) =
      -\frac{1}{32 \times 2!} tr( \Gamma_{\mu_{1} \mu_{2}} \Gamma^{\mu_{1}
       \mu_{2}} ) =  
      -\frac{1}{32 \times 3!} tr( \Gamma_{\mu_{1} \mu_{2} \mu_{3}}
       \Gamma^{\mu_{1} \mu_{2} \mu_{3} } )  \nonumber \\
  &=&  \frac{1}{32 \times 4!} tr( \Gamma_{\mu_{1} \cdots \mu_{4}}
       \Gamma^{\mu_{1} \cdots \mu_{4}}) = 
       \frac{1}{32 \times 5!} tr( \Gamma_{\mu_{1} \cdots \mu_{5}}
       \Gamma^{\mu_{1} \cdots \mu_{5}}) = 1,
   \end{eqnarray} 
  where the duplicate indices do not give a summation.
  And if we 
  take different indices $X,Y$, the trace is given by
     \begin{eqnarray}
       tr( \Gamma^{X} \Gamma^{Y} ) = 0 \textrm{ for } X \neq Y.
     \end{eqnarray}
  These results give the orthogonality of the gamma matrices with
  respect to the $32 \times 32$ trace. Suppose an arbitrary $32 \times 
  32$ matrix ${\cal W}$ is expressed by
   \begin{eqnarray}
    {\cal W} = A {\bf 1} + A_{\mu} \Gamma^{\mu} + \frac{1}{2!}
    A_{\mu_{1} \mu_{2}} \Gamma^{\mu_{1} \mu_{2}} + \frac{1}{3!}
    A_{\mu_{1} \mu_{2} \mu_{3}} \Gamma^{\mu_{1} \mu_{2} \mu_{3}} +
    \frac{1}{4!} A_{\mu_{1} \cdots \mu_{4}} \Gamma^{\mu_{1} \cdots
    \mu_{4}} + \frac{1}{5!} A_{\mu_{1} \cdots \mu_{5}} \Gamma^{\mu_{1} 
    \cdots \mu_{5}}.
   \end{eqnarray}
  Then, the orthogonality of the gamma matrices enables one to pick up 
  the coefficients of the gamma matrices as follows: \\
   \shadowbox{ \parbox{16cm}{
   \begin{eqnarray}
  & &  A = \frac{1}{32} tr( {\cal W}),  \hspace{3mm} 
       A_{\mu} = \frac{1}{32} tr( {\cal W} \Gamma_{\mu}), \hspace{3mm}
       A_{\mu_{1} \mu_{2}} = - \frac{1}{32} tr({\cal W}
       \Gamma_{\mu_{1} \mu_{2}} ) , \hspace{3mm}
       A_{\mu_{1} \mu_{2} \mu_{3}} = - \frac{1}{32} tr( {\cal W}
       \Gamma_{\mu_{1} \mu_{2} \mu_{3}}),  \nonumber \\
  & &  A_{\mu_{1} \cdots \mu_{4}} = \frac{1}{32} tr( {\cal W}
       \Gamma_{\mu_{1} \cdots \mu_{4}}), \hspace{3mm}
       A_{\mu_{1} \cdots \mu_{5}} = \frac{1}{32} tr( {\cal W}
       \Gamma_{\mu_{1} \cdots \mu_{5}} ). \label{AZMAdec7777} 
   \end{eqnarray} }}

  And we present the proof of the formula of the Fierz transformation
  (\ref{AZM31fierz}), as an example of the utility of this
  decomposition. However, we note that the denominator is {\it not 32 
  but 16} because we {\it limit the following argument to the case of
  the Weyl fermion in 10
  dimensions}. When we write explicitly the $16 \times 16$ indices,
  the left-hand side is 
   \begin{eqnarray}
  {\bar \epsilon}_{1} \Gamma_{j} \psi \Gamma^{ij} \epsilon_{2} 
  =  ({\bar \epsilon})^{\alpha} {(\Gamma_{j})_{\alpha}}^{\beta} 
     \psi_{\beta} {(\Gamma^{ij})_{\gamma}}^{\delta} \epsilon_{2
  \delta} 
  = - {(\Gamma^{ij})_{\gamma}}^{\delta} {({\bar \epsilon}_{1}
  \epsilon_{2})_{\delta}}^{\alpha} {(\Gamma_{j})_{\alpha}}^{\beta}
  \psi_{\beta}. \label{AZMAf2}
   \end{eqnarray} 
  Here, the minus sign emerges because we have flipped the order of
  the Grassmann odd fermions. And ${({\bar \epsilon}_{1}
  \epsilon_{2})_{\delta}}^{\alpha}$ is a $16 \times 16$ matrices and
  we expand this in terms of the Gamma matrices as
  \begin{eqnarray}
   {({\bar \epsilon}_{1} \epsilon_{2})_{\delta}}^{\alpha} &=& 
    \frac{1}{16} ({\bar \epsilon}_{1} \epsilon_{2}) {\bf 1}
  + \frac{1}{16} ({\bar \epsilon}_{1} \Gamma_{k} \epsilon_{2})
    \Gamma^{k} 
  - \frac{1}{16 \times 2!} ({\bar \epsilon}_{1} \Gamma_{k_{1} k_{2}}
    \epsilon_{2}) \Gamma^{k_{1} k_{2}} 
  -  \frac{1}{16 \times 3!} ({\bar \epsilon}_{1} \Gamma_{k_{1} k_{2}
    k_{3}} \epsilon_{2}) \Gamma^{k_{1} k_{2} k_{3}} \nonumber \\ 
  &+&  \frac{1}{16 \times 4!} ({\bar \epsilon}_{1} \Gamma_{k_{1}
    \cdots k_{4}}  \epsilon_{2}) \Gamma^{k_{1} \cdots k_{4}} 
  +  \frac{1}{16 \times 5!} ({\bar \epsilon}_{1} \Gamma_{k_{1} \cdots k_{5}}
    \epsilon_{2}) \Gamma^{k_{1} \cdots k_{5}}. \label{AZMAf}
  \end{eqnarray}
  We note that only the gamma matrices of rank 1 and 5 survive for the 
  following reasoning:
  \begin{itemize}
   \item{The gamma matrices of even rank vanish because the fermions
       are Weyl ones. We set all the fermions to be of the left chirality,
       and $\epsilon_{1} = \frac{1 + \Gamma^{\sharp}}{2} \epsilon_{1O}$: 
    \begin{eqnarray}
    {\bar \epsilon_{1}} \Gamma^{k_{1} \cdots k_{2m}} \epsilon_{2} 
  &=& {\bar \epsilon}_{1O}
   \frac{1 - \Gamma^{\sharp}}{2}  \Gamma^{k_{1} \cdots k_{2m}} 
   \frac{1 + \Gamma^{\sharp}}{2} \epsilon_{2O}
   =  {\bar \epsilon}_{1O} \frac{1 - \Gamma^{\sharp}}{2}  
   \frac{1 + (-1)^{2m} \Gamma^{\sharp}}{2} \Gamma^{k_{1} \cdots
   k_{2m}}  \epsilon_{2O} \nonumber \\
   &=& {\bar \epsilon}_{1O} \frac{1 - \Gamma^{\sharp}}{2}   
   \frac{1 + \Gamma^{\sharp}}{2} \Gamma^{k_{1} \cdots
   k_{2m}}  \epsilon_{2O} = 0.
    \end{eqnarray}}
   \item{We ignore the terms of rank 3 because of the flipping
       properties described in Appendix. \ref{AZCflipping}. Since we
       are now looking at the difference 
       $ {\bar \epsilon}_{1} \Gamma_{j} \psi \Gamma^{ij} \epsilon_{2}
       -  {\bar \epsilon}_{2} \Gamma_{j} \psi \Gamma^{ij}
       \epsilon_{1}$, this contribution cancels.} 
  \end{itemize}

  And we perform the computation of the gamma matrices as 
  \begin{eqnarray}
   \Gamma^{ij} \Gamma_{k} \Gamma_{j} &=& \Gamma^{ij} (\Gamma_{kj} +
   \eta_{kj}) 
   = (- {\eta_{j}}^{j} {\Gamma^{i}}_{k} + 2 {\Gamma^{i}}_{k}
     - {\eta_{j}}^{j} {\eta^{i}}_{k} + {\eta^{i}}_{k} ) +
   {\Gamma^{i}}_{k}
   \stackrel{{\eta^{j}}_{j} = 10}{=} 7 {\Gamma_{k}}^{i} - 9
   {\eta^{i}}_{k} \nonumber \\ 
   &=& 7 ( \Gamma_{k} \Gamma^{i} - {\eta^{i}}_{k} ) -  9 {\eta^{i}}_{k}
   = 7 \Gamma_{k} \Gamma^{i} - 16 {\eta^{i}}_{k}, \label{AZMArank1} \\
  \Gamma^{ij} \Gamma_{k_{1} \cdots k_{5}} \Gamma_{j} 
   &=& \Gamma^{ij} ( - \Gamma_{j k_{1} \cdots k{5}} + 5 \eta_{j [k_{1}}
   \Gamma_{k_{2} \cdots k_{5}]} ) \nonumber \\
  &=& - ( - \eta^{i j} \Gamma_{j k_{1} \cdots k_{5}} +
   {\eta^{j}}_{j} {\Gamma^{i}}_{k_{1} \cdots k_{5}}
   - 5 \eta_{j [k_{1}} {\Gamma^{i j}}_{k_{2} \cdots k_{5}]} )
      + 5 ( {\Gamma^{i}}_{k_{1} \cdots k_{5}} 
        - {\eta^{i}}_{[ k_{2}} \Gamma_{k_{1} k_{3} k_{4} k_{5}]} )
   \nonumber \\  
  &=& - {\Gamma_{k_{1} \cdots k_{5}}}^{i} + 5 {\eta^{i}}_{[k_{1}}
   \Gamma_{k_{2} \cdots k_{5}]}
   = \Gamma_{k_{1} \cdots k_{5}} \Gamma^{i}. \label{AZMArank5}
  \end{eqnarray}
  When we substitute (\ref{AZMAf}), (\ref{AZMArank1}) and (\ref{AZMArank5})
  into (\ref{AZMAf2}), the computation of the Fierz transformation
  goes as follows:
  \begin{eqnarray}
 {\bar \epsilon}_{1} \Gamma_{j} \psi \Gamma^{ij} \epsilon_{2} 
  &=&  - {(\Gamma^{ij})_{\gamma}}^{\delta} {({\bar \epsilon}_{1}
  \epsilon_{2})_{\delta}}^{\alpha} {(\Gamma_{j})_{\alpha}}^{\beta}
  \psi_{\beta} \nonumber \\
  &=& - \frac{1}{16} ({\bar \epsilon}_{1} \Gamma_{k} \epsilon_{2}) 
    (\Gamma^{ij} \Gamma^{k} \Gamma_{j}) \psi
    - \frac{1}{16 \times 5!} ({\bar \epsilon}_{1} \Gamma_{k_{1} \cdots
  k_{5}} \epsilon_{2}) (\Gamma^{ij} \Gamma^{k_{1} \cdots k_{5}}
  \Gamma_{j} ) \psi  +  (\textrm{rank 3 term}) \nonumber \\
  &=&     {\bar \epsilon_{1}} \Gamma^{i} \epsilon_{2} \psi - \frac{7}{16}
      {\bar \epsilon_{1}} \Gamma^{k} \epsilon_{2} \Gamma_{k}
      \Gamma^{i} \psi - \frac{1}{16 \times 5!} {\bar \epsilon_{1}}
      \Gamma^{k_{1} \cdots k_{5}} \epsilon_{2} \Gamma_{k_{1} \cdots
      k_{5}} \Gamma^{i} \psi + (\textrm{rank 3 term}), \label{AZMAQED}
  \end{eqnarray}
 and we complete the proof of (\ref{AZM31fierz}).

\subsection{Supermatrices} \label{AZMA091441}
 This section is devoted to introducing the definitions of the notion
 of supermatrices. In treating supermatrices, there are many points we 
 should be careful about, because what holds true of ordinary matrices 
 is not applicable to the supermatrices. 

 \subsubsection{Transpose} 
  We first introduce a notion of the transpose, emphasizing on the
 difference from the ordinary matrices. In considering such objects,
 it is extremely important to settle the starting point, because the
 other notions are defined so that they are consistent with this
 starting point. 
 
 \paragraph{ Transpose of Vector} . \\ 
 The guiding principle in considering the transpose of the
 supermatrices is {\it the transpose of the vector}. \\
 \shadowbox{ \parbox{16cm}{
  \begin{itemize}
   \item{ The guiding principle is that {\it the transpose of a
     vector} is defined as 
     \begin{eqnarray}
         { \left( \begin{array}{c}  x_{1} \\ \vdots  \\  x_{n}  \end{array}
         \right)}^{T} = ( x_{1}  ,\cdots , x_{n}  ). 
     \end{eqnarray}      
     We denote $\{ x_{i} \}$ as the  components of $v$, and these
     components mean  {\sf both bosons and fermions}.}
   \item{ We define the vector as $v = \left( \begin{array}{c}  \eta
     \\ b  \end{array} \right) $ where $\eta$ and $b$ are fermionic and
     bosonic {\it real}  fields respectively. }
  \end{itemize} }}

  \paragraph{Transpose of Supermatrices} . \\ 
  The transpose of supermatrices must be defined so that the
  definition is consistent with the transpose of a vector. Therefore the 
  transpose of a supermatrix must satisfy 
   \begin{eqnarray}
     {^{T}(Mv)} = {^{T}v} {^{T}M},
   \end{eqnarray}
  where $M$ is a supermatrix and $v$ is a vector. Following this rule, 
  the transpose of a supermatrix is defined as follows.\\
  \shadowbox{ \parbox{16cm}{
  \begin{eqnarray}
     For \hspace{2mm} M =  \left( \begin{array}{cc} a & \beta \\ \gamma 
    & d \end{array} \right) \hspace{2mm} and  \hspace{2mm} v = 
    \left( \begin{array}{c}   \eta \\ b 
    \end{array} \right)  ,\hspace{4mm} {^{T}{M}} =  \left
    ( \begin{array}{cc} {^{T}{a}} & - {^{T}{\gamma}} \\ {^{T}{\beta}}
    & {^{T}{d}} \end{array} \right). \label{defsupertrace}
  \end{eqnarray}
   \begin{itemize}
    \item{ $a$ and $d$ are bosonic (i.e. Grassmann even) $m \times
        m$ and $n \times n$ matrices, respectively.}
    \item{ $\beta$ and $\gamma$ are $M \times n$ and $n \times m$
        fermionic (i.e. Grassmann odd) matrices , respectively.} 
    \item{ $\eta$ ($b$) denote the upper $m$ (lower $n$) 
        bosonic(fermionic) components of the supervector respectively.}
   \end{itemize}  
    }} \\
 {\sf (Proof)  This can be verified using the very definition of the
 transpose.  
  \begin{eqnarray} Mv =   \left( \begin{array}{c}   a\eta + \beta b  \\
       \gamma \eta + d b  \end{array} \right).
  \end{eqnarray}
  Then the  transpose of this vector is by definition 
   \begin{eqnarray} 
    {^{T}(Mv)} = ( {^{T}\eta} {^{T}a} + {^{T}b} {^{T}\beta} , - {^{T}\eta}
    {^{T}\gamma} + {^{T}b} {^{T}d} ) = {^{T}v} {^{T}M}.
    \end{eqnarray}
   The point is that {\sf the sign of } $\gamma \eta$ {\sf has changed
   because these are Grassmann odd}.   Noting this fact, we can read
   off the result (\ref{defsupertrace}). (Q.E.D.) }\\

  We have one caution about the transpose of the supermatrix. {\sf
  The transpose of the transpose does not give an original matrix.}
  This 'anomalous' property can be immediately read off from the
  definition of the supermatrix (\ref{defsupertrace}):
  \begin{eqnarray}
   {^{T} ({^{T} \left( \begin{array}{cc} a & \beta \\ \gamma  & d
   \end{array} \right)} ) } = {^{T} \left( \begin{array}{cc} {^{T}{a}}
   & - {^{T}{\gamma}} \\ {^{T}{\beta}} & {^{T}{d}} \end{array} \right) 
    } = \left( \begin{array}{cc} a & - \beta \\ - \gamma
    & d \end{array} \right).
  \end{eqnarray}

  \paragraph{Transpose of Transverse Vector} . \\
  We have seen an important fact that the transpose of the transpose
  of a supermatrix does not give the original supermatrix. In fact,
  the same holds true of the transpose of the transpose of a vector. 
   Conclusion coming first, the definition is \\
   \shadowbox{ \parbox{16cm}{
    \begin{eqnarray}
     {^{T} y} = {^{T} ( \eta ,b)} = \left( \begin{array}{c} - {^{T} \eta} \\
     {^{T}b} \end{array} \right).
    \end{eqnarray} }} \\
   We confirm that this is actually a well-defined settlement. This
   notion is defined so that 
    \begin{eqnarray}
      {^{T} (yM)} = {^{T}M } {^{T}y},
    \end{eqnarray}    
   where $y$ is a transverse vector $y = (\eta ,b)$ and $M$ is a
   supermatrix $M = 
   \left( \begin{array}{cc} a & \beta \\ \gamma & d \end{array}
   \right)$.  We
   compute both the L.H.S and the R.H.S and verify that they actually
   match if we follow the above definition.
   \begin{itemize}
    \item{L.H.S. :  ${^{T} (yM) } = {^{T} ( \eta a + b \gamma , \eta
         \beta + bd)} =  
         \left( \begin{array}{c} - {^{T} (\eta a)} - {^{T} (b \gamma)} \\
         {^{T}( \eta \beta)} +  
         {^{T} (bd)} \end{array} \right) = \left( \begin{array}{c}
         - {^{T}a} {^{T} \eta} - {^{T} \gamma} {^{T}b} \\ -
         {^{T}\beta} {^{T} \eta} + 
         {^{T}d} {^{T}b} \end{array} \right)$. \\
         We have used the fact that in the transpose of $fF$, we must
         multiply $-1$ because a fermion jumps over another fermion.}
    \item{ R.H.S. : ${^{T}M} {^{T}y} = \left( \begin{array}{cc} {^{T}a} 
            & - {^{T}\gamma} \\ {^{T} \beta} & {^{T} d} \end{array} \right) 
            \left( \begin{array}{c} - {^{T}\eta } \\ {^{T}b}
            \end{array} \right) = \left( \begin{array}{c} - {^{T}a} 
            {^{T} \eta} - {^{T}\gamma} {^{T}b} \\ - {^{T}\beta} {^{T}
            \eta} + {^{T}\gamma}  
            {^{T}b} \end{array} \right)$. }
   \end{itemize} 
  Thus we have verified that the above definition of the transpose is
  consistent with the condition ${^{T} (yM)} = {^{T}M } {^{T}y}$. 
  As we have mentioned before, the transpose of the transpose of a
  vector does not give the original vector, because 
   \begin{eqnarray}
    {^{T} ( {^{T} \left( \begin{array}{c} \eta \\ b \end{array} \right)}
    )} = {^{T} ( {^{T} \eta}, {^{T}b} )} = \left( \begin{array}{c}
    -\eta \\ b \end{array} \right).
   \end{eqnarray}

  \subsubsection{Hermitian Conjugate}
  We introduce a notion of hermitian conjugate of supermatrix. This
  notion is much simpler than the transpose, and we do not see 
  anomalous properties as emerged in the transpose. The starting
  point of this notion is \\
   \shadowbox{ \parbox{16cm}{
    \begin{itemize}
     \item{ For a fermionic {\it number} $\alpha, \beta$, the complex
         conjugate is $(\alpha \beta)^{\dagger} = (\beta)^{\dagger}
         (\alpha)^{\dagger}$.}
    \item{ For a vector $v = \left( \begin{array}{c} \eta \\ b
          \end{array} \right)$, the complex conjugate is 
        $ \left( \begin{array}{c} \eta \\ b \end{array}
         \right)^{\dagger} = ( \eta^{\dagger}, b^{\dagger})$. }
    \end{itemize} }} \\

  Under this definition, the hermitian conjugate of a supermatrix is
  defined as \\
  \shadowbox{ \parbox{16cm}{
   \begin{eqnarray}
    \textrm{ For } M = \left( \begin{array}{cc} a & \beta \\ \gamma &
    d \end{array} \right), \hspace{4mm}  M^{\dagger} = \left
    ( \begin{array}{cc} a^{\dagger} & \gamma^{\dagger} \\ \beta^{\dagger} & 
    d^{\dagger} \end{array} \right).
   \end{eqnarray} }} \\
 {\sf
 (Proof) The guiding principle to determine the hermitian conjugate of 
 a supermatrix is the condition 
  \begin{eqnarray}
   ( M v)^{\dagger} = (v^{\dagger}) (M^{\dagger}).
  \end{eqnarray}
  For $M = \left( \begin{array}{cc} a & \beta \\ \gamma &
    d \end{array} \right)$ and $v = \left( \begin{array}{c} \eta \\ b
    \end{array} \right)$, $(Mv)^{\dagger}$ is computed to be,
    utilizing the definition for the vector, 
     \begin{eqnarray}
        (Mv)^{\dagger} = \left( \begin{array}{c} a \eta + \beta b \\
        \gamma \eta + d b \end{array} \right)^{\dagger} = 
        ( (a \eta)^{\dagger} + (\beta b)^{\dagger} , (\gamma
        \eta)^{\dagger} + (d b)^{\dagger} ) = ( {\eta}^{\dagger}
        a^{\dagger} + b^{\dagger} {\beta}^{\dagger} ,
        {\eta}^{\dagger} {\gamma}^{\dagger} + b^{\dagger} d^{\dagger} ). 
     \label{AZMA134718} 
    \end{eqnarray}
    The explicit form of the hermitian conjugate of a supermatrix can
    be read off from (\ref{AZMA134718}), and this completes the
    proof. (Q.E.D.)}

    The definition of the hermitian conjugate of a transverse vector
    is now straightforward. This is given by \\
    \shadowbox{ \parbox{16cm}{
    \begin{eqnarray}
       y^{\dagger} = ( \eta , b)^{\dagger} = \left( \begin{array}{c}
       \eta^{\dagger} \\ b^{\dagger} \end{array} \right).  
    \end{eqnarray}  }} \\
    It is straightforward to verify that this definition is consistent
    with the guiding principle 
     \begin{eqnarray}
       (y M)^{\dagger} = M^{\dagger} y^{\dagger},
     \end{eqnarray}
   because $(l.h.s.) = (r.h.s.) = \left( \begin{array}{c} a^{\dagger}
   \eta^{\dagger} + \gamma^{\dagger} b^{\dagger} \\ \beta^{\dagger}
   \eta^{\dagger} + d^{\dagger} b^{\dagger} \end{array} \right)$.

  \subsubsection{Complex Conjugate} \label{AZCaho}
  We define a notion of complex conjugate for a supermatrix. The
  guiding principle to define the complex conjugate is to require the
  matrices and the vectors to satisfy the condition 
   \begin{eqnarray}
       (Mv)^{\ast} = M^{\ast} v^{\ast}. \label{AZMA1guidingcc}
   \end{eqnarray}
  In order to satisfy this condition, we define the complex conjugate
  of the vectors and the matrices as follows\footnote{  Be careful
  about the fact that ${^{T}(M^{\dagger})}$ {\it is 
  different from } $({^{T} M})^{\dagger} = M^{\ast}$. They are
  computed to be ${^{T}(M^{\dagger})} = \left( \begin{array}{cc}
  a^{\ast} & - \beta^{\ast} \\ \gamma^{\ast} & d^{\ast} \end{array}
  \right)$, ${^{T}(v^{\dagger})} = \left( \begin{array}{c}
  - \eta^{\ast} \\  b^{\ast} \end{array} \right) $ and ${^{T}
  (y^{\dagger}) } =  (\eta^{\ast}, b^{\ast})$ and these are {\it
  not} the complex conjugate of the vectors or the matrices.}. \\
   \shadowbox{ \parbox{16cm}{
   \begin{eqnarray}
      v^{\ast} \stackrel{def}{=} ({^{T} v})^{\dagger}, \hspace{4mm}
      M^{\ast} \stackrel{def}{=} ({^{T} M})^{\dagger}. \label{AZMA3defcc}
   \end{eqnarray} }} \\
  It is straightforward to verify that this definition is consistent
  with the guiding principle (\ref{AZMA1guidingcc}).
   \begin{eqnarray}
    (Mv)^{\ast} = (({^{T}v})({^{T}M}))^{\dagger} = ({^{T}
    M})^{\dagger} ({^{T} v})^{\dagger} = M^{\ast} v^{\ast}.
   \end{eqnarray}
  Combining the results obtained in the previous section, the explicit 
  form of the complex conjugate of the vectors and the matrices are\\
   \shadowbox{ \parbox{16cm}{
    \begin{eqnarray}
      M^{\ast} = \left( \begin{array}{cc} a & \beta \\ \gamma & d
      \end{array} \right)^{\ast} = \left( \begin{array}{cc}
        a^{\ast} & \beta^{\ast} \\ - \gamma^{\ast} & d^{\ast}
        \end{array}  \right), \hspace{2mm} 
      v^{\ast} = \left( \begin{array}{c} \eta \\ b \end{array}
        \right)^{\ast} = \left( \begin{array}{c} \eta^{\ast} \\
        b^{\ast} \end{array} \right), \hspace{2mm}
      y^{\ast} = ( \eta, b)^{\ast}  = ( - \eta^{\ast}, b^{\ast}).  
    \label{AZMA1misccc} \end{eqnarray} }} \\
  This is clearly consistent with the guiding principle $(Mv)^{\ast}
  = M^{\ast} v^{\ast}$ because $(l.h.s.) = (r.h.s.) = 
  \left( \begin{array}{c} a^{\ast} \eta^{\ast} + \beta^{\ast} b^{\ast} \\
   - \gamma^{\ast} \eta^{\ast} + d^{\ast} b^{\ast} \end{array}
  \right) $. \\ 

   We have following properties which relates the transpose, hermitian
  conjugate and the complex conjugate. \\
  \shadowbox{ \parbox{16cm}{
   {\sf (Prop) (1)${^{T}M} = (M^{\ast})^{\dagger}$, (2)$M^{\dagger} =
  {^{T} (M^{\ast})}$.  }}} \\
  {\sf
  (Proof) These properties can be verified by noting that the
  hermitian conjugate of the hermitian conjugate gives back the
  original quantity, which can be readily verified by definition.
  \begin{enumerate}
   \item{ $(M^{\ast})^{\dagger} = (({^{T} M})^{\dagger} )^{\dagger} = 
       {^{T} M}$.}
   \item{ ${^{T} (M^{\ast})} \stackrel{(1)}{=}
       ((M^{\ast})^{\ast})^{\dagger} = M^{\dagger}$. In the last
       equality, we have utilized the fact that, for a supermatrix,
       $(M^{\ast})^{\ast} = M$, which can be readily verified from
       the explicit form of the complex conjugate
       (\ref{AZMA1misccc}).  }
  \end{enumerate}
  This completes the proof of the above properties. (Q.E.D.)
  }\\

  Now we are ready to answer the question : {\it what do we mean by 'a
  supermatrix is real ?'}. In considering physics, we must
  take into account the reality condition. We utilize supermatrices in 
  the context of expressing the action of superstring theory, the
  action must be real, and we are required to solidify the definition
  of the reality of a supermatrix. \\
  \shadowbox{ \parbox{16cm}{
   {\sf (Def) A supermatrix $M$ is real
  $\stackrel{def}{\Leftrightarrow}$ $M$ is a mapping from {\it a real
  vector} to {\it a real vector}}. }} \\
   This statement is equivalent to, for the above definition of
  complex conjugate, 
    \begin{eqnarray}
      M^{\ast} = M. \label{AZMA3defreality}
    \end{eqnarray}
  This can be verified by noting the starting guiding principle that
  $M$ is designed to satisfy $(Mv)^{\ast} = 
  M^{\ast} v^{\ast}$. If $M^{\ast} = M$ is satisfied,
  $(Mv)^{\ast}$ is a real vector if $v$ is real, because
   \begin{eqnarray}
     (Mv)^{\ast} = M^{\ast} v^{\ast} \stackrel{v \textrm{ is
     real}}{=} M^{\ast} v \stackrel{M^{\ast} = M}{=} M v.
   \end{eqnarray}
    The relationship  (\ref{AZMA3defreality}) tells us the conditions
  for the components of $M$ to satisfy. Noting the explicit form of
  complex conjugate (\ref{AZMA1misccc}), we can derive $a^{\ast} =
  a$, $d^{\ast} =  d$ , $\beta^{\ast} = \beta$ and $\gamma^{\ast} =
  - \gamma$,  id est, \\ 
  \shadowbox{ \parbox{16cm}{
  \begin{itemize}
   \item{$a, \beta, d$ should be real.}
   \item{$\gamma$ should be pure imaginary.}
  \end{itemize} }}
  
 \subsection{The Properties of $su(N)$ Lie algebra.} \label{AZCsun}
  This section is devoted to the introduction of the properties of the 
  $su(N)$ algebra. $su(N)$ is a $N^{2} -1 $ dimensional Lie algebra
  composed of anti-hermitian $N \times N$ matrices. The generators of this
  gauge group are denoted by \footnote{ In considering a large N reduced
  model, $N$ is large enough that $N^{2} -1 $ is nearly $N^{2}$. And
  in this sense we often write the dimension of $su(N)$ algebra as
  $N^{2}$.}
   \begin{eqnarray}
    T^{a} \textrm{ (with } a= 1,2, \cdots N^{2}-1 \textrm{)},
   \end{eqnarray} 
  where $T^{a}$ are all {\it hermitian} matrices and 
   \begin{eqnarray}
     \exp( i t T^{a}) = 1 + i t T^{a} + {\cal O}(t^{2})
   \end{eqnarray}
  belongs to the $SU(N)$ Lie group (of course, $t$ is a real number).
  Before arguing the
  properties of $su(N)$, let us pause and investigate the properties
  of general hermitian and anti-hermitian matrices.\footnote{ Since we
  concentrate on the bosonic matrices here, not the supertrace, we do
  not see a complexity as seen in the previous section.}
   \begin{itemize}
    \item{Hermitian matrices: ${\bf H} = \{ M \in M_{N \times N}({\bf
            C}) | M^{\dagger} = M \} $. }
    \item{Anti-hermitian matrices : ${\bf A} = \{ M \in M_{N \times N}
          ({\bf C}) | M^{\dagger} = -M \} $.}
   \end{itemize}
  Let us have a look at the following properties.\\
  \shadowbox{ \parbox{16cm}{
   Let the matrices $h, h_{1}, h_{2}$ and $a, a_{1}, a_{2}$ be
  hermitian and anti-hermitian matrices respectively. 
  \begin{eqnarray}
  & & (1) [h_{1}, h_{2} ] \in {\bf A}, \hspace{3mm}
      (2) [h    , a     ] \in {\bf H},    \hspace{3mm}
      (3) [a_{1}, a_{2} ] \in {\bf A}, \nonumber \\
  & & (4)\{h_{1}, h_{2}\} \in {\bf H},    \hspace{3mm}
      (5)\{h    , a    \} \in {\bf A}, \hspace{3mm}
      (6)\{a_{1}, a_{2}\} \in {\bf H}.
 \end{eqnarray} }}  \\
 {\sf
 (Proof) These properties can be verified by taking the hermitian
 conjugates one by one, noting the fact that $(XY)^{\dagger} =
 Y^{\dagger} X^{\dagger}$.
  \begin{enumerate}
   \item{$ ([h_{1}, h_{2} ])^{\dagger} = (h_{1} h_{2} - h_{2}
       h_{1})^{\dagger} = h_{2}^{\dagger}
       h_{1}^{\dagger} -   h_{1}^{\dagger} h_{2}^{\dagger} = -
       [h_{1}^{\dagger}, h_{2}^{\dagger}] =  - [h_{1}, h_{2}] $, }
   \item{$ ([h,a])^{\dagger} = (h a - a h )^{\dagger} = a^{\dagger}
       h^{\dagger} - h^{\dagger} 
       a^{\dagger} = - [h^{\dagger}, a^{\dagger}] = [h,a] $,}
   \item{$ ([a_{1}, a_{2} ])^{\dagger} = (a_{1} a_{2} - a_{2}
       a_{1})^{\dagger} = a_{2}^{\dagger}
       a_{1}^{\dagger} - a_{1}^{\dagger} a_{2}^{\dagger} = -
       [a_{1}^{\dagger}, a_{2}^{\dagger}] = -  [a_{1}, a_{2}]$,}
   \item{$ (\{h_{1}, h_{2}\})^{\dagger} = (h_{1} h_{2} + h_{2}
       h_{1})^{\dagger} = h_{2}^{\dagger}
       h_{1}^{\dagger} +   h_{1}^{\dagger} h_{2}^{\dagger} = 
       \{ h_{1}^{\dagger}, h_{2}^{\dagger} \} = \{ h_{1}, h_{2} \}$,} 
  \item{$ (\{h,a\})^{\dagger} = (ha + ah)^{\dagger} = a^{\dagger}
       h^{\dagger} + h^{\dagger} 
       a^{\dagger} = \{ h^{\dagger}, a^{\dagger} \} = - \{ h, a \}$,}
   \item{$ (\{ a_{1}, a_{2} \})^{\dagger} = (a_{1} a_{2} + a_{2} a_{1} 
       )^{\dagger} = a_{2}^{\dagger}
       a_{1}^{\dagger} + a_{1}^{\dagger} a_{2}^{\dagger} = 
       \{ a_{1}^{\dagger}, a_{2}^{\dagger}  
       \} = \{a_{1}, a_{2} \} $.}
  \end{enumerate}
  Comparing the original quantity and their complex conjugates, we
 verify the above statement. (Q.E.D.) }\\

 As we have seen in the above commutation relations, it is not ${\bf
 H}$, but ${\bf A}$ that constitutes a closed Lie algebra. However,
 ${\bf H}$, to which the basis $\{ T^{a} \}$ belong, is regarded as a
 representation of the Lie algebra $su(N)$ because of the commutation
 relation $[h,a] \in {\bf H}$. This means that the generators $T^{a}$
 remain hermitian after the infinitesimal transformation by the
 anti-hermitian matrices.\\

 Having these properties in mind, let us consider the structure
 constant of $su(N)$ Lie algebra. The commutator and the
 anticommutators of the generators are usually described by the
 structure constant $f_{abc}$ and $d_{abc}$ respectively.
   \begin{itemize}
    \item{ Commutator: The structure constant $f_{abc}$ is defined 
        so that\footnote{ We are
        sloppy in this paper about whether the indices of the color
        should be on the  upper or lower side.} $[T^{a}, T^{b}] = i
        f_{abc} T^{c}$. 
       \begin{itemize} 
        \item{
        $f_{abc}$ is {\it
        per se} a {\it real} quantity. As we have seen in the above
        properties (1), the commutator of the generators of $su(N)$ is 
        {\it an anti-hermitian} matrix because $T^{a}$ and $T^{b}$ are 
        hermitian. That is why we must multiply $i$ in the
        R.H.S.}
        \item{ Another natural but important property is that
      $f_{abc}$ possess a cyclic symmetry $f_{abc} = f_{bca} =
      f_{cab}$ while they are anti-symmetric with respect to the
      exchange of two indices : $f_{abc} = - f_{bac}$.}
  \end{itemize}
      }
    \item{Anti-commutator: The structure constant $d_{abc}$ is
        defined so that $\{ T^{a}, T^{b} \} = d_{abc} T^{c}$.  
       \begin{itemize}
        \item{ The anticommutator of two hermitian matrices
            is hermitian, so that $d_{abc}$ are
            real quantity and we do not need to multiply $i$.}
        \item{ The structure constant $d_{abc}$ not only possesses
            cyclic symmetry $d_{abc} = d_{bca} = d_{cab}$ but also are 
            symmetric in that $d_{abc} = d_{bac}$.}
       \end{itemize}
     }
   \item{ Trace : It is clear from the definition of $su(N)$ Lie
       algebra that $tr(T^{a})=0$. And the trace of the product 
       of these generators are known to be $Tr(T^{a} T^{b}) =
       \frac{\delta^{ab}}{2} $.}
   \end{itemize}

 \subsection{Tensor Product} \label{AZCtensorpr}
  In considering the gauged theory according to the proposal of
  L. Smolin\cite{0006137}, we need to introduce a tensor product of
  the two Lie algebras. The tensor product is trivially defined for
  two matrices: for $a \in M_{m} ({\bf C})$ and $b \in M_{n}({\bf
  C})$, the tensor product $a \otimes b$ is a well-defined
  notion. Here, we consider the extension of this notion to the Lie
  algebras. In this discussion, we limit the Lie algebras to {\it
  linear Lie algebras}\footnote{The term {\it linear Lie group} is
  defined as the closed subgroup of $GL(N,{\bf C})$ group. For example,
  such groups as $GL(N, {\bf C})$ itself and $SU(N)$ are linear Lie
  groups.\\  The term {\it linear Lie algebra} is defined as {\it a Lie
  algebra of a linear Lie group}, id est, the subalgebra of
  $M_{N}({\bf C})$. }. 

  \paragraph{Tensor Products of two Sets of Matrices $(\otimes)$}. \\
  The tensor product $\otimes$ is defined for two sets of
  matrices, regardless of whether they are closed Lie algebras or
  not.  Let ${\cal A}$ and ${\cal B}$ be a set of matrices which
  are not necessarily closed Lie algebras. Let $\{ a_{i} \}$ and 
  $\{ b_{j} \}$ be the bases of ${\cal A}$ and ${\cal B}$,
  respectively.\\
   \shadowbox{ \parbox{16cm}{
    \begin{eqnarray}
    ( \textrm{The tensor product } {\cal A} \otimes {\cal B})
    \stackrel{def}{=}  
    \textrm{The linear space spanned by the bases } a_{i} \otimes
    b_{j}.  
   \end{eqnarray} }}
 
  The important point is that ${\cal A} \otimes {\cal B}$ {\it does not
  necessarily constitute a closed Lie algebra}, even if each of ${\cal
  A}$ and ${\cal B}$ is a closed Lie algebra. If the tensor product
  ${\cal A} \otimes {\cal B}$ is to be a closed Lie algebra, this must
  close with respect to the commutator of the tensor product. For
  $a_{1}, a_{2} \in {\cal A}$ and $b_{1}, b_{2} \in {\cal B}$, the
  commutator of the tensor product of the matrices is
   \begin{eqnarray}
    [a_{1} \otimes b_{1}, a_{2} \otimes b_{2}] = \frac{1}{2} ( 
   \{ a_{1}, a_{2} \} \otimes  [b_{1}, b_{2} ] + 
    [ a_{1}, a_{2} ]  \otimes \{b_{1}, b_{2} \}).
   \end{eqnarray}
 However, the tensor product ${\cal A} \otimes {\cal B}$ does not close
 with respect to this commutator. For example, consider the tensor
 product of two Lie algebras ${\cal H}$ and ${\bf H}$, defined in
 Sec. 5. The tensor product ${\cal H} \otimes {\bf H}$ does not
 constitute a closed Lie algebra. This can be seen by the fact that 
     \begin{eqnarray}
      [ ({\cal H} \otimes {\bf H}) , ({\cal H} \otimes {\bf H}) ] = 
     (\{ {\cal H}, {\cal H} \} \otimes [{\bf H},{\bf H}]) \oplus 
     ( [ {\cal H}, {\cal H}] \otimes \{ {\bf H}, {\bf H} \})      =
    ({\cal A}' \otimes {\bf A}) \oplus ({\cal H} \otimes {\bf H}),
     \end{eqnarray} 
 with the definition of ${\cal A}'$ and ${\bf A}$, and the reasoning
 given in Sec. 5. Thus, the tensor product ${\cal H} \otimes {\bf H}$
 is not a closed Lie algebra.

 \paragraph{Tensor Products of two Lie algebras $(\check{\otimes})$
 }. \\
 We introduce another notion of the tensor product
 $\check{\otimes}$. This is a notion limited to the linear Lie
 algebras, unlike the tensor product $\otimes$. Let ${\cal A}$ and
 ${\cal B}$ now be two linear Lie algebras. \\
  \shadowbox{ \parbox{16cm}{
   \begin{eqnarray}
      ( \textrm{The tensor product } {\cal A} \check{\otimes} {\cal B}) 
    \stackrel{def}{=} \textrm{the smallest closed Lie algebra
    containing } {\cal A} \otimes {\cal B}. 
   \end{eqnarray} }}

  The above example ${\cal H}$ and ${\bf H}$ teaches us clearly that
  ${\cal H} \otimes {\bf H}$ and ${\cal H} \check{\otimes} {\bf H}$
  are completely different. From the commutation relations $[ ({\cal
  H} \otimes {\bf H}) , ({\cal H} \otimes {\bf H}) ] =  
      ({\cal A}' \otimes {\bf A}) \oplus ({\cal H} \otimes {\bf H})$
  and $     [ ( {\cal A}' \otimes {\bf A}) , ({\cal A}' \otimes {\bf A}) ] = 
     ( {\cal A}' \otimes {\bf A} ) \oplus ( {\cal H} \otimes {\bf H}
  )$, the smallest Lie algebra containing ${\cal H} \otimes {\bf H}$
  is
  \begin{eqnarray}
   {\cal H} \check{\otimes} {\bf H} = ({\cal H} \otimes {\bf H})
   \oplus ({\cal A}' \otimes {\bf A}).
  \end{eqnarray} 
  
 \paragraph{Enhancement of the Gauge Symmetry}. \\
  We next explain how the Smolin's proposal for the gauged action
 \cite{0006137} enhances the gauge symmetry. Smolin's original
 proposal is to alter the gauge symmetry from the Lie algebra of
 $OSp(1|32,R) \times SU(N)$ to the tensor product of the Lie algebras
 $osp(1|32,R) \check{\otimes} su(N)$\footnote{In our discussion, we
 consider the analytic continuation of  $osp(1|32,R) \check{\otimes}
 su(N)$ Lie algebra: $gl(1|32,R) \otimes gl(N,R) (=gl(1|32,R)
 \check{\otimes}  gl(N,R))$. }. We grasp the structure of the
 enhancement of the gauge symmetry by considering the toy
 model for simplicity.\\ 

  The Lie algebra $su(6)$ is known as the tensor product of the Lie
  algebras $su(3)$ and $su(2)$:
   \begin{eqnarray}
     su(6) = su(3) \check{\otimes} su(2).
   \end{eqnarray}
  This fact is discerned as follows. Let $\lambda^{a}$ and
  $\sigma^{i}$ be the basis of the Lie algebra $su(3)$ and $su(2)$,
  respectively. $su(3)$ and $su(2)$ are 8 and 3 dimensional Lie
  algebras respectively, and the indices run $a = 1, \cdots, 8$ and $i 
  = 1, 2, 3$. The tensor product $su(3) \check{\otimes} su(2)$
  consists of the following elements.
   \begin{itemize}
    \item{ $\lambda^{a} \otimes \sigma^{i}$: These are the elements of 
        the tensor product as a set of matrices: $su(3) \otimes
        su(2)$. This set, per se, does not constitute a closed Lie
        algebra.} 
   \item{ $\lambda^{a} \otimes {\bf 1}$ and ${\bf 1} \otimes
       {\sigma^{i}}$: These are the generators of the
       group $SU(3) \times SU(2)$.}
   \end{itemize}   
  These elements are known to constitute the algebra of $su(6)$. 
  This example shows in a pedagogical way how the notion of 'gauged
  theory' enhances the gauge symmetry. While the Lie algebra of the
  gauge group $SU(3) \times SU(2)$ is a $8+3=11$ dimensional algebra,
  the tensor product $su(3) \check{\otimes} su(2)$ is a $8+3+24=35$
  dimensional Lie algebra. This is the structure of the enhancement of 
  the gauge symmetry. Note that the similar enhancement of the gauge
  symmetry is seen  in Smolin's proposal\cite{0006137}.

\section{Miscellaneous Calculations}
 \subsection{Proof of (\ref{2-42-}) } \label{proofofmeasure}
   We give a proof of the measure of matrix integration
 (\ref{2-42-}). The measure $d^{N^{2}}M$ means an integration  
 with respect to all components of the matrix $M$, that is $d^{N^{2}}
 M = \prod^{N}_{i,j=1} dM_{ij}$. It is easier to analyze the matrix measure in 
 terms of the eigenvalues $\{ \lambda_{i} \}$, and we diagonalize the
 matrix $M$. Since $M$ is an hermitian matrix, there exists a unitary
 matrix $U$ such that 
  \begin{eqnarray}
   U M U^{\dagger} = diag(\lambda_{1} , \lambda_{2} , \cdots ,
   \lambda_{N}). 
  \end{eqnarray}
  The integral is then divided into that of eigenvalues and that of
  the unitary matrix:
   \begin{eqnarray}
    d^{N^{2}}M = d \lambda_{1} \cdots d \lambda_{N} h(\lambda_{1},
    \cdots \lambda_{N}) dU_{ij}, 
   \end{eqnarray}
  where $h(\lambda_{1} ,\cdots ,\lambda_{N})$ is a function we
  determine immediately.
  The degree of freedom of the eigenvalue is $N$, where as that of the 
  unitary matrix is $N^{2}-N$ ( because this excludes the freedom of
  the eigenvalues). 
  \begin{figure}[htbp]
   \begin{center}
    \scalebox{.5}{\includegraphics{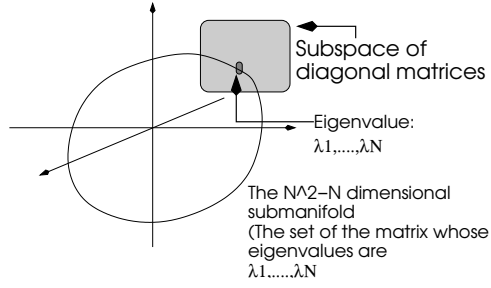} }
   \end{center}
   \caption{This figure shows the space of the hermitian $N^{2}$
     dimensional matrix. The submanifold show the orbit of the unitary 
     transformation of the diagonal matrix $diag(\lambda_{1}, \cdots,
     \lambda_{N})$. }
  \label{measure}
  \end{figure}
  Now, we consider the $N^{2}-N$ dimensional submanifold, depicted by
  the 'loop' in the above picture,  of the manifold of the entire
  $N^{2}$ dimensional matrix $M$. This manifold can be expressed by
  $N$ parameters, $\lambda_{1} \cdots \lambda_{N}$. We consider
  the effect of the integration with respect to the unitary matrix. In 
  order to understand this effect, we consider the infinitesimal
  transformation with respect to the unitary matrix. Let $U$ be $U =1
  + i \epsilon$. The bases of infinitesimal matrix (id est, the Lie algebra of
  the unitary group) are $u_{ij} = \epsilon_{ij} E_{ij} +
  \epsilon^{\dagger}_{ij} E_{ji}$ (where $E_{ij}$ is a matrix whose
  $(i,j)$ component is $1$ and other components are 0). Taking the
  commutator, we obtain
  \begin{eqnarray}
   [ diag(\lambda_{1}, \cdots, \lambda_{N}) , u_{ij} ] =
   -\epsilon_{ij} E_{ij} (\lambda_{i} - \lambda_{j}) +
   \epsilon^{\dagger}_{ij} E_{ji} (\lambda_{i} - \lambda_{j}),  
  \end{eqnarray}
   Therefore, the contribution to the measure is $(\lambda_{i} -
   \lambda_{j})^{2}$. We take all the products of $i,j (i<j)$, then the 
   integration of the unitary matrix gives $h(\lambda_{1}, \cdots,
   \lambda_{N}) = \prod_{i<j} (\lambda_{i} - 
   \lambda_{j})^{2}$. We finally obtain our desired result \\
 \shadowbox{ \parbox{16cm}{ 
    \begin{eqnarray}
       d^{N^{2}} M = \prod_{i} d \lambda_{i} \prod_{i<j} (\lambda_{i} -
     \lambda_{j})^{2} dU_{ij}.
    \end{eqnarray} }}

 \subsection{Proof of (\ref{AZM52actionpmb}) }
    \label{AZCospactionres} 
   This appendix is devoted to the computation of the action of
 $osp(1|32,R)$ cubic matrix model. The computation of the bosonic part 
 $tr(m^{a} m^{b} m^{c}) Tr(T^{a} [T^{b}, T^{c}]) $ is especially
 tedious, and this  appendix provides the  
 technique to deal with this computation. In this computation, it is
 easier to rely on the color indices, rather than to adopt large $N$
 representation. The bosonic $32 \times 32$ matrices $m^{a}$ are
 expressed in terms of the basis of the gamma matrices
  \begin{eqnarray}
    m^{a} = u^{a}_{\mu} \Gamma^{\mu} + \frac{1}{2} u^{a}_{\mu_{1}
    \mu_{2}} \Gamma^{\mu_{1} \mu_{2}} + \frac{1}{5!} u^{a}_{\mu_{1}
    \cdots \mu_{5}} \Gamma^{\mu_{1} \cdots \mu_{5}}.
  \end{eqnarray}
 We work out the computation of the trace $tr(m^{a} m^{b} m^{c})$ in
  terms of the coefficients of gamma matrix representation, and the
  properties in the Appendix. \ref{AZCgammaproperty} play an essential
  role in the analysis. 
  \begin{itemize}
   \item{ The property in Appendix. \ref{AZCvanish} indicates that all we
       have to consider is the coefficients of ${\bf 1}_{32 \times
       32}$ in the product $m^{a} m^{b} m^{c}$, because
       \begin{eqnarray} 
       tr(\Gamma^{\mu})=tr(\Gamma^{\mu_{1} \mu_{2}}) = tr(\Gamma^{\mu_{1}
       \mu_{2} \mu_{3}}) = tr(\Gamma^{\mu_{1} \cdots \mu_{4}}) =
       tr(\Gamma^{\mu_{1} \cdots \mu_{5}}) = 0.
       \end{eqnarray}
      }
   \item{ When we consider the product of the gamma matrices, we
       utilize the formula in Sec. \ref{AZCproductgamma}. We have a
       table of what rank of gamma matrices emerge in the product of 
       two gamma matrices. For example, The product $\Gamma^{\mu_{1}
       \mu_{2}} \Gamma^{\nu_{1} \nu_{2}}$ produces the 
       gamma matrices of rank 0,2,4. Multiplying another elements $m^{c} \in
       sp(32)$, only the term $\frac{1}{2!} u^{c}_{\mu_{1} \mu_{2}}
       \Gamma^{\mu_{1} \mu_{2}}$ contribute to the trace $tr(m^{a}
       m^{b} m^{c})$ because of the property mentioned in the previous 
       item.}   
  \end{itemize}

   We adopt a new abbreviation $(xyz)$, which means
    \begin{eqnarray}
     (xyz) = \textrm{the terms emerging from } \frac{1}{x!y!z!}
     u^{a}_{\mu_{1} \cdots \mu_{x}} u^{b}_{\nu_{1} \cdots \nu_{y}}
     u^{c}_{\rho_{1} \cdots \rho_{z}} tr(\Gamma^{\mu_{1} \cdots
     \mu_{x}} \Gamma^{\nu_{1} \cdots \nu_{y}} \Gamma^{\rho_{1} \cdots
     \rho_{z}}), \nonumber 
    \end{eqnarray}
   where $x,y,z = 1,2,5$ because we are considering $osp(1|32,R)$
   matrix model. Based on the above prescription, the candidate of the 
   nonvanishing terms are as follows:
    \begin{eqnarray}
     (112), (155), (222), (255), (555).
    \end{eqnarray}
   The bosonic part of the action can be written as 
    \begin{eqnarray}
     I_{b} &=& - \frac{f_{abc}}{2g^{2}} tr(m^{a} m^{b} m^{c}) \nonumber
     \\    &=& - \frac{f_{abc}}{2g^{2}} tr( (112) + (155) + (222) +
     (255) + (555) ). \label{AZM42actioncomp4}  
    \end{eqnarray}
  We compute these terms one by one.

  \paragraph{(112) terms} .\\
  In the computation of these terms, we must
  be cautions about the following point: in picking up the term
  $(xyz)$, we must take into account where we have picked up the $x$
  (or $y$, $z$) term. In this case, we must consider the sum 
   \begin{eqnarray}
    (112) &=& \frac{1}{2!}( u^{a}_{\mu_{1}} u^{b}_{\nu_{1}}
    u^{c}_{\rho_{1} \rho_{2}} tr(\Gamma^{\mu_{1}} \Gamma^{\nu_{1}}
    \Gamma^{\rho_{1} \rho_{2}}) + u^{a}_{\mu_{1}} u^{b}_{\nu_{1}
    \nu_{2}} u^{c}_{\rho_{1}} tr(\Gamma^{\mu_{1}} \Gamma^{\nu_{1}
    \nu_{2}} \Gamma^{\rho_{1}})  
       + u^{a}_{\mu_{1} mu_{2}} u^{b}_{\nu_{1}} u^{c}_{\rho_{1}}
    tr(\Gamma^{\mu_{1} \mu_{2}} \Gamma^{\nu_{1}} \Gamma^{\rho_{1}}) )
    \nonumber \\
      &=& \frac{1}{2!} u^{a}_{\mu_{1}} u^{b}_{\nu_{1}} u^{c}_{\rho_{1} 
    \rho_{2}} tr(\Gamma^{\mu_{1}} \Gamma^{\nu_{1}} \Gamma^{\rho_{1}
    \rho_{2}} + \Gamma^{\nu_{1}} \Gamma^{\rho_{1} \rho_{2}}
    \Gamma^{\mu_{1}} + \Gamma^{\rho_{1} \rho_{2}} \Gamma^{\mu_{1}}
    \Gamma^{\nu_{1}} ) \nonumber \\ 
    &=&  \frac{3}{2!} u^{a}_{\mu_{1}} u^{b}_{\nu_{1}} u^{c}_{\rho_{1} 
    \rho_{2}} tr(\Gamma^{\mu_{1}} \Gamma^{\nu_{1}} \Gamma^{\rho_{1}
    \rho_{2}} ), \label{AZ42a111}
   \end{eqnarray}
   where we have utilized the cyclic symmetry of the trace with
   respect to the gamma matrices. The formula of the product of the
   gamma matrices (\ref{AZproduct}) indicates that 
    \begin{eqnarray}
     tr(\Gamma^{\mu_{1}} \Gamma^{\nu_{1}} \Gamma^{\rho_{1}
    \rho_{2}} ) = tr( (\Gamma^{\mu_{1} \mu_{2}} + \eta^{\mu_{1}
    \mu_{2}}) \Gamma^{\rho_{1} \rho_{2}}) = tr(-2 \eta^{\mu_{1}
    \rho_{1}} \eta^{\nu_{1} \rho_{2}} ) = -64 \eta^{\mu_{1}
    \rho_{1}} \eta^{\nu_{1} \rho_{2}}.
    \end{eqnarray}
  Thus we complete the computation of the (112) term 
   \begin{eqnarray}
    (112) = -96 u^{a}_{\mu_{1}} u^{b}_{\nu_{1}} u^{c \mu_{1} \nu_{1}}.
    \label{AZ42f112} \end{eqnarray}

  \paragraph{(155) terms} . \\
  We perform a computation similar to the case of the (112) term:
  \begin{eqnarray}
  (155) &=& \frac{1}{(5!)^{2}} ( u^{a}_{\mu_{1}}
  u^{b}_{\nu_{1} \cdots \nu_{5}}  u^{c}_{\rho_{1} \cdots \rho_{5}}
    tr( \Gamma^{\mu_{1}} \Gamma^{\nu_{1} \cdots \nu_{5} }
    \Gamma^{\rho_{1} \cdots \rho_{5}} ) + u^{a}_{\mu_{1}
    \cdots \mu_{5}} u^{b}_{\nu_{1}} u^{c}_{\rho_{1} \cdots \rho_{5}}
    tr( \Gamma^{\mu_{1} \cdots \mu_{5}} \Gamma^{\nu_{1}}
    \Gamma^{\rho_{1} \cdots \rho_{5}} ) \nonumber \\ &+& 
     u^{a}_{\mu_{1} \cdots \mu_{5}} u^{b}_{\nu_{1} \cdots \nu_{5}}
    u^{c}_{\rho_{1}} tr( \Gamma^{\mu_{1} \cdots \mu_{5}} 
     \Gamma^{\nu_{1} \cdots \nu_{5}} \Gamma^{\rho_{1}} ) )
    \nonumber \\ &=&   \frac{1}{(5!)^{2}}   u^{a}_{\mu_{1}}
    u^{b}_{\nu_{1} \cdots \nu_{5}} u^{c}_{\rho_{1}
     \cdots  \rho_{5}} tr(  \Gamma^{\mu_{1}}
    \Gamma^{\nu_{1}\cdots \nu_{5}} \Gamma^{ \rho_{1} \cdots \rho_{5}}
    + \Gamma^{\rho_{1} \cdots \rho_{5} } \Gamma^{\mu_{1}}  
    \Gamma^{\nu_{1} \cdots \nu_{5}} + \Gamma^{\nu_{1} \cdots  \nu_{5}}
    \Gamma^{\rho_{1} \cdots \rho_{5}} \Gamma^{\mu_{1} } ) \nonumber
    \\ &=&
     \frac{3}{(5!)^{2}} u^{a}_{\mu_{1}}
    u^{b}_{\nu_{1} \cdots \nu_{5}} u^{c}_{\rho_{1}
     \cdots  \rho_{5}} tr( \Gamma^{\mu_{1}}
    \Gamma^{\nu_{1}\cdots \nu_{5}} \Gamma^{ \rho_{1} \cdots \rho_{5}}
  ). \label{AZ42a155}
    \end{eqnarray} 
  The biggest difference from the previous case is that the
  nonvanishing term emerges due to the dual of the gamma matrices: 
   \begin{eqnarray}  & & tr(\Gamma^{\mu_{1}} \Gamma^{\nu_{1}\cdots \nu_{5}}
     \Gamma^{ \rho_{1} \cdots \rho_{5}}) =  tr(\Gamma^{\mu_{1}
     \nu_{1}\cdots \nu_{5}}  \Gamma^{ \rho_{1} \cdots \rho_{5}}) =
     \frac{-1}{5!} tr(\epsilon^{\mu_{1} \nu_{1}\cdots \nu_{5} \chi_{1} 
      \cdots \chi_{5}} \Gamma_{\chi_{1} \cdots \chi_{5}}  \Gamma^
     { \rho_{1} \cdots \rho_{5}} ) \\ \nonumber &=&
    \frac{-1}{5!} tr(\epsilon^{\mu_{1} \nu_{1}\cdots \nu_{5} \chi_{1} 
      \cdots \chi_{5}} 5! \eta_{\chi_{1}}^{\rho_{1}} \cdots
     \eta_{\chi_{5}}^{\rho_{5}})  = - 32 \epsilon^{\mu_{1} \nu_{1}
     \cdots \nu_{5} \rho_{1} \cdots \rho_{5}}.
  \end{eqnarray}
  Therefore the final result is obtained by
     \begin{eqnarray}
    (155) = -\frac{1}{150}  u^{a}_{\mu_{1}}
    u^{b}_{\nu_{1} \cdots \nu_{5}} u^{c}_{\rho_{1}
     \cdots  \rho_{5}}  \epsilon^{\mu_{1} \nu_{1}
     \cdots \nu_{5} \rho_{1} \cdots \rho_{5} }.  \label{AZ42f155}
    \end{eqnarray} 

  \paragraph{(222) terms} . \\
  This term emerges from the choice that we pick up $\frac{1}{2!}
  u_{\mu_{1} \mu_{2}} \Gamma^{\mu_{1} \mu_{2}}$ from all of $m^{a},
  m^{b}, m^{c}$. The difference from the previous two cases is that we 
  do not have to consider the number of the ways to pick up the term
  (222). The contribution is therefore 
   \begin{eqnarray} \frac{1}{(2!)^{3}}  u^{a}_{\mu_{1} \mu_{2}}
    u^{b}_{\nu_{1} \nu_{2}} u^{c}_{\rho_{1}  \rho_{2}} tr(
    \Gamma^{\mu_{1} \mu_{2}} \Gamma^{\nu_{1}  \nu_{2}} \Gamma^
    { \rho_{1} \rho_{2} } ). \label{AZ42a222}
   \end{eqnarray}
  We likewise compute the gamma matrices, so that we obtain 
   \begin{eqnarray}
  & &  tr(\Gamma^{\mu_{1} \mu_{2}} \Gamma^{\nu_{1} \nu_{2}}
    \Gamma^{\rho_{1} \rho_{2}} ) = tr( (\Gamma^{\mu_{1} \mu_{2}
    \nu_{1} \nu_{2}} - 4 {\eta^{[\mu_{1}}}^{[\nu_{1}}
    {\Gamma^{\mu_{2}]}}^{\nu_{2}]} - 2 {\eta^{[\mu_{1}}}^{[\nu_{1}}
    {\eta^{\mu_{2}]}}^{\nu_{2}]} )\Gamma^{\rho_{1} \rho_{2}} )
    \nonumber \\
 &=& tr(8 {\eta^{[\mu_{1}}}^{[\nu_{1}} {\eta^{\mu_{2}]}}^{[\rho_{1}}
    {\eta^{\nu_{2}]} }^{\rho_{2}]} ),
   \end{eqnarray}
  where the bracket $[\mu_{1} \mu_{2}\cdots ]$ represents the
  anti-symmetry of the 
  indices (for example $P_{[\mu_{1} \mu_{2}]} = \frac{1}{2} ( P_{\mu_{1}
  \mu_{2}} - P_{\mu_{2} \mu_{1}})$), but it is not necessary to write
  explicitly  this awkward mark because the anti-symmetry is obvious.
  These  indices contract with the antisymmetric indices of
  $u^{a}_{\mu_{1} \cdots }$.  The final result is obtained by 
   \begin{eqnarray}
    (222) = 32 u^{a}_{\mu_{1} \mu_{2}} {u^{b \mu_{1}}}_{\nu_{1}}
    u^{c \mu_{2} \nu_{1}}.
   \end{eqnarray}

 \paragraph{(255) terms} . \\
  We only repeat the similar procedure, and we proceed rather
  quickly. The contribution of these terms is 
     \begin{eqnarray}
    \frac{3}{2!(5!)^{2}} u^{a}_{\mu_{1} \mu_{2} } u^{b}_{\nu_{1} \cdots 
    \nu_{5}} u^{c}_{\rho_{1} \cdots  \rho_{5}}  
    tr ( \Gamma^{\mu_{1} \mu_{2}} \Gamma^{\nu_{1} \cdots \nu_{5}}
    \Gamma^{\rho_{1} \cdots  \rho_{5}} ). \label{AZ42a255}
  \end{eqnarray} 
    Computing the gamma matrices similarly,
  \begin{eqnarray}
  & &   \Gamma^{\mu_{1} \mu_{2}} \Gamma^{\nu_{1} \cdots \nu_{5}}
  \Gamma^{\rho_{1} \cdots  \rho_{5}} = -10 \eta^{\mu_{1} \nu_{1}}
  \Gamma^{  \mu_{2} \nu_{2} \cdots \nu_{5} }  \Gamma^{\rho_{1} \cdots
  \rho_{5}} \nonumber \\ &=& -1200  \eta^{\mu_{1} \nu_{1}}
  \eta^{\mu_{2} \rho_{1}} 
  \eta^{\nu_{2} \rho_{2}}  \eta^{\nu_{3} \rho_{3}}  \eta^{\nu_{4}
  \rho_{4}}  \eta^{\nu_{5} \rho_{5}}.  \end{eqnarray}
  Therefore, the final result is
   \begin{eqnarray}
    (255) = -4  u^{a}_{\mu_{1} \mu_{2} } {u^{b \mu_{1}}}_{\nu_{1} \cdots 
    \nu_{4}} u^{c \mu_{2} \nu_{1} \cdots  \nu_{4}}. \label{AZ42f255}
   \end{eqnarray} 

 \paragraph{(555) terms}  . \\
  The contribution is obtained by 
  \begin{eqnarray}
      \frac{1}{(5!)^{3}} u^{a}_{\mu_{1} \cdots \mu_{5} }
    u^{b}_{\nu_{1} \cdots \nu_{5}} u^{c}_{\rho_{1} \cdots
    \rho_{5}} tr( \Gamma^{\mu_{1} \cdots \mu_{5}}
    \Gamma^{\nu_{1} \cdots \nu_{5}} \Gamma^{\rho_{1} 
    \cdots  \rho_{5}} ). \label{AZ42a555}
  \end{eqnarray}
  The product of the gamma matrices is computed to be 
  \begin{eqnarray}
   & & \Gamma^{\mu_{1} \cdots \mu_{5}} \Gamma^{\nu_{1} \cdots \nu_{5}}
    \Gamma^{\rho_{1} \cdots  \rho_{5}} = \frac{200}{5!} \eta^{\mu_{1} \nu_{1}}
    \eta^{\mu_{2} \nu_{2}} \epsilon^{\mu_{3} \mu_{4} \mu_{5} \nu_{3}
    \nu_{4} \nu_{5} \chi_{1} \cdots \chi_{5} } \Gamma_{\chi_{1} \cdots 
    \chi_{5}} \Gamma^{\rho_{1} \cdots \rho_{5}} \nonumber \\ &=& 200
    \eta^{\mu_{1}  \nu_{1}}  \eta^{\mu_{2} \nu_{2}} \epsilon_{\mu_{3}
    \mu_{4} \mu_{5} \nu_{3} \nu_{4} \nu_{5} \rho_{1} \cdots \rho_{5} }. 
  \end{eqnarray} 
  The result is therefore,
   \begin{eqnarray}
   (555) = \frac{1}{270} u^{a}_{\mu_{1} \cdots \mu_{5} }
    {u^{b \mu_{1} \mu_{2}}}_{\nu_{3} \nu_{4} \nu_{5}} u^{c}_{\rho_{1}
    \cdots  \rho_{5}} \epsilon^{\mu_{3} \mu_{4} \mu_{5} \nu_{3}
    \nu_{4} \nu_{5} \rho_{1} \cdots \rho_{5} }. \label{AZ42f555}
  \end{eqnarray} 
 
  The conclusion of this lengthy computation is summarized as 
   \begin{eqnarray}
      (112) &=& -96 u^{a}_{\mu_{1}} u^{b}_{\nu_{1}} u^{c \mu_{1} \nu_{1}},
    \nonumber \\
      (155) &=&  -\frac{1}{150}  u^{a}_{\mu_{1}}
    u^{b}_{\nu_{1} \cdots \nu_{5}} u^{c}_{\rho_{1}
     \cdots  \rho_{5}}  \epsilon^{\mu_{1} \nu_{1}
     \cdots \nu_{5} \rho_{1} \cdots \rho_{5} },  \nonumber \\
      (222) &=& 32 u^{a}_{\mu_{1} \mu_{2}} {u^{b \mu_{1}}}_{\nu_{1}}
    u^{c \mu_{2} \nu_{1}}, \nonumber \\
      (255) &=&  -4  u^{a}_{\mu_{1} \mu_{2} } {u^{b \mu_{1}}}_{\nu_{1} \cdots 
    \nu_{4}} u^{c \mu_{2} \nu_{1} \cdots  \nu_{4}}, \nonumber \\
      (555) &=&  \frac{1}{270} u^{a}_{\mu_{1} \cdots \mu_{5} }
    {u^{b \mu_{1} \mu_{2}}}_{\nu_{3} \nu_{4} \nu_{5}} u^{c}_{\rho_{1}
    \cdots  \rho_{5}} \epsilon^{\mu_{3} \mu_{4} \mu_{5} \nu_{3}
    \nu_{4} \nu_{5} \rho_{1} \cdots \rho_{5} }. \nonumber 
    \end{eqnarray}

  We next examine the fermionic part of this
  action. However, the computation of its contribution is trivial:
   \begin{eqnarray}
    I_{f} &=& - \frac{f_{abc}}{2g^{2}} (-3i {\bar \psi^{a}} m^{b}
    \psi^{c} ) \nonumber \\
          &=&   \frac{3i f_{abc}}{2g^{2}} ( 
   {\bar \psi^{a}} \Gamma^{\mu_{1}} u^{b}_{\mu_{1}} \psi^{c}
 + \frac{1}{2!} {\bar \psi^{a}} \Gamma^{\mu_{1} \mu_{2}}
    u^{b}_{\mu_{1} \mu_{2}} \psi^{c} 
 + \frac{1}{5!} {\bar \psi^{a}} \Gamma^{\mu_{1} \cdots \mu_{5}}
    u^{b}_{\mu_{1} \cdots \mu_{5}} \psi^{c} ). 
   \end{eqnarray}
 
  We are now finished with the computation of the action in terms of
  the components $\{ u_{X} \}$. This is a description in terms of 11
  dimensional indices. One of the problems we should tackle with is
  the correspondence between this cubic matrix theory and the existing 
  10 dimensional matrix theory such as IKKT model. For this purpose,
  it is more convenient to express the indices for 10
  dimensions. Rewriting the above terms in terms of the variables
  (\ref{AZ43fieldredef}), we obtain a final result.

 \subsection{Propagator of the action (\ref{AZ44partial}) $\sim$
   (\ref{AZ44fermother}) } 
  \label{AZCkillac}
    This section is devoted to the investigation of the Feynman rules of 
  the action (\ref{AZ44partial}) $\sim$ (\ref{AZ44fermother}). Most of
  the Feynman rules are trivially read off from the Lagrangian,
  however the non-trivial problem is the propagator among the fields
  $W$, $A_{i}$ and $C_{ij}$. This is due to the kinetic term
   \begin{eqnarray}
    \frac{1}{g^{2}} Tr_{N \times N}( 96 W \partial_{i} A^{(-) i} - 96
    A^{(-)}_{i} \partial_{j} C^{ij} ),
   \end{eqnarray}
  in which the vector fields $A^{(-)}_{i}$ are connected with both $W$ 
  and $C_{ij}$ fields. We would like to investigate the Feynman
  diagram among the three fields $W$, $A^{(-)}_{i}$ and $C_{ij}$. For
  simplicity,  we omit an awkward coefficient
  $\frac{96}{g^{2}}$. Then, the Lagrangian in question is 
    \begin{eqnarray}
      I_{WAC} = - W \partial_{i} A^{(-)}_{i} + A^{(-)}_{i}
      \partial_{j} C^{ij}.  \label{AZMBwacac}
    \end{eqnarray}

  \begin{itemize}
   \item{First, we disprove the existence of the propagator $\langle
       A^{(-)}_{i} C^{ij} \rangle$. This stems from the gauge
       freedom of the field $C^{ij}$. As is easily read off from the
       action (\ref{AZMBwacac}), $C^{ij}$ is invariant under the
       transformation 
        \begin{eqnarray}
          C^{ij} \to C^{ij} + \partial_{k} \chi^{ijk},
        \end{eqnarray}
      where the quantity $\chi^{ijk}$ is anti-symmetric with
      respect to the exchange of the indices. Therefore, the propagators 
      $\langle A^{(-)}_{i} C^{ij} \rangle$ do not exist in this
      theory.}
   \item{We next consider the propagator $\langle W A^{(-)}_{i}
       \rangle$. This is understood by first integrating out the
       fields $C^{ij}$. Since this action is linear with respect to
       $C^{ij}$, this is tantamount to solving the classical
       Euler-Lagrange equation with respect to $C^{ij}$:
        \begin{eqnarray}
         \partial_{k} ( \frac{ \partial I_{WAC}}{\partial (\partial_{k} 
         C^{ij}) } ) - \frac{\partial I_{WAC}}{\partial C^{ij}}  =
         \partial_{[j} A^{(-)}_{i]} = 0.
        \end{eqnarray}
      Note that the indices are anti-symmetrized because we have
      solved the Euler-Lagrange equation  with respect to $C^{ij}$.
      The gauge field $A^{(-)}_{i}$ is thus constrained to be
       \begin{eqnarray}
        \partial_{i} A^{(-)}_{j} - \partial_{j} A^{(-)}_{i} = 0,
        \Rightarrow {^{\exists} \lambda} \textrm{ s.t. } A^{(-)}_{i} = 
        \partial_{i} \lambda. \label{AZMB4aminus}
       \end{eqnarray}
      Substituting (\ref{AZMB4aminus}) into the action
      (\ref{AZMBwacac}), the action is 
      \begin{eqnarray}
        I_{WAC} = - W \partial_{i} \partial^{i} \lambda.
      \end{eqnarray} 
      This indicates that the propagator for $W$ and $\lambda$ is
      $\langle W \lambda \rangle \sim \frac{1}{p^{2}}$. Therefore, we
      obtain a propagator for $W$ and $A^{(-)}_{i}$:
       \begin{eqnarray}
        \langle W A^{(-)}_{i} \rangle \sim \frac{p_{i}}{p^{2}}.
       \end{eqnarray}
     }
    \end{itemize}

 \subsection{Disproof of the existence of the perturbative
   propagators $\langle WW \rangle$ and $\langle A^{(-)}_{i}
   A^{(-)}_{j}  \rangle$  }
  \label{AZCkillwwaa}
   In this subsection, we develop a disproof of such propagators as
 $\langle WW \rangle$ and $\langle A^{(-)}_{i} A^{(-)}_{j} \rangle$ at 
 a perturbative level. This statement is verified more generally, and
 we use a rather abstract expression. The bosonic terms in
 $osp(1|32,R)$ cubic   matrix theory is distinguished into the
 following two groups.
  \begin{itemize}
   \item{ $U_{e}$ denotes the fields of even rank with respect to 10
       dimensional indices. Id est, we mean the fields $W$, $C_{ij}$ and
       $H_{ijkl}$\footnote{Originally the terms $W$ and $H_{ijkl}$
       comes from the fields $u_{\mu \cdots}$ of rank 1 and 5
       respectively. These mean $W = u_{\sharp}$ and $H_{ijkl} =
       u_{ijkl\sharp}$ respectively, and in terms of the 10
       dimensional indices, these are fields of even rank.}. }
   \item{ $U^{(\pm)}_{o}$ denotes the fields of odd rank in terms of 10
       dimensional indices with plus (or minus) chirality. 
         \begin{itemize}
          \item{$U^{(+)}_{i}$ denotes $A^{(+)}_{i}$ or $I^{(+)}_{i_{1} 
                \cdots i_{5} }$. }
          \item{$U^{(-)}_{i}$ denotes $A^{(-)}_{i}$ or $I^{(-)}_{i_{1} 
                \cdots i_{5} }$. }
         \end{itemize} }
  \end{itemize}
  We have expanded this cubic matrix theory around the classical 
  solution (\ref{AZ44cs}) and considered the following mapping
   \begin{eqnarray}
     [ A^{(+)}_{i}, X ] \Rightarrow - i \partial_{i} X + [a^{(+)}_{i},X]. 
   \end{eqnarray}
  Therefore, the action with the kinetic terms introduced by this
  mapping is abstractly expressed as follows:
  \begin{eqnarray}
   I = \{ (\partial U_{e}) U^{(-)}_{o} + (\partial {\bar \psi_{L}})
   \psi_{L} \} + ( U_{e} U^{(+)}_{o} U^{(-)}_{o}
   + {\bar \psi_{L}} U^{(+)}_{o} \psi_{L} + {\bar \psi_{R}}
   U^{(-)}_{o} \psi_{R} + U^{3}_{e} + {\bar \psi_{L}} U_{e} \psi_{R}).
  \end{eqnarray}
  \begin{itemize}
   \item{ We can in general verify that the product of plural fields
       of the same chirality (such as $U^{(+)}_{o} U^{(+)}_{o}$ or
       $U^{(-)}_{o} U^{(-)}_{o}$) never emerges in this cubic
       action. This can be verified by noting the structure of gamma
       matrices. For example, suppose that  such terms as
       $U^{(+)}_{o} U^{(+)}_{o} U^{X}$ emerges in the bosonic part of
       the cubic action. This would appear in the action,
        \begin{eqnarray}
          Tr_{N \times N}( U^{(+)}_{o} U^{(+)}_{o} U^{X})
          tr(\Gamma^{o} ( 1 + \Gamma^{\sharp}) \Gamma^{o} (1 +
          \Gamma^{\sharp}) \Gamma^{X}),
        \end{eqnarray}
       where $\Gamma^{o}$ denotes the gamma matrices of odd rank in
       terms of 10 dimensional indices, such as $\Gamma^{i}$ or
       $\Gamma^{ijklm}$.  Be careful about the structure of the gamma
       matrices. Noting the Clifford algebra $\Gamma^{i}
       \Gamma^{\sharp} + \Gamma^{\sharp} \Gamma^{i} =0$, this gamma
       matrix is shown to vanish:
        \begin{eqnarray}
         \Gamma^{o} ( 1 + \Gamma^{\sharp}) \Gamma^{o} (1 +
         \Gamma^{\sharp}) \Gamma^{X} = \Gamma^{o} \Gamma^{o}
         (1-\Gamma^{\sharp}) (1+\Gamma^{\sharp}) \Gamma^{X} = 0.
        \end{eqnarray} 
       Therefore, The bosonic part is restricted to be
       $U^{(+)}_{o} U^{(-)}_{o} U_{e}$ or $U_{e}^{3}$. }
     \item{ For the same reason, the kinetic term is restricted to be
       $(\partial 
       U^{(-)}_{o}) U_{e}$ since the kinetic terms emerge from the
       expansion around the classical solution (\ref{AZ44cs})
       \footnote{ Note that $(\partial U^{(-)}_{o}) U_{e}$ is
       identical to  $(\partial U_{e}) U^{(-)}_{o}$ up to the partial 
       integration.} }
   \item{ The fermionic terms is restricted to be of the form ${\bar
         \psi_{L}} U^{(+)}_{o} \psi_{L}$, ${\bar \psi_{R}} U^{(-)}_{o}
         \psi_{R}$ or ${\bar \psi_{L}} U_{e} \psi_{R}$, which is
         understood from the properties (\ref{AZMA21fermvanish}). }
  \end{itemize}
  In order to develop  perturbative argument, we assign a following
  charge, and perform the correspondence rescaling.
    \begin{itemize}
     \item{$U_{e}$, $U^{(+)}_{o}$ and $U^{(-)}_{o}$ are assigned the
         charge $1$, $0$ and $-1$ respectively. Id est, the scaling is $U^{e}
         \to e^{\lambda} U_{e}$, $U^{(+)}_{o} \to U^{(+)}_{o}
         $ and $U^{(-)}_{o} \to U^{(-)} e^{-\lambda} $.}
     \item{$\psi_{L}$ and $\psi_{R}$ are assigned charge 0 and
         $\frac{1}{2}$ respectively. Id est, the scaling is 
         $\psi_{L} \to \psi_{L}$ and $\psi_{R} \to
         \psi_{R} e^{\frac{\lambda}{2}}$.} 
    \end{itemize}
  Following these scaling rules, this action is thus rescaled as follows:
    \begin{eqnarray}
   I = \{ (\partial U_{e}) U^{(-)}_{o} + (\partial {\bar \psi_{L}})
   \psi_{L} \} + ( U_{e} U^{(+)}_{o} U^{(-)}_{o}
   + {\bar \psi_{L}} U^{(+)}_{o} \psi_{L} + {\bar \psi_{R}}
   U^{(-)}_{o} \psi_{R}) + (U^{3}_{e} e^{3 \lambda})  + ({\bar \psi_{L}}
   U_{e} \psi_{R} e^{\frac{3 \lambda}{2}}).
  \end{eqnarray}

  This action possesses only the propagators $\langle \psi_{L} \psi_{L}
  \rangle$ and $\langle U^{(-)}_{o} U_{e} \rangle$ at tree
  level. Other propagators are induced by the multi-loop effect.
  And the vertices in this action and their charges are given below.
    \begin{figure}[htbp]
   \begin{center}
    \scalebox{.5}{\includegraphics{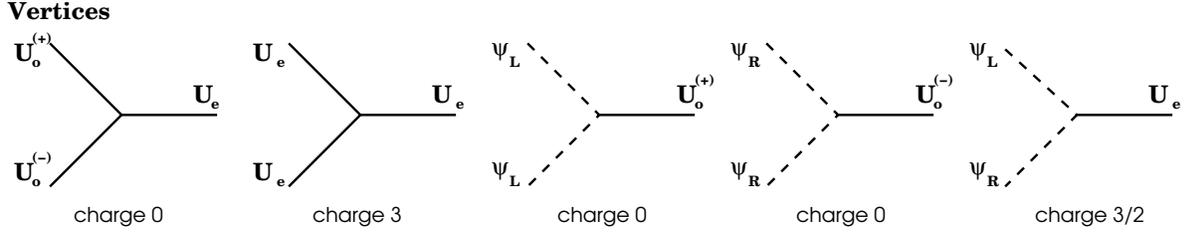} }
   \end{center}
    \caption{The possible vertices of this Lagrangian and their charge.}
   \label{bvertex}
  \end{figure}
 If the induced propagators of $\langle U_{e} U_{e} \rangle$, $\langle
 U^{(-)}_{o} U^{(-)}_{o} \rangle$ and $\langle \psi_{R} \psi_{R}
 \rangle$ are to exist, they can be constructed by the combination of
 those vertices. The charge of $U_{e} U_{e}$, $U^{(-)}_{o}
 U^{(-)}_{o}$ and $\psi_{R} \psi_{R}$ is $+2, -2, +1$,
 respectively. Therefore, the possible charges of these induced
 propagators are listed below.
     \begin{center} \begin{tabular}{|c||c|c|c|} \hline
  propagator &  $\langle U_{e} U_{e} \rangle$ & $\langle
 U^{(-)}_{o} U^{(-)}_{o} \rangle$ & $\psi_{R} \psi_{R}$ \\ \hline
  charge     &  $2 + 3n + \frac{3m}{2}$ & $-2 + 3n + \frac{3m}{2}$ & $1 + 
 3n + \frac{3m}{2}$ \\  \hline
  \end{tabular} \end{center}
  where $n$ and $m$ are the number of the vertices $U_{e}-U_{e}-U_{e}$ 
  and $\psi_{L}-\psi_{R}-U_{e}$. Since the propagators at tree level
  is of charge 0, the propagators do not contribute to the charge of
  the induced propagators.   The point is that, even if we
  consider the multi-loop effect, the charge of these propagators
  increases by the multiple of $3$ or $\frac{3}{2}$. Thus, it is
  impossible to construct the propagators of these fields with charge
  0 no matter how we combine the propagators and the vertices.\\

  If the charge of the propagators is nonzero, these induced
  propagators vanishes when we integrate out the parameter
  $\lambda$. Therefore, such propagators as $\langle W W \rangle$ or
  $\langle A^{(-)}_{i} A^{(-)}_{j} \rangle$ cannot emerge by the
  perturbative multi-loop effect. 

 \subsection{Explicit form of the $gl(1|32,R) \otimes gl(N,R)$ gauged
   cubic action} 
  \label{AZCgaugedactionres}
     This section is devoted to executing the tedious computation of the
    $gl(1|32,R) 
   \otimes gl(N,R)$ gauged cubic matrix theory. Especially we focus on
    the bosonic part $tr(m^{a} m^{b} m^{c})$, because this is the most
    difficult and  
    cumbersome to  compute. We rely on the expression of color
    indices. We first express the trace in terms of the basis of the gamma
   matrices: 
  \begin{eqnarray}
   m^{a} = u^{a} {\bf 1} + u^{a}_{\mu_{1}} \Gamma^{\mu_{1}}
     + \frac{1}{2!} u^{a}_{\mu_{1} \mu_{2}} \Gamma^{\mu_{1} \mu_{2}} 
     + \frac{1}{3!} u^{a}_{\mu_{1} \mu_{2} \mu_{3}} \Gamma^{\mu_{1}
     \mu_{2} \mu_{3}}
     + \frac{1}{4!} u^{a}_{\mu_{1} \cdots \mu_{4}} \Gamma^{\mu_{1} \cdots
     \mu_{4} }
     + \frac{1}{5!} u^{a}_{\mu_{1} \cdots \mu_{5}} \Gamma^{\mu_{1} \cdots
     \mu_{5} }.
  \end{eqnarray}

  And whether these coefficients  
  appear as a commutator or anti-commutator affects the physics
  crucially. Let us investigate the structure of the action from
  another point of view. Considering the correspondence of the super
  Lie algebra $gl(1|32,R)$ with $u(1|16,16)$, we have in the
  previous section noted that $gl(1|32,R)$ can be expressed by the
  direct sum $gl(1|32,R) = {\cal H} \oplus {\cal A}$. The following
  properties  play  an essential role in the
 investigation of the action, especially the cumbersome bosonic
 part.
 For $H_{1}, H_{2}, H_{3} \in {\cal H}$ and $A_{1}, A_{2}, A_{3}
 \in {\cal A}$,
 \begin{eqnarray}
  & &(1)Str(H_{1} H_{2} H_{3}) = \frac{1}{2} Str( H_{1}[H_{2},
  H_{3}]), \hspace{2mm}
     (2)Str(H_{1} H_{2} A_{3}) = \frac{1}{2} Str( H_{1} \{ H_{2},
  A_{3} \} ), \hspace{2mm} \nonumber \\
  & &(3)Str(H_{1} A_{2} A_{3}) = \frac{1}{2} Str( H_{1} [A_{2},
  A_{3}]) \hspace{2mm}
     (4)Str(A_{1} A_{2} A_{3}) = \frac{1}{2} Str( A_{1} \{A_{2}, A_{3} 
 \} ). \label{AZ53distinguish} 
 \end{eqnarray}
  {\sf(Proof)
   This relationship is readily verified by noting the commutation and 
   anti-commutation relationships of the two  different representations of
   $osp(1|32,R)$, as we have seen in (\ref{AZB1comha}). Noting these
   relationships, we readily obtain the following property: 
    \begin{eqnarray}
   \textrm{For } H \in {\cal H} \textrm{ and } A \in {\cal A},
   Str(HA)=0. \label{AZB1tracevanish}
  \end{eqnarray}
   This can be verified by noting that $G^{2}=-1$ and that the cyclic
 rule holds true of the supertrace:
 \begin{eqnarray}
  Str(HA) = Str({^{T}A} {^{T}H}) = - Str({^{T}A}G^{2}{^{T}H}) =
  Str({^{T}A}G H G) = Str(GAHG) = - Str(AH) = -Str(HA). \nonumber
  \end{eqnarray}
 $Str(HA)$ satisfies $Str(HA) = - Str(HA)$, and thus this is shown to
 vanish.\\

  Utilizing the relationships (\ref{AZB1comha}) and
  (\ref{AZB1tracevanish}), the proof of (\ref{AZ53distinguish}) goes
  as follows:
    \begin{enumerate}
   \item{$Str(H_{1} H_{2} H_{3}) = \frac{1}{2} Str(H_{1} ( [H_{2},
       H_{3}] + \{ H_{2}, H_{3} \}) )
      = \frac{1}{2} Str( H_{1}[H_{2}, H_{3}]) + \frac{1}{2}Str(H_{1}
       {\cal A}) 
      = \frac{1}{2} Str(H_{1} [H_{2}, H_{3}])$,}
   \item{$Str(H_{1} H_{2} A_{3}) = \frac{1}{2} Str(H_{1} ( [H_{2},
       A_{3}] + \{H_{2}, A_{3}\})) 
     =  \frac{1}{2} Str(H_{1} {\cal A}) + \frac{1}{2}Str( H_{1} 
        \{ H_{2}, A_{3} \}) 
     =  \frac{1}{2} Str(H_{1} \{ H_{2}, A_{3} \})$,}
   \item{$Str(H_{1} A_{2} A_{3}) = \frac{1}{2} Str( H_{1} ( [A_{2},
       A_{3}] + \{ A_{2}, A_{3} \})) 
     = \frac{1}{2} Str( H_{1} [A_{2}, A_{3}]) + \frac{1}{2} Str(H_{1}
       {\cal A}) 
     = \frac{1}{2} Str(H_{1} [A_{2}, A_{3}])$,}
   \item{$Str(A_{1} A_{2} A_{3}) = \frac{1}{2} Str( A_{1} ( [A_{2},
       A_{3}] + \{ A_{2} , A_{3}\})) 
     = \frac{1}{2} Str(A_{1} {\cal H}) + \frac{1}{2} Str(A_{1} 
      \{ A_{2}, A_{3} \} ) 
     = \frac{1}{2} Str(A_{1} \{ A_{2}, A_{3} \} )$.}
  \end{enumerate}
 These completes the proof of the properties (\ref{AZ53distinguish}).
 (Q.E.D.) }\\

  Now let us separate the components of $gl(1|32,R)$ into the
  subalgebras ${\cal H}$ and ${\cal A}$. The element $M^{a} \in
  gl(1|32,R)$ can be separated as 
   \begin{eqnarray}
    M^{a} = H^{a} + A^{a} = 
    \left( \begin{array}{cc} m^{a}_{1} & \psi^{a}_{1} \\ i {\bar
    \psi_{1}}^{a} & 0  \end{array} \right) + 
    \left( \begin{array}{cc} m^{a}_{2} & \psi^{a}_{2} \\ -i {\bar
    \psi_{2}}^{a} & v^{a} \end{array} \right), 
   \end{eqnarray}
  where $m_{1}, m_{2}, \psi_{1}, \psi_{2}$ are defined in the previous 
  section. Then, the action is rewritten as
   \begin{eqnarray}
     I &=& \frac{1}{g^{2}} Str(M^{a} M^{b} M^{c}) Tr(T^{a} T^{b}
     T^{c})
        =  \frac{1}{2 g^{2}} Str((H^{a} + A^{a}) (H^{b} + A^{b}) (H^{c}
     + A^{c})) ( Tr(T^{a}[T^{b}, T^{c}]) + Tr(T^{a} \{ T^{b}, T^{c}
     \})) \nonumber \\
   &\stackrel{(\ref{AZ53distinguish})}{=}& 
       \frac{1}{4g^{2}} ( Str(H^{a}[H^{b}, H^{c}]) + Str(H^{a}[A^{b},
     A^{c}])) Tr(T^{a} [T^{b}, T^{c}])  \nonumber \\
   &+&  \frac{1}{4g^{2}} ( Str(H^{a} \{ H^{b}, A^{c} \} ) + Str(A^{a}
     \{ A^{b}, A^{c} \}) ) Tr(T^{a} \{ T^{b}, T^{c} \}) \nonumber \\
   &\stackrel{(\ref{AZ52structureconst})}{=}&
      \frac{f_{abc}}{4g^{2}} ( Str(H^{a} [H^{b}, H^{c}]) + Str(H^{a},
     [ A^{b}, A^{c}] )) 
    + \frac{d_{abc}}{4g^{2}} ( Str(H^{a} \{ H^{b}, A^{c} \}) + Str
     ( A^{a} \{ A^{b}, A^{c} \} ) )  \label{AZ53disaction1}
     \\ 
   &=&  \frac{f_{abc}}{2g^{2}} ( Str(H^{a} H^{b} H^{c}) + Str(H^{a}
     A^{b} A^{c}) ) 
    + \frac{d_{abc}}{2g^{2}} ( Str(H^{a} H^{b} A^{c}) + Str
     ( A^{a} A^{b} A^{c}) ).  \label{AZ53disaction2}
   \end{eqnarray}

 We compute the trace $tr(m^{a} m^{b} m^{c} )$ according to the same
 principle as in 
 $osp(1|32,R)$ bosonic part. Since the trace of the gamma matrices is
  \begin{eqnarray}
    tr(\Gamma^{\mu}) = tr(\Gamma^{\mu_{1} \mu_{2}}) =
    tr(\Gamma^{\mu_{1} \mu_{2} \mu_{3}}) = 
    tr(\Gamma^{\mu_{1} \cdots \mu_{4}} ) = 
    tr(\Gamma^{\mu_{1} \cdots \mu_{5}} ) = 0,
  \end{eqnarray}
 the goal of the computation of $tr(mmm)$ is to extract the
 coefficient of ${\bf 1}_{32 \times 32}$, and we utilize the formula
 (\ref{AZproduct}) for the computation of the gamma matrices. 
  We separate the contribution of the bosonic part $tr(mmm)$ into the 
  following four parts, and we inherit the abbreviation $(xyz)$ from
  the argument of $osp(1|32,R)$:
  \begin{eqnarray}
   (xyz) \stackrel{def}{=} \textrm{the terms emerging from }
   \frac{1}{x!y!z!} u^{a}_{\mu_{1} \cdots \mu_{x}} u^{b}_{\mu_{1}
   \cdots \mu_{y}} u^{c}_{\mu_{1} \cdots \mu_{z}} tr(\Gamma^{\mu_{1}
   \cdots \mu_{x}} \Gamma^{\mu_{1} \cdots \mu_{y}} \Gamma^{\mu_{1}
   \cdots \mu_{z}} ),
  \end{eqnarray}
 where now $x,y,z=0, 1, \cdots, 5$. The technique of the following
 computation is totally the same as in $osp(1|32,R)$ cubic model, as
 was discussed in Appendix. \ref{AZCospactionres}

 \subsubsection{(F1) terms}
  We nane the terms emerging from $\frac{f_{abc}}{2g^{2}} Str(H^{a}
  H^{b} H^{c})$ (F1) terms. We utilize the color indices notation
  for simplicity, and switch to the large $N$ matrix representation. 
  These terms are attached to the commutator of the basis of
  $gl(N,R)$, and therefore anti-symmetric with respect to the exchange
  of the color indices $a \leftrightarrow b$. As we have seen, the
  bosonic part $m_{1}$ is 
   \begin{eqnarray}
    m^{a} = u^{a}_{\mu_{1}} \Gamma^{\mu_{1}} + \frac{1}{2!}
    u^{b}_{\mu_{1} \mu_{2}} \Gamma^{\mu_{1} \mu_{2} }
    + \frac{1}{5!} u^{a}_{\mu_{1} \cdots \mu_{5}} \Gamma^{\mu_{1}
    \cdots \mu_{5}}.  
   \end{eqnarray}
  Therefore, the nonvanishing trace is 
 \begin{eqnarray}
 (112), (155), (222), (255), (555),
 \end{eqnarray}
  whose computation we do not repeat because we are finished
  with these computations in the investigation of $osp(1|32,R)$. 

 \subsubsection{(F2) terms}
  We refer to the terms $\frac{f_{abc}}{2g^{2}} Str(H^{a} A^{b}
  A^{c})$ as (F2) terms. The nonvanishing trace comes from
 \begin{eqnarray}
  (134), (233), (244), (335), (345), (445).
 \end{eqnarray}
 The fundamental techniques and cautions are already introduced in the 
 analysis of $osp(1|32,R)$, and we proceed rather quickly.

  \paragraph{(134) terms}
  There are 6 ways as to where we should pick up which terms, and  the 
  contribution is
   \begin{eqnarray}
   (134) = \frac{6}{3! 4!} u^{a}_{\mu_{1}} u^{b}_{\nu_{1} \nu_{2}
   \nu_{3}} u^{c}_{\rho_{1} \cdots \rho_{4}} tr(\Gamma^{\mu_{1}}
   \Gamma^{\nu_{1} \nu_{2} \nu_{3}} \Gamma^{\rho_{1} \cdots \rho_{4}}).
   \end{eqnarray}
 On the other hand, the trace of the gamma matrices are computed to be
  \begin{eqnarray}
   tr(\Gamma^{\mu_{1}} \Gamma^{\nu_{1} \nu_{2} \nu_{3}}
   \Gamma^{\rho_{1} \cdots \rho_{4}}) =
   tr(\Gamma^{\mu_{1} \nu_{1} \nu_{2} \nu_{3}} \Gamma^{\rho_{1} \cdots
   \rho_{4}}) = 24 \eta^{\mu_{1} \rho_{1}} \eta^{\nu_{1} \rho_{2}}
   \eta^{\nu_{2} \rho_{3}} \eta^{\nu_{3} \rho_{4}}. 
  \end{eqnarray}
 Therefore, the contribution is 
  \begin{eqnarray}
    (134) = 32 u^{a}_{\mu_{1}} u^{b}_{\nu_{1} \nu_{2}  \nu_{3}}
   u^{c \mu_{1} \nu_{1} \nu_{2}  \nu_{3}}.
  \end{eqnarray}

 \paragraph{(233) terms} 
   The contribution of (233) term is 
  \begin{eqnarray} \frac{3}{2!(3!)^{2}}  tr( u^{a}_{\mu_{1} \mu_{2}}
    u^{b}_{\nu_{1} \cdots \nu_{3}} u^{c}_{\rho_{1}  \cdots \rho_{3}}
    f_{abc}  \Gamma^{\mu_{1} \mu_{2}} \Gamma^{\nu_{1} \cdots  \nu_{3}} \Gamma^
    { \rho_{1} \cdots \rho_{3} } ).    
   \end{eqnarray}
  Again, using the formula of the gamma matrices,
  \begin{eqnarray} 
    \Gamma^{\mu_{1} \mu_{2}} \Gamma^{\nu_{1} \cdots  \nu_{3}} \Gamma^
    { \rho_{1} \cdots \rho_{3} } = -6 \eta^{\mu_{1} \nu_{1}}
    \Gamma^{\mu_{2} \nu_{2} \nu_{3}} \Gamma^{ \rho_{1} \cdots \rho_{3}
    } = 36  \eta^{\mu_{1} \nu_{1}}  \eta^{\mu_{2} \rho_{1}}
    \eta^{\nu_{2} \rho_{2}}  \eta^{\nu_{3} \rho_{3}}.
  \end{eqnarray}
  Then, the final result is 
   \begin{eqnarray}
    (233) = 48  u^{a}_{\mu_{1} \mu_{2}} {u^{b \mu_{1}}}_{\nu_{2} \nu_{3}}
    u^{c \mu_{2} \nu_{2} \nu_{3}}.
   \end{eqnarray} 

 \paragraph{(244) terms}
    Again, there are 3 choices with respect to where we should pick up
  $u_{\mu_{1} \mu_{2}}$ from $m^{a}, m^{b} , m^{c}$. Then, the
  contribution is 
    \begin{eqnarray} \frac{3}{2!(4!)^{2}}  tr( u^{a}_{\mu_{1} \mu_{2}}
    u^{b}_{\nu_{1} \cdots \nu_{4}} u^{c}_{\rho_{1}  \cdots \rho_{4}}
    f_{abc}  \Gamma^{\mu_{1} \mu_{2}} \Gamma^{\nu_{1} \cdots  \nu_{4}} \Gamma^
    { \rho_{1} \cdots \rho_{4} } ).    
   \end{eqnarray}
  The gamma matrices are computed to be
    \begin{eqnarray} 
    \Gamma^{\mu_{1} \mu_{2}} \Gamma^{\nu_{1} \cdots  \nu_{4}} \Gamma^
    { \rho_{1} \cdots \rho_{4} } = -8 \eta^{\mu_{1} \nu_{1}}
    \Gamma^{\mu_{2} \nu_{2} \nu_{3} \nu_{4} } \Gamma^{ \rho_{1} \cdots
    \rho_{4} } = -192  \eta^{\mu_{1} \nu_{1}}  \eta^{\mu_{2} \rho_{1}}
    \eta^{\nu_{2} \rho_{2}}  \eta^{\nu_{3} \rho_{3}}  \eta^{\nu_{4}
    \rho_{4}}.  
  \end{eqnarray}
  Then we obtain,
   \begin{eqnarray}
    (244) = -16 u^{a}_{\mu_{1} \mu_{2}} {u^{b \mu_{1}}}_{\nu_{2} \nu_{3}
    \nu_{4}}  u^{c \mu_{2} \nu_{2} \nu_{3} \nu_{4}}. 
   \end{eqnarray}   

 \paragraph{(335) terms}
   The contribution of this term is 
     \begin{eqnarray} \frac{3}{(3!)^{2}5!}  tr( u^{a}_{\mu_{1} \cdots
         \mu_{3}}  u^{b}_{\nu_{1} \cdots \nu_{3}} u^{c}_{\rho_{1}
    \cdots \rho_{5}}  
    f_{abc}  \Gamma^{\mu_{1} \cdots \mu_{3}} \Gamma^{\nu_{1} \cdots
    \nu_{3}} \Gamma^{ \rho_{1} \cdots \rho_{5} } ).
   \end{eqnarray}
  The gamma matrices are computed to be 
   \begin{eqnarray}
  & &   \Gamma^{\mu_{1} \cdots \mu_{3}} \Gamma^{\nu_{1} \cdots
    \nu_{3}} \Gamma^{ \rho_{1} \cdots \rho_{5} } = \Gamma^{\mu_{1}
    \cdots \mu_{3} \nu_{1} \cdots \nu_{3}} \Gamma^{ \rho_{1}
    \cdots \rho_{5} }  = -\frac{1}{5!} \epsilon^{\mu_{1}
    \cdots \mu_{3} \nu_{1} \cdots \nu_{3} \chi_{1} \cdots \chi_{5} }
    \Gamma_{\chi^{1} \cdots \chi^{5}} \Gamma^{ \rho_{1} \cdots
    \rho_{5} } \nonumber \\ 
  &=&  - \epsilon^{\mu_{1} \cdots \mu_{3} \nu_{1} \cdots
    \nu_{3} \rho_{1} \cdots \rho_{5}}. \end{eqnarray}
  Therefore, the final result is 
    \begin{eqnarray}
    (335) = - \frac{1}{45} (u^{a}_{\mu_{1} \cdots
         \mu_{3}}  u^{b}_{\nu_{1} \cdots \nu_{3}} u^{c}_{\rho_{1}
    \cdots \rho_{5}}  \epsilon^{\mu_{1} \cdots \mu_{3} \nu_{1} \cdots
    \nu_{3} \rho_{1} \cdots \rho_{5}} ). 
   \end{eqnarray} 

 \paragraph{(345) terms}
   The contribution to this term is 
      \begin{eqnarray}
   \frac{6}{3!4!5!} u^{a}_{\mu_{1} \cdots \mu_{3} } u^{b}_{\nu_{1}
    \cdots  \nu_{4}} u^{c}_{\rho_{1} \cdots  \rho_{5}} f_{abc}  
    tr ( \Gamma^{\mu_{1} \cdots \mu_{3}} \Gamma^{\nu_{1} \cdots \nu_{4}}
    \Gamma^{\rho_{1} \cdots  \rho_{5}} ).
  \end{eqnarray} 
  On the other hand, the gamma matrices are evaluated to be
    \begin{eqnarray} 
    \Gamma^{\mu_{1} \cdots \mu_{3}} \Gamma^{\nu_{1} \cdots \nu_{4}}
    \Gamma^{\rho_{1} \cdots  \rho_{5}}  =
    12 \eta^{\mu_{1} \nu_{1}} \Gamma^{\mu_{2} \mu_{3} \nu_{2} \nu_{3} 
    \nu_{4} } \Gamma^{\rho_{1} \cdots \rho_{5}}   =
    1440  \eta^{\mu_{1} \nu_{1}}  \eta^{\mu_{2} \rho_{1}}
    \eta^{\mu_{3} \rho_{2}}   \eta^{\nu_{2} \rho_{3}} \eta^{\nu_{3}
    \rho_{4} }  \eta^{\nu_{4} \rho_{5}}.
    \end{eqnarray}  
  Therefore, the final result is 
  \begin{eqnarray}
   (345) = 16 u^{a}_{\mu_{1} \cdots \mu_{3} } {u^{b \mu_{1}}}_{\nu_{2} 
        \nu_{3}  \nu_{4}} u^{c \mu_{2} \mu_{3} \nu_{2} \nu_{3} \nu_{4}}. 
  \end{eqnarray} 

 \paragraph{(445) terms}
   The contribution is expressed as   
    \begin{eqnarray}
      \frac{3}{(4!)^{2}5!} tr(u^{a}_{\mu_{1} \cdots \mu_{4} }
    u^{b}_{\nu_{1} \cdots \nu_{4}} u^{c}_{\rho_{1} \cdots
    \rho_{5}} f_{abc}  ( \Gamma^{\mu_{1} \cdots \mu_{4}}
    \Gamma^{\nu_{1} \cdots \nu_{4}} \Gamma^{\rho_{1} 
    \cdots  \rho_{5}} ) ).
  \end{eqnarray}
  The multiplication of the gamma matrices is 
   \begin{eqnarray}
      \Gamma^{\mu_{1} \cdots \mu_{4}} \Gamma^{\nu_{1} \cdots \nu_{4}}
      \Gamma^{\rho_{1} \cdots  \rho_{5}} = \frac{16}{5!}
      \eta^{\mu_{1} \nu_{1}} \epsilon^{\mu_{2} \mu_{3} \mu_{4} \nu_{2} 
      \nu_{3} \nu_{4} \chi_{1} \cdots \chi_{5}} \Gamma_{\chi_{1}
      \cdots \chi_{5}}  \Gamma^{\rho_{1} \cdots  \rho_{5}} = 16
      \eta^{\mu_{1} \nu_{1}} \epsilon^{\mu_{2} \mu_{3} \mu_{4} \nu_{2} 
      \nu_{3} \nu_{4} \rho_{1} \cdots  \rho_{5} }. 
   \end{eqnarray}
  Therefore, the contribution is seen to be 
  \begin{eqnarray}
    \frac{1}{45} u^{a}_{\mu_{1} \cdots \mu_{4} } {u^{b
  \mu_{1}}}_{\nu_{2} \nu_{3}  \nu_{4}} u^{c}_{\rho_{1} \cdots
  \rho_{5}}  \epsilon^{\mu_{2} \mu_{3} \mu_{4} \nu_{2}  \nu_{3}
  \nu_{4} \rho_{1} \cdots  \rho_{5} }.
  \end{eqnarray}

 \subsubsection{(D1) terms}
 We next investigate the terms coming from the contribution
 $\frac{d_{abc}}{2g^{2}} Str( A^{a} A^{b} A^{c}) $. The difference
 from the previous two cases is that these terms are attached to the
 anti-commutator of the gamma matrices, so that the result is {\it
 symmetric} with respect to the exchange of the color indices $a
 \leftrightarrow b$. This term is named '(D1) term' after the
 structure constant $d_{abc}$. The nonvanishing trace of these this
 term stems from the following terms:
  \begin{eqnarray}
   (000), (033), (044), (334), (344), (444).
  \end{eqnarray}

 \paragraph{(000) terms}
 The computation of this contribution is trivial:
  \begin{eqnarray} (000) = u^{a} u^{b} u^{c} tr( {\bf 1}^{3}) = 32
    u^{a} u^{b} u^{c}.
 \end{eqnarray}

 \paragraph{(033) terms}
  This is also trivial, paying attention to where we should pick up
  the $u {\bf 1}$:
  \begin{eqnarray}
   (033) &=& \frac{3}{(3!)^{2}} u^{a} u^{b}_{\nu_{1} \nu_{2} \nu_{3}}
   u^{c}_{\rho_{1} \rho_{2} \rho_{3}} tr( \Gamma^{\nu_{1} \nu_{2}
   \nu_{3}} \Gamma^{\rho_{1} \rho_{2} \rho_{3}} ) \nonumber \\
   &=& \frac{3}{(3!)^{2}} u^{a} u^{b}_{\nu_{1} \nu_{2} \nu_{3}}
   u^{c}_{\rho_{1} \rho_{2} \rho_{3}} (-32 \times 6 \eta^{\mu_{1}
   \nu_{1}} \eta^{\mu_{2} \nu_{2}} \eta^{\mu_{3} \nu_{3}} ) = -16
   u^{a} u^{b}_{\mu_{1} \mu_{2} \mu_{3}} u^{c \mu_{1} \mu_{2} \mu_{3}}. 
  \end{eqnarray}

 \paragraph{(044) terms} Again, we can compute the contribution with
   ease:
   \begin{eqnarray}
    (044) &=& \frac{3}{(4!)^{2}} u^{a} u^{b}_{\nu_{1} \cdots \nu_{4}}
    u^{c}_{\rho_{1} \cdots \rho_{4}} tr( \Gamma^{\nu_{1} \cdots
    \nu_{4}} \Gamma^{\rho_{1} \cdots \rho_{4}} ) \nonumber \\
     &=&  \frac{3}{(4!)^{2}} u^{a} u^{b}_{\nu_{1} \cdots \nu_{4}}
    u^{c}_{\rho_{1} \cdots \rho_{4}} ( 32 \times 24 \eta^{\mu_{1}
    \nu_{1}} \cdots \eta^{\mu_{4} \nu_{4}} ) = 4 u^{a} u^{b}_{\nu_{1}
    \cdots \nu_{4}} u^{c \nu_{1} \cdots \nu_{4}}. 
   \end{eqnarray}
 
 \paragraph{(334) terms}
  This term can be computed in the similar fashion to the previous
  cases. The contribution is
   \begin{eqnarray}
     \frac{3}{(3!)^{2} 4!} u^{a}_{\mu_{1} \mu_{2} \mu_{3}}
     u^{b}_{\nu_{1} \nu_{2} \nu_{3}} u^{c}_{\rho_{1} \cdots \rho_{4}}
     tr( \Gamma^{\mu_{1} \mu_{2} \mu_{3}} \Gamma^{\nu_{1} \nu_{2}
     \nu_{3}} \Gamma^{\rho_{1} \cdots \rho_{4}} ). 
   \end{eqnarray}
  The trace of the gamma matrices are now
   \begin{eqnarray}
    tr( \Gamma^{\mu_{1} \mu_{2} \mu_{3}} \Gamma^{\nu_{1} \nu_{2}
     \nu_{3}} \Gamma^{\rho_{1} \cdots \rho_{4}} ) =
     9 \eta^{\mu_{1} \nu_{1}} tr( \Gamma^{\mu_{2} \mu_{3} \nu_{2}
     \nu_{3}} \Gamma^{\rho_{1} \cdots \rho_{4}} ) 
   = 216 \times 32 \eta^{\mu_{1} \nu_{1}} \eta^{\mu_{2} \rho_{1}}
     \eta^{\mu_{3} \rho_{2}} \eta^{\nu_{2} \rho_{3}} \eta^{\nu_{3}
     \rho_{4}}. 
   \end{eqnarray}
  The contribution is now obtained by
  \begin{eqnarray}
   (334) &=& 24 u^{a}_{\mu_{1} \mu_{2} \mu_{3}} {u^{b
   \mu_{1}}}_{\nu_{2} \nu_{3}} u^{c \mu_{2} \mu_{3} \nu_{2} \nu_{3}}. 
  \end{eqnarray}

 \paragraph{(344) terms}
  We can pick up the following terms 
   \begin{eqnarray}
    \frac{3}{(3!)(4!)^{2}} u^{a}_{\mu_{1} \mu_{2} \mu_{3}}
    u^{b}_{\nu_{1} \cdots \nu_{4}} u^{c}_{\rho_{1} \cdots \rho_{4}}
    tr( \Gamma^{\mu_{1} \mu_{2} \mu_{3}} \Gamma^{\nu_{1} \cdots
    \nu_{4}} \Gamma^{\rho_{1} \cdots \rho_{4}} ).  
   \end{eqnarray}
  The trace of the gamma matrices are now
   \begin{eqnarray}
    tr( \Gamma^{\mu_{1} \mu_{2} \mu_{3}} \Gamma^{\nu_{1} \cdots
    \nu_{4}} \Gamma^{\rho_{1} \cdots \rho_{4}} ) =
    tr( \Gamma^{\mu_{1} \mu_{2} \mu_{3} \nu_{1} \cdots
    \nu_{4}} \Gamma^{\rho_{1} \cdots \rho_{4}} ) 
   = - \epsilon^{ \mu_{1} \mu_{2} \mu_{3} \nu_{1} \cdots
    \nu_{4} \rho_{1} \cdots \rho_{4}}.
   \end{eqnarray}
  Therefore, we obtain the following contribution
   \begin{eqnarray}
    (344) &=& - \frac{1}{36} u^{a}_{\mu_{1} \mu_{2} \mu_{3}}
    u^{b}_{\nu_{1} \cdots \nu_{4}} u^{c}_{\rho_{1} \cdots \rho_{4}}
    \epsilon^{ \mu_{1} \mu_{2} \mu_{3} \nu_{1} \cdots \nu_{4} \rho_{1}
    \cdots \rho_{4} }.
   \end{eqnarray}

 \paragraph{(444) terms}
  Here we do not have to multiply 3, because there is only one way to
  pick up the (444) term from $m^{a} m^{b} m^{c}$ 
   \begin{eqnarray}
    \frac{1}{(4!)^{3}} u^{a}_{\mu_{1} \cdots \mu_{4}} u^{b}_{\nu_{1}
    \cdots \nu_{4}} u^{c}_{\rho_{1} \cdots \rho_{4}} 
    tr( \Gamma^{\mu_{1} \cdots \mu_{4}} \Gamma^{\nu_{1} \cdots
    \nu_{4}} \Gamma^{\rho_{1} \cdots \rho_{4}} ).
   \end{eqnarray} 
  On the other hand, the nonvanishing trace of the gamma matrices are 
 \begin{eqnarray}
  & &  tr( \Gamma^{\mu_{1} \cdots \mu_{4}} \Gamma^{\nu_{1} \cdots
    \nu_{4}} \Gamma^{\rho_{1} \cdots \rho_{4}} ) =
    - 72 \eta^{\mu_{1} \nu_{1}} \eta^{\mu_{2}
    \nu_{2}} tr( \Gamma^{\mu_{3} \mu_{4} \nu_{3} \nu_{4}}
    \Gamma^{\rho_{1} \cdots \rho_{4}} ) \nonumber \\ 
 &=& - 72 \times 24 \times 32 \eta^{\mu_{1} \nu_{1}} \eta^{\mu_{2}
    \nu_{2}} \eta^{\mu_{3} \rho_{1}} \eta^{\mu_{4} \rho_{2}}
    \eta^{\nu_{3} \rho_{3}} \eta^{\nu_{4} \rho_{4}}.
 \end{eqnarray}
  The contribution of this term is now obtained by
  \begin{eqnarray}
  (444) &=& -4 u^{a}_{\mu_{1} \cdots \mu_{4}} {u^{b \mu_{1}
  \mu_{2}}}_{\nu_{3} \nu_{4}} u^{c \mu_{3} \mu_{4} \nu_{3} \nu_{4}}.
  \end{eqnarray}

 \subsubsection{(D2) terms}
  This is a term stemming from the contribution of
  $\frac{d_{abc}}{2g^{2}} Str( H^{a} H^{b} A^{c})$, which are also
  attached to the anti-commutator of the generators. This term also
  serves as an anti-commutator of the large $N$ field matrices. The
  nonvanishing terms are now
   \begin{eqnarray}
    (011), (022), (055), (123), (145), (224), (235), (245), (355),
    (455). 
   \end{eqnarray}
  which can be known from the table in Appendix. \ref{AZCproductgamma}. 
  Two of the terms are picked up from $m_{1}$, while the rest of one
  term stems from $m_{2}$. 
 
  \paragraph{(011), (022), (055) terms}
    The computation of these terms is trivial. There are 3 ways as to 
    where we should pick up $u^{a}$ from $tr(m^{a} m^{b} m^{c})$. 
     \begin{eqnarray}
     (011) &=& 3 u^{a} u^{b}_{\nu_{1}} u^{c}_{\rho_{1}}
     tr(\Gamma^{\nu_{1}} \Gamma^{\rho_{1}} ) = 96  u^{a}
     u^{b}_{\nu_{1}} u^{c \nu_{1}}, \\
     (022) &=& \frac{3}{(2!)^{2}} u^{a} u^{b}_{\nu_{1} \nu_{2}}
     u^{c}_{\rho_{1} \rho_{2}} tr( \Gamma^{\nu_{1} \nu_{2}}
     \Gamma^{\rho_{1} \rho_{2}} ) = -48 u^{a} u^{b}_{\nu_{1} \nu_{2}}
     u^{c \nu_{1} \nu_{2}}, \\
     (055) &=& \frac{3}{(5!)^{2}} u^{a} u^{b}_{\nu_{1} \cdots \nu_{5}} 
     u^{c}_{\rho_{1} \cdots \rho_{5}} tr( \Gamma^{\nu_{1} \cdots
     \nu_{5}} \Gamma^{\rho_{1} \cdots \rho_{5}} ) = \frac{4}{5} u^{a}
     u^{b}_{\nu_{1} \cdots \nu_{5}} u^{c \nu_{1} \cdots \nu_{5}}.
     \end{eqnarray}

  \paragraph{(123), (145) terms}
  There are $3! = 6$ ways as to the way to pick up the terms. And the
  nonvanishing traces are computed respectively as
   \begin{eqnarray}
    (123) &=& \frac{6}{(2!)(3!)} u^{a}_{\mu_{1}} u^{b}_{\nu_{1}
    \nu_{2}} u^{c}_{\rho_{1} \rho_{2} \rho_{3}} tr(\Gamma^{\mu_{1}}
    \Gamma^{\nu_{1} \nu_{2}} \Gamma^{\rho_{1} \rho_{2} \rho_{3}}) = -
    \frac{6 \times 6}{(2!)(3!)}  u^{a}_{\mu_{1}} u^{b}_{\nu_{1}
    \nu_{2}} u^{c}_{\rho_{1} \rho_{2} \rho_{3}} tr(\eta^{\mu_{1}
    \rho_{1}} \eta^{\nu_{1} \rho_{2}} \eta^{\nu_{2} \rho_{3}}) 
    \nonumber \\ 
    &=& -96 u^{a}_{\mu_{1}} u^{b}_{\nu_{1} \nu_{2}} u^{c \mu_{1}
    \nu_{1} \nu_{2}}, \\ 
    (145) &=& \frac{6}{(4!)(5!)} u^{a}_{\mu_{1}} u^{b}_{\nu_{1} \cdots 
    \nu_{4}} u^{c}_{\rho_{1} \cdots \rho_{5}} tr( \Gamma^{\mu_{1}}
    \Gamma^{\nu_{1} \cdots \nu_{4}} \Gamma^{\rho_{1} \cdots \rho_{5}}
    ) = \frac{6 \times 120}{(4!)(5!)} u^{a}_{\mu_{1}} u^{b}_{\nu_{1}
    \cdots \nu_{4}} u^{c}_{\rho_{1} \cdots \rho_{5}} tr( \eta^{\mu_{1} 
    \rho_{1}} \eta^{\nu_{1} \rho_{2}} \cdots \eta^{\nu_{4} \rho_{5}} ) 
   \nonumber \\ 
   &=& 8 u^{a}_{\mu_{1}} u^{b}_{\nu_{1} \cdots \nu_{4}} u^{c \mu_{1}
    \nu_{1} \cdots \nu_{4}}. 
   \end{eqnarray}

  \paragraph{(224) terms}
  The contribution of this term to the non-gauged action is
   \begin{eqnarray}
    \frac{3}{(2!)^{2} 4!} u^{a}_{\mu_{1} \mu_{2}} u^{b}_{\nu_{1}
    \nu_{2}} u^{c}_{\rho_{1} \cdots \rho_{4}} tr( \Gamma^{\mu_{1}
    \mu_{2}} \Gamma^{\nu_{1} \nu_{2}} \Gamma^{\rho_{1} \cdots
    \rho_{4}}),
   \end{eqnarray}
  where the trace of the gamma matrices is 
   \begin{eqnarray}
    tr( \Gamma^{\mu_{1} \mu_{2}} \Gamma^{\nu_{1} \nu_{2}}
    \Gamma^{\rho_{1} \cdots \rho_{4}}) = tr(\Gamma^{ \mu_{1} \mu_{2}
    \nu_{1} \nu_{2}} \Gamma^{\rho_{1} \cdots \rho_{4}}) = 
    24 \times 32 \eta^{\mu_{1} \rho_{1}} \eta^{\mu_{2} \rho_{2}} \eta^{\nu_{1}
    \rho_{3}} \eta^{\nu_{2} \rho_{4}}.
   \end{eqnarray}
  Therefore, this term gives the contribution
   \begin{eqnarray}
   (224) =  24 u^{a}_{\mu_{1} \mu_{2}} u^{b}_{\nu_{1} \nu_{2}} u^{c \mu_{1}
    \mu_{2} \nu_{1} \nu_{2}}.
   \end{eqnarray}

  \paragraph{(235) terms}
   We repeat the same procedure as has been performed.
   \begin{eqnarray}
    (235) &=& \frac{6}{(2!)(3!)(5!)} u^{a}_{\mu_{1} \mu_{2}}
    u^{b}_{\nu_{1} \nu_{2} \nu_{3}} u^{c}_{\rho_{1} \cdots \rho_{5}}
    tr( \Gamma^{\mu_{1} \mu_{2}} \Gamma^{\nu_{1} \nu_{2} \nu_{3}}
    \Gamma^{\rho_{1} \cdots \rho_{5}} ) \nonumber \\
    &=& \frac{6 \times 120 \times
    32}{(2!)(3!)(5!)}  u^{a}_{\mu_{1} \mu_{2}}  u^{b}_{\nu_{1} \nu_{2}
    \nu_{3}} u^{c}_{\rho_{1} \cdots \rho_{5}} \eta^{\mu_{1} \rho_{1}}
    \eta^{\mu_{2} \rho_{2}} \eta^{\nu_{1} \rho_{3}} \eta^{\nu_{2}
    \rho_{4}} \eta^{\nu_{3} \rho_{5}} 
    = 16  u^{a}_{\mu_{1} \mu_{2}}  u^{b}_{\nu_{1} \nu_{2}
    \nu_{3}} u^{c \mu_{1} \mu_{2} \nu_{1} \nu_{2} \nu_{3}}.
   \end{eqnarray} 

  \paragraph{(245) terms}
  \begin{eqnarray}
   (245) = \frac{6}{(2!)(4!)(5!)} u^{a}_{\mu_{1} \mu_{2}}
   u^{b}_{\nu_{1} \cdots \nu_{4}} u^{c}_{\rho_{1} \cdots \rho_{5}} 
   tr(\Gamma^{\mu_{1} \mu_{2}} \Gamma^{\nu_{1} \cdots \nu_{4}}
   \Gamma^{\rho_{1} \cdots \rho_{5}} ) = - \frac{1}{30} u^{a}_{\mu_{1}
   \mu_{2}}  u^{b}_{\nu_{1} \cdots \nu_{4}} u^{c}_{\rho_{1} \cdots
   \rho_{5}} \epsilon^{\mu_{1} \mu_{2} \nu_{1} \cdots \nu_{4} \rho_{1} \cdots
   \rho_{5}}.
  \end{eqnarray}
   
  \paragraph{(355) terms}
   This term gives the following contribution to the action:
   \begin{eqnarray}
    \frac{3}{(3!)(5!)^{2}} u^{a}_{\mu_{1} \mu_{2} \mu_{3}}
    u^{b}_{\nu_{1} \cdots \nu_{5}} u^{c}_{\rho_{1} \cdots \rho_{5}}
    tr( \Gamma^{\mu_{1} \mu_{2} \mu_{3}} \Gamma^{\nu_{1} \cdots
    nu_{5}} \Gamma^{\rho_{1} \cdots \rho_{5}}).
   \end{eqnarray}
  Computing the trace of the gamma matrices, we obtain
  \begin{eqnarray}
    tr( \Gamma^{\mu_{1} \mu_{2} \mu_{3}} \Gamma^{\nu_{1} \cdots
    nu_{5}} \Gamma^{\rho_{1} \cdots \rho_{5}}) = 25 \eta^{\nu_{1}
    \rho_{1}} tr( \Gamma^{\mu_{1} \mu_{2} \mu_{3} \nu_{2} \cdots \nu_{5}}
    \Gamma^{\rho_{2} \cdots \rho_{5}} ) = -25 \times 32 \eta^{\nu_{1}
    \rho_{1}} \epsilon^{\mu_{1} \mu_{2} \mu_{3} \nu_{2} \cdots \nu_{5}
    \rho_{2} \cdots \rho_{5} }.
  \end{eqnarray}
    Therefore, the final result is
   \begin{eqnarray}
   (355) &=&  - \frac{1}{36}  u^{a}_{\mu_{1} \mu_{2} \mu_{3}}
   u^{b}_{\nu_{1} \cdots \nu_{5}} {u^{c \nu_{1}}}_{\rho_{2} \cdots
   \rho_{5}} \epsilon^{\mu_{1} \mu_{2} \mu_{3} \nu_{2} \cdots \nu_{5}
    \rho_{2} \cdots \rho_{5} }.
   \end{eqnarray}

  \paragraph{(455) terms}
  This term gives the contribution to the action
   \begin{eqnarray}
    \frac{3}{(4!)(5!)^{2}} u^{a}_{\mu_{1} \cdots \mu_{4}}
    u^{b}_{\nu_{1} \cdots \nu_{5}} u^{c}_{\rho_{1} \cdots \rho_{5}}
    tr( \Gamma^{\mu_{1} \cdots \mu_{4}} \Gamma^{\nu_{1} \cdots
    \nu_{5}} \Gamma^{\rho_{1} \cdots \rho_{5}}).
   \end{eqnarray}
  We compute the gamma matrices again to obtain
   \begin{eqnarray}
   & &  tr( \Gamma^{\mu_{1} \cdots \mu_{4}} \Gamma^{\nu_{1} \cdots
    \nu_{5}} \Gamma^{\rho_{1} \cdots \rho_{5}}) = - 6 \times 10 \times 
    (2!) \eta^{\mu_{1} \nu_{1}} \eta^{\mu_{2} \nu_{2}}
    tr( \Gamma^{\mu_{3} \mu_{4} \nu_{3} \nu_{4} \nu_{5}} \Gamma^{\rho_{1}
    \cdots \rho_{5}} ) \nonumber \\
  &=& -120 \times 120 \times 32    \eta^{\mu_{1} \nu_{1}}
    \eta^{\mu_{2} \nu_{2}} \eta^{\mu_{3} \rho_{1}} \eta^{\mu_{4}
    \rho_{2}} \eta^{\nu_{3} \rho_{3}} \eta^{\nu_{4} \rho_{4}}
    \eta^{\nu_{5} \rho_{5}}.
   \end{eqnarray}
  Therefore, the result is
   \begin{eqnarray}
    (455) &=& -4 u^{a}_{\mu_{1} \cdots \mu_{4}} {u^{b \mu_{1}
    \mu_{2}}}_{\nu_{3} \nu_{4} \nu_{5}} u^{c \mu_{3} \mu_{4} \nu_{3}
    \nu_{4} \nu_{5}}. 
   \end{eqnarray}
  
 This completes the lengthy computation of the bosonic part $tr(m^{a}
 m^{b} m^{c})$. The next job is to express these terms utilizing the
 fields with the $x^{10} = x^{\sharp}$ direction specified. 
   Expressing this action in terms of the variables defined in
  (\ref{AZ43fieldredef}), the bosonic part $I_{b}$ and the fermionic
  part $I_{f}$ is rewritten as follows.
   \begin{eqnarray}
     I = \frac{1}{g^{2}} Tr_{N \times N} ( I_{b} + I_{f} ), \textrm{ where}
   \end{eqnarray}
   \begin{eqnarray}
    I_{b} &=& (F1) + (F2) + (D1) + (D2) + (vvv) \nonumber \\
  &=& v^{3} + 32 Z^{3} + 96 ZW^{2} + 48 Z \{  A^{(+) i_{1}} ,
   A^{(-)}_{i_{1}} \} -24 Z \{ C_{i_{1} i_{2}} , C^{i_{1} i_{2}} \} -
   24 Z \{ D_{i_{1} i_{2}} , D^{i_{1} i_{2}} \} \nonumber \\
  &-& 8Z \{ E^{(+)}_{i_{1} i_{2} i_{3}} , E^{(-) i_{1} i_{2} i_{3}} \} 
   + 2Z \{ G_{i_{1} \cdots i_{4}} , G^{i_{1} \cdots i_{4}} \} 
   + 2Z \{ H_{i_{1} \cdots i_{4}} , H^{i_{1} \cdots i_{4}} \}
   + \frac{2}{5} Z \{ I^{(+)}_{i_{1} \cdots i_{5}} , I^{(-) i_{1}
   \cdots i_{5}} \} \nonumber \\
   &-& 48 W [ A^{(+)}_{i}, A^{(-)i} ]
    -  48 W \{ C_{ij} , D^{ij} \}
    +   8 W [ E^{(+)}_{i_{1} i_{2} i_{3}} , E^{(-) i_{1} i_{2} i_{3}}]
    +   4 W \{ G_{i_{1} \cdots i_{4}} , H^{i_{1} \cdots i_{4}} \}
    - \frac{4}{5} W [ I^{(+)}_{i_{1} \cdots i_{5}} , I^{(-)i_{1} \cdots
   i_{5}} ] \nonumber \\
    &-& 48 [ A^{(+)}_{i_{1}} , {A^{(-)}}_{i_{2}}] C^{i_{1} i_{2}} 
     +  48 \{  A^{(+)}_{i_{1}} , {A^{(-)}}_{i_{2}} \} D^{i_{1} i_{2}} 
     -  24 ( \{ A^{(+)}_{i_{1}} , C_{i_{2} i_{3}} \} E^{(-)i_{1} i_{2} 
   i_{3}} + \{ A^{(-)}_{i_{1}} , C_{i_{2} i_{3}} \} E^{(+) i_{1} i_{2} 
   i_{3}} ) \nonumber \\
   &+&  24 ( - A^{(+)}_{i_{1}} [ D_{i_{2} i_{3}} , E^{(-) i_{1} i_{2}
   i_{3}}]  + A^{(-)}_{i_{1}} [ D_{i_{2} i_{3}} , E^{(+) i_{1} i_{2}
   i_{3}}] )
    +    8 ( [ A^{(+)}_{i_{1}} , E^{(-)}_{i_{2} i_{3} i_{4}} ] +
   [ A^{(-)}_{i_{1}} , E^{(+)}_{i_{2} i_{3} i_{4}} ] ) G^{i_{1} \cdots 
   i_{4}}  \nonumber \\ 
   &+&   8 ( \{ A^{(+)}_{i_{1}} , E^{(-)}_{i_{2} i_{3} i_{4}} \}
   H^{i_{1} \cdots i_{4}} + \{ A^{(-)}_{i_{1}} , E^{(+)}_{i_{2} i_{3}
   i_{4}} \}   H^{i_{1} \cdots i_{4}} ) 
    +    4 ( \{ A^{(+)}_{i_{1}} , G_{i_{2} \cdots i_{5}} \} I^{(-)
   i_{1} \cdots i_{5}} + \{ A^{(-)}_{i_{1}} , G_{i_{2} \cdots i_{5}}
   \} I^{(+)  i_{1} \cdots i_{5}} ) \nonumber \\
   &+&    4 ( A^{(+)}_{i_{1}} [ H_{i_{2} \cdots i_{5}} , I^{(-) i_{1}
   \cdots i_{5}} ] - A^{(-)}_{i_{1}} [  H_{i_{2} \cdots i_{5}} ,
   I^{(+) i_{1} \cdots i_{5}} ] ) \nonumber \\
  &+& 16 C_{i_{1} i_{2}} [ {C^{i_{1}}}_{i_{3}} , C^{i_{2} i_{3}}] 
   +  12 \{ C_{i_{1} i_{2}} , C_{i_{3} i_{4}} \} G^{i_{1} \cdots
   i_{4}} 
   +  48 C_{i_{1} i_{2}} [ {D^{i_{1}}}_{i_{3}} , D^{i_{2} i_{3}}]
   +  24 \{ C_{i_{1} i_{2}}, D_{i_{3} i_{4}} \} H^{i_{1} \cdots i_{4}} 
   \nonumber \\
  &+& 24 C_{i_{1} i_{2}} [ {E^{(+) i_{1}}}_{i_{3} i_{4}} , E^{(-) i_{2} 
   i_{3} i_{4}} ] 
   +   8 C_{i_{1} i_{2}} ( \{ E^{(+)}_{i_{3} i_{4} i_{5}} , I^{(-)
   i_{1} \cdots i_{5}} \} + \{ E^{(-)}_{i_{3} i_{4} i_{5}} , I^{(+)
   i_{1} \cdots i_{5}} \}  )
   -   8 C_{i_{1} i_{2}} [ {G^{i_{1}}}_{i_{3} \cdots i_{5}}, G^{i_{2}
   \cdots i_{5}} ] \nonumber \\
  &-& \frac{1}{12} C_{i_{1} i_{2}} \{ G_{i_{3} \cdots i_{6}} ,
   H_{i_{7} \cdots i_{10} } \}  \epsilon^{i_{1} \cdots i_{10} \sharp }
   -   8 C_{i_{1} i_{2}} [ {H^{i_{1}}}_{i_{3} i_{4} i_{5}} , H^{i_{2}
   \cdots i_{5}} ]
   -   4 C_{i_{1} i_{2}} [ {I^{(+) I_{1}}}_{i_{3} \cdots i_{6}} ,
   I^{(-) i_{2} \cdots i_{6}} ] \nonumber \\
  &+& 12 \{ D_{i_{1} i_{2}} , D_{i_{3} i_{4}} \} G^{i_{1} \cdots
   i_{4}} 
   -  24 D_{i_{1} i_{2}} \{ {E^{(+) i_{1}}}_{i_{3} i_{4}} , E^{(-)
   i_{2} i_{3} i_{4}} \} 
   +   4 D_{i_{1} i_{2}} ( - [ E^{(+)}_{i_{3} i_{4} i_{5}} , I^{(-)
   i_{1} \cdots i_{5}} ] +  [ E^{(-)}_{i_{3} i_{4} i_{5}} , I^{(+)
   i_{1} \cdots i_{5}} ] ) \nonumber \\
  &-& \frac{1}{24} D_{i_{1} i_{2}} \{ G_{i_{3} \cdots i_{6}} ,
   G_{i_{7} \cdots i_{10}} \} \epsilon^{i_{1} \cdots i_{10} \sharp} 
   -  16 [ D_{\rho i_{1}} , {G^{\rho}}_{i_{2} i_{3} i_{4}} ] H^{i_{1}
   \cdots i_{4}} 
   +  \frac{1}{24} D_{i_{1} i_{2}} \{ H_{i_{3} \cdots i_{6}} ,
   H_{i_{7} \cdots i_{10}} \} \epsilon^{i_{1} \cdots i_{10} \sharp}  
 \nonumber \\
  &+& \frac{1}{24}  D_{i_{1} i_{2}} ( \{ I^{(+)}_{\rho i_{3} \cdots
   i_{6}}  , {I^{(-) \rho}}_{i_{7} \cdots i_{10}} \} + 
   \{ I^{(-)}_{\rho i_{3} \cdots i_{6}}  , {I^{(+) \rho}}_{i_{7} \cdots
   i_{10}} \} )
   \epsilon^{i_{1} \cdots i_{10} \sharp} \nonumber \\
  &+& \frac{1}{18} \{ E^{(+)}_{i_{1} i_{2} i_{3}} , E^{(-)}_{i_{4}
   i_{5} i_{6}} \} G_{i_{7} \cdots i_{10}} \epsilon^{i_{1} \cdots
   i_{10} \sharp}
   + 6 \{ E^{(+)}_{\rho i_{1} i_{2}} , {E^{(-) \rho}}_{i_{3} i_{4}} \}
   G^{i_{1} \cdots i_{4}} \nonumber \\
  &-& 12 [ E^{(+)}_{\rho i_{1} i_{2}} , {E^{(-) \rho}}_{i_{3} i_{4}}]
   H^{i_{1} \cdots i_{4}} 
   - \frac{1}{36} [ E^{(+)}_{i_{1} i_{2} i_{3}} , E^{(-)}_{ i_{4} i_{5}
   i_{6}} ] H_{i_{7} \cdots i_{10}} \epsilon^{i_{1} \cdots i_{10}
   \sharp} \nonumber \\
  &+& 8 ( [ {E^{(+) \rho}}_{i_{1} i_{2}} , G_{\rho i_{3} i_{4} i_{5}}]
   I^{(-) i_{1} \cdots i_{5}} + [ {E^{(-) \rho}}_{i_{1} i_{2}} ,
   G_{\rho i_{3} i_{4} i_{5}}] I^{(+) i_{1} \cdots i_{5}} ) \nonumber \\ 
  &+& \frac{1}{18} ( E^{(+)}_{i_{1} i_{2} i_{3}}  \{ H_{\rho i_{4}
   i_{5} i_{6}} ,  {I^{(-) \rho}}_{i_{7} \cdots i_{10}} \} +
    E^{(-)}_{i_{1} i_{2} i_{3}}  \{ H_{\rho i_{4}
   i_{5} i_{6}} ,  {I^{(+) \rho}}_{i_{7} \cdots i_{10}} \})
 \epsilon^{ i_{1} \cdots i_{10} \sharp} \nonumber \\
  &+&  4 ( E^{(+)}_{\rho i_{1} i_{2}}  \{ {H^{\rho}}_{i_{3} i_{4}
   i_{5}} , I^{(-) i_{1} \cdots i_{5}}  \}
   -  E^{(-)}_{\rho i_{1} i_{2}}  \{ {H^{\rho}}_{i_{3} i_{4}
   i_{5}} , I^{(+) i_{1} \cdots i_{5}}  \}) \nonumber \\
  &-& 2 G_{i_{1} \cdots i_{4}} \{ {G^{i_{1} i_{2}}}_{j_{1} j_{2}} ,
   G^{i_{3} i_{4} j_{1} j_{2}} \} 
   + \frac{1}{18} [ {G^{\rho}}_{i_{1} i_{2} i_{3}} , G_{\rho i_{4}
   i_{5} i_{6}} ] H_{i_{7} \cdots i_{10}} \epsilon^{i_{1} \cdots
   i_{10} \sharp} 
   -  6 G_{i_{1} \cdots i_{4}} \{ {H^{i_{1} i_{2}}}_{j_{1} j_{2}} ,
   H^{i_{3} i_{4} j_{1} j_{2}} \} \nonumber \\
  &-& 2 G_{i_{1} \cdots i_{4}} ( \{ {I^{(+) i_{1} i_{2}}}_{j_{1} j_{2}
   j_{3}} ,  I^{(-) i_{3} i_{4} j_{1} j_{2} j_{3}} \} + \{ {I^{(-)
   i_{1} i_{2}}}_{j_{1} j_{2} j_{3}} ,  I^{(+) i_{3} i_{4} j_{1} j_{2}
   j_{3}} \} ) \nonumber \\
  &+& \frac{1}{54} [ {H^{\rho}}_{i_{1} i_{2} i_{3}} , H_{\rho i_{4}
   i_{5} i_{6}} ]  H_{i_{7} \cdots i_{10}} \epsilon^{i_{1} \cdots
   i_{10} \sharp} 
   + \frac{4}{3} {H^{\rho \sigma}}_{i_{1} i_{2}} ( [ I^{(+)}_{\rho
   \sigma i_{3} i_{4} i_{5}} , I^{(-) i_{1} \cdots i_{5}} ] -
   [ I^{(-)}_{\rho \sigma i_{3} i_{4} i_{5}} , I^{(+)  i_{1} \cdots
   i_{5}} ] ) \nonumber \\
  &+& \frac{1}{54} [ {I^{\rho \sigma (+)}}_{i_{1} i_{2} i_{3}} ,
   I^{(-)}_{\rho \sigma i_{4} i_{5} i_{6}} ] H_{i_{7} \cdots i_{10}}
   \epsilon^{i_{1} \cdots i_{10} \sharp}. \label{AZMBbospm} \\
  I_{f} &=&  - 3i ( {\bar \phi_{L}} (v+Z-W) \psi_{R} + 
         {\bar \phi_{R}} (v+Z+W) \psi_{L} ) 
    - 3i ( {\bar \phi_{L}} \Gamma^{i} A^{(+)}_{i} \psi_{L} + 
          {\bar \phi_{R}} \Gamma^{i} A^{(-)}_{i} \psi_{R} )
                  \nonumber \\
   &-&  \frac{3i}{2!} ( {\bar \phi_{L}} \Gamma^{i_{1} i_{2}}
                  (C_{i_{1} i_{2}} - D_{i_{1} i_{2}} ) \psi_{R}
                + {\bar \phi_{R}} \Gamma^{i_{1} i_{2}}
                  (C_{i_{1} i_{2}} +  D_{i_{1} i_{2}}) \psi_{L} )
      - \frac{3i}{3!} ( {\bar \phi_{L}} \Gamma^{i_{1} i_{2}
                  i_{3}} E^{(+)}_{i_{1} i_{2} i_{3}} \psi_{L}
                +  {\bar \phi_{R}} \Gamma^{i_{1} i_{2}
                  i_{3}} E^{(-)}_{i_{1} i_{2} i_{3}} \psi_{R} )
  \nonumber  \\
   &-&  \frac{3i}{4!} ( {\bar \phi_{L}} \Gamma^{i_{1} \cdots i_{4}}
                  (G_{i_{1} \cdots i_{4}} - H_{i_{1} \cdots i_{4}} ) \psi_{R} 
                +  {\bar \phi_{R}} \Gamma^{i_{1} \cdots i_{4}} 
                  (G_{i_{1} \cdots i_{4}} + H_{i_{1} \cdots i_{4}} )
                \psi_{L} ) \nonumber \\
   &-&  \frac{3i}{5!} ( 2 {\bar \phi_{L}} \Gamma^{i_{1} \cdots
                  i_{5}} I^{(+)}_{i_{1} \cdots i_{5}} \psi_{L} + 
                       2 {\bar \phi_{R}} \Gamma^{i_{1} \cdots
                  i_{5}} I^{(-)}_{i_{1} \cdots i_{5}} \psi_{R} ).
  \label{AZMBfermpm} 
   \end{eqnarray}

 \subsection{Proof of (\ref{AZ55cocoro})} \label{AZCcocoro}
     In this section,
  we verify that other terms than are given  in (\ref{AZ55cocoro})
  vanish from this action.  
  \begin{itemize}
   \item{First, we investigate the first term $tr(m^{2}
       \Gamma^{\sharp})$. Using the decomposition $m = m_{e} + m_{o}$, 
       this term is $tr((m^{2}_{e} + m^{2}_{o} + m_{e} m_{o} + m_{o}
       m_{e}) \Gamma^{\sharp} )$. 
       \begin{itemize}
        \item{The cross terms $tr((m_{e} m_{o} + m_{o} m_{e})
            \Gamma^{\sharp} )$ are easily understood to vanish. The
            formula of the product of gamma matrices in
            Appendix. \ref{AZCproductgamma}. readily shows that the
            product of $\Gamma^{e}$ and $\Gamma^{o}$ is of the odd
            rank. Since the trace of the gamma matrix survives only
            for the rank 0 (id est, the unit matrix ${\bf 1}_{32 \times 
            32}$), these cross terms do not affect the action.}
       \item{Next, we verify that the term $tr(m_{o}^{2}
           \Gamma^{\sharp})$ also vanishes. This is less
           trivial than in the previous case. Consider the product of
           two gamma matrices of odd rank $\Gamma^{i_{1} \cdots
           i_{2k+1}}$ and $\Gamma^{j_{1} \cdots j_{2l+1}}$. Utilizing
           the formula in Appendix. \ref{AZCvanish}. only one of them
           is required to  possess the index $\sharp$. Another
           requirement is that, of course, $k=l$, if we are to extract 
           the terms of rank 0. In this case, the only surviving term
           is 
          \begin{eqnarray}
        & &    \Gamma^{i_{1} \cdots i_{2k+1}} \Gamma^{j_{1} \cdots
           j_{2k+1} \sharp} \Gamma^{\sharp} = X (-1)^{2k + \cdots + 1 }
           {\eta^{[i_{1}}}^{[j_{1}} \cdots
           {\eta^{i_{2k+1}]}}^{j_{2k+1}]} (\Gamma^{\sharp})^{2},
           \label{AZ53moocomp1} \\
       & & \Gamma^{j_{1} \cdots j_{2k+1} \sharp} \Gamma^{i_{1} \cdots
       i_{2k+1}} \Gamma^{\sharp} = X (-1)^{(2k+1) + 2k + \cdots 1}
       {\eta^{[i_{1}}}^{[j_{1}} \cdots 
           {\eta^{i_{2k+1}]}}^{j_{2k+1}]} (\Gamma^{\sharp})^{2},
       \label{AZ53moocomp2} 
          \end{eqnarray}
       where $X$ is a coefficient read off from the formula
       (\ref{AZproduct}). Now, it is not the coefficient $X$ but the
       relative sign of these two terms that matters.  Since $2k+1$ is 
       an odd number, $(-1)^{2k + \cdots + 1}$ and $(-1)^{(2k+1) + 2k
       + \cdots 1}$ is {\it different} in sign. This means that {\it
       the anti-commutators $\{ m_{o}, m'_{o} \}$ all vanish, and that 
       only the commutators $[m_{o}, m'_{o}]$ survive. } The trace
       $tr(R [m_{o}, m'_{o}] \Gamma^{\sharp})$ vanish,
       because  of the cyclic property of the trace\footnote{ Completely
       likewise, we can understand that the terms $tr(m^{2}_{e}
       \Gamma^{\sharp})$ only survive as an anti-commutator.}. 
      We have thus justified that the term $tr(m^{2}_{o}
       \Gamma^{\sharp})$ is excluded from this action.
      }
     \end{itemize}
      }
   \item{We next have a look at the cubic term with respect to the
       fluctuation. The discussion is much easier than
       before. Originally, the cubic term involves the following terms: 
       \footnote{We are sloppy about the order of the matrix, because
       we only mind the rank of the gamma matrices.} 
        \begin{eqnarray}
         tr(m^{3}) = tr(m_{e}^{3}) + 3 tr(m_{e}^{2} m_{o}) + 3
         tr(m_{e} m_{o}^{2}) + tr(m_{o}^{3}).
        \end{eqnarray}
       However, the terms $tr(m_{e}^{2} m_{o})$ and $tr(m_{o}^{3})$
       clearly vanish, because these terms includes only the terms of
       the odd rank, considering the product formula
       (\ref{AZproduct}). Therefore, the trace vanishes from
       Appendix. \ref{AZCvanish}.}
   \item{The chirality of the fermionic terms are determined by the
       properties (\ref{AZMA21fermvanish}). }
  \end{itemize}
     
\section{Wigner In{\"o}n{\"u} Contraction} \label{AZCwicont}
  We introduce a notion named the 'Wigner-In{\"o}n{\"u} contraction'. This
is a technique to produce the Poincar{\'e} algebra in ${\bf
  R}^{d-1,1}$ from the algebra of $AdS$ algebra. The $d$ dimensional anti-de
Sitter(AdS) space can be represented as the hyperboloid 
  \begin{eqnarray}
   - X_{0}^{2} - X_{d}^{2} + \sum_{i=1}^{d-1} X_{i}^{2} = - R^{2},
  \end{eqnarray} 
with the metric of the space 
 \begin{eqnarray}
 \eta^{{\hat \mu} {\hat \nu}} = diag(-1,1, \cdots, 1, -1),
 \end{eqnarray}
 where the indices ${\hat \mu}$ runs from $0, 1, \cdots, d-1, d$. The
 symmetry of $AdS_{d}$ space is equal to $d-1$ dimensional conformal
 field theory, as is well known in the context of AdS/CFT correspondence
 \cite{9711200}. 
  \begin{eqnarray}
 & &  [M_{{\hat \mu} {\hat \nu}}, M_{{\hat \rho} {\hat \sigma}} ] =
   \eta_{{\hat \nu} {\hat \rho}} M_{{\hat \mu} {\hat \sigma}}
 + \eta_{{\hat \mu} {\hat \sigma}} M_{{\hat \nu} {\hat \rho}}
 - \eta_{{\hat \mu} {\hat \rho}} M_{{\hat \nu} {\hat \sigma}}
 - \eta_{{\hat \nu} {\hat \sigma}} M_{{\hat \mu} {\hat \rho}}.
   \label{AZC1MM} 
  \end{eqnarray}
 Then, we specify the $x^{d}$ direction, and we define an operator
 $P_{\mu}$, where $\mu$ runs $\mu = 0, 1, \cdots d-1$, as 
  \begin{eqnarray}
   P_{\mu} = \frac{1}{R} M_{d \mu} ( \textrm{with } R \to \infty ).
  \end{eqnarray}
 The algebra of $P_{\mu}$ and $M_{\mu \nu}$ is now obtained by
  \begin{eqnarray}
& & [M_{\mu \nu}, M_{\rho \sigma} ] =
   \eta_{\nu \rho} M_{\mu \sigma}
 + \eta_{\mu \sigma} M_{\nu \rho}
 - \eta_{\mu \rho} M_{\nu \sigma}
 - \eta_{\nu \sigma} M_{\mu \rho}, 
   \label{AZMC1MM+} \\
& & [P_{\mu}, M_{\rho \sigma}] = - \eta_{\mu \sigma} P_{\rho} +
  \eta_{\mu \rho} P_{\sigma}, \label{AZMC1MP+} \\
& & [P_{\mu}, P_{\nu}] = \frac{1}{R^{2}} M_{\mu \nu}. \label{AZMC1PP+}
  \end{eqnarray}
  Now, the generators $P_{\mu}$ and $M_{\mu \nu}$ should be regarded
  as the translation and the Lorentz transformation in $d-1$
  dimensional spacetime, 
  respectively. The virtue of taking the radius of $AdS$ space to be
  infinitely large lies in the fact that the two operation of the
  translations (\ref{AZMC1PP+}) commute with each other. The results
  (\ref{AZMC1MM+}), (\ref{AZMC1MP+}) and (\ref{AZMC1PP+}) indicates
  the algebra of $AdS_{d}$ space in the limit of the infinitely large
  radius $R \to \infty$ becomes the Poincar{\'e} algebra in ${\bf
  R}^{d-1,1}$ spacetime.\\  

  Note that the set of 11 dimensional gamma matrices of rank 1 and 2
  constitutes the closed algebra, and this set satisfy totally the
  same algebra as $d-1$ dimensional conformal field theory. Let us
  investigate the commutator of these gamma matrices.
   \begin{eqnarray}
 & &   [\frac{1}{2} \Gamma^{\mu \nu}, \frac{1}{2} \Gamma^{\rho \sigma}] =
     \frac{1}{2} \eta^{\nu \rho} \Gamma^{\mu \sigma} 
  +  \frac{1}{2} \eta^{\mu \sigma} \Gamma^{\nu \rho}
  -  \frac{1}{2} \eta^{\mu \rho} \Gamma^{\nu \sigma}
  -  \frac{1}{2} \eta^{\nu \sigma} \Gamma^{\mu \rho}, \label{AZC122+} \\
 & &  [ \Gamma^{\mu}, \frac{1}{2} \Gamma^{\rho \sigma} ] =
     \frac{1}{2} \eta^{\mu \rho} \Gamma^{\sigma} 
 -   \frac{1}{2} \eta^{\mu \sigma} \Gamma^{\rho}, \label{AZMC112+} \\
 & &  [ \Gamma^{\mu}, \Gamma^{\nu}] = 2 \Gamma^{\mu \nu}. \label{AZMC111+}
   \end{eqnarray} 
 Therefore, these gamma matrices are identified with the
 transformation and the  
 Lorentz transformation of ${\bf R}^{d-1,1}$ Minkowski spacetime, and the 
 correspondence is as follows:
  \begin{eqnarray}
   P^{\mu} = \frac{1}{R} \Gamma^{\mu}, \hspace{3mm} 
   M^{\mu \nu} = \frac{1}{2} \Gamma^{\mu \nu}.
  \end{eqnarray}
 
  In order to grasp the intuitive image of the Wigner-In{\"o}n{\"u}
  contraction, let us have a look at the case of the sphere
  $S^{2}$. Although this is not an $AdS$ space, the argument is
  similar, and serves to understand the Wigner-In{\"o}n{\"u}
  contraction in a pedagogical way.
  We are considering the $S^{2}$ sphere embedded in the 3 dimensional
  Euclidean space, and the metric is now $\delta^{ij} = diag(1,1,1)$. 
  The generator of the rotation around the $x^{k}$ direction is obtained by
  \begin{eqnarray}
   {\bf J}= {\bf x} \times {\bf p} \Rightarrow 
   J^{k}  = J^{ij} = -i (x^{i} \nabla^{j} - x^{j} \nabla^{i} ),
  \end{eqnarray}
 where $(ijk) = (xyz), (yzx), (zxy)$. This generator satisfy 
 the following commutation relation, which is the same as that of
 $AdS$ space, up to $i$: 
  \begin{eqnarray}
    [J^{ij}, J^{kl}] = -i (\delta^{jk} J^{il} + \delta^{il} J^{jk} -
    \delta^{ik} J^{jl} - \delta^{jl} J^{ik} ).
  \end{eqnarray}
 This is translated into the well-known $SU(2)$ algebra $[J^{i},
 J^{j}] = i \epsilon^{ijk} J^{k}$, if we adopt the expression $J^{k} = 
 {\epsilon_{ij}}^{k} J^{ij}$. Now let us consider a sphere whose
 radius is large enough. And let us have a look at the transformations 
 $J^{i}$ from the viewpoint of the observer at the north pole. 

   \begin{figure}[htbp]
   \begin{center}
    \scalebox{.7}{\includegraphics{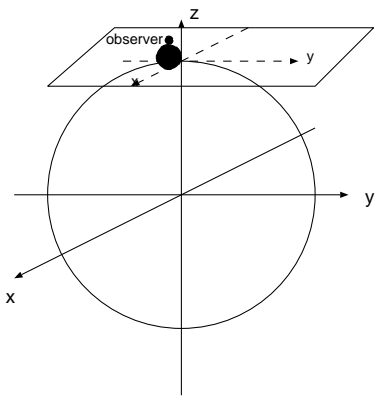} }
   \end{center}
   \caption{An example of the large enough $S^{2}$ sphere. The
     rotation around $x,y$ axis is perceived as the translation in $x,y$
  direction, from the viewpoint of the observer.}
  \label{PicWignerinonu}
  \end{figure}

  Let the  operator $P_{x}$ and $P_{y}$ be defined as
  \begin{eqnarray}
   P_{x} = \frac{1}{R} J_{x}, \hspace{3mm} P_{y} = \frac{1}{R} J_{y}.
  \end{eqnarray}
  The Wigner-In{\"o}n{\"u} contraction means that the rotation around the
  $x,y$ axis can be perceived as the translation in $x,y$ direction for the
  observer at the north pole. This can be algebraically seen by the
  following commutation relation:
  \begin{eqnarray}
   [P_{x}, P_{y}] = \frac{i}{R^{2}} J_{z}.
  \end{eqnarray}
  This commutation relation corresponds to the fact that we
  perceive the earth as a flat space even though the fact is that the
  earth is a sphere whose radius is tremendously large. And there is
  no difference for us residing on the earth even if we step east and
  then step north, or we step north and after that we step east. Even
  though our ancestors, except Pythagoras who
  believed that our world should be a round sphere for aesthetic
  reason, have believed that our world should be a flat plane
  supported by a giant Atlas, they have overlooked the possibility
  that the $S^{2}$ sphere becomes a flat plane $E^{2}$ by the
  Wigner-In{\"o}n{\"u} contraction.

\end{document}